%% file: GrIP.tex
\documentclass[11pt,a4paper,oneside]{article}
\pdfoutput=1
\usepackage{jheppub}

\usepackage{verbatim}
\usepackage{mathrsfs}
\usepackage{appendix}
\usepackage{caption}
\usepackage{float}
\usepackage{subfig}
\usepackage[vcentermath]{youngtab}
\usepackage{alltt}
\usepackage{array,multirow}

\usepackage{amssymb,amsmath,amsfonts,mathrsfs,enumerate,wrapfig,latexsym,wasysym}

\usepackage{graphicx}
\usepackage{adjustbox}
\usepackage{xcolor}  
\usepackage{slashed}    
\usepackage{url}
\usepackage{hyperref}
\usepackage{framed}
\usepackage[numbers,sort&compress]{natbib}
\usepackage{fancyvrb}
\allowdisplaybreaks
\usepackage{multirow}
\usepackage{framed}
\RequirePackage{flushend}
\usepackage{makecell}
\usepackage{color}
\usepackage{pifont}
\usepackage{cancel}
\usepackage{import}
\usepackage{subfiles}
\usepackage{filecontents}
\usepackage{threeparttable,tablefootnote}
\usepackage{mathtools}
\usepackage{pbox}
\usepackage{setspace}
\usepackage{lscape}

\definecolor{mmaLabel}{RGB}{70,70,153}            
\definecolor{mmaLink}{RGB}{20,40,153}
\definecolor{mmaUndefined}{RGB}{0,44,195}         
\definecolor{mmaFunctionLocal}{RGB}{60,125,145}   
\definecolor{mmaLocal}{RGB}{67,137,88}           
\definecolor{mmaMessage}{RGB}{129,43,38}
\definecolor{mmaError}{RGB}{255,51,51}
\definecolor{mmaSyntaxError}{RGB}{194,85,204}
\definecolor{mmaEmphasizedError}{RGB}{204,0,0}
\definecolor{mmaEmphasizedErrorBackground}{RGB}{255,225,130}
\definecolor{mmaFormattingError}{RGB}{255,85,85}
\definecolor{mmaFormattingErrorBackground}{RGB}{255,230,230}
\definecolor{mmaString}{gray}{.4}  
\definecolor{mmaComment}{gray}{.6}

\title{Characters and Group Invariant Polynomials of (Super)fields: Road to ``Lagrangian"}

\author{Upalaparna Banerjee, Joydeep Chakrabortty, Suraj Prakash} \author{and Shakeel Ur Rahaman} \author{\\}

\affiliation{Indian Institute of Technology Kanpur, Kalyanpur, Kanpur 208016. INDIA. \\}

\emailAdd{upalab, joydeep, surajprk, shakel@iitk.ac.in}

\abstract
{
The dynamics of the subatomic fundamental particles, represented by quantum fields, and their interactions are determined uniquely by the assigned transformation properties, i.e., the quantum numbers associated with the underlying symmetry of the model under consideration. These fields constitute a finite number of group invariant operators which are assembled to build a polynomial, known as the Lagrangian of that particular model. The order of the polynomial is determined by the mass dimension.  In this paper, we have introduced an automated $\text{\texttt{Mathematica}}^{\tiny\textregistered}$ package, \textbf{GrIP}, that computes the complete set of operators that form a basis at each such order for a model containing any number of fields transforming under connected compact groups. The spacetime symmetry is restricted to the Lorentz group. 
The first part of the paper is dedicated to formulating the algorithm of \textbf{GrIP}. In this context, the detailed and explicit construction of the characters of different representations corresponding to connected compact groups and respective Haar measures have been discussed in terms of the coordinates of their respective maximal torus. In the second part, we have documented the user manual of \textbf{GrIP} that captures the generic features of the main program and guides to prepare the input file.  We have attached a sub-program  \textbf{CHaar} to compute characters and  Haar measures for  $SU(N), SO(2N), SO(2N+1), Sp(2N)$. This program works very efficiently to find out the higher mass (non-supersymmetric) and canonical (supersymmetric) dimensional operators relevant to the Effective Field Theory (EFT). We have demonstrated the working principles with two examples:- the Standard Model (SM) and the Minimal Supersymmetric Standard Model (MSSM).  We have further highlighted important features of  \textbf{GrIP}, e.g., identification of effective operators leading to specific rare processes linked with the violation of baryon and lepton numbers,  using several Beyond Standard Model (BSM) scenarios. We have also tabulated a complete set of dimension-6 operators for each such model. Some of the operators possess rich flavour structures which are discussed in detail. This work paves the way towards BSM-EFT.
}

\begin{document}
	\maketitle

%%%%%%%%%%%%%%%%%%%%%%%%%%%%%%%%%%%%%%%%%%%%%%%%%%%%%%%%%%%%%%%%%

\input{Introduction.tex}

%%%%%%%%%%%%%%%%%%%%%%%%%%%%%%%%%%%%%%%%%%%%%%%%%%%%%%%%%%%%%%%%%

\input{HilbertSeries.tex}

%%%%%%%%%%%%%%%%%%%%%%%%%%%%%%%%%%%%%%%%%%%%%%%%%%%%%%%%%%%%%%%%%

\clearpage

%%%%%%%%%%%%%%%%%%%%%%%%%%%%%%%%%%%%%%%%%%%%%%%%%%%%%%%%%%%%%%%%%

\input{KnownLagrangians.tex} 

%%%%%%%%%%%%%%%%%%%%%%%%%%%%%%%%%%%%%%%%%%%%%%%%%%%%%%%%%%%%%%%%%

\input{GrIPCodeDescription.tex}

%%%%%%%%%%%%%%%%%%%%%%%%%%%%%%%%%%%%%%%%%%%%%%%%%%%%%%%%%%%%%%%%%

\input{GrIPExamples.tex}

%%%%%%%%%%%%%%%%%%%%%%%%%%%%%%%%%%%%%%%%%%%%%%%%%%%%%%%%%%%%%%%%%

\input{OperatorCategorization.tex}

%%%%%%%%%%%%%%%%%%%%%%%%%%%%%%%%%%%%%%%%%%%%%%%%%%%%%%%%%%%%%%%%%

\input{GrIPpheno.tex}

%%%%%%%%%%%%%%%%%%%%%%%%%%%%%%%%%%%%%%%%%%%%%%%%%%%%%%%%%%%%%%%%%

\input{BSMeftBasis.tex}

%%%%%%%%%%%%%%%%%%%%%%%%%%%%%%%%%%%%%%%%%%%%%%%%%%%%%%%%%%%%%%%%%

\input{FlavourStructure.tex}

%%%%%%%%%%%%%%%%%%%%%%%%%%%%%%%%%%%%%%%%%%%%%%%%%%%%%%%%%%%%%%%%%

\clearpage

%%%%%%%%%%%%%%%%%%%%%%%%%%%%%%%%%%%%%%%%%%%%%%%%%%%%%%%%%%%%%%%%%

\input{conclusion.tex}

%%%%%%%%%%%%%%%%%%%%%%%%%%%%%%%%%%%%%%%%%%%%%%%%%%%%%%%%%%%%%%%%%

\appendix
\input{appendix_theory.tex}
\input{appendix_code.tex}

\section*{}

\providecommand{\href}[2]{#2}
\addcontentsline*{toc}{section}{}
\bibliographystyle{JHEP}
\bibliography{GrIP}

\end{document}

%% file: Introduction.tex
\section{Introduction}\label{sec:intro}
%%%%%%%%%%%%%%%%%%%%%%%%%%%%%%%%%%%%%%%%%%%%%%%%%%%%%%%%%%%%%%%%%%%%%%%%%%%%%%%
Particle physics, an intricate medley between theory and experiment, aims to provide an accurate description of the dynamics and interactions of the subatomic particles. The experimental results are quantified by a set of  observables, e.g., decay widths and the scattering cross-sections. The symbiotic relationship between theory and experiment implies that each measurement lends credence to some theoretically calculated number. To calculate the theoretical values of these observables we need to rely on Feynman vertices which are derived from the expanded form of the Lagrangian density. Therefore, it is the Lagrangian density\footnote{In what follows we will use the terms Lagrangian and Lagrangian density interchangeably even though their specific usage depends on whether we are working with discrete or continuum theories.} that is the \textit{fons et origo} of any justifiable or falsifiable claim that we can attempt to make based on the theory. 

Now, this begs a couple of questions, the first being which terms are allowed in the Lagrangian that would ultimately determine the characteristics of the interactions and the structure of the Feynman vertices. The second is whether we can backtrack further, i.e., are there more rudimentary aspects below the level of the Lagrangian from which the complete theory can be built procedurally. One can ask if the Lagrangian is the true genesis of the theory or if we can probe its anatomy further. The answers to these questions are affirmative. There are well-defined, mathematically sound guidelines that determine what interactions are allowed and which ones are not and in a nutshell, these are the interplay of the conservation and violation of certain symmetries. Also, it is quite evident that the minimum information that we require for building a Lagrangian and in turn constructing a model is the quantum fields representing the particles and their transformation properties guided by the underlying symmetries of the model. The concept that we endeavour to forge an understanding of is how to build a full-fledged theoretical model, whose predictions could be corroborated using ingeniously designed high energy experiments, using nothing but this minimal piece of information.

We need to do meticulous scrutiny of the eccentric features of a general Lagrangian. For the sake of our analysis, we will treat the Lagrangian as a polynomial of certain spurion variables which are nothing but the quantum fields representing the actual particles.  Just as we can define the order of a polynomial in terms of the powers of the variables, analogously we can define the order of the Lagrangian density in terms of certain parameters associated to the fields. The mass dimension of the terms of the Lagrangian in natural units (where $c=1$, $\hbar=1$ and consequently $\left[L\right]=\left[T\right]=\left[M\right]^{-1}$) is customarily used to define this order. We restrict ourselves to $d=3+1$ space-time dimensions. Here, the action is defined as:

\vspace{-0.5cm}
{\small\begin{eqnarray}\label{eq:action}
\mathcal{S} = \int d^4x\hspace{0.1cm} \mathcal{L}.
\end{eqnarray}}
Since $\mathcal{S}$ is dimensionless ($[\mathcal{S}] \equiv 0$) and the integration measure possesses a mass dimension of ``-$4$" ($[d^4x] \equiv -4$), the Lagrangian density $(\mathcal{L})$ must have a mass dimension ``+$4$" ($ [\mathcal{L}]\equiv 4$). This has two significant consequences. First, this fixes the mass dimensions of bosonic $(\phi,A_{\mu})$ and fermionic ($\psi$) fields in $3+1$ dimensions based on their kinetic terms. We can also obtain the mass dimension for field strength tensors $F_{\mu\nu}$ using the gauge kinetic terms. Thus to summarize,

\vspace{-0.5cm}
{\small\begin{eqnarray}
[\mathcal{L}]\equiv 4,\hspace{0.2cm} [\mathcal{D}_{\mu}]\equiv 1 \hspace{0.1cm} \implies \hspace{0.1cm} [\phi]\equiv [A_{\mu}] \equiv 1, \hspace{0.2cm} [\psi] \equiv \frac{3}{2}, \hspace{0.2cm} [F_{\mu\nu}] \equiv 2.
\end{eqnarray}}
The second major consequence is that even though we may add terms of any mass dimension to the Lagrangian density, they all must be suitably multiplied by coefficients of suitable mass dimensions to be successfully accommodated in $\mathcal{L}$. Thus, a schematic form of the Lagrangian density can be written as: 
\begin{eqnarray}
\mathcal{L} = \alpha^{(1)}\mathcal{O}^{(1)} + \alpha^{(2)}\mathcal{O}^{(2)} + \alpha^{(3)}\mathcal{O}^{(3)} + \alpha^{(4)}\mathcal{O}^{(4)} + \alpha^{(5)}\mathcal{O}^{(5)} + \alpha^{(6)}\mathcal{O}^{(6)} + \cdots ,
\end{eqnarray}
where $\mathcal{O}^{(i)}$'s are operators of mass dimension $i$ and $\alpha^{(i)}$'s are the coefficients of mass dimension $(4-i)$. Now, since we can have operators of mass dimension $>4$, this implies that certain coupling constants will have negative mass dimensions. Then from power counting arguments and taking into account the issue of superficial renormalizability we divide the full Lagrangian density into two parts: the renormalizable Lagrangian and the effective Lagrangian:
\vspace{-0.5cm}
\begin{eqnarray}
\mathcal{L} = \mathcal{L}_{\text{renorm}} + \sum_{i=5}^{n}\sum_{j=1}^{N_i}\frac{\mathcal{C}_j^{(i)}}{\Lambda^{i-4}}\mathcal{O}_j^{(i)}.
\end{eqnarray}
Here, $i$ denotes the mass dimension of the operators and since there can be more than one operator at a particular mass dimension, therefore we have a sum over all such operators $(\sum_j)$. The total number of operators at a given mass dimension has been denoted by $N_i$. $\Lambda$ has dimensions of mass and $\mathcal{C}_j^{(i)}$'s are dimensionless coefficients known as the Wilson coefficients. The second term on the RHS is called the effective Lagrangian ($\mathcal{L}_{EFT}$) \cite{Georgi:1994qn,Manohar:1996cq,Kaplan:1995uv,Burgess:2007pt}.  

Having established the form of the Lagrangian density the next question is given some quantum fields, can we include all possible combinations of these fields in the Lagrangian density or are there certain restrictions. In other words, how does one fix $N_i$ for a specific theory? 

\begin{figure}[h!]
	\centering
	{
		\includegraphics[scale=0.4]{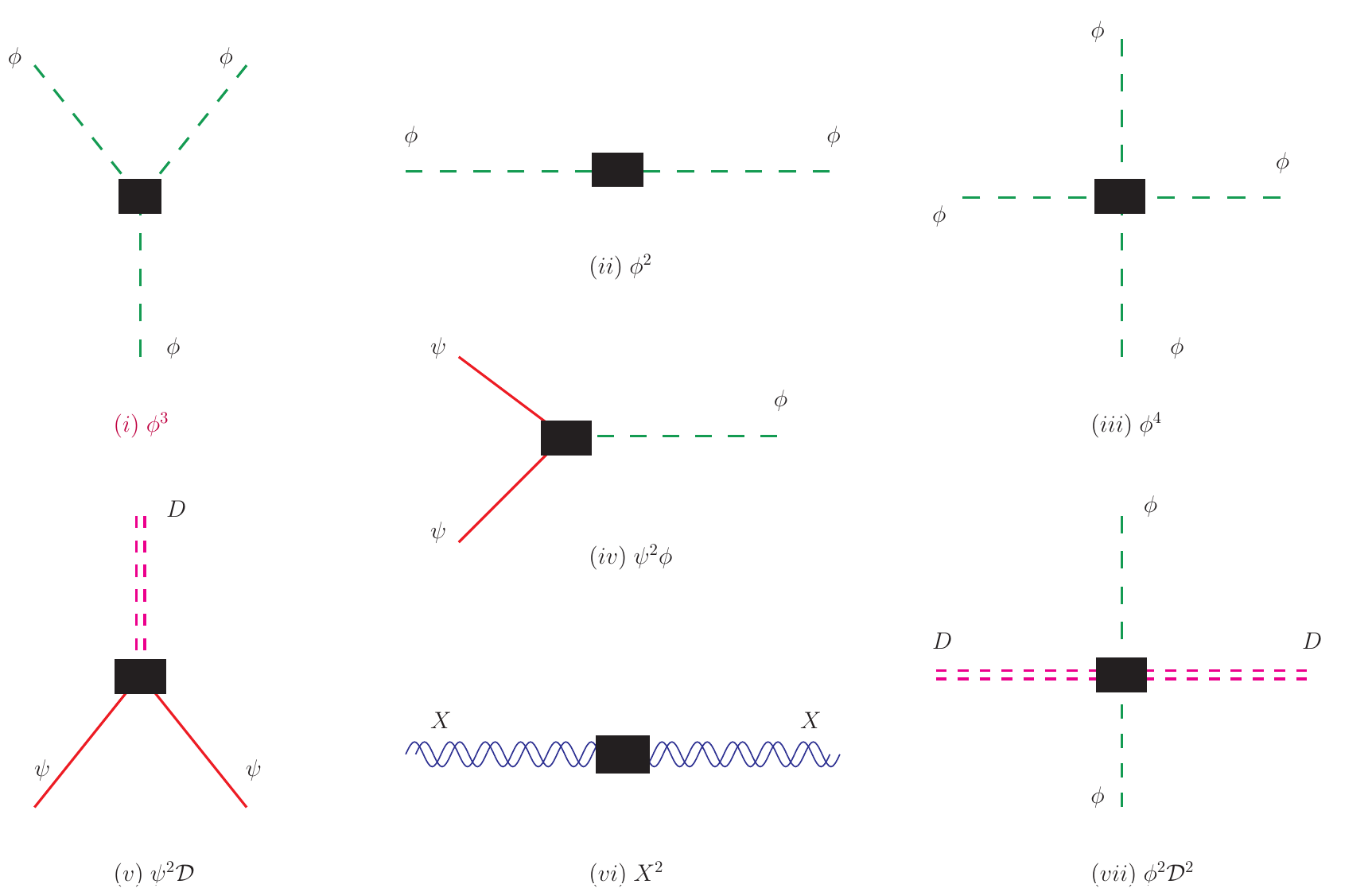}
		\includegraphics[scale=0.4]{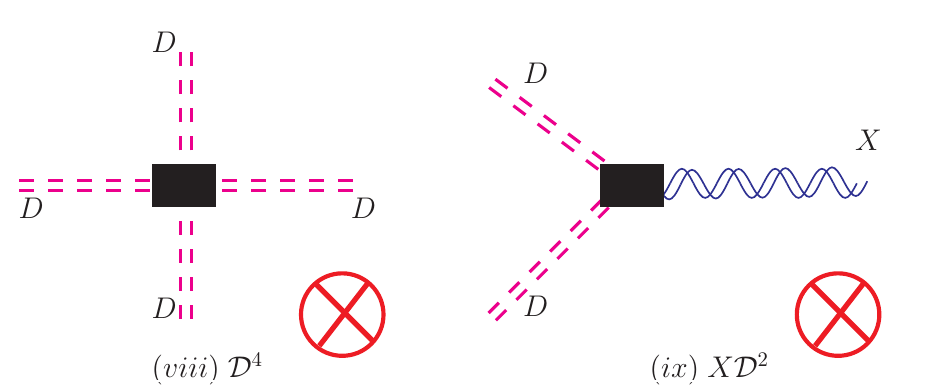}
	}
	\caption{\small Diagrammatic representation of operator classes at mass dimension-4 (for $3+1$ space-time dimensions) or less constituted by combining spin-0 ($\phi$), spin-1/2 ($\psi$) and Field Strength Tensor ($X$) of spin-1 fields and the covariant derivative ($\mathcal{D}$). Of these, \textcolor{purple}{$(i)$} does not appear in the case of SM hence we have highlighted its caption in colour. Also, the last two structures being total derivative terms are excluded from the Lagrangian.}
	\label{fig:op-structure-dim4}
\end{figure}

\begin{figure}[h!]
	\centering
	{
		\includegraphics[scale=0.4]{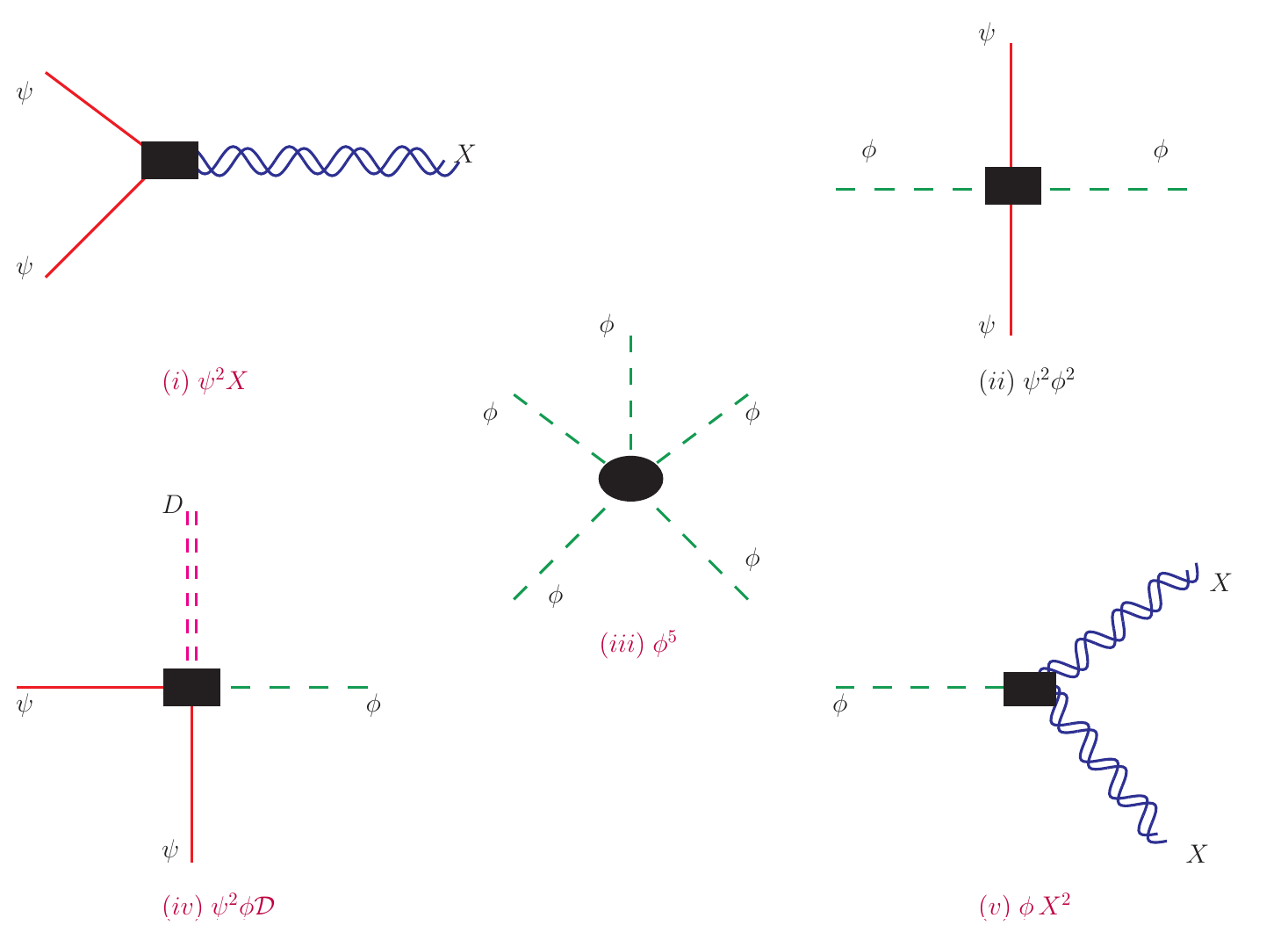}
	}
	\caption{\small Diagrammatic representation of  mass dimension-5 (for $3+1$ space-time dimensions) operator classes constituted by combining spin-0 ($\phi$), spin-1/2 ($\psi$) and Field Strength Tensor ($X$) of spin-1 fields and the covariant derivative ($\mathcal{D}$). Only one of these (diagram $(ii)$) appears in the case of SM. The other structures with red captions \textcolor{purple}{$(i),\,(iii),\,(iv)$} and \textcolor{purple}{$(v)$} may appear for other models with different particle content and symmetry.}
	\label{fig:op-structure-dim5}
\end{figure}

\begin{figure}[h!]
	\centering
	{
		\includegraphics[scale=0.7]{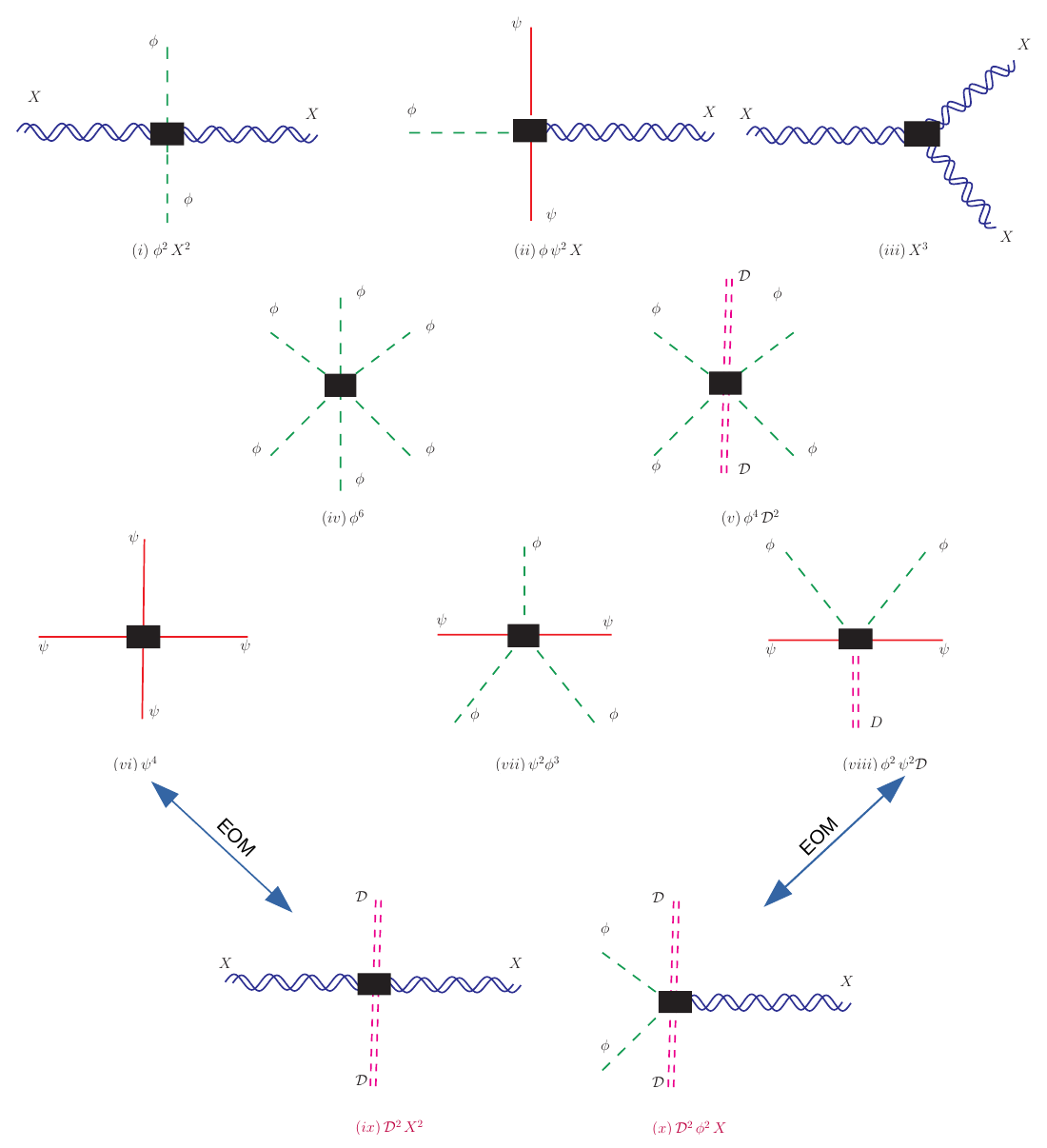}
	}
	\caption{\small Diagrammatic representation of  mass dimension-6 (for $3+1$ space-time dimensions) SM operators classes constituted by combining spin-0 ($\phi$), spin-1/2 ($\psi$) and Field Strength Tensor ($X$) of spin-1 fields and the covariant derivative ($\mathcal{D}$). Operators described by \textcolor{purple}{$(ix)$} and \textcolor{purple}{$(x)$} are related to $(vi)$ and $(viii)$ respectively through the equation of motion of the gauge fields. Based on which terms are included in the operator set we have two popular operator bases. The Warsaw basis \cite{Grzadkowski:2010es} includes the operator classes $(i)-(viii)$ and forms a complete set. While, the SILH \cite{Giudice:2007fh} basis trades off $(ii)$, $(vi)-(viii)$ (the operators composed of fermionic fields) in favour of \textcolor{purple}{$(ix)$} and \textcolor{purple}{$(x)$}, This forms an under-complete set.}
	\label{fig:op-structure-dim6}
\end{figure}

Again, the answer comes from looking at Eq.~\eqref{eq:action}, since $\mathcal{S}$ is invariant w.r.t. spacetime symmetry as well as any internal symmetry (and so does $d^4x$), the Lagrangian density $\mathcal{L}$ must be invariant as well under the same set of symmetries. We further demand that the operators must form a complete and independent set. Below we shall illustrate the role of symmetry in restricting the inclusion of arbitrary operators with a few examples. For a theory consisting only of a real scalar field $\phi$, the renormalizable Lagrangian is given as: 

\vspace{-0.5cm}
{\small\begin{eqnarray}
\mathcal{L}_{\phi} = \frac{1}{2}(\partial_{\mu}\phi)(\partial^{\mu}\phi) - \mathcal{V}(\phi), \hspace{0.8cm} \mathcal{V}(\phi) = \mathcal{M}^3\phi-\frac{1}{2}m^2\phi^2+\frac{\mu}{3!}\phi^3+\frac{\lambda}{4!}\phi^4.
\end{eqnarray}}
Now, for a theory consisting of a scalar field $\rho$ which possesses a discrete symmetry $\rho \rightarrow -\rho$, the Lagrangian becomes:

\vspace{-0.5cm}
{\small\begin{eqnarray}
\mathcal{L}_{\rho} = \frac{1}{2}(\partial_{\mu}\rho)(\partial^{\mu}\rho) - \mathcal{V}(\rho), \hspace{0.8cm} \mathcal{V}(\rho) = -\frac{1}{2}m^2\rho^2+\frac{\lambda}{4!}\rho^4.
\end{eqnarray}}
It is evident that this discrete symmetry ($\mathbb{Z}_2$) rules out the linear and cubic terms as these are no longer invariant. As a second example, we consider the case of a complex scalar field and its conjugate which transform under a global $U(1)$ symmetry:

\vspace{-0.5cm}
{\small\begin{eqnarray}
\phi = \phi_1 + i \phi_2,\hspace{0.4cm} \phi^{*} = \phi_1 - i \phi_2; \hspace{1.4cm} \phi \rightarrow e^{i\theta}\phi, \hspace{0.4cm}
\phi^{*} \rightarrow \phi^{*}e^{-i\theta}.
\end{eqnarray}}
Even in this case, we see that if the Lagrangian has to be invariant w.r.t the $U(1)$ symmetry we cannot have terms linear, quadratic or trilinear in only one of the fields. Hence, the permissible Lagrangian looks like: 
 
\vspace{-0.5cm}
{\small\begin{eqnarray}
\mathcal{L}_{\phi,\phi^{*}} = (\partial_{\mu}\phi^{*})(\partial^{\mu}\phi) - \mathcal{V}(\phi), \hspace{0.8cm} \mathcal{V}(\phi) = -m^2(\phi^{*}\phi)+\lambda(\phi^{*}\phi)^2.
\end{eqnarray}}
The situation becomes more involved if we have a gauge symmetry and when the number of degrees of freedom (DOF) is large. For most cases constructing an independent set of operators is a painstaking task not only for higher dimensions but even at the renormalizable level. Let us look at the most popular model, i.e., the Standard Model (SM) of particle physics where in addition to spacetime symmetry we also have local $SU(3)_C\otimes SU(2)_L \otimes U(1)_Y$ symmetry, then the renormalizable Lagrangian is:
 
\vspace{-0.5cm}
{\small\begin{eqnarray}
\mathcal{L}_{SM} &=& (\mathcal{D}_{\mu}\phi)^{\dagger}(\mathcal{D}^{\mu}\phi)+\frac{1}{2}m^2(\phi^{\dagger}\phi)-\frac{\lambda}{4!}(\phi^{\dagger}\phi)^2 -\frac{1}{4}B^{\mu\nu}B_{\mu\nu}-\frac{1}{4}G^{a\mu\nu}G^{a}_{\mu\nu}-\frac{1}{4}W^{I \mu\nu}W^{I}_{\mu\nu}\nonumber\\
& & -i\left(\bar{L}_L^{p}\slashed{\mathcal{D}}L^{p}_L+
\bar{Q}_L^{p}\slashed{\mathcal{D}}Q^{p}_L+\bar{u}_R^{p}\slashed{\mathcal{D}}u^{p}_R+\bar{d}_R^{p}\slashed{\mathcal{D}}d^{p}_R+\bar{e}_R^{p}\slashed{\mathcal{D}}e^{p}_R\right)\nonumber\\
& & -\left(y_e^{rs}\bar{L}^r_L \phi e^s_R + y_d^{rs} \bar{Q}^r_L \phi d^s_R + y_u^{rs} \bar{Q}^r_L \tilde{\phi} u^s_R\right) + h.c.
\end{eqnarray}}
Here, $\lambda, y_e, y_d, y_u$ are dimensionless couplings while $m$ is the mass parameter. We can neatly categorize each of these terms as in Table~\ref{table:sm-renorm-operator-class}. Different terms, i.e., the operators in the Lagrangian can have diagrammatic representations. The operator classes corresponding to the renormalizable Lagrangian have been depicted in Fig.~\ref{fig:op-structure-dim4}\footnote{All these diagrams have been generated using \texttt{JaxoDraw} \cite{Binosi:2003yf}.}. We have shown 2 more invariant operator structures ($\mathcal{D}^4, \hspace{0.1cm} \mathcal{D}^2X$) at mass dimension-4 which are excluded from the Lagrangian as they are total derivative terms and therefore do not affect the dynamics. We have also displayed the Feynman diagrams representing processes encapsulated in operators of dimensions-5 and -6 (Figs.~\ref{fig:op-structure-dim5} and \ref{fig:op-structure-dim6}). Here, in addition to the operator classes in which the SM operators can be categorized into, we have also identified the classes which appear for general theories with fields having spins-0, -1/2, and -1.

\begin{table}[h]
	\centering
	\renewcommand{\arraystretch}{1.8}
	{\scriptsize\begin{tabular}{|c|c|c|}
			\hline
			\textbf{Category}&
			\textbf{Constitution}&
			\textbf{Operators (for SM)}\\
			\hline
			Scalar Potential&
			$\phi^n, \hspace{0.2cm}n\leq 4$&
			$(\phi^{\dagger}\phi),\hspace{0.1cm} (\phi^{\dagger}\phi)^2$\\
			\hline
			Scalar Kinetic Term&
			$\phi^2\mathcal{D}^2$& 
			$(\mathcal{D}_{\mu}\phi)^{\dagger}(\mathcal{D}^{\mu}\phi)$\\
			\hline
			Fermion Kinetic Term&
			$\psi^2\mathcal{D}$& 
			$\bar{L}_L^{p}\slashed{\mathcal{D}}L^{p}_L,\hspace{0.1cm} 
			\bar{Q}_L^{p}\slashed{\mathcal{D}}Q^{p}_L,\hspace{0.1cm} \bar{u}_R^{p}\slashed{\mathcal{D}}u^{p}_R,\hspace{0.1cm} \bar{d}_R^{p}\slashed{\mathcal{D}}d^{p}_R,\hspace{0.1cm} \bar{e}_R^{p}\slashed{\mathcal{D}}e^{p}_R$\\
			\hline	
			Gauge Kinetic Term&
			$(F_{\mu\nu})^2$& 
			$B^{\mu\nu}B_{\mu\nu},\hspace{0.1cm} G^{a\mu\nu}G^{a}_{\mu\nu},\hspace{0.1cm} W^{I \mu\nu}W^{I}_{\mu\nu}$\\
			\hline	
			Yukawa Interaction Term&
			$\psi^2\phi$& 
			$\bar{L}^r_L \phi e^s_R,\hspace{0.1cm} \bar{Q}^r_L \phi d^s_R,\hspace{0.1cm} \bar{Q}^r_L \tilde{\phi} u^s_R$\\
			\hline	
	\end{tabular}}
	\caption{\small Operator classification for the renormalizable Lagrangian composed of scalars, spinors, gauge field strength tensors, and covariant derivatives.}
	\label{table:sm-renorm-operator-class}
\end{table}

\noindent
The fields under consideration have a dynamical nature. This is substantiated by the presence of the covariant derivative ($\mathcal{D}_{\mu}$). The $\mathcal{D}_{\mu}$ is a singlet under the internal symmetries but transforms non-trivially under the Lorentz group. As we go to higher mass dimensions, we encounter operators with multiple derivatives. The presence of $\mathcal{D}_{\mu}$ leads to redundancy in the operator set, see \cite{Grzadkowski:2010es,Henning:2015daa,Henning:2017fpj,Lehman:2015coa}. Essentially, two operators containing the covariant derivative can be related to each other through integration by parts (IBP) and removal of a total derivative from the Lagrangian density. Also, two classes of operators could be related through the equations of motion (EOM) of one of the fields, e.g., the operators described by  Figs.~\ref{fig:op-structure-dim6}\,$(ix)$ and $(x)$ are related to those described by Figs.~\ref{fig:op-structure-dim6}\,$(vi)$ and $(viii)$ respectively through the equation of motion of the gauge fields. We need to ensure that the operators which are a part of our set at a given dimension are invariant w.r.t. spacetime as well as internal symmetries and also form a complete and independent set, i.e., a basis at a given order of the polynomial. To do so any IBP and EOM redundancies in the operator set must be taken care of. 

Now, the positive thing is that the guiding principle behind this sequence of steps is not entirely an unfathomable, esoteric mathematical artifact. In fact, it can be elegantly described in terms familiar to a physicist \cite{Henning:2017fpj,Feng:2007ur}. The centerpiece of this construction is the Hilbert Series \cite{Lehman:2015via,Hanany:2008sb,Hanany:2010vu,Hanany:2014dia,Henning:2015alf} which can be generated from group theoretic principles. Before performing phenomenological analysis on any proposed model, the most important task is to write down the correct Lagrangian. Keeping that in mind we have developed a $\text{\texttt{Mathematica}}^{\tiny\textregistered}$ \cite{Mathematica} based package, \textbf{GrIP} which automatizes the myriad of steps involved in constructing \textbf{Gr}oup \textbf{I}nvariant \textbf{P}olynomials, i.e., the Lagrangian for any given model based on the very minimal input, the field content of the model and their transformation properties. We are sure that \textbf{GrIP} will be an indispensable addition to the phenomenologist's EFT toolbox \cite{Brivio:2019irc} along with other ingenious computational packages, like {\texttt CoDEx} \cite{Bakshi:2018ics}, {\texttt DsixTools} \cite{Celis:2017hod}, {\texttt FlavorKit} \cite{Porod:2014xia,Vicente:2014xda}, {\texttt FormFlavor} \cite{Evans:2016lzo}, {\texttt Wilson} \cite{Aebischer:2018bkb}, {\texttt SMEFT-FR} \cite{Dedes:2019uzs}, {\texttt SMEFTsim} \cite{Brivio:2017btx}, {\texttt SPheno} \cite{Porod:2003um,Porod:2011nf}, {\texttt WCxf-python} \cite{Aebischer:2017ugx}, {\texttt Sym2Int} \cite{Fonseca:2019yya} and {\texttt ECO} \cite{1791986}.

We have divided this work into two broad parts. The first part highlights the theoretical principles behind group invariant polynomial construction and its necessity in particle physics model building. To start with, in section~\ref{sec:group-theory-hilbert-series},  we have outlined the detailed mathematics behind the computation of the basic ingredients of the Hilbert Series, i.e., the characters corresponding to the representations under given groups and the Haar measures of various groups. We have delineated the explicit calculations for the connected compact groups $SU(N), SO(2N), SO(2N+1)$ and $Sp(2N)$. 
This is followed by a brief discussion on the non-triviality associated with the Lorentz group. Then in section~\ref{sec:hilbert-series-examples}, we have employed the Hilbert Series approach to build the operator sets for a few known models. We have revisited the Two Higgs Doublet Model and unveiled the detailed intermediate steps. Then we have introduced the Pati-Salam Model, etc. and performed a comparative analysis with the existing literature to underline the power of this method. 

The second half of this work sheds light on the salient features of \textbf{GrIP}. We start with section~\ref{sec:grip-description} where we have described the chronological steps to elaborate (i) the  installation of the package, (ii) preparation of  a general input file and interfacing it with the main program,  and (iii) generating specific as well as generic output in the form of operators of different mass dimensions.  We have provided specific illustrations in section~\ref{sec:grip-examples} using two example models: the Standard Model and the Minimal Supersymmetric Standard Model. We have also drawn attention towards  certain \textbf{GrIP} functions that help us to filter out the operators leading to rare processes. 

In section~\ref{sec:bsmeft} we have discussed the bottom-up approach to formulate Effective Field Theory \cite{Georgi:1994qn,Manohar:1996cq,Kaplan:1995uv,Burgess:2007pt} and further paved the way to construct Beyond Standard Model Effective Field Theory (BSM-EFT). We have demonstrated this idea through a few examples where SM is extended by different choices of  infrared degrees of freedoms (IR-DOFs). For each such scenario, we have computed the additional (beyond the SM-EFT ones) operators of dimensions-5 and -6 using \textbf{GrIP}. We have also outlined other possible features of this code to generate unique effective operators based on the specific phenomenological demands. Our program \textbf{GrIP}  provides the operators for arbitrary number of fermion flavours $(N_f)$ keeping the provision to analyse the explicit flavour dependence.   In section~\ref{sec:flavour-structure}, we have reasoned the origin of different $N_f$-dependent factors that appear for similar structures across various phenomenological models. We have also tabulated the operator sets for a few more models and some necessary group-theoretic information in the appendices.   

%% file: HilbertSeries.tex
\section{Hilbert Series: The Underlying Theoretical Framework for GrIP}\label{sec:group-theory-hilbert-series}

The object of our inquiry in this section is the Hilbert Series (HS) method  \cite{Henning:2015daa,Kobach:2017xkw,Henning:2015alf,Lehman:2015coa,Trautner:2018ipq,Henning:2017fpj,Hanany:2010vu,Feng:2007ur,Lehman:2015via} based on which \textbf{GrIP} has been developed. In the context of particle physics models, we can define a set of quantum fields representing particles that posses certain transformation properties under the symmetries of the model. The Hilbert Series is an infinite series consisting of all possible symmetry group  invariant clusters of the quantum fields and is built on two necessary ingredients: (i) the Plethystic Exponential (PE) and (ii) the Haar measure. The relevant generic form of the Hilbert Series is given as \cite{Hanany:2010vu,Feng:2007ur,Henning:2017fpj,Henning:2015alf}:

{\small\begin{eqnarray}\label{eq:HS}
	\mathcal{H}[\varphi] = \prod^{n}_{j=1} \int_{\mathcal{G}_j} \hspace{0.1cm}\underbrace{d\mu_j}_{\text{Haar}\hspace{0.1cm}\text{Measure}} \hspace{0.1cm}\underbrace{PE[\varphi,  R]}_{\hspace{0.1cm}\text{Plethystic}\hspace{0.1cm}\text{Exponential}}\hspace{-0.5cm},
	\end{eqnarray}}
where $\varphi$ is a  spurion variable that represents either a scalar ($\phi$) or a fermion ($\psi$),  or a gauge field ($A_{\mu}$). With the aid of the Haar measure, the PEs are integrated on the symmetry group space. The Plethystic Exponentials for fields having integer and half-integer spins can be depicted as \cite{Hanany:2010vu,Feng:2007ur,Henning:2017fpj,Henning:2015alf}:
{\small\begin{eqnarray}\label{eq:PE}
	PE[\phi, R] &=& \exp\left[\sum_{r=1}^{\infty}\frac{\phi^r\chi_R(z^r_j)}{r}\right],\\
	PE[\psi, R] &=& \exp\left[\sum_{r=1}^{\infty}(-1)^{r+1}\frac{\psi^r\chi_R(z^r_j)}{r}\right],
	\end{eqnarray}}
respectively.
Here, $R$ denotes the representation of the symmetry group $\mathcal{G}_j$ under which the fields ($\phi,\psi$) transform and $\chi_R(z^r_j)$ the  corresponding ``Weyl" character.

\begin{figure}[h]
	\centering
	{
		\includegraphics[scale=0.7]{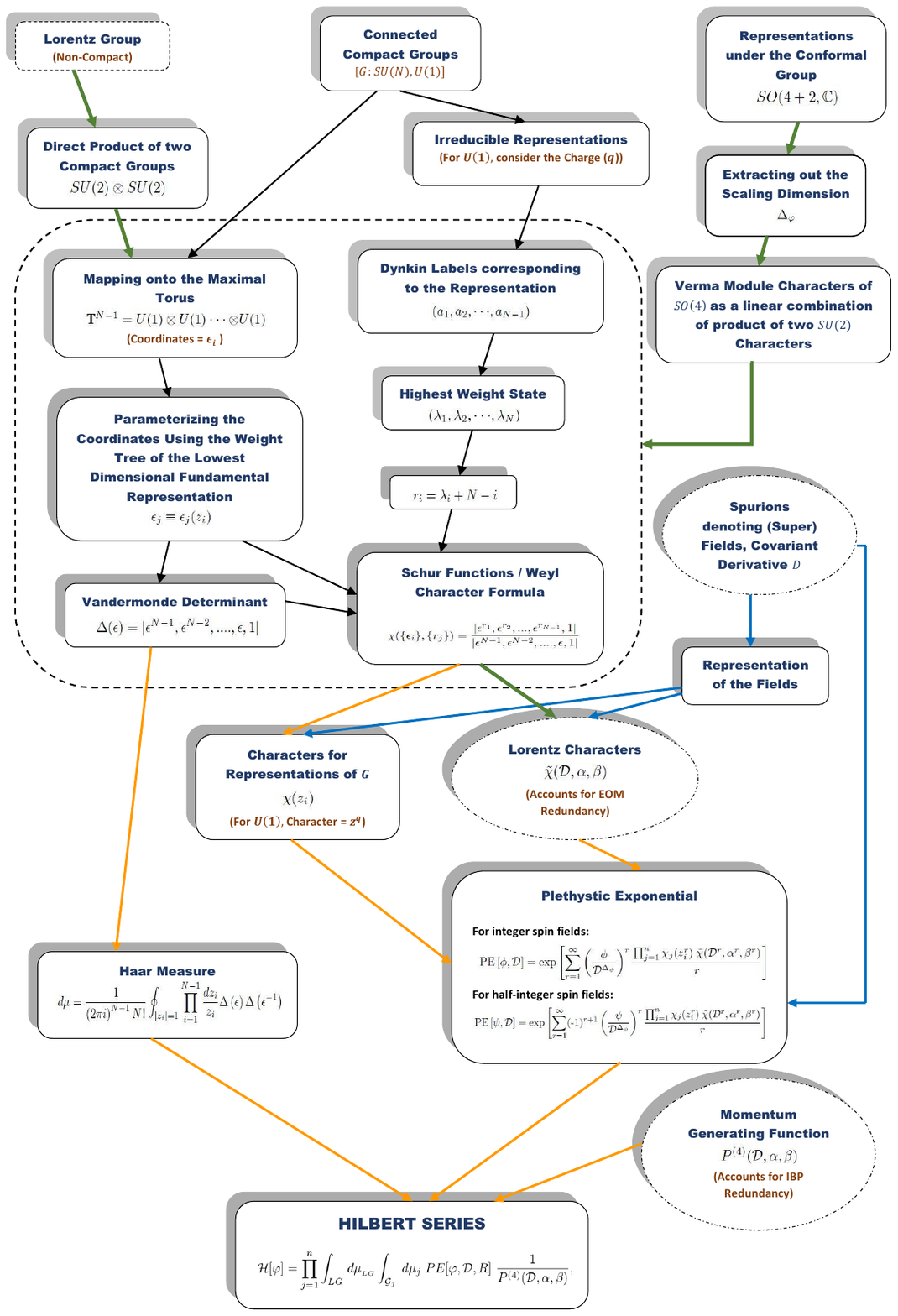}
	}
	\caption{Flow chart outlining the mathematical steps followed for computing the Hilbert Series.}
	\label{fig:flowchart-math}
\end{figure}

We are specifically interested in studying the representations of connected compact Lie groups which encapsulates the internal symmetry of  the particle physics models. In addition, the non-compact Lorentz group which describes the space-time transformations of the fields also attracts our attention. We have summarized the complete scheme of building the Hilbert Series through the explicit computation of characters and Haar measures in Fig.~\ref{fig:flowchart-math}. We have started by explicitly calculating the Haar measures of the groups ($SU(N), SO(2N+1), SO(2N), Sp(2N)$) \cite{Weyl} for small values of $N$ and built characters of some example representations in subsection~\ref{subsec:char-haar-compute}. Then we have briefly examined the non-triviality associated with constructing characters and Haar measure for the non-compact Lorentz group in subsection~\ref{subsec:hs-lorentz-group}. 

\subsection{Characters and Haar measures of connected compact Lie groups}\label{subsec:char-haar-compute}

We are interested in both abelian and non-abelian Lie groups. The procedures followed for computing the characters and Haar measures for each of these groups have been described below.

\subsubsection*{\textbf{\underline{\large Abelian Group~-~$U(1)$}}}\label{subsubsec:u1-char-haar-measure}

\textbf{Characters}\\
The $U(1)$ characters depend on the associated charge of a field. For a field having charge $q$ the $U(1)$ character is simply
\vspace{-0.5cm}
\begin{eqnarray}\label{eq:u1-char}
\chi_{({U(1)})_{q}}(z) = z^q. 
\end{eqnarray}
\textbf{Haar Measure}\\
The maximal torus of $U(1)$ is simply the unit circle. So, the group space integral is equivalent to the integral over some $\theta$ from $\theta = 0$ to $\theta = 2\pi$. With the parametrization as $z = e^{i\theta}$, this turns into a contour integral over $z$. 
Thus, the $U(1)$ Haar measure can be written as: 

\vspace{-0.5cm}
{\scriptsize\begin{eqnarray}\label{eq:u1-haar-measure}
	\int_{0}^{2\pi} d\theta &\equiv& \frac{1}{2\pi i} \oint_{|z|=1} \frac{dz}{z}, \hspace{0.5cm} \text{where}\hspace{0.2cm} 
	z = e^{i\theta}. \nonumber \\
	\therefore \hspace{0.2cm} \int d\mu_{U(1)} &=& \frac{1}{2\pi i} \oint_{|z|=1} \frac{dz}{z}.
	\end{eqnarray}}
\noindent
\textbf{\underline{\large Non-Abelian Groups}}
\subsubsection*{$1.~SU(N)$}\label{subsubsec:sun-haar-char}
\textbf{Characters}\\
\\
For $SU(N)$ the Weyl character formula \cite{Balantekin:2001id,Plymen:1976,koike1987,littlewood1977theory,rossmann2006lie} is given as:
{\scriptsize\begin{eqnarray}
	\chi_{r_1,r_2,...,r_{N-1}}^{(M(\epsilon))} = \frac{|\epsilon^{r_1},\epsilon^{r_2},...,\epsilon^{r_{N-1}},1|}{|\epsilon^{N-1},\epsilon^{N-2},....,\epsilon,1|},
	\end{eqnarray}}
where $M(\epsilon)$ = diag$(\epsilon_1,\epsilon_2,...,\epsilon_N)$ identifies a particular representation of $SU(N)$ with $\prod^{N}_{a=1}\epsilon_{a}$ = $1$ and $r_1, r_2,..., r_{N-1}$ are integers such that $r_1>r_2>\cdots r_{N-1}>0$ and these are obtained from the Dynkin labels of a particular representation. We can write the numerator in expanded form as:

\vspace{-0.5cm}
{\scriptsize\begin{eqnarray}\label{eq:char-numerator}
	|\epsilon^{r_1},\epsilon^{r_2},...,\epsilon^{r_{N-1}},1| = 
	\begin{vmatrix}
	\epsilon_1^{r_1}&\epsilon_1^{r_2} & . . . & \epsilon_1^{r_{N-1}}& 1\\
	\vspace{0.4cm}\epsilon_2^{r_1} &\epsilon_2^{r_2} & . . . & \epsilon_2^{r_{N-1}}&  1\\
	\vdots & \vdots &  \ddots &\vdots & \vdots \\ 
	\epsilon_N^{r_1}&\epsilon_N^{r_2} & . . . & \epsilon_N^{r_{N-1}}& 1\\
	\end{vmatrix},
	\end{eqnarray}}

\clearpage

\noindent while the denominator is the Vandermonde determinant as given in Eq.~\eqref{eq:vandermonde}:

\vspace{-0.5cm}
{\scriptsize\begin{eqnarray}\label{eq:vandermonde}
|\epsilon^{N-1},\epsilon^{N-2},....,\epsilon,1| = 
\begin{vmatrix}
\epsilon_1^{N-1}& \epsilon_1^{N-2}& . . . &\epsilon_1^{2}& \epsilon_1& 1\\
\epsilon_2^{N-1}& \epsilon_2^{N-2}& . . . &\epsilon_2^{2}& \epsilon_2& 1\\
\vdots & \vdots &  \ddots & \vdots & \vdots & \vdots \\ 
\epsilon_N^{N-1}& \epsilon_N^{N-2}& . . . & \epsilon_N^{2}& \epsilon_N& 1\\
\end{vmatrix} = \prod_{1\leq a<b\leq N} \left(\epsilon_a-\epsilon_b\right).
\end{eqnarray}}

\vspace{-0.5cm}
The $\epsilon_a$'s define coordinates on the maximal torus of $SU(N)$ which is the group $\mathbb{T}^{N-1} = U(1)\otimes U(1)\cdots\otimes U(1)$ \cite{Dieck}. It can be described by the matrix shown in Eq.~\eqref{eq:torus-suN}: 

\vspace{-0.5cm}
{\scriptsize\begin{eqnarray}\label{eq:torus-suN}
\mathbb{T}^{N-1}:
\begin{pmatrix}
e^{i\theta_1} & 0 & 0 &\cdots& 0\\
0 & e^{i(\theta_2\text{-}\theta_1)} & 0 &\cdots& 0\\
0 & 0 & e^{i(\theta_3\text{-}\theta_2)} &\cdots& 0\\
\vdots & \vdots & \vdots & \ddots & \vdots \\
0 & 0 & 0 & \cdots & e^{\text{-}i\theta_{N-1}}\\
\end{pmatrix}_{N\times N}\hspace{-0.7cm}.
\end{eqnarray}}

\vspace{-0.5cm}
Here $\theta_i$'s parametrize the points on the torus. The co-ordinates of these points on the torus can be reparametrized in terms of the complex variables $z_i$'s as $z_i=e^{i\theta_i}$. The $\epsilon_a$'s are functions of these $z_i$'s, i.e., $\epsilon_a = \epsilon_a\left(z_1, z_2,..., z_{N-1}\right)$ and they follow the properties:

\vspace{-0.5cm}
{\scriptsize\begin{eqnarray}
\prod_{a=1}^{N}\epsilon_a = 1, \hspace{0.5cm} |\epsilon_a| = 1, \hspace{0.7cm} a = 1,2,...,N.
\end{eqnarray}}	
This relationship between $z_i$'s and $\epsilon_a$ is determined using 
the weight tree \cite{Foster:2016lectures} with respect to the lowest dimension fundamental (LDF) representation\footnote{For a particular group, the  fundamental representations are those whose Dynkin labels ($N-1$ tuples for $SU(N)$) have a single entry as unity while all other entries are 0s. Among these the representations denoted by $\left(1,0,0,...,0,0\right)$ and $\left(0,0,0,...,0,1\right)$ (which are conjugate to each other) have the lowest dimension equal to $N$. For example, for $SU(3)$, the fundamental representations are $\left(1,0\right)$ and $\left(0,1\right)$ which are conjugate to each other and each of them have dimension 3. While for $SU(4)$, the fundamental representations are $\left(1,0,0\right)$, $\left(0,1,0\right)$ and $\left(0,0,1\right)$. Among these $\left(1,0,0\right)$ and $\left(0,0,1\right)$ have the lowest dimension 4 whereas $\left(0,1,0\right)$ has dimension 6.} corresponding to the group. The weight tree can be constructed starting from the respective Dynkin label  $\left(1,0,..,0\right)$ by successively subtracting the rows ($\alpha_i$) of the Cartan matrix shown below:

{\scriptsize\begin{eqnarray}\label{eq:sun-cartan matrix}
\mathcal{A}_{SU(N)} = 
\begin{pmatrix}
- -\alpha_1- -\\
- -\alpha_2- -\\
\vline \\
- -\alpha_{N-2}- -\\
- -\alpha_{N-1}- -
\end{pmatrix} = 
\begin{pmatrix}
2 & \text{-}1& 0& \cdots& 0& 0 & 0\\
\text{-}1&  2& \text{-}1& \cdots& 0& 0 & 0\\
\vdots & \vdots & \vdots & \ddots& \vdots & \vdots & \vdots  \\
0 & 0& 0& \cdots & \text{-}1& 2 & \text{-}1\\
0 &  0&  0& \cdots& 0& \text{-}1 & 2\\
\end{pmatrix}_{(N-1)\times(N-1)}\hspace{-1.8cm}.
\end{eqnarray}}
\noindent	
The weight tree corresponding to the LDF representation of $SU(N)$ is shown below:

\vspace{-0.5cm}	
{\scriptsize\begin{eqnarray}
L_1 &=& \underbrace{(1,0,0,...,0,0)}_{(\text{$N$-1 tuple})}, \nonumber\\
L_2 = L_1 - \alpha_1 &=& (-1,1,0,...,0,0),\nonumber\\
&\vdots&             \nonumber\\
L_k = L_{k-1}-\alpha_{k-1} &=& (0,..,-1,1,.,0),\nonumber\\
&\vdots&             \nonumber\\
L_{N-1} = L_{N-2}-\alpha_{N-2}&=& (0,0,...,-1,1),\nonumber\\
L_N = L_{N-1}-\alpha_{N-1}&=& (0,0,...,0,-1).
\end{eqnarray}}
Then, if the $(N-1)$ tuple $L_i$ is denoted as $(l_i^{(1)},l_i^{(2)},l_i^{(3)},...,l_i^{(N-1)})$, a general formula for $\epsilon_a$ can be written in terms of $z_i$'s as:

\vspace{-0.5cm}
{\scriptsize\begin{eqnarray}
\epsilon_i = z_1^{l_i^{(1)}}\times z_2^{l_i^{(2)}}\times z_3^{l_i^{(3)}}\times . . . \times z_{N-1}^{l_i^{(N-1)}},
\end{eqnarray}}

\vspace{-0.7cm}
\noindent which enables us to write:

\vspace{-0.6cm}
{\scriptsize\begin{eqnarray}\label{eq:eps-z-reln}
\epsilon_1 &=& z_1^{1}\times z_2^{0}\times z_3^{0}\times . . . \times z_{N-1}^{0} = z_1, \hspace{1.8cm}
\epsilon_2 = z_1^{-1}\times z_2^{1}\times z_3^{0}\times . . . \times z_{N-1}^{0} = z_1^{-1}z_2, \nonumber \\
\epsilon_k &=&  z_1^{0}\times . . . z_{k-1}^{-1}\times z_k^{1}\times ... \times z_{N-1}^{0} = z_{k-1}^{-1}z_k, \hspace{0.5cm}
\epsilon_N =  z_1^{0}\times z_2^{0}\times z_3^{0}\times . . . \times z_{N-1}^{-1} =  z_{N-1}^{-1}.
\end{eqnarray}}

\noindent
\underline{\textbf{Calculating the $r_i$'s}}
\\
A particular representation of $SU(N)$ of dimension $d$ can be uniquely identified by its Dynkin label $(a_1,a_2,.....,a_{N-1})$ and it is represented by the Young diagram consisting of $N-1$ rows with boxes. To find the $r_i$'s we first need to obtain the $\lambda_i$'s, which can be obtained as solutions of the following equation in terms of the Dynkin label and the fundamental weight tree of the LDF representation.

{\scriptsize\begin{eqnarray}
(a_1,a_2,...,a_{N-1}) = \lambda_1 (1,0,...,0,0) + \lambda_2 (-1,1,0,...,0) + \cdots +\lambda_{N-1} (0,...,0,-1,1) + \lambda_{N}(0,0,...,0,-1).\nonumber
\end{eqnarray}}
\noindent
Note that the equation contains $N$-unknowns in $\lambda_1,\lambda_2,...,\lambda_N$ while the Dynkin label and the fundamental weights are ($N-1$) tuples. Thus we are required to solve ($N-1$) equations in $N$-unknowns but this difficulty is remedied by making an association between the $\lambda_i$ and the Young diagrams. It turns out that $\lambda_i$ equals the number of boxes in the $i$-th row of the Young diagrams for the particular representation of $SU(N)$ and for non-trivial representations $\lambda_N$ = 0. Using this we get ($N-1$) equations in ($N-1$) unknowns:

\vspace{-0.4cm}
{\scriptsize\begin{eqnarray}
a_k &=& \lambda_k-\lambda_{k+1}, \hspace{0.7cm} k = 1,\cdots,N-1 \hspace{0.6cm} \text{and} \hspace{0.4cm} \lambda_{N}=0.  
\end{eqnarray}}

\vspace{-0.5cm}
\noindent Solving this we get: 

\vspace{-0.5cm}
{\scriptsize\begin{eqnarray}\label{eq:lambda-recursion}
\lambda_{N} &=& 0, \nonumber\\
\lambda_{N-1} &=& a_{N-1} =\left(\sum_{i=1}^{N-1}a_i\right) -\left(\sum_{j=1}^{N-2}a_j\right),\nonumber\\
\lambda_k &=& a_k+a_{k+1}+\cdots+a_{N-1} =\left(\sum_{i=1}^{N-1}a_i\right) -\left(\sum_{j=1}^{k-1}a_j\right).
\end{eqnarray}}

\vspace{-0.5cm}
\noindent
The $r_i$'s are related to the $\lambda_i$'s through the following equation:

\vspace{-0.4cm}
{\scriptsize\begin{eqnarray}
\vec{r} = \vec{\lambda}+\vec{\rho} \hspace{0.5cm}\text{where}\hspace{0.5cm}\rho_i = N-i, \hspace{0.5cm} i=1,2,..,N.
\end{eqnarray}}

\begin{figure}[h]
\centering
\includegraphics[clip,trim=3cm 5cm 1cm 13cm,scale=0.3]{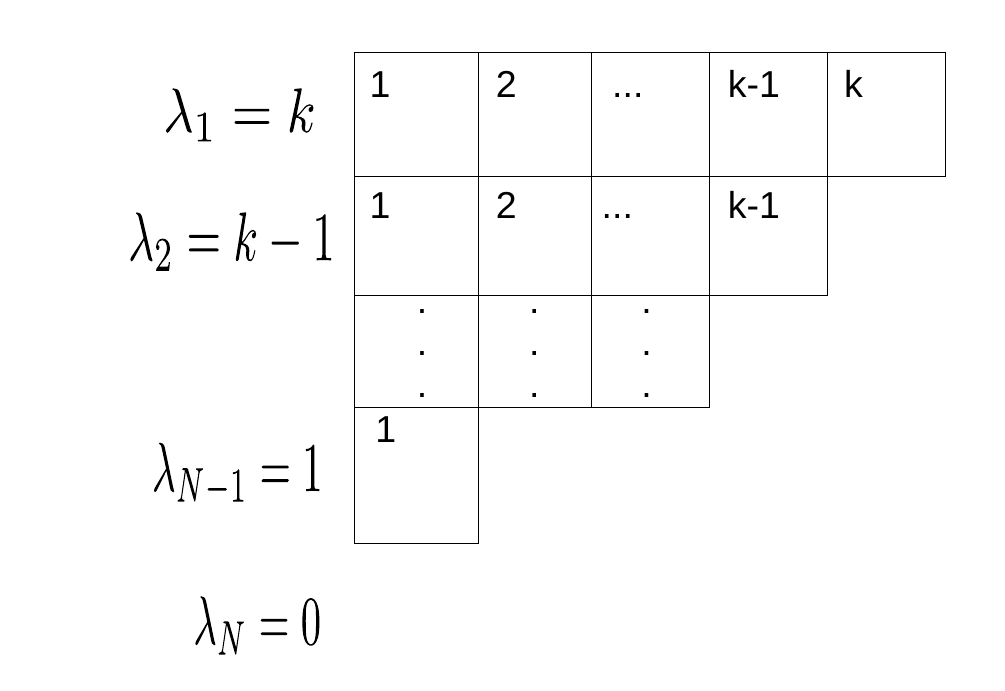}
\caption{A schematic form of the Young diagram corresponding to an arbitrary representation of  $SU(N)$. $\lambda_i$ is equal to the number of boxes in the $i$-th row of the diagram. From here one can immediately infer that $\lambda_N$ = 0. }
\label{fig:young}
\end{figure}

\vspace{-0.6cm}
\noindent
Since $\lambda_N = 0$ and $\rho_N = N-N = 0$, therefore $r_N = 0$. Now, having obtained $r_1, r_2,..., r_{N-1}$, the numerator can be computed using Eq.~\eqref{eq:char-numerator} corresponding to the given representation and subsequently the full character as well. We must mention that the $\lambda_i$'s and hence the $r_i$'s can be directly obtained from the Young diagram corresponding to the representation as shown in Fig.~\ref{fig:young}.
\newpage
\vspace{-0.5cm}

\noindent
\textbf{Haar Measure}
\\
The general formula can be written as \cite{Gray:2008yu,Hanany:2014dia,Henning:2017fpj}:

\vspace{-0.7cm} 
{\scriptsize\begin{eqnarray}\label{eq:haar-measure}
\int_{SU(N)} d\mu_{SU(N)} = \frac{1}{\left(2\pi i\right)^{N-1}N!}\oint_{|z_l|=1} \prod_{l=1}^{N-1}\frac{dz_l}{z_l}\Delta\left(\epsilon\right)\Delta\left(\epsilon^{-1}\right).
\end{eqnarray}}
$\Delta\left(\epsilon\right)$ is the Vandermonde determinant as given in Eq.~\eqref{eq:vandermonde} which can be evaluated in terms of the $\epsilon_a(z_l)$'s, $\Delta(\epsilon^{-1})$ can similarly be computed by substituting $\epsilon^{-1}_a$ in the place of $\epsilon_a$  in the expression of $\Delta\left(\epsilon\right)$. Finally, substituting for known quantities in Eq.~\eqref{eq:haar-measure} the Haar measure for a given group can be obtained.

Next, the characters for certain representations \cite{Slansky:1981yr,Yamatsu:2015npn} of $SU(2)$, $SU(3)$, $SU(4)$ and $SU(5)$ have been computed and the Haar measures of these groups have also been explicitly calculated.

\begin{center}
\rule{15cm}{1.25pt}	
\end{center}
\noindent
\fbox{$SU(2)$}\label{examples:su2-char-haar}\\
\\
\textbf{Haar Measure}
\\
Using Eq.~\eqref{eq:eps-z-reln}, we can obtain for $SU(2)$, $\epsilon_1 = z_1$ and $\epsilon_2$ = $z_1^{-1}$. The Vandermonde determinant for this case is: 

\vspace{-0.7cm}
{\scriptsize\begin{eqnarray}\label{eq:vandermonde-su2}
	\Delta\left(\epsilon\right) = \begin{vmatrix}
	\epsilon_1\ \ 1\\
	\epsilon_2\ \ 1\\
	\end{vmatrix} = \left(\epsilon_1-\epsilon_2\right) \xRightarrow{\epsilon_i(z_j)} \left(z_1-\frac{1}{z_1}\right).
	\end{eqnarray}}

\vspace{-0.7cm}
\noindent
$\Delta(\epsilon^{-1})$ is obtained by replacing $\epsilon_i$ by $\epsilon_i^{-1}$ in the above expression. Then using Eq.~\eqref{eq:haar-measure},

\vspace{-0.4cm}
{\scriptsize\begin{eqnarray}\label{eq:su2-haar-measure}
	d\mu_{SU(2)} = \frac{1}{2!\left(2\pi i\right)} \frac{dz_1}{z_1}\Delta\left(\epsilon\right)\Delta\left(\epsilon^{-1}\right) = \frac{1}{2\left(2\pi i\right)} \frac{dz_1}{z_1}\left(1-z_1^2\right)\left(1-\frac{1}{z_1^2}\right).
	\end{eqnarray}}

\noindent
\textbf{Characters}
\begin{itemize}
	\item The Singlet Representation: $1 \equiv (0) \equiv  {\scriptsize\yng(1,1)}$ 
	
	For the singlet representation, we can directly get $\vec{\lambda}$ = $(1,1)$ from the Young diagram. Now, since $\vec{\rho}$ = $(1,0)$ therefore, $\vec{r} = \vec{\lambda}+\vec{\rho}= (2,1)$ and the character is obtained as:
	
	\vspace{-0.7cm}
	{\scriptsize\begin{eqnarray}\label{eq:su2-1-char}
	\chi(\epsilon_1,\epsilon_2) = \frac{|		\epsilon^2,\hspace{0.1cm}\epsilon|}{|\epsilon,\hspace{0.1cm}1|} = \frac{1}{\left(\epsilon_1-\epsilon_2\right)}
	\begin{vmatrix}
	\epsilon_1^2\hspace{0.1cm}\ \ \epsilon_1\\
	\epsilon_2^2\hspace{0.1cm}\ \ \epsilon_2\\
	\end{vmatrix}=\epsilon_1\cdot\epsilon_2, \hspace{0.5cm}	
	\implies\;\chi_{({SU(2)})_{1}}(z_1) = z_1\times \frac{1}{z_1}  = 1.
	\end{eqnarray}} 
	Now, this result is easily generalized for the case of general $SU(N)$, i.e., the character of the singlet representation for any $SU(N)$ is simply
	
	\vspace{-0.7cm}
	{\scriptsize\begin{eqnarray}\label{eq:su-n-1-char}
	\chi_{({SU(N)})_{1}} = \prod_{i=1}^{N}\epsilon_i = z_1\times \frac{z_2}{z_1}\times \cdots \times \frac{z_{N-1}}{z_{N-2}}\times \frac{1}{z_{N-1}}  = 1.
	\end{eqnarray}}
	
	\vspace{-0.8cm}
	\item The (Anti-)Fundamental Representation: $2\equiv\overline{2} \equiv (1) \equiv {\scriptsize\yng(1)}$ 
	
	Using Eq.~\eqref{eq:lambda-recursion}, we get $\vec{\lambda}$ = $(1,0)$. Now, since $\vec{\rho}$ = $(1,0)$ therefore, $\vec{r} = \vec{\lambda}+\vec{\rho} = (2,0)$ and the character is obtained as:
	
	\vspace{-0.6cm}
	{\scriptsize\begin{eqnarray}\label{eq:su2-2-char}
	\chi(\epsilon_1,\epsilon_2) = \frac{|\epsilon^2,\hspace{0.1cm}1|}{|\epsilon,\hspace{0.1cm}1|} = \frac{1}{\left(\epsilon_1-\epsilon_2\right)}
	\begin{vmatrix}
	\epsilon_1^2\hspace{0.1cm}\ \ 1\\
	\epsilon_2^2\hspace{0.1cm}\ \ 1\\
	\end{vmatrix} =  \epsilon_1+\epsilon_2, \hspace{0.5cm}
	\implies\;\chi_{({SU(2)})_{2(\overline{2})}}(z_1) = z_1+\frac{1}{z_1}.
	\end{eqnarray}}   
	\item The Adjoint Representation: $3 \equiv (2) \equiv {\scriptsize\yng(2)}$ 
	
	We obtain $\vec{\lambda}$ = $(2,0)$ and $\vec{r} = \vec{\lambda}+\vec{\rho}= (3,0)$ and the character is obtained as:
	
	\vspace{-0.5cm}
	{\scriptsize\begin{eqnarray}\label{eq:su2-3-char}
	\chi(\epsilon_1,\epsilon_2) &=& \frac{|\epsilon^3,\hspace{0.1cm}1|}{|\epsilon,\hspace{0.1cm}1|} = \frac{1}{\left(\epsilon_1-\epsilon_2\right)}
	\begin{vmatrix}
	\epsilon_3^2\hspace{0.1cm}\ \ 1\\
	\epsilon_3^2\hspace{0.1cm}\ \ 1\\
	\end{vmatrix} = \frac{\epsilon_1^3-\epsilon_2^3}{\epsilon_1-\epsilon_2} = \epsilon_1^2+\epsilon_1\epsilon_2+\epsilon_2^2. \nonumber\\ 
	\therefore\;\chi_{({SU(2)})_{3}}(z_1) &=&  z_1^2+1+\frac{1}{z_1^2}.
	\end{eqnarray}} 
\end{itemize}
It must be noted that the character of the fundamental representation of any $SU(N)$ is simply $\sum_{i=1}^{N}\epsilon_i = z_1 + \sum_{k=2}^{k={N-1}}\frac{z_k}{z_{k-1}} + \frac{1}{z_{N-1}}$ and the character of the anti-fundamental representation is obtained by making the substitution $z_i\leftrightarrow\frac{1}{z_i}$. The character of adjoint representation can be computed using that for the fundamental and anti-fundamental representations as we know:

\vspace{-0.7cm}
\begin{eqnarray}
& &\text{fundamental}\otimes\text{anti-fundamental} = \text{adjoint}\oplus\text{singlet}\nonumber\\
& &\implies \chi_{\text{f}}\times\chi_{\text{af}} = \chi_{\text{adj}}+\chi_{\text{singl}} .\hspace{0.4cm}
\therefore\;\chi_{\text{adj}} = \chi_{\text{f}}\times\chi_{\text{af}} - 1. \nonumber
\end{eqnarray}
So, the true utility of the formalism outlined above lies in the character computation of other representations of a given group. Below, a few more examples of character computation for irreducible representations of $SU(2)$ have been elucidated.

\begin{itemize}
	\item The Quadruplet Representation: $4 \equiv (3) \equiv {\scriptsize\yng(3)}$ 
	
	We obtain $\vec{\lambda}$ = $(3,0)$ and $\vec{r} = \vec{\lambda}+\vec{\rho} = (4,0)$ and the character is obtained as:
	
	\vspace{-0.7cm}
	{\scriptsize\begin{eqnarray}\label{eq:su2-4-char}
		\chi(\epsilon_1,\epsilon_2) &=& \frac{|\epsilon^4,\hspace{0.1cm}1|}{|\epsilon,\hspace{0.1cm}1|} = \frac{1}{\left(\epsilon_1-\epsilon_2\right)}
		\begin{vmatrix}
		\epsilon_4^2\hspace{0.1cm}\ \ 1\\
		\epsilon_4^2\hspace{0.1cm}\ \ 1\\
		\end{vmatrix} = \frac{\epsilon_1^4-\epsilon_2^4}{\epsilon_1-\epsilon_2} = \epsilon_1^3+\epsilon_1^2\epsilon_2+\epsilon_1\epsilon_2^2 + \epsilon_2^3. \nonumber\\ 
		\therefore\;\chi_{({SU(2)})_{4}}(z_1) &=&  z_1^3+z_1+\frac{1}{z_1}+\frac{1}{z_1^3}.
		\end{eqnarray}}
	\item The Quintuplet Representation: $5 \equiv (4) \equiv {\scriptsize\yng(4)}$ 
	
	We obtain $\vec{\lambda}$ = $(4,0)$ and $\vec{r} = \vec{\lambda}+\vec{\rho} = (5,0)$ and the character is obtained as:
	
	\vspace{-0.7cm}
	{\scriptsize\begin{eqnarray}\label{eq:su2-5-char}
		\chi(\epsilon_1,\epsilon_2) &=& \frac{|\epsilon^5,\hspace{0.1cm}1|}{|\epsilon,\hspace{0.1cm}1|} = \frac{1}{\left(\epsilon_1-\epsilon_2\right)}
		\begin{vmatrix}
		\epsilon_5^2\hspace{0.1cm}\ \ 1\\
		\epsilon_5^2\hspace{0.1cm}\ \ 1\\
		\end{vmatrix} = \frac{\epsilon_1^5-\epsilon_2^5}{\epsilon_1-\epsilon_2} = \epsilon_1^4+\epsilon_1^3\epsilon_2+\epsilon_1^2\epsilon_2^2+\epsilon_1\epsilon_2^3 + \epsilon_2^4. \nonumber\\ 
		\therefore\;\chi_{({SU(2)})_{5}}(z_1) &=&  z_1^4+z_1^2+1+\frac{1}{z_1^2}+\frac{1}{z_1^4}.
		\end{eqnarray}}
\end{itemize} 
\noindent \rule{15cm}{1.25pt}
\\
\\
\noindent
\fbox{$SU(3)$}\label{examples:su3-char-haar}\\
\\
\textbf{Haar Measure}	\renewcommand{\arraystretch}{1.2}
\\
Using Eq.~\eqref{eq:eps-z-reln}, we can obtain for $SU(3)$, $\epsilon_1 = z_1$, $\epsilon_2$ = $z_1^{-1}z_2$ and $\epsilon_3$ = $z_2^{-1}$. The Vandermonde determinant for this case is:

\vspace{-0.6cm}
{\scriptsize\begin{eqnarray}\label{eq:vandermonde-su3}
	\Delta\left(\epsilon\right) &=& \begin{vmatrix}
	\epsilon_1^{2} \ \ \epsilon_1\ \ 1\\
	\epsilon_2^{2} \ \ \epsilon_2\ \ 1\\
	\epsilon_3^{2} \ \ \epsilon_3\ \ 1\\
	\end{vmatrix} = \prod_{1\leq i<j\leq 3} \left(\epsilon_i-\epsilon_j\right) \xRightarrow{\epsilon_i(z_j)} \left(z_1-\frac{z_2}{z_1}\right)\left(z_1-\frac{1}{z_2}\right)\left(\frac{z_2}{z_1}-\frac{1}{z_2}\right).
	\end{eqnarray}}

\vspace{-0.4cm}
\noindent $\Delta(\epsilon^{-1})$ is obtained by replacing $\epsilon_i$ by $\epsilon_i^{-1}$ in the above expression. Then using Eq.~\eqref{eq:haar-measure},

\vspace{-0.6cm}
{\scriptsize\begin{eqnarray}\label{eq:su3-haar-measure}
d\mu_{SU(3)} &=& \frac{1}{3!\left(2\pi i\right)^{2}} \frac{dz_1}{z_1}\frac{dz_2}{z_2}\Delta\left(\epsilon\right)\Delta\left(\epsilon^{-1}\right) 
\nonumber \\ 
&=& \frac{1}{6\left(2\pi i\right)^{2}} \frac{dz_1}{z_1}\frac{dz_2}{z_2}\left(1-z_1z_2\right)\left(1-\frac{1}{z_1z_2}\right)\left(1-\frac{z_1}{z_2^2}\right)\left(1-\frac{z_2^2}{z_1}\right)\left(1-\frac{z_2}{z_1^2}\right)\left(1-\frac{z_1^2}{z_2}\right).
\end{eqnarray}}
\noindent
\textbf{Characters}
\vspace{-0.3cm}
\begin{itemize}
	\item The Fundamental Representation: $3 \equiv (1,0) \equiv {\scriptsize\yng(1)}$ 
	
	Using Eq.~\eqref{eq:lambda-recursion}, we get $\vec{\lambda}$ = $(1,0,0)$. Now, since $\vec{\rho}$ = $(2,1,0)$ therefore,
	$\vec{r} = \vec{\lambda}+\vec{\rho} = (3,1,0)$ and the character is obtained as:
	
	\vspace{-0.9cm}
	{\scriptsize\begin{eqnarray}\label{eq:su3-3-char}
	\chi(\epsilon_1,\epsilon_2,\epsilon_3) &=& \frac{|\epsilon^3,\hspace{0.1cm}
	\epsilon,\hspace{0.1cm}1|}{|\epsilon^2,\hspace{0.1cm}\epsilon,\hspace{0.1cm}1|} = \frac{1}{\prod_{1\leq i<j\leq 3}\left(\epsilon_i-\epsilon_j\right)}
	\begin{vmatrix}
	\epsilon_1^{3}\hspace{0.1cm} \ \ \epsilon_1\hspace{0.1cm}\ \ 1\\
	\epsilon_2^{3}\hspace{0.1cm} \ \ \epsilon_2\hspace{0.1cm}\ \ 1\\
	\epsilon_3^{3}\hspace{0.1cm} \ \ \epsilon_3\hspace{0.1cm}\ \ 1\\
	\end{vmatrix} = \epsilon_1+\epsilon_2+\epsilon_3, \nonumber\\
	\therefore\;\chi_{({SU(3)})_{3}}(z_1, z_2) &=& z_1+\frac{z_2}{z_1} + \frac{1}{z_2}.
	\end{eqnarray}}

	\vspace{-0.9cm}
	\item The Anti-fundamental Representation: $\overline{3} \equiv (0,1)\equiv {\scriptsize\yng(1,1)}$ 
	
	We obtain $\vec{\lambda}$ = $(1,1,0)$ and $\vec{r} = \vec{\lambda}+\vec{\rho} = (3,2,0)$ and the character is obtained as:
	
	\vspace{-0.8cm}
	{\scriptsize\begin{eqnarray}\label{eq:su3-3-bar-char}
	\chi(\epsilon_1,\epsilon_2,\epsilon_3) &=& \frac{|\epsilon^3,\hspace{0.1cm}
	\epsilon^2,\hspace{0.1cm}1|}{|\epsilon^2,\hspace{0.1cm}\epsilon,\hspace{0.1cm}1|} = \frac{1}{\prod_{1\leq i<j\leq 3}\left(\epsilon_i-\epsilon_j\right)}
	\begin{vmatrix}
	\epsilon_1^{3}\hspace{0.1cm} \ \ \epsilon_1^{2}\hspace{0.1cm}\ \ 1\\
	\epsilon_2^{3}\hspace{0.1cm} \ \ \epsilon_2^{2}\hspace{0.1cm}\ \ 1\\
	\epsilon_3^{3}\hspace{0.1cm} \ \ \epsilon_3^{2}\hspace{0.1cm}\ \ 1\\
	\end{vmatrix}= \epsilon_1\epsilon_2+\epsilon_1\epsilon_3+\epsilon_2\epsilon_3. \nonumber\\
	\therefore\;\chi_{(SU(3))_{\overline{3}}}(z_1, z_2) &=&  z_2+\frac{z_1}{z_2} + \frac{1}{z_1}.
	\end{eqnarray}} 

	\vspace{-0.8cm}
	\item The Adjoint Representation: $8 \equiv (1,1) \equiv {\scriptsize\yng(2,1)}$ 
	
	We obtain $\vec{\lambda}$ = $(2,1,0)$ and $\vec{r} = \vec{\lambda}+\vec{\rho} = (4,2,0)$ and the character is obtained as:
	
	\vspace{-0.8cm}
	{\scriptsize\begin{eqnarray}\label{eq:su3-8-char}
		\chi(\epsilon_1,\epsilon_2,\epsilon_3) &=& \frac{|\epsilon^4,\hspace{0.1cm}
		\epsilon^2,\hspace{0.1cm}1|}{|\epsilon^2,\hspace{0.1cm}\epsilon,\hspace{0.1cm}1|} = \frac{1}{\prod_{1\leq i<j\leq 3}\left(\epsilon_i-\epsilon_j\right)}
		\begin{vmatrix}
		\epsilon_1^{4}\hspace{0.1cm} \ \ \epsilon_1^{2}\hspace{0.1cm}\ \ 1\\
		\epsilon_2^{4}\hspace{0.1cm} \ \ \epsilon_2^{2}\hspace{0.1cm}\ \ 1\\
		\epsilon_3^{4}\hspace{0.1cm} \ \ \epsilon_3^{2}\hspace{0.1cm}\ \ 1\\
		\end{vmatrix}\nonumber\\
		&=&
		\epsilon_1^2\epsilon_2+\epsilon_2^2\epsilon_1+\epsilon_1^2\epsilon_3+\epsilon_3^2\epsilon_1+\epsilon_2^2\epsilon_3+\epsilon_3^2\epsilon_2+2\epsilon_1\epsilon_2\epsilon_3. \nonumber\\
		\therefore\;\chi_{(SU(3))_{8}}(z_1, z_2) &=& z_1z_2 + \frac{1}{z_1z_2}+2+\frac{z_1}{z_2^2}+\frac{z_1^2}{z_2}+\frac{z_2}{z_1^2}+\frac{z_2^2}{z_1}. 
		\end{eqnarray} }
		
		\vspace{-0.8cm}
	    \item The Sextet Representation: $6 \equiv (2,0) \equiv {\scriptsize\yng(2)}$ 
		
		We obtain $\vec{\lambda}$ = $(2,0,0)$ and $\vec{r} = \vec{\lambda}+\vec{\rho} = (4,1,0)$	and the character is obtained as:
		
		\vspace{-0.8cm}
		{\scriptsize\begin{eqnarray}\label{eq:su3-6-char}
		\chi(\epsilon_1,\epsilon_2,\epsilon_3) &=& \frac{|\epsilon^4,\hspace{0.1cm}
		\epsilon,\hspace{0.1cm}1|}{|\epsilon^2,\hspace{0.1cm}\epsilon,\hspace{0.1cm}1|} = \frac{1}{\prod_{1\leq i<j\leq 3}\left(\epsilon_i-\epsilon_j\right)}
		\begin{vmatrix}
		\epsilon_1^{4}\hspace{0.1cm} \ \ \epsilon_1\hspace{0.1cm}\ \ 1\\
		\epsilon_2^{4}\hspace{0.1cm} \ \ \epsilon_2\hspace{0.1cm}\ \ 1\\
		\epsilon_3^{4}\hspace{0.1cm} \ \ \epsilon_3\hspace{0.1cm}\ \ 1\\
		\end{vmatrix} =
		\epsilon_1^2+\epsilon_2^2+\epsilon_3^2+\epsilon_1\epsilon_2+\epsilon_2\epsilon_3+\epsilon_1\epsilon_3. \nonumber\\
		\therefore\;\chi_{(SU(3))_{6}}(z_1, z_2) &=& z_1^2 + \frac{z_2^2}{z_1^2}+\frac{1}{z_2^2}+\frac{1}{z_1}+\frac{z_1}{z_2}+z_2. 
		\end{eqnarray} }
		\item The 27-dimensional Representation: $27 \equiv (2,2) \equiv {\scriptsize\yng(4,2)}$ 
		
		We obtain $\vec{\lambda}$ = $(4,2,0)$ and $\vec{r} = \vec{\lambda}+\vec{\rho} = (6,3,0)$ and the character is obtained as:
		
		\vspace{-0.8cm}
		{\scriptsize\begin{eqnarray}\label{eq:su3-10-char}
			\chi(\epsilon_1,\epsilon_2,\epsilon_3) &=& \frac{|\epsilon^6,\hspace{0.1cm}
			\epsilon^3,\hspace{0.1cm}1|}{|\epsilon^2,\hspace{0.1cm}\epsilon,\hspace{0.1cm}1|} = \frac{1}{\prod_{1\leq i<j\leq 3}\left(\epsilon_i-\epsilon_j\right)}
			\begin{vmatrix}
			\epsilon_1^{6}\hspace{0.1cm} \ \ \epsilon_1^{3}\hspace{0.1cm}\ \ 1\\
			\epsilon_2^{6}\hspace{0.1cm} \ \ \epsilon_2^{3}\hspace{0.1cm}\ \ 1\\
			\epsilon_3^{6}\hspace{0.1cm} \ \ \epsilon_3^{3}\hspace{0.1cm}\ \ 1\\
			\end{vmatrix} =
			\frac{\left(\epsilon_1^3-\epsilon_2^3\right)}{\left(\epsilon_1-\epsilon_2\right)}\; \frac{\left(\epsilon_1^3-\epsilon_3^3\right)}{\left(\epsilon_1-\epsilon_3\right)}\; \frac{\left(\epsilon_2^3-\epsilon_3^3\right)}{\left(\epsilon_2-\epsilon_3\right)} \nonumber\\ 
			&=&
			\left(\epsilon_1^2+\epsilon_1\epsilon_2+\epsilon_2^2\right) \left(\epsilon_1^2+\epsilon_1\epsilon_3+\epsilon_3^2\right) \left(\epsilon_2^2+\epsilon_2\epsilon_3+\epsilon_3^2\right). \nonumber\\
			\therefore\;\chi_{(SU(3))_{27}}(z_1, z_2) &=& 3 + z_1^3 + \frac{1}{z_1^3} + z_2^3 + \frac{1}{z_2^3} + \frac{z_1^3}{z_2^3} + \frac{z_2^3}{z_1^3} + \frac{z_1^4}{z_2^2} + \frac{z_2^2}{z_1^4} + \frac{z_2^4}{z_1^2} + \frac{z_1^2}{z_2^4} \nonumber\\
			& & + \frac{2z_1^2}{z_2} + \frac{2z_2}{z_1^2} + \frac{2z_2^2}{z_1} + \frac{2z_1}{z_2^2} + z_1^2z_2^2 + \frac{1}{z_1^2z_2^2} + 2z_1z_2 + \frac{2}{z_1z_2}.
			\end{eqnarray} }
\end{itemize}
\noindent \rule{15cm}{1.25pt}
\\
\\
\noindent
\fbox{$SU(4)$}\label{examples:su4-char-haar}\\
\\
\textbf{Haar Measure}
\\
Using Eq.~\eqref{eq:eps-z-reln}, we can obtain for $SU(4)$, $\epsilon_1 = z_1$, $\epsilon_2$ = $z_1^{-1}z_2$, $\epsilon_3$ = $z_2^{-1}z_3$ and $\epsilon_4 = z_3^{-1}$. The Vandermonde determinant for this case is:

\vspace{-0.6cm} 
{\scriptsize\begin{eqnarray}\label{eq:vandermonde-su4}
	\Delta\left(\epsilon\right) = \begin{vmatrix}
	\epsilon_1^{3}\hspace{0.1cm} \ \ \epsilon_1^{2}\hspace{0.1cm} \ \ \epsilon_1\hspace{0.1cm}\ \ 1\\
	\epsilon_2^{3}\hspace{0.1cm} \ \ \epsilon_2^{2}\hspace{0.1cm} \ \ \epsilon_2\hspace{0.1cm}\ \ 1\\
	\epsilon_3^{3}\hspace{0.1cm} \ \ \epsilon_3^{2}\hspace{0.1cm} \ \ \epsilon_3\hspace{0.1cm}\ \ 1\\
	\epsilon_4^{3}\hspace{0.1cm} \ \ \epsilon_4^{2}\hspace{0.1cm} \ \ \epsilon_4\hspace{0.1cm}\ \ 1\\
	\end{vmatrix} = \prod_{1\leq i<j\leq 4}\left(\epsilon_i-\epsilon_j\right).
	\end{eqnarray}}
$\Delta(\epsilon^{-1})$ is obtained by replacing $\epsilon_i$ by $\epsilon_i^{-1}$ in the above expression. Then using Eq.~\eqref{eq:haar-measure},

\vspace{-0.5cm}
{\scriptsize\begin{eqnarray}\label{eq:su4-haar-measure}
	d\mu_{SU(4)} &=& \frac{1}{4!\left(2\pi i\right)^{3}} \frac{dz_1}{z_1}\frac{dz_2}{z_2}\frac{dz_3}{z_3}\Delta\left(\epsilon\right)\Delta\left(\epsilon^{-1}\right)\nonumber\\
	&=& \frac{1}{24\left(2\pi i\right)^{3}} \frac{dz_1}{z_1}\frac{dz_2}{z_2}\frac{dz_3}{z_3}\left(1-z_1z_3\right)\left(1-\frac{1}{z_1z_3}\right)\left(1-\frac{z_1^2}{z_2}\right)\left(1-\frac{z_2}{z_1^2}\right)\left(1-\frac{z_1z_2}{z_3}\right)\left(1-\frac{z_3}{z_1z_2}\right) \nonumber\\
	& & \left(1-\frac{z_2^2}{z_1z_3}\right)\left(1-\frac{z_1z_3}{z_2^2}\right)\left(1-\frac{z_2z_3}{z_1}\right)\left(1-\frac{z_1}{z_2z_3}\right)\left(1-\frac{z_3^2}{z_2}\right)\left(1-\frac{z_2}{z_3^2}\right).
	\end{eqnarray}}
\textbf{Characters}
\begin{itemize}
	\item The Fundamental Representation: $4 \equiv (1,0,0) \equiv {\scriptsize\yng(1)}$ 
	
	Using Eq.~\eqref{eq:lambda-recursion}, we get $\vec{\lambda}$ = $(1,0,0,0)$. Now, since $\vec{\rho}$ = $(3,2,1,0)$ therefore,\\ 
	$\vec{r} = \vec{\lambda}+\vec{\rho} = (4,2,1,0)$ and the character is obtained as:
	
	\vspace{-0.6cm}
	{\scriptsize\begin{eqnarray}\label{eq:su4-4-char}
		\chi(\epsilon_1,\epsilon_2,\epsilon_3,\epsilon_4) &=& \frac{|\epsilon^4,\hspace{0.1cm}\epsilon^2,\hspace{0.1cm}
		\epsilon^1,\hspace{0.1cm}1|}{|\epsilon^3,\hspace{0.1cm}\epsilon^2,\hspace{0.1cm}\epsilon,\hspace{0.1cm}1|} = \frac{1}{\prod_{1\leq i<j\leq 4}\left(\epsilon_i-\epsilon_j\right)}
		\begin{vmatrix}
		\epsilon_1^{4}\hspace{0.1cm} \ \ \epsilon_1^{2}\hspace{0.1cm} \ \ \epsilon_1\hspace{0.1cm}\ \ 1\\
		\epsilon_2^{4}\hspace{0.1cm} \ \ \epsilon_2^{2}\hspace{0.1cm} \ \ \epsilon_2\hspace{0.1cm}\ \ 1\\
		\epsilon_3^{4}\hspace{0.1cm} \ \ \epsilon_3^{2}\hspace{0.1cm} \ \ \epsilon_3\hspace{0.1cm}\ \ 1\\
		\epsilon_4^{4}\hspace{0.1cm} \ \ \epsilon_4^{2}\hspace{0.1cm} \ \ \epsilon_4\hspace{0.1cm}\ \ 1\\
		\end{vmatrix} = \epsilon_1 + \epsilon_2 + \epsilon_3 + \epsilon_4. \nonumber\\
		\therefore\;\chi_{({SU(4)})_{4}}(z_1, z_2, z_3) &=& z_1 + \frac{z_2}{z_1} + \frac{z_3}{z_2} + \frac{1}{z_3}.
		\end{eqnarray}}
	
	\vspace{-0.9cm}
	\item The Anti-fundamental Representation: $\overline{4} \equiv (0,0,1) \equiv {\scriptsize\yng(1,1,1)}$ 
	
	We get $\vec{\lambda}$ = $(1,1,1,0)$ and $\vec{r} = \vec{\lambda}+\vec{\rho} = (4,3,2,0)$ and the character is obtained as:
	
	\vspace{-0.9cm}
	{\scriptsize\begin{eqnarray}\label{eq:su4-4-bar-char}
		\chi(\epsilon_1,\epsilon_2,\epsilon_3,\epsilon_4) &=& \frac{|\epsilon^4,\hspace{0.1cm}\epsilon^3,\hspace{0.1cm}
		\epsilon^2,\hspace{0.1cm}1|}{|\epsilon^3,\hspace{0.1cm}\epsilon^2,\hspace{0.1cm}\epsilon,\hspace{0.1cm}1|} = \frac{1}{\prod_{1\leq i<j\leq 4}\left(\epsilon_i-\epsilon_j\right)}
		\begin{vmatrix}
		\epsilon_1^{4}\hspace{0.1cm} \ \ \epsilon_1^{3}\hspace{0.1cm} \ \ \epsilon_1^{2}\hspace{0.1cm}\ \ 1\\
		\epsilon_2^{4}\hspace{0.1cm} \ \ \epsilon_2^{3}\hspace{0.1cm} \ \ \epsilon_2^{2}\hspace{0.1cm}\ \ 1\\
		\epsilon_3^{4}\hspace{0.1cm} \ \ \epsilon_3^{3}\hspace{0.1cm} \ \ \epsilon_3^{2}\hspace{0.1cm}\ \ 1\\
		\epsilon_4^{4}\hspace{0.1cm} \ \ \epsilon_4^{3}\hspace{0.1cm} \ \ \epsilon_4^{2}\hspace{0.1cm}\ \ 1\\
		\end{vmatrix} \nonumber\\
		&=& \epsilon_1\epsilon_2\epsilon_3 + \epsilon_1\epsilon_2\epsilon_4 + \epsilon_1\epsilon_3\epsilon_4 + \epsilon_2\epsilon_3\epsilon_4. \nonumber\\
		\therefore\;\chi_{({SU(4)})_{\overline{4}}}(z_1, z_2, z_3) 
		&=& z_3 + \frac{z_2}{z_3} + \frac{z_1}{z_2} + \frac{1}{z_1}.
		\end{eqnarray}}
	
	\vspace{-0.9cm}
	\item The Decuplet Representation: $10 \equiv (2,0,0) \equiv {\scriptsize\yng(2)}$ 
	
	We get $\vec{\lambda}$ = $(2,0,0,0)$ and $ \vec{r} = \vec{\lambda}+\vec{\rho} = (5,2,1,0)$ and the character is obtained as:
	
	\vspace{-0.8cm}
	{\scriptsize\begin{eqnarray}\label{eq:su4-10-char}
		\chi(\epsilon_1,\epsilon_2,\epsilon_3,\epsilon_4) &=& \frac{|\epsilon^5,\hspace{0.1cm}\epsilon^2,\hspace{0.1cm}
		\epsilon,\hspace{0.1cm}1|}{|\epsilon^3,\hspace{0.1cm}\epsilon^2,\hspace{0.1cm}\epsilon,\hspace{0.1cm}1|} = \frac{1}{\prod_{1\leq i<j\leq 4}\left(\epsilon_i-\epsilon_j\right)}
		\begin{vmatrix}
		\epsilon_1^{5}\hspace{0.1cm} \ \ \epsilon_1^{2}\hspace{0.1cm} \ \ \epsilon_1\hspace{0.1cm}\ \ 1\\
		\epsilon_2^{5}\hspace{0.1cm} \ \ \epsilon_2^{2}\hspace{0.1cm} \ \ \epsilon_2\hspace{0.1cm}\ \ 1\\
		\epsilon_3^{5}\hspace{0.1cm} \ \ \epsilon_3^{2}\hspace{0.1cm} \ \ \epsilon_3\hspace{0.1cm}\ \ 1\\
		\epsilon_4^{5}\hspace{0.1cm} \ \ \epsilon_4^{2}\hspace{0.1cm} \ \ \epsilon_4\hspace{0.1cm}\ \ 1\\
		\end{vmatrix}\nonumber\\
		&=&  \epsilon_1^2 + \epsilon_2^2 + \epsilon_3^2 + \epsilon_4^2 + \epsilon_1\epsilon_2 + \epsilon_1\epsilon_3 + \epsilon_1\epsilon_4 + \epsilon_2\epsilon_3 + \epsilon_2\epsilon_4 + \epsilon_3\epsilon_4.\nonumber\\
		\therefore\; \chi_{({SU(4)})_{10}}(z_1, z_2, z_3) &=& z_1^2 + \frac{z_2^2}{z_1^2} + \frac{z_3^2}{z_2^2} + \frac{1}{z_3^2} + z_2 + \frac{1}{z_2} + \frac{z_1}{z_3} + \frac{z_3}{z_1} + \frac{z_2}{z_1z_3} + \frac{z_1z_3}{z_2}.
		\end{eqnarray}}
	
	\vspace{-0.9cm}
	\item The Anti-decuplet Representation: $\overline{10} \equiv (0,0,2) \equiv {\scriptsize\yng(2,2,2)}$ 
	
	We get $\vec{\lambda}$ = $(2,2,2,0)$ and $\vec{r} = \vec{\lambda}+\vec{\rho} = (5,4,3,0)$ and the character is obtained as:
	
	\vspace{-0.8cm}
	{\scriptsize\begin{eqnarray}\label{eq:su4-10-bar-char}
		\chi(\epsilon_1,\epsilon_2,\epsilon_3,\epsilon_4) &=& \frac{|\epsilon^5,\hspace{0.1cm}\epsilon^4,\hspace{0.1cm}
		\epsilon^3,\hspace{0.1cm}1|}{|\epsilon^3,\hspace{0.1cm}\epsilon^2,\hspace{0.1cm}\epsilon,\hspace{0.1cm}1|} = \frac{1}{\prod_{1\leq i<j\leq 4}\left(\epsilon_i-\epsilon_j\right)}
		\begin{vmatrix}
		\epsilon_1^{5}\hspace{0.1cm} \ \ \epsilon_1^{4}\hspace{0.1cm} \ \ \epsilon_1^{3}\hspace{0.1cm}\ \ 1\\
		\epsilon_2^{5}\hspace{0.1cm} \ \ \epsilon_2^{4}\hspace{0.1cm} \ \ \epsilon_2^{3}\hspace{0.1cm}\ \ 1\\
		\epsilon_3^{5}\hspace{0.1cm} \ \ \epsilon_3^{4}\hspace{0.1cm} \ \ \epsilon_3^{3}\hspace{0.1cm}\ \ 1\\
		\epsilon_4^{5}\hspace{0.1cm} \ \ \epsilon_4^{4}\hspace{0.1cm} \ \ \epsilon_4^{3}\hspace{0.1cm}\ \ 1\\
		\end{vmatrix}\nonumber\\
		&=&  \epsilon_1^2\epsilon_2^2\epsilon_3^2 + \epsilon_1^2\epsilon_2^2\epsilon_4^2 + \epsilon_1^2\epsilon_3^2\epsilon_4^2 + \epsilon_2^2\epsilon_3^2\epsilon_4^2 + \epsilon_1^2\epsilon_2^2\epsilon_3\epsilon_4 + \epsilon_1^2\epsilon_2\epsilon_3^2\epsilon_4 \nonumber\\
		& & + \epsilon_1^2\epsilon_2\epsilon_3\epsilon_4^2 + \epsilon_1\epsilon_2^2\epsilon_3^2\epsilon_4 + \epsilon_1\epsilon_2^2\epsilon_3\epsilon_4^2 + \epsilon_1\epsilon_2\epsilon_3^2\epsilon_4^2. \nonumber\\
		\therefore\;\chi_{({SU(4)})_{\overline{10}}}(z_1, z_2, z_3) 
		&=& z_3^2 + \frac{z_2^2}{z_3^2} + \frac{z_1^2}{z_2^2} + \frac{1}{z_1^2} + \frac{1}{z_2} + z_2 + \frac{z_3}{z_1} + \frac{z_1}{z_3} + \frac{z_1z_3}{z_2} + \frac{z_2}{z_1z_3}.
		\end{eqnarray}}
	
	\vspace{-0.8cm}
	\item The Adjoint Representation: $15 \equiv (1,0,1) \equiv {\scriptsize\yng(2,1,1)}$ 
	
	We get $\vec{\lambda}$ = $(2,1,1,0)$ and $\vec{r} = \vec{\lambda}+\vec{\rho} = (5,3,2,0)$ and the character is obtained as:
	
	\vspace{-0.8cm}
	{\scriptsize\begin{eqnarray}\label{eq:su4-15-char}
		\chi(\epsilon_1,\epsilon_2,\epsilon_3,\epsilon_4) &=& \frac{|\epsilon^5,\hspace{0.1cm}\epsilon^3,\hspace{0.1cm}
		\epsilon^2,\hspace{0.1cm}1|}{|\epsilon^3,\hspace{0.1cm}\epsilon^2,\hspace{0.1cm}\epsilon,\hspace{0.1cm}1|} = \frac{1}{\prod_{1\leq i<j\leq 4}\left(\epsilon_i-\epsilon_j\right)}
		\begin{vmatrix}
		\epsilon_1^{5}\hspace{0.1cm} \ \ \epsilon_1^{3}\hspace{0.1cm} \ \ \epsilon_1^{2}\hspace{0.1cm}\ \ 1\\
		\epsilon_2^{5}\hspace{0.1cm} \ \ \epsilon_2^{3}\hspace{0.1cm} \ \ \epsilon_2^{2}\hspace{0.1cm}\ \ 1\\
		\epsilon_3^{5}\hspace{0.1cm} \ \ \epsilon_3^{3}\hspace{0.1cm} \ \ \epsilon_3^{2}\hspace{0.1cm}\ \ 1\\
		\epsilon_4^{5}\hspace{0.1cm} \ \ \epsilon_4^{3}\hspace{0.1cm} \ \ \epsilon_4^{2}\hspace{0.1cm}\ \ 1\\
		\end{vmatrix}\nonumber\\
		&=&  \epsilon_1^2\epsilon_2\epsilon_3 + \epsilon_1\epsilon_2^2\epsilon_3 + \epsilon_1\epsilon_2\epsilon_3^2 + \epsilon_1^2\epsilon_2\epsilon_4 + \epsilon_1\epsilon_2^2\epsilon_4 + \epsilon_1\epsilon_2\epsilon_4^2  + \epsilon_1^2\epsilon_3\epsilon_4 \nonumber\\
		& & + \epsilon_1\epsilon_3^2\epsilon_4 + \epsilon_1\epsilon_3\epsilon_4^2 + \epsilon_2^2\epsilon_3\epsilon_4 + \epsilon_2\epsilon_3^2\epsilon_4 + \epsilon_2\epsilon_3\epsilon_4^2 + 3\epsilon_1\epsilon_2\epsilon_3\epsilon_4. \nonumber\\
		\therefore\;\chi_{({SU(4)})_{15}}(z_1, z_2, z_3) 
		&=& \frac{z_1^2}{z_2}+\frac{z_2}{z_1^2}+\frac{z_1 z_3}{z_2^2}+\frac{z_2^2}{z_1 z_3}+\frac{z_1 z_2}{z_3}+\frac{z_1}{z_2 z_3}+\frac{z_3}{z_1 z_2}+\frac{z_2 z_3}{z_1}\nonumber\\ 
		& &+z_1 z_3+\frac{1}{z_1 z_3}+\frac{z_3^2}{z_2}+\frac{z_2}{z_3^2}+3.
		\end{eqnarray}}
\end{itemize}
\noindent \rule{15cm}{1.25pt}

\subsection*{$2.~SO(2N+1)$}\label{subsec:soodd-haar-char}
\textbf{Characters}\\
The Weyl character formula for $SO(2N+1)$ representations  \cite{Balantekin:2001id,koike1987,littlewood1977theory,rossmann2006lie} can be  written as:

\vspace{-0.6cm}
{\scriptsize\begin{eqnarray}\label{eq:soodd-char-formula}
\chi_{r_1,r_2,...,r_{N}}^{(M(\epsilon))} = \frac{|\epsilon^{r_1}-\epsilon^{-r_1},\epsilon^{r_2}-\epsilon^{-r_2},...,\epsilon^{r_N}-\epsilon^{-r_N}|}{|\epsilon^{N-\frac{1}{2}}-\epsilon^{-N+\frac{1}{2}},\epsilon^{N-\frac{3}{2}}-\epsilon^{-N+\frac{3}{2}},....,\epsilon^{\frac{1}{2}}-\epsilon^{-\frac{1}{2}}|},
\end{eqnarray}}
where $M(\epsilon)$ = diag$(\epsilon_1,\epsilon_2,...,\epsilon_N)$ identifies a particular representation of $SO(2N+1)$ and the $r_i$'s are obtained from the Dynkin labels of a particular representation.
The $\epsilon_i$'s can again be obtained by inspecting the matrix form of the maximal torus of $SO(2N+1)$ \cite{Dieck}:

\vspace{-0.6cm}
{\scriptsize\begin{eqnarray}\label{eq:torus-soodd}
\mathbb{T}^N:
\begin{pmatrix}
\cos\,\theta_1 & \text{-}\sin\,\theta_1 &\cdots & 0 & 0 & 0\\
\sin\,\theta_1 & \cos\,\theta_1 &\cdots & 0 & 0 & 0\\
\vdots & \vdots & \ddots & \vdots & \vdots & \vdots \\
0 & 0 & \cdots & \cos\,\theta_N & \text{-}\sin\,\theta_N & 0\\
0 & 0 & \cdots & \sin\,\theta_N & \cos\,\theta_N & 0\\
0 & 0 & \cdots & 0 & 0 & 1\\
\end{pmatrix}_{(2N+1)\times (2N+1)}\hspace{-1.9cm}.
\end{eqnarray}}
Each $2\times 2$ block can be written in diagonal form as: 
{\scriptsize$\begin{pmatrix}
\cos\,\theta_i \ \ \text{-}\sin\,\theta_i\\
\sin\,\theta_i \ \ \cos\,\theta_i
\end{pmatrix} \rightarrow   
\begin{pmatrix}
e^{i\theta_i} \ \ 0\\
0 \ \ e^{\text{-}i\theta_i}
\end{pmatrix}$}. Then, by defining $\epsilon_i = e^{i\theta_i}$, we find $(2N+1)$ parameters $\{\epsilon_i, \epsilon_i{}^{\text{-}1}, 1\}$ where $i=1,2,\cdots N$. The numerator of Eq.~\eqref{eq:soodd-char-formula} can be recast as: 

\vspace{-0.6cm}
{\scriptsize\begin{eqnarray}\label{eq:sooddchar-numerator}
|\epsilon^{r_1}-\epsilon^{-r_1},\epsilon^{r_2}-\epsilon^{-r_2},...,\epsilon^{r_N}-\epsilon^{-r_N}| = 
\begin{vmatrix}
\epsilon_1^{r_1}-\epsilon_1^{-r_1} & \epsilon_1^{r_2}-\epsilon_1^{-r_2} & \cdots &\epsilon_1^{r_N}-\epsilon_1^{-r_N} \\
\epsilon_2^{r_1}-\epsilon_2^{-r_1} & \epsilon_2^{r_2}-\epsilon_2^{-r_2} & \cdots & \epsilon_2^{r_N}-\epsilon_2^{-r_N} \\
\vdots & \vdots &  \ddots &  \vdots \\ 
\epsilon_N^{r_1}-\epsilon_N^{-r_1} & \epsilon_N^{r_2}-\epsilon_N^{-r_2} & \cdots & \epsilon_N^{r_N}-\epsilon_N^{-r_N} \\
\end{vmatrix},
\end{eqnarray}}
\noindent
where the denominator is expressed as: 

\vspace{-0.6cm}
{\scriptsize\begin{eqnarray}\label{eq:sooddvandermonde}
|\epsilon^{N-\frac{1}{2}}-\epsilon^{-N+\frac{1}{2}},...,\epsilon^{\frac{1}{2}}-\epsilon^{-\frac{1}{2}}| = \begin{vmatrix}
\epsilon_1^{N-\frac{1}{2}}-\epsilon_1^{-N+\frac{1}{2}} & \epsilon_1^{N-\frac{3}{2}}-\epsilon_1^{-N+\frac{3}{2}}  & \cdots & \epsilon_1^{\frac{1}{2}}-\epsilon_1^{-\frac{1}{2}} \\
\epsilon_2^{N-\frac{1}{2}}-\epsilon_2^{-N+\frac{1}{2}} & \epsilon_2^{N-\frac{3}{2}}-\epsilon_2^{-N+\frac{3}{2}} & \cdots & \epsilon_2^{\frac{1}{2}}-\epsilon_2^{-\frac{1}{2}} \\
\vdots &  \vdots & \ddots &  \vdots \\ 
\epsilon_N^{N-\frac{1}{2}}-\epsilon_N^{-N+\frac{1}{2}} & \epsilon_N^{N-\frac{3}{2}}-\epsilon_N^{-N+\frac{3}{2}} & \cdots & \epsilon_N^{\frac{1}{2}}-\epsilon_N^{-\frac{1}{2}} \\
\end{vmatrix}.
\end{eqnarray}}	
\underline{\textbf{Calculating the $r_i$'s}}
\\
The Cartan matrix corresponding to $SO(2N+1)$ is an $N\times N$ matrix

\vspace{-0.7cm}
{\scriptsize\begin{eqnarray}\label{eq:sooddcartan matrix}
\mathcal{A}_{SO(2N+1)} = 
\begin{pmatrix}
- -\alpha_1- -\\
- -\alpha_2- -\\
\vline \\
- -\alpha_{N-1}- -\\
- -\alpha_{N}- -
\end{pmatrix} = 
\begin{pmatrix}
2 & \text{-}1&  0& \cdots & 0 & 0 & 0\\
\text{-}1 &  2 & \text{-}1 & \cdots & 0 & 0 & 0\\
\vdots & \vdots & \vdots & \ddots & \vdots & \vdots & \vdots\\
0 &  0 &  0 & \cdots & \text{-}1 & 2 & \text{-}2\\
0 &  0 &  0 & \cdots & 0 & \text{-}1 & 2\\
\end{pmatrix}_{N\times N}\hspace{-0.7cm}.
\end{eqnarray}}
To find the $r_i$'s, the weight tree \cite{Foster:2016lectures} corresponding to the LDF representation of $SO(2N+1)$ needs to be constructed first. This can be done by successively subtracting the rows of the Cartan matrix ($\mathcal{A}_{SO(2N+1)}$), given in Eq.~\eqref{eq:sooddcartan matrix}, from the respective Dynkin label  $\left(1,0,..,0\right)$. The general structure of the weight tree is given as:

\vspace{-0.6cm}
{\scriptsize\begin{eqnarray}
L_1 &=& \underbrace{(1,0,0,...,0,0)}_{(\text{$N$-tuple})}, \nonumber\\
L_2 = L_1 - \alpha_1 &=& (-1,1,0,...,0,0),\nonumber\\
&\vdots&             \nonumber\\
L_k = L_{k-1}-\alpha_{k-1} &=& (0,..,-1,1,.,0),\nonumber\\
&\vdots&             \nonumber\\
L_{N-1} = L_{N-2}-\alpha_{N-2}&=& (0,0,...,-1,1,0),\nonumber\\
L_N = L_{N-1}-\alpha_{N-1}&=& (0,0,...,0,-1,2).
\end{eqnarray}}

\vspace{-0.8cm}
\noindent A particular representation of $SO(2N+1)$ can be uniquely identified by its Dynkin label $(a_1,a_2,.....,a_{N})$. To find the $r_i$'s, first, we  solve the following equation in terms of the Dynkin label and the fundamental weight tree of LDF representation to obtain $\lambda_i$'s:

\vspace{-0.6cm}
{\scriptsize\begin{eqnarray}
(a_1,a_2,....,a_{N}) = \lambda_1 (1,0,...,0,0) + \lambda_2 (-1,1,0,...,0) + \cdots +\lambda_{N-1} (0,...,-1,1,0) + \lambda_{N}(0,0,...,-1,2).\nonumber
\end{eqnarray}}

\vspace{-0.8cm}
\noindent
Here, we have $N$-unknowns in $\{ \lambda_1,\lambda_2,...,\lambda_N \}$ and there are N-equations in $\{ a_1,a_2,....,a_{N} \}$ as:

\vspace{-0.6cm}
{\scriptsize\begin{eqnarray}
a_N = 2\lambda_{N} \hspace{0.4cm} \text{and} \hspace{0.4cm} a_k &=& \lambda_k-\lambda_{k+1}, \hspace{0.7cm} k = 1,\cdots,N-1 \hspace{0.1cm}.  
\end{eqnarray}}

\vspace{-0.9cm}
\noindent
Thus, we find unique solutions of $\lambda_i$'s as:

\vspace{-0.6cm}
{\scriptsize\begin{eqnarray}\label{eq:sooddlambda-recursion}
\lambda_{N} = \frac{a_N}{2}, \hspace{0.4cm}
\lambda_{N-1} = a_{N-1}+\frac{a_N}{2}, \hspace{0.6cm}
\lambda_k = a_k+a_{k+1}+\cdots+a_{N-1}+\frac{a_N}{2}.
\end{eqnarray}}
The $r_i$'s are related to the $\lambda_i$'s through the following equation:

\vspace{-0.6cm}
{\scriptsize\begin{eqnarray}
\vec{r} = \vec{\lambda}+\vec{\rho} \hspace{0.5cm}\text{where}\hspace{0.5cm}\rho_i = N-i+\frac{1}{2}, \hspace{0.5cm} i=1,2,..,N.
\end{eqnarray}}

\vspace{-0.8cm}
\noindent
Having obtained $r_i$'s, we can compute the numerator given in  Eq.~\eqref{eq:sooddchar-numerator} and subsequently the full character.
\\
\\
\textbf{Haar Measure}
\\
The general form of the Haar measure can be written as \cite{Gray:2008yu,Hanany:2014dia,Henning:2017fpj}:

\vspace{-0.5cm}
{\scriptsize\begin{eqnarray}\label{eq:sooddhaar-measure}
\int_{SO(2N+1)} d\mu_{SO(2N+1)} = \frac{1}{\left(2\pi i\right)^{N} 2^{N} N!}\oint_{|\epsilon_l|=1} \prod_{l=1}^{N}\frac{d\epsilon_l}{\epsilon_l}\Delta\left(\epsilon\right)\Delta\left(\epsilon^{-1}\right).
\end{eqnarray}}
$\Delta\left(\epsilon\right)$ is given in Eq.~\eqref{eq:sooddvandermonde}, $\Delta(\epsilon^{-1})$ can similarly be computed by substituting $\epsilon^{-1}_i$ in the place of $\epsilon_i$  in the expression of $\Delta\left(\epsilon\right)$. Finally, substituting for known quantities in Eq.~\eqref{eq:sooddhaar-measure} we can obtain the Haar measure for given group.
	
Next, the characters for certain representations \cite{Slansky:1981yr,Yamatsu:2015npn} of $SO(7)$ and $SO(9)$ have been computed and the explicit computation of the respective Haar measures of these groups have also been shown.	

\begin{center}
	\rule{15cm}{1.25pt}	
\end{center}

\noindent
\fbox{$SO(7)$}\label{examples:so7-char-haar}\\
\\
\textbf{Haar Measure}
\\
The denominator for the character formula in this case is: 

\vspace{-0.5cm}
{\scriptsize\begin{eqnarray}\label{eq:vandermonde-so7}
	\Delta\left(\epsilon\right) = \begin{vmatrix}
	\epsilon_1^{\frac{5}{2}}-\epsilon_1^{-\frac{5}{2}}\hspace{0.1cm} \ \ \epsilon_1^{\frac{3}{2}}-\epsilon_1^{-\frac{3}{2}} \ \ \epsilon_1^{\frac{1}{2}}-\epsilon_1^{-\frac{1}{2}}\\
	\epsilon_2^{\frac{5}{2}}-\epsilon_2^{-\frac{5}{2}}\hspace{0.1cm} \ \ \epsilon_2^{\frac{3}{2}}-\epsilon_2^{-\frac{3}{2}} \ \ \epsilon_2^{\frac{1}{2}}-\epsilon_2^{-\frac{1}{2}}\\
	\epsilon_3^{\frac{5}{2}}-\epsilon_3^{-\frac{5}{2}}\hspace{0.1cm} \ \ \epsilon_3^{\frac{3}{2}}-\epsilon_3^{-\frac{3}{2}} \ \ \epsilon_3^{\frac{1}{2}}-\epsilon_3^{-\frac{1}{2}}\\
	\end{vmatrix}.
	\end{eqnarray}}

\vspace{-0.5cm}
\noindent $\Delta(\epsilon^{-1})$ is obtained by replacing $\epsilon_i$ by $\epsilon_i^{-1}$ in the above expression. Then using Eq.~\eqref{eq:sooddhaar-measure},

\vspace{-0.6cm}
{\scriptsize\begin{eqnarray}\label{eq:so7-haar-measure}
	d\mu_{SO(7)} &=& \frac{1}{3! 2^{3}\left(2\pi i\right)^{3}} \frac{d\epsilon_1}{\epsilon_1}\frac{d\epsilon_2}{\epsilon_2}\frac{d\epsilon_3}{\epsilon_3}\Delta\left(\epsilon\right)\Delta\left(\epsilon^{-1}\right)\nonumber\\
	&=& -\frac{1}{48\left(2\pi i\right)^{3}} \frac{d\epsilon_1 \hspace{1mm} d\epsilon_2 \hspace{1mm} d\epsilon_3}{\epsilon_1^{6} \epsilon_2^{6} \epsilon_3^{6}} \left(-1+\epsilon_{1}\right)^{2}\left(\epsilon_{1}-\epsilon_{2}\right)^{2}\left(-1+\epsilon_{2}\right)^{2}\left(-1+\epsilon_{1} \epsilon_{2}\right)^{2}\left(\epsilon_{1}-\epsilon_{3}\right)^{2}\left(\epsilon_{2}-\epsilon_{3}\right)^{2} \nonumber \\
	& & \left(-1+\epsilon_{3}\right)^{2} \left(-1+\epsilon_{1} \epsilon_{3}\right)^{2}\left(-1+\epsilon_{2} \epsilon_{3}\right)^{2}.
	\end{eqnarray}}

\vspace{-0.9cm}
\noindent
\textbf{Characters}

\vspace{-0.3cm}
\begin{itemize}
	\item The Fundamental Representation: $7 \equiv (1,0,0)$ 
	
	Using Eq.~\eqref{eq:sooddlambda-recursion}, we get $\vec{\lambda}$ = $(1,0,0)$. Now, since $\vec{\rho}$ = $(\frac{5}{2},\frac{3}{2},\frac{1}{2})$, therefore,
	$\vec{r} = \vec{\lambda}+\vec{\rho} = (\frac{7}{2},\frac{3}{2},\frac{1}{2})$ and the character is obtained as:
	
	\vspace{-1.1cm}
	{\scriptsize\begin{eqnarray}\label{eq:so7-7-char}
		\chi_{({SO(7)})_{7}}(\epsilon_1,\epsilon_2,\epsilon_3) &=& \frac{|\epsilon^{\frac{7}{2}}-\epsilon^{-\frac{7}{2}},\hspace{0.1cm}\epsilon^{\frac{3}{2}}-\epsilon^{-\frac{3}{2}},\hspace{0.1cm}\epsilon^{\frac{1}{2}}-\epsilon^{-\frac{1}{2}}|} {|\epsilon^{\frac{5}{2}}-\epsilon^{-\frac{5}{2}},\hspace{0.1cm}\epsilon^{\frac{3}{2}}-\epsilon^{-\frac{3}{2}},\hspace{0.1cm}\epsilon^{\frac{1}{2}}-\epsilon^{-\frac{1}{2}}|} = \frac{
			\begin{vmatrix}
			\epsilon_1^{\frac{7}{2}}-\epsilon_1^{-\frac{7}{2}}\hspace{0.1cm} \ \ \epsilon_1^{\frac{3}{2}}-\epsilon_1^{-\frac{3}{2}}\hspace{0.1cm} \ \ \epsilon_1^{\frac{1}{2}}-\epsilon_1^{-\frac{1}{2}}\\
			\epsilon_2^{\frac{7}{2}}-\epsilon_2^{-\frac{7}{2}}\hspace{0.1cm} \ \ \epsilon_2^{\frac{3}{2}}-\epsilon_2^{-\frac{3}{2}}\hspace{0.1cm} \ \ \epsilon_2^{\frac{1}{2}}-\epsilon_2^{-\frac{1}{2}}\\
			\epsilon_3^{\frac{7}{2}}-\epsilon_3^{-\frac{7}{2}}\hspace{0.1cm} \ \ \epsilon_3^{\frac{3}{2}}-\epsilon_3^{-\frac{3}{2}}\hspace{0.1cm} \ \ \epsilon_3^{\frac{1}{2}}-\epsilon_3^{-\frac{1}{2}}\\
			\end{vmatrix}}{
			\begin{vmatrix}
			\epsilon_1^{\frac{5}{2}}-\epsilon_1^{-\frac{5}{2}}\hspace{0.1cm} \ \ \epsilon_1^{\frac{3}{2}}-\epsilon_1^{-\frac{3}{2}}\hspace{0.1cm} \ \ \epsilon_1^{\frac{1}{2}}-\epsilon_1^{-\frac{1}{2}}\\
			\epsilon_2^{\frac{5}{2}}-\epsilon_2^{-\frac{5}{2}}\hspace{0.1cm} \ \ \epsilon_2^{\frac{3}{2}}-\epsilon_2^{-\frac{3}{2}}\hspace{0.1cm} \ \ \epsilon_2^{\frac{1}{2}}-\epsilon_2^{-\frac{1}{2}}\\
			\epsilon_3^{\frac{5}{2}}-\epsilon_3^{-\frac{5}{2}}\hspace{0.1cm} \ \ \epsilon_3^{\frac{3}{2}}-\epsilon_3^{-\frac{3}{2}}\hspace{0.1cm} \ \ \epsilon_3^{\frac{1}{2}}-\epsilon_3^{-\frac{1}{2}}\\
			\end{vmatrix}}\nonumber\\
		&=& 1+\epsilon_{1}+\frac{1}{\epsilon_{1}}+\epsilon_{2}+\frac{1}{\epsilon_{2}}+\epsilon_{3}+\frac{1}{\epsilon_{3}} .
		\end{eqnarray}} 	
	
	\vspace{-0.9cm}
	\item The Spinor Representation: $8 \equiv (0,0,1)$ 
	
	We obtain $\vec{\lambda}$ = $(\frac{1}{2},\frac{1}{2},\frac{1}{2})$ and $\vec{r} = \vec{\lambda}+\vec{\rho} = (3,2,1)$	and the character is obtained as:
	
	\vspace{-0.7cm}
	{\scriptsize\begin{eqnarray}\label{eq:so7-8-char}
		\hspace{-0.8cm}\chi_{({SO(7)})_{8}}(\epsilon_1,\epsilon_2,\epsilon_3) &=& \frac{|\epsilon^{3}-\epsilon^{-3},\hspace{0.1cm}\epsilon^{2}-\epsilon^{-2},\hspace{0.1cm}\epsilon^{1}-\epsilon^{-1}|} {|\epsilon^{\frac{5}{2}}-\epsilon^{-\frac{5}{2}},\hspace{0.1cm}\epsilon^{\frac{3}{2}}-\epsilon^{-\frac{3}{2}},\hspace{0.1cm}\epsilon^{\frac{1}{2}}-\epsilon^{-\frac{1}{2}}|} = \frac{
			\begin{vmatrix}
			\epsilon_1^{3}-\epsilon_1^{-3}\hspace{0.1cm} \ \ \epsilon_1^{2}-\epsilon_1^{-2}\hspace{0.1cm} \ \ \epsilon_1^{1}-\epsilon_1^{-1}\\
			\epsilon_2^{3}-\epsilon_2^{-3}\hspace{0.1cm} \ \ \epsilon_2^{2}-\epsilon_2^{-2}\hspace{0.1cm} \ \ \epsilon_2^{1}-\epsilon_2^{-1}\\
			\epsilon_3^{3}-\epsilon_3^{-3}\hspace{0.1cm} \ \ \epsilon_3^{2}-\epsilon_3^{-2}\hspace{0.1cm} \ \ \epsilon_3^{1}-\epsilon_3^{-1}\\
			\end{vmatrix}}{
			\begin{vmatrix}
			\epsilon_1^{\frac{5}{2}}-\epsilon_1^{-\frac{5}{2}}\hspace{0.1cm} \ \ \epsilon_1^{\frac{3}{2}}-\epsilon_1^{-\frac{3}{2}}\hspace{0.1cm} \ \ \epsilon_1^{\frac{1}{2}}-\epsilon_1^{-\frac{1}{2}}\\
			\epsilon_2^{\frac{5}{2}}-\epsilon_2^{-\frac{5}{2}}\hspace{0.1cm} \ \ \epsilon_2^{\frac{3}{2}}-\epsilon_2^{-\frac{3}{2}}\hspace{0.1cm} \ \ \epsilon_2^{\frac{1}{2}}-\epsilon_2^{-\frac{1}{2}}\\
			\epsilon_3^{\frac{5}{2}}-\epsilon_3^{-\frac{5}{2}}\hspace{0.1cm} \ \ \epsilon_3^{\frac{3}{2}}-\epsilon_3^{-\frac{3}{2}}\hspace{0.1cm} \ \ \epsilon_3^{\frac{1}{2}}-\epsilon_3^{-\frac{1}{2}}\\
			\end{vmatrix}}\\
		&=& \sqrt{\epsilon_{1} \epsilon_{2} \epsilon_{3}}+\sqrt{\frac{\epsilon_{1} \epsilon_{2}}{\epsilon_{3}}}+\sqrt{\frac{\epsilon_{1} \epsilon_{3}}{\epsilon_{2}}}+\sqrt{\frac{\epsilon_{2} \epsilon_{3}}{\epsilon_{1}}}+\frac{1}{\sqrt{\epsilon_{1} \epsilon_{2} \epsilon_{3}}}+\sqrt{\frac{\epsilon_{3}}{\epsilon_{1} \epsilon_{2}}}+\sqrt{\frac{\epsilon_{2}}{\epsilon_{3} \epsilon_{1}}}+ \sqrt{\frac{\epsilon_{1}}{\epsilon_{2} \epsilon_{3}}}. \nonumber
		\end{eqnarray}} 
	
	\vspace{-1.0cm}		
	\item The Adjoint Representation: $21 \equiv (0,1,0)$ 
	
	We obtain $\vec{\lambda}$ = $(1,1,0)$ and $\vec{r} = \vec{\lambda}+\vec{\rho} = (\frac{7}{2},\frac{5}{2},\frac{1}{2})$ and the character is obtained as:
	
	\vspace{-0.7cm}
	{\scriptsize\begin{eqnarray}\label{eq:so7-21-char}
		\chi_{({SO(7)})_{21}}(\epsilon_1,\epsilon_2,\epsilon_3) &=& \frac{|\epsilon^{\frac{7}{2}}-\epsilon^{-\frac{7}{2}},\hspace{0.1cm}\epsilon^{\frac{5}{2}}-\epsilon^{-\frac{5}{2}},\hspace{0.1cm}\epsilon^{\frac{1}{2}}-\epsilon^{-\frac{1}{2}}|} {|\epsilon^{\frac{5}{2}}-\epsilon^{-\frac{5}{2}},\hspace{0.1cm}\epsilon^{\frac{3}{2}}-\epsilon^{-\frac{3}{2}},\hspace{0.1cm}\epsilon^{\frac{1}{2}}-\epsilon^{-\frac{1}{2}}|} = \frac{
			\begin{vmatrix}
			\epsilon_1^{\frac{7}{2}}-\epsilon_1^{-\frac{7}{2}}\hspace{0.1cm} \ \ \epsilon_1^{\frac{5}{2}}-\epsilon_1^{-\frac{5}{2}}\hspace{0.1cm} \ \ \epsilon_1^{\frac{1}{2}}-\epsilon_1^{-\frac{1}{2}}\\
			\epsilon_2^{\frac{7}{2}}-\epsilon_2^{-\frac{7}{2}}\hspace{0.1cm} \ \ \epsilon_2^{\frac{5}{2}}-\epsilon_2^{-\frac{5}{2}}\hspace{0.1cm} \ \ \epsilon_2^{\frac{1}{2}}-\epsilon_2^{-\frac{1}{2}}\\
			\epsilon_3^{\frac{7}{2}}-\epsilon_3^{-\frac{7}{2}}\hspace{0.1cm} \ \ \epsilon_3^{\frac{5}{2}}-\epsilon_3^{-\frac{5}{2}}\hspace{0.1cm} \ \ \epsilon_3^{\frac{1}{2}}-\epsilon_3^{-\frac{1}{2}}\\
			\end{vmatrix}}{
			\begin{vmatrix}
			\epsilon_1^{\frac{5}{2}}-\epsilon_1^{-\frac{5}{2}}\hspace{0.1cm} \ \ \epsilon_1^{\frac{3}{2}}-\epsilon_1^{-\frac{3}{2}}\hspace{0.1cm} \ \ \epsilon_1^{\frac{1}{2}}-\epsilon_1^{-\frac{1}{2}}\\
			\epsilon_2^{\frac{5}{2}}-\epsilon_2^{-\frac{5}{2}}\hspace{0.1cm} \ \ \epsilon_2^{\frac{3}{2}}-\epsilon_2^{-\frac{3}{2}}\hspace{0.1cm} \ \ \epsilon_2^{\frac{1}{2}}-\epsilon_2^{-\frac{1}{2}}\\
			\epsilon_3^{\frac{5}{2}}-\epsilon_3^{-\frac{5}{2}}\hspace{0.1cm} \ \ \epsilon_3^{\frac{3}{2}}-\epsilon_3^{-\frac{3}{2}}\hspace{0.1cm} \ \ \epsilon_3^{\frac{1}{2}}-\epsilon_3^{-\frac{1}{2}}\\
			\end{vmatrix}}\nonumber\\
		&=& 3+\epsilon_{1}+\epsilon_{2}+\epsilon_{3}+\frac{1}{\epsilon_{1}}+\frac{1}{\epsilon_{2}}+\frac{1}{\epsilon_{3}}+\epsilon_{1}\epsilon_{2}+\epsilon_{2}\epsilon_{3}+\epsilon_{1}\epsilon_{3}+\frac{1}{\epsilon_1 \epsilon_2}+\frac{1}{\epsilon_2 \epsilon_3}+\frac{1}{\epsilon_1 \epsilon_3} \nonumber \\
		& & +\frac{\epsilon_1}{\epsilon_2}+\frac{\epsilon_2}{\epsilon_1}+ \frac{\epsilon_1}{\epsilon_3}+\frac{\epsilon_3}{\epsilon_1}+\frac{\epsilon_2}{\epsilon_3}+\frac{\epsilon_3}{\epsilon_2}.
		\end{eqnarray}} 	
\end{itemize}
\noindent \rule{15.5cm}{1.25pt}
\\
\\
\noindent
\fbox{$SO(9)$}\label{examples:so9-char-haar}\\
\\
\textbf{Haar Measure}
\\
The denominator for the character formula in this case is: 

\vspace{-0.5cm}
{\scriptsize\begin{eqnarray}\label{eq:vandermonde-so9}
	\Delta\left(\epsilon\right) = \begin{vmatrix}
	\epsilon_1^{\frac{7}{2}}-\epsilon_1^{-\frac{7}{2}}\hspace{0.1cm} \ \		\epsilon_1^{\frac{5}{2}}-\epsilon_1^{-\frac{5}{2}}\hspace{0.1cm} \ \ \epsilon_1^{\frac{3}{2}}-\epsilon_1^{-\frac{3}{2}}\hspace{0.1cm} \ \ \epsilon_1^{\frac{1}{2}}-\epsilon_1^{-\frac{1}{2}}\\
	\epsilon_2^{\frac{7}{2}}-\epsilon_2^{-\frac{7}{2}}\hspace{0.1cm} \ \		\epsilon_2^{\frac{5}{2}}-\epsilon_2^{-\frac{5}{2}}\hspace{0.1cm} \ \ \epsilon_2^{\frac{3}{2}}-\epsilon_2^{-\frac{3}{2}}\hspace{0.1cm} \ \ \epsilon_2^{\frac{1}{2}}-\epsilon_2^{-\frac{1}{2}}\\
	\epsilon_3^{\frac{7}{2}}-\epsilon_3^{-\frac{7}{2}}\hspace{0.1cm} \ \		\epsilon_3^{\frac{5}{2}}-\epsilon_3^{-\frac{5}{2}}\hspace{0.1cm} \ \ \epsilon_3^{\frac{3}{2}}-\epsilon_3^{-\frac{3}{2}}\hspace{0.1cm} \ \ \epsilon_3^{\frac{1}{2}}-\epsilon_3^{-\frac{1}{2}}\\
	\epsilon_4^{\frac{7}{2}}-\epsilon_4^{-\frac{7}{2}}\hspace{0.1cm} \ \		\epsilon_4^{\frac{5}{2}}-\epsilon_4^{-\frac{5}{2}}\hspace{0.1cm} \ \ \epsilon_4^{\frac{3}{2}}-\epsilon_4^{-\frac{3}{2}}\hspace{0.1cm} \ \ \epsilon_4^{\frac{1}{2}}-\epsilon_4^{-\frac{1}{2}}\\
	\end{vmatrix}.
	\end{eqnarray}}
$\Delta(\epsilon^{-1})$ is obtained by replacing $\epsilon_i$ by $\epsilon_i^{-1}$ in the above expression. Then using Eq.~\eqref{eq:sooddhaar-measure},
{\scriptsize\begin{eqnarray}\label{eq:so9-haar-measure}
	d\mu_{SO(9)} &=& \frac{1}{4! 2^{4}\left(2\pi i\right)^{4}} \frac{d\epsilon_1}{\epsilon_1}\frac{d\epsilon_2}{\epsilon_2}\frac{d\epsilon_3}{\epsilon_3}\frac{d\epsilon_4}{\epsilon_4}\Delta\left(\epsilon\right)\Delta\left(\epsilon^{-1}\right).
	\end{eqnarray}}
\textbf{Characters}
\begin{itemize}
	\item The Fundamental Representation: $9 \equiv (1,0,0,0)$ 
	
	Using Eq.~\eqref{eq:sooddlambda-recursion}, we get $\vec{\lambda}$ = $(1,0,0,0)$. Now, since $\vec{\rho}$ = $(\frac{7}{2},\frac{5}{2},\frac{3}{2},\frac{1}{2})$, therefore,\\
	$\vec{r} = \vec{\lambda}+\vec{\rho} = (\frac{9}{2},\frac{5}{2},\frac{3}{2},\frac{1}{2})$ and the character is obtained as:
	
	\vspace{-0.7cm}
	{\scriptsize\begin{eqnarray}\label{eq:so9-9-char}
		\chi_{({SO(9)})_{9}}(\epsilon_1,\epsilon_2,\epsilon_3,\epsilon_4) &=& \frac{|\epsilon^{\frac{9}{2}}-\epsilon^{-\frac{9}{2}},\hspace{0.1cm}\epsilon^{\frac{5}{2}}-\epsilon^{-\frac{5}{2}},\hspace{0.1cm}\epsilon^{\frac{3}{2}}-\epsilon^{-\frac{3}{2}},\hspace{0.1cm}\epsilon^{\frac{1}{2}}-\epsilon^{-\frac{1}{2}}|} {|\epsilon^{\frac{7}{2}}-\epsilon^{-\frac{7}{2}},\hspace{0.1cm}\epsilon^{\frac{5}{2}}-\epsilon^{-\frac{5}{2}},\hspace{0.1cm}\epsilon^{\frac{3}{2}}-\epsilon^{-\frac{3}{2}},\hspace{0.1cm}\epsilon^{\frac{1}{2}}-\epsilon^{-\frac{1}{2}}|} 
		= \frac{
			\begin{vmatrix}
			\epsilon_1^{\frac{9}{2}}-\epsilon_1^{-\frac{9}{2}}\hspace{0.1cm} \ \		\epsilon_1^{\frac{5}{2}}-\epsilon_1^{-\frac{5}{2}}\hspace{0.1cm} \ \ \epsilon_1^{\frac{3}{2}}-\epsilon_1^{-\frac{3}{2}}\hspace{0.1cm} \ \ \epsilon_1^{\frac{1}{2}}-\epsilon_1^{-\frac{1}{2}}\\
			\epsilon_2^{\frac{9}{2}}-\epsilon_2^{-\frac{9}{2}}\hspace{0.1cm} \ \		\epsilon_2^{\frac{5}{2}}-\epsilon_2^{-\frac{5}{2}}\hspace{0.1cm} \ \ \epsilon_2^{\frac{3}{2}}-\epsilon_2^{-\frac{3}{2}}\hspace{0.1cm} \ \ \epsilon_2^{\frac{1}{2}}-\epsilon_2^{-\frac{1}{2}}\\
			\epsilon_3^{\frac{9}{2}}-\epsilon_3^{-\frac{9}{2}}\hspace{0.1cm} \ \		\epsilon_3^{\frac{5}{2}}-\epsilon_3^{-\frac{5}{2}}\hspace{0.1cm} \ \ \epsilon_3^{\frac{3}{2}}-\epsilon_3^{-\frac{3}{2}}\hspace{0.1cm} \ \ \epsilon_3^{\frac{1}{2}}-\epsilon_3^{-\frac{1}{2}}\\
			\epsilon_4^{\frac{9}{2}}-\epsilon_4^{-\frac{9}{2}}\hspace{0.1cm} \ \		\epsilon_4^{\frac{5}{2}}-\epsilon_4^{-\frac{5}{2}}\hspace{0.1cm} \ \ \epsilon_4^{\frac{3}{2}}-\epsilon_4^{-\frac{3}{2}}\hspace{0.1cm} \ \ \epsilon_4^{\frac{1}{2}}-\epsilon_4^{-\frac{1}{2}}\\
			\end{vmatrix}}{
			\begin{vmatrix}
			\epsilon_1^{\frac{7}{2}}-\epsilon_1^{-\frac{7}{2}}\hspace{0.1cm} \ \		\epsilon_1^{\frac{5}{2}}-\epsilon_1^{-\frac{5}{2}}\hspace{0.1cm} \ \ \epsilon_1^{\frac{3}{2}}-\epsilon_1^{-\frac{3}{2}}\hspace{0.1cm} \ \ \epsilon_1^{\frac{1}{2}}-\epsilon_1^{-\frac{1}{2}}\\
			\epsilon_2^{\frac{7}{2}}-\epsilon_2^{-\frac{7}{2}}\hspace{0.1cm} \ \		\epsilon_2^{\frac{5}{2}}-\epsilon_2^{-\frac{5}{2}}\hspace{0.1cm} \ \ \epsilon_2^{\frac{3}{2}}-\epsilon_2^{-\frac{3}{2}}\hspace{0.1cm} \ \ \epsilon_2^{\frac{1}{2}}-\epsilon_2^{-\frac{1}{2}}\\
			\epsilon_3^{\frac{7}{2}}-\epsilon_3^{-\frac{7}{2}}\hspace{0.1cm} \ \		\epsilon_3^{\frac{5}{2}}-\epsilon_3^{-\frac{5}{2}}\hspace{0.1cm} \ \ \epsilon_3^{\frac{3}{2}}-\epsilon_3^{-\frac{3}{2}}\hspace{0.1cm} \ \ \epsilon_3^{\frac{1}{2}}-\epsilon_3^{-\frac{1}{2}}\\
			\epsilon_4^{\frac{7}{2}}-\epsilon_4^{-\frac{7}{2}}\hspace{0.1cm} \ \		\epsilon_4^{\frac{5}{2}}-\epsilon_4^{-\frac{5}{2}}\hspace{0.1cm} \ \ \epsilon_4^{\frac{3}{2}}-\epsilon_4^{-\frac{3}{2}}\hspace{0.1cm} \ \ \epsilon_4^{\frac{1}{2}}-\epsilon_4^{-\frac{1}{2}}\\
			\end{vmatrix}}\nonumber\\
		&=& 1+\epsilon_{1}+\frac{1}{\epsilon_{1}}+\epsilon_{2}+\frac{1}{\epsilon_{2}}+\epsilon_{3}+\frac{1}{\epsilon_{3}}+\epsilon_{4}+\frac{1}{\epsilon_{4}} .
		\end{eqnarray}} 
	\item The Spinor Representation: $16 \equiv (0,0,0,1)$ 
	
	With $\vec{\lambda}$ = $(\frac{1}{2},\frac{1}{2},\frac{1}{2},\frac{1}{2})$ and $\vec{r} = \vec{\lambda}+\vec{\rho}$ = $(4,3,2,1)$, the character is obtained as:
	
	\vspace{-0.5cm}
	{\scriptsize\begin{eqnarray}\label{eq:so7-16-char}
		\chi_{({SO(9)})_{16}}(\epsilon_1,\epsilon_2,\epsilon_3,\epsilon_4) &=& \frac{|\epsilon^{4}-\epsilon^{-4},\hspace{0.1cm}\epsilon^{3}-\epsilon^{-3},\hspace{0.1cm}\epsilon^{2}-\epsilon^{-2},\hspace{0.1cm}\epsilon^{1}-\epsilon^{-1}|} {|\epsilon^{\frac{7}{2}}-\epsilon^{-\frac{7}{2}},\hspace{0.1cm}\epsilon^{\frac{5}{2}}-\epsilon^{-\frac{5}{2}},\hspace{0.1cm}\epsilon^{\frac{3}{2}}-\epsilon^{-\frac{3}{2}},\hspace{0.1cm}\epsilon^{\frac{1}{2}}-\epsilon^{-\frac{1}{2}}|} 
		=\frac{
			\begin{vmatrix}
			\epsilon_1^{4}-\epsilon_1^{-4}\hspace{0.1cm} \ \ \epsilon_1^{3}-\epsilon_1^{-3}\hspace{0.1cm} \ \ \epsilon_1^{2}-\epsilon_1^{-2}\hspace{0.1cm} \ \ \epsilon_1^{1}-\epsilon_1^{-1}\\
			\epsilon_2^{4}-\epsilon_2^{-4}\hspace{0.1cm} \ \ \epsilon_2^{3}-\epsilon_2^{-3}\hspace{0.1cm} \ \ \epsilon_2^{2}-\epsilon_2^{-2}\hspace{0.1cm} \ \ \epsilon_2^{1}-\epsilon_2^{-1}\\
			\epsilon_3^{4}-\epsilon_3^{-4}\hspace{0.1cm} \ \ \epsilon_3^{3}-\epsilon_3^{-3}\hspace{0.1cm} \ \ \epsilon_3^{2}-\epsilon_3^{-2}\hspace{0.1cm} \ \ \epsilon_3^{1}-\epsilon_3^{-1}\\
			\epsilon_4^{4}-\epsilon_4^{-4}\hspace{0.1cm} \ \ \epsilon_4^{3}-\epsilon_4^{-3}\hspace{0.1cm} \ \ \epsilon_4^{2}-\epsilon_4^{-2}\hspace{0.1cm} \ \ \epsilon_4^{1}-\epsilon_4^{-1}\\
			\end{vmatrix}}{
			\begin{vmatrix}
			\epsilon_1^{\frac{7}{2}}-\epsilon_1^{-\frac{7}{2}}\hspace{0.1cm} \ \		\epsilon_1^{\frac{5}{2}}-\epsilon_1^{-\frac{5}{2}}\hspace{0.1cm} \ \ \epsilon_1^{\frac{3}{2}}-\epsilon_1^{-\frac{3}{2}}\hspace{0.1cm} \ \ \epsilon_1^{\frac{1}{2}}-\epsilon_1^{-\frac{1}{2}}\\
			\epsilon_2^{\frac{7}{2}}-\epsilon_2^{-\frac{7}{2}}\hspace{0.1cm} \ \		\epsilon_2^{\frac{5}{2}}-\epsilon_2^{-\frac{5}{2}}\hspace{0.1cm} \ \ \epsilon_2^{\frac{3}{2}}-\epsilon_2^{-\frac{3}{2}}\hspace{0.1cm} \ \ \epsilon_2^{\frac{1}{2}}-\epsilon_2^{-\frac{1}{2}}\\
			\epsilon_3^{\frac{7}{2}}-\epsilon_3^{-\frac{7}{2}}\hspace{0.1cm} \ \		\epsilon_3^{\frac{5}{2}}-\epsilon_3^{-\frac{5}{2}}\hspace{0.1cm} \ \ \epsilon_3^{\frac{3}{2}}-\epsilon_3^{-\frac{3}{2}}\hspace{0.1cm} \ \ \epsilon_3^{\frac{1}{2}}-\epsilon_3^{-\frac{1}{2}}\\
			\epsilon_4^{\frac{7}{2}}-\epsilon_4^{-\frac{7}{2}}\hspace{0.1cm} \ \		\epsilon_4^{\frac{5}{2}}-\epsilon_4^{-\frac{5}{2}}\hspace{0.1cm} \ \ \epsilon_4^{\frac{3}{2}}-\epsilon_4^{-\frac{3}{2}}\hspace{0.1cm} \ \ \epsilon_4^{\frac{1}{2}}-\epsilon_4^{-\frac{1}{2}}\\
			\end{vmatrix}}\nonumber\\
		&=& \sqrt{\epsilon_{1} \epsilon_{2} \epsilon_{3}\epsilon_{4}}+\sqrt{\frac{\epsilon_{1} \epsilon_{2}\epsilon_{3}}{\epsilon_{4}}}+\sqrt{\frac{\epsilon_{1} \epsilon_{3}\epsilon_{4}}{\epsilon_{2}}}+\sqrt{\frac{\epsilon_{1} \epsilon_{2}\epsilon_{4}}{\epsilon_{3}}}+\sqrt{\frac{\epsilon_{2} \epsilon_{3}\epsilon_{4}}{\epsilon_{1}}}+\sqrt{\frac{\epsilon_{1}\epsilon_{2}}{\epsilon_{3} \epsilon_{4}}} +\sqrt{\frac{\epsilon_{1}\epsilon_{4}}{\epsilon_{3} \epsilon_{2}}}+\sqrt{\frac{\epsilon_{1}\epsilon_{3}}{\epsilon_{2} \epsilon_{4}}}+ \nonumber \\
		& & \sqrt{\frac{\epsilon_{3}\epsilon_{4}}{\epsilon_{1} \epsilon_{2}}} +\sqrt{\frac{\epsilon_{2}\epsilon_{3}}{\epsilon_{1} \epsilon_{4}}}+\sqrt{\frac{\epsilon_{2}\epsilon_{4}}{\epsilon_{1} \epsilon_{3}}}  +\sqrt{\frac{\epsilon_{4}}{\epsilon_{3} \epsilon_{2} \epsilon_{1}}}+\sqrt{\frac{\epsilon_{3}}{\epsilon_{2} \epsilon_{1} \epsilon_{4}}}+\sqrt{\frac{\epsilon_{2}}{\epsilon_{1} \epsilon_{3} \epsilon_{4}}}+\sqrt{\frac{\epsilon_{1}}{\epsilon_{2} \epsilon_{3} \epsilon_{4}}}  +\frac{1}{\sqrt{\epsilon_{1} \epsilon_{2} \epsilon_{3} \epsilon_{4}}}.
		\end{eqnarray}} 		
	\item The Adjoint Representation: $36 \equiv (0,1,0,0)$ 
	
	With $\vec{\lambda}$ = $(1,1,0,0)$ and $\vec{r} = \vec{\lambda}+\vec{\rho}$ = $(\frac{9}{2},\frac{7}{2},\frac{3}{2},\frac{1}{2})$, the character is obtained as:
	
	\vspace{-0.5cm}
	{\scriptsize\begin{eqnarray}\label{eq:so9-36-char}
		\chi_{({SO(9)})_{36}}(\epsilon_1,\epsilon_2,\epsilon_3,\epsilon_4) &=& \frac{|\epsilon^{\frac{9}{2}}-\epsilon^{-\frac{9}{2}},\hspace{0.1cm}\epsilon^{\frac{7}{2}}-\epsilon^{-\frac{7}{2}},\hspace{0.1cm}\epsilon^{\frac{3}{2}}-\epsilon^{-\frac{3}{2}},\hspace{0.1cm}\epsilon^{\frac{1}{2}}-\epsilon^{-\frac{1}{2}}|} {|\epsilon^{\frac{7}{2}}-\epsilon^{-\frac{7}{2}},\hspace{0.1cm}\epsilon^{\frac{5}{2}}-\epsilon^{-\frac{5}{2}},\hspace{0.1cm}\epsilon^{\frac{3}{2}}-\epsilon^{-\frac{3}{2}},\hspace{0.1cm}\epsilon^{\frac{1}{2}}-\epsilon^{-\frac{1}{2}}|} 
		=\frac{
			\begin{vmatrix}
			\epsilon_1^{\frac{9}{2}}-\epsilon_1^{-\frac{9}{2}}\hspace{0.1cm} \ \ \epsilon_1^{\frac{7}{2}}-\epsilon_1^{-\frac{7}{2}}\hspace{0.1cm} \ \ \epsilon_1^{\frac{3}{2}}-\epsilon_1^{-\frac{3}{2}}\hspace{0.1cm} \ \ \epsilon_1^{\frac{1}{2}}-\epsilon_1^{-\frac{1}{2}}\\
			\epsilon_2^{\frac{9}{2}}-\epsilon_2^{-\frac{9}{2}}\hspace{0.1cm} \ \ \epsilon_2^{\frac{7}{2}}-\epsilon_2^{-\frac{7}{2}}\hspace{0.1cm} \ \ \epsilon_2^{\frac{3}{2}}-\epsilon_2^{-\frac{3}{2}}\hspace{0.1cm} \ \ \epsilon_2^{\frac{1}{2}}-\epsilon_2^{-\frac{1}{2}}\\
			\epsilon_3^{\frac{9}{2}}-\epsilon_3^{-\frac{9}{2}}\hspace{0.1cm} \ \ \epsilon_3^{\frac{7}{2}}-\epsilon_3^{-\frac{7}{2}}\hspace{0.1cm} \ \ \epsilon_3^{\frac{3}{2}}-\epsilon_3^{-\frac{3}{2}}\hspace{0.1cm} \ \ \epsilon_3^{\frac{1}{2}}-\epsilon_3^{-\frac{1}{2}}\\
			\epsilon_4^{\frac{9}{2}}-\epsilon_4^{-\frac{9}{2}}\hspace{0.1cm} \ \ \epsilon_4^{\frac{7}{2}}-\epsilon_4^{-\frac{7}{2}}\hspace{0.1cm} \ \ \epsilon_4^{\frac{3}{2}}-\epsilon_4^{-\frac{3}{2}}\hspace{0.1cm} \ \ \epsilon_4^{\frac{1}{2}}-\epsilon_4^{-\frac{1}{2}}\\
			\end{vmatrix}}{
			\begin{vmatrix}
			\epsilon_1^{\frac{7}{2}}-\epsilon_1^{-\frac{7}{2}}\hspace{0.1cm} \ \		\epsilon_1^{\frac{5}{2}}-\epsilon_1^{-\frac{5}{2}}\hspace{0.1cm} \ \ \epsilon_1^{\frac{3}{2}}-\epsilon_1^{-\frac{3}{2}}\hspace{0.1cm} \ \ \epsilon_1^{\frac{1}{2}}-\epsilon_1^{-\frac{1}{2}}\\
			\epsilon_2^{\frac{7}{2}}-\epsilon_2^{-\frac{7}{2}}\hspace{0.1cm} \ \		\epsilon_2^{\frac{5}{2}}-\epsilon_2^{-\frac{5}{2}}\hspace{0.1cm} \ \ \epsilon_2^{\frac{3}{2}}-\epsilon_2^{-\frac{3}{2}}\hspace{0.1cm} \ \ \epsilon_2^{\frac{1}{2}}-\epsilon_2^{-\frac{1}{2}}\\
			\epsilon_3^{\frac{7}{2}}-\epsilon_3^{-\frac{7}{2}}\hspace{0.1cm} \ \		\epsilon_3^{\frac{5}{2}}-\epsilon_3^{-\frac{5}{2}}\hspace{0.1cm} \ \ \epsilon_3^{\frac{3}{2}}-\epsilon_3^{-\frac{3}{2}}\hspace{0.1cm} \ \ \epsilon_3^{\frac{1}{2}}-\epsilon_3^{-\frac{1}{2}}\\
			\epsilon_4^{\frac{7}{2}}-\epsilon_4^{-\frac{7}{2}}\hspace{0.1cm} \ \		\epsilon_4^{\frac{5}{2}}-\epsilon_4^{-\frac{5}{2}}\hspace{0.1cm} \ \ \epsilon_4^{\frac{3}{2}}-\epsilon_4^{-\frac{3}{2}}\hspace{0.1cm} \ \ \epsilon_4^{\frac{1}{2}}-\epsilon_4^{-\frac{1}{2}}\\
			\end{vmatrix}}\nonumber\\
		&=& 4+\epsilon_{1}+\epsilon_{2}+\epsilon_{3}+\epsilon_{4}+\epsilon_{1}\epsilon_{2}+\epsilon_{2}\epsilon_{3}+\epsilon_{1}\epsilon_{3}+\epsilon_{1}\epsilon_{4}+\epsilon_{2}\epsilon_{4}+\epsilon_{3}\epsilon_{4}+\frac{1}{\epsilon_{1}}+\frac{1}{\epsilon_{2}}+\frac{1}{\epsilon_{3}}
		\nonumber \\
		& & +\frac{1}{\epsilon_{4}}+\frac{\epsilon_1}{\epsilon_2}+\frac{\epsilon_2}{\epsilon_1}+\frac{\epsilon_1}{\epsilon_3}+\frac{\epsilon_3}{\epsilon_1}+\frac{\epsilon_2}{\epsilon_3}+\frac{\epsilon_3}{\epsilon_2}+\frac{\epsilon_1}{\epsilon_4}+\frac{\epsilon_4}{\epsilon_1}+\frac{\epsilon_2}{\epsilon_4}+\frac{\epsilon_4}{\epsilon_2}+\frac{\epsilon_3}{\epsilon_4}+\frac{\epsilon_4}{\epsilon_3}\nonumber\\
		& &+\frac{1}{\epsilon_1 \epsilon_2}+\frac{1}{\epsilon_2 \epsilon_3}+\frac{1}{\epsilon_1 \epsilon_3}+\frac{1}{\epsilon_1 \epsilon_4}+\frac{1}{\epsilon_2 \epsilon_4}+\frac{1}{\epsilon_4 \epsilon_3}.
		\end{eqnarray}} 
\end{itemize}
\noindent \rule{18cm}{1.25pt}

\subsection*{$3.~SO(2N)$}\label{subsec:soeven-haar-char}
\textbf{Characters}\\
\\
 The character computation of a particular representation for $SO(2N)$ needs special attention as it possesses the notion of  simple and double characters. First, we compute two character functions \cite{Balantekin:2001id,koike1987,littlewood1977theory} 

\vspace{-0.6cm}
{\scriptsize\begin{eqnarray}\label{eq:soeven-CS-formula}
\mathcal{C}_{r_1,r_2,...,r_{N}}^{(M(\epsilon))} &=& \frac{|\epsilon^{r_1}+\epsilon^{-r_1},\epsilon^{r_2}+\epsilon^{-r_2},...,\epsilon^{r_N}+\epsilon^{-r_N}-\delta_{\lambda_{N} 0}|}{|\epsilon^{N-1}+\epsilon^{-N+1},\epsilon^{N-2}+\epsilon^{-N+2},....,1|}, \nonumber\\
\mathcal{S}_{r_1,r_2,...,r_{N}}^{(M(\epsilon))} &=& \frac{|\epsilon^{r_1}-\epsilon^{-r_1},\epsilon^{r_2}-\epsilon^{-r_2},...,\epsilon^{r_N}-\epsilon^{-r_N}|}{|\epsilon^{N-1}+\epsilon^{-N+1},\epsilon^{N-2}+\epsilon^{-N+2},....,1|}.
\end{eqnarray}}
For $\lambda_{N}=0$, the simple character is given as:

\vspace{-0.7cm}
{\scriptsize\begin{eqnarray}\label{eq:soeven-char-vec-formula}	
\chi^{(M(\epsilon))}_{r_1,r_2,...,r_{N}}=\mathcal{C}_{r_1,r_2,...,r_{N}}^{(M(\epsilon))}.
\end{eqnarray}}
Otherwise, if  $\lambda_{N}\neq0$,  the simple character computation is redefined as:

\vspace{-0.6cm}	
{\scriptsize\begin{eqnarray}\label{eq:soeven-char-spinor-formula}
\chi^{(M(\epsilon))}_{r_1,r_2,...,r_{N}}=\frac{1}{2}(\mathcal{C}_{r_1,r_2,...,r_{N}}^{(M(\epsilon))}+\mathcal{S}_{r_1,r_2,...,r_{N}}^{(M(\epsilon))}).
\end{eqnarray}}
Here $M(\epsilon)$ = diag$(\epsilon_1,\epsilon_2,...,\epsilon_N)$ identifies a particular representation of $SO(2N)$. The $r_i$'s can be obtained from the Dynkin labels of the representation while the $\epsilon_a$'s can be obtained by examining the matrix form of the maximal torus of $SO(2N)$ \cite{Dieck} given below:

\vspace{-0.5cm}
{\scriptsize\begin{eqnarray}\label{eq:torus-soeven}
\mathbb{T}^N:
\begin{pmatrix}
\cos\,\theta_1 & \text{-}\sin\,\theta_1 &\cdots & 0 & 0 \\
\sin\,\theta_1 & \cos\,\theta_1 &\cdots & 0 & 0 \\
\vdots & \vdots & \ddots & \vdots & \vdots \\
0 & 0 & \cdots & \cos\,\theta_N & \text{-}\sin\,\theta_N \\
0 & 0 & \cdots &\sin\,\theta_N & \cos\,\theta_N\\
\end{pmatrix}_{2N\times 2N}\hspace{-0.8cm}.
\end{eqnarray}}
Each $2\times 2$ block can be written in diagonal form  as: 
{\scriptsize$\begin{pmatrix}
\cos\,\theta_i \ \ \text{-}\sin\,\theta_i\\
\sin\,\theta_i \ \ \cos\,\theta_i
\end{pmatrix} \rightarrow   
\begin{pmatrix}
e^{i\theta_i} \ \ 0\\
0 \ \ e^{\text{-}i\theta_i}
\end{pmatrix}$}. Then, by defining $\epsilon_i = e^{i\theta_i}$ we get $2N$ parameters $\{\epsilon_i, \epsilon_i{}^{\text{-}1}\}$ where $i=1,2,\cdots N$. The respective numerators on the RHS of Eq.~\eqref{eq:soeven-CS-formula} can be explicitly written  as: 

\vspace{-0.5cm}
{\scriptsize\begin{eqnarray}\label{eq:soevenchar-numerator}
|\epsilon^{r_1}+\epsilon^{-r_1},\epsilon^{r_2}+\epsilon^{-r_2},...,\epsilon^{r_N}+\epsilon^{-r_N}-\delta_{\lambda_{N} 0}| &=& 
\begin{vmatrix}
\epsilon_1^{r_1}+\epsilon_1^{-r_1} & \epsilon_1^{r_2}+\epsilon_1^{-r_2} & \cdots & \epsilon_1^{r_N}+\epsilon_1^{-r_N}-\delta_{\lambda_{N} 0} \\
\epsilon_2^{r_1}+\epsilon_2^{-r_1} & \epsilon_2^{r_2}+\epsilon_2^{-r_2} & \cdots & \epsilon_2^{r_N}+\epsilon_2^{-r_N}-\delta_{\lambda_{N} 0} \\
\vdots & \vdots & \ddots & \vdots \\ 
\epsilon_N^{r_1}+\epsilon_N^{-r_1} & \epsilon_N^{r_2}+\epsilon_N^{-r_2} & \cdots & \epsilon_N^{r_N}+\epsilon_N^{-r_N}-\delta_{\lambda_{N} 0} \\
\end{vmatrix}, \nonumber \\
\vspace{2mm}
|\epsilon^{r_1}-\epsilon^{-r_1},\epsilon^{r_2}-\epsilon^{-r_2},...,\epsilon^{r_N}-\epsilon^{-r_N}| &=& 
\begin{vmatrix}
\epsilon_1^{r_1}-\epsilon_1^{-r_1} & \epsilon_1^{r_2}-\epsilon_1^{-r_2} & \cdots & \epsilon_1^{r_N}-\epsilon_1^{-r_N} \\
\epsilon_2^{r_1}-\epsilon_2^{-r_1} & \epsilon_2^{r_2}-\epsilon_2^{-r_2} & \cdots & \epsilon_2^{r_N}-\epsilon_2^{-r_N} \\
\vdots & \vdots & \ddots & \vdots \\ 
\epsilon_N^{r_1}-\epsilon_N^{-r_1} & \epsilon_N^{r_2}-\epsilon_N^{-r_2} & \cdots & \epsilon_N^{r_N}-\epsilon_N^{-r_N} \\
\end{vmatrix},
\end{eqnarray}}
while the denominator is 
	
\vspace{-0.7cm}
{\scriptsize\begin{eqnarray}\label{eq:soevenvandermonde}
|\epsilon^{N-1}+\epsilon^{-N+1},\epsilon^{N-2}+\epsilon^{-N+2},....,1| = \begin{vmatrix}
\epsilon_1^{N-1}+\epsilon_1^{-N+1} & \epsilon_1^{N-2}+\epsilon_1^{-N+2} & \cdots & 1 \\
\epsilon_2^{N-1}+\epsilon_2^{-N+1} & \epsilon_2^{N-2}+\epsilon_2^{-N+2} & \cdots & 1 \\
\vdots & \vdots & \ddots & \vdots \\ 
\epsilon_N^{N-1}+\epsilon_N^{-N+1} & \epsilon_N^{N-2}+\epsilon_N^{-N+2} & \cdots & 1 \\
\end{vmatrix}.
\end{eqnarray}}
\underline{\textbf{Calculating the $r_i$'s}}
\\
Cartan matrix for $SO(2N)$ group is an $N\times N$ matrix

\vspace{-0.7cm}
	{\scriptsize\begin{eqnarray}\label{eq:soevencartan matrix}
			\mathcal{A}_{SO(2N)} = 
			\begin{pmatrix}
				- -\alpha_1- -\\
				- -\alpha_2- -\\
				\vline \\
				- -\alpha_{N-2}- -\\
				- -\alpha_{N-1}- -\\
				- -\alpha_{N}- -
			\end{pmatrix} = 
			\begin{pmatrix}
				2 & \text{-}1 &  0 & \cdots & 0 & 0 & 0\\
				\text{-}1 & 2 & \text{-}1 & \cdots & 0 & 0 & 0\\
				\vdots & \vdots & \vdots & \ddots & \vdots & \vdots & \vdots \\
				0 &  0 &  0 & \cdots & 2 & \text{-}1 & \text{-}1\\
				0 &  0 &  0 & \cdots & \text{-}1 & 2 & 0\\
				0 &  0 &  0 & \cdots & \text{-}1 & 0 & 2\\
			\end{pmatrix}_{N\times N}\hspace{-0.7cm}.
	\end{eqnarray}}
	Once again our first step in calculating the $r_i$'s is the construction of the weight tree \cite{Foster:2016lectures} corresponding to the LDF representation of $SO(2N)$. We start with the Dynkin label  $\left(1,0,..,0\right)$ and successively subtract rows of the Cartan matrix shown in Eq.~\eqref{eq:soevencartan matrix}. The weight tree can be expressed as:
	
	\vspace{-0.6cm}
	{\scriptsize\begin{eqnarray}
			L_1 &=& \underbrace{(1,0,0,...,0,0)}_{(\text{$N$-tuple})}, \nonumber\\
			L_2 = L_1 - \alpha_1 &=& (-1,1,0,...,0,0),\nonumber\\
			&\vdots&             \nonumber\\
			L_k = L_{k-1}-\alpha_{k-1} &=& (0,..,-1,1,.,0),\nonumber\\
			&\vdots&             \nonumber\\
			L_{N-1} = L_{N-2}-\alpha_{N-2}&=& (0,0,...,-1,1,1),\nonumber\\
			L_N = L_{N-1}-\alpha_{N-1}&=& (0,0,...,0,-1,1).
	\end{eqnarray}}
	A particular representation of $SO(2N)$ is uniquely identified by its Dynkin label $(a_1,\cdots,a_{N})$. The elements of Dynkin label $a_i$'s can be written  in terms of the fundamental weight tree of LDF representation and  the $\lambda_i$'s as:
	
	\vspace{-0.5cm}
	{\scriptsize\begin{eqnarray}
			(a_1,a_2,....,a_{N}) &=& \lambda_1 (1,0,...,0,0) + \lambda_2 (-1,1,0,...,0) + \cdots +\lambda_{N-1} (0,...,-1,1,1) + \lambda_{N}(0,0,...,-1,1).\nonumber
	\end{eqnarray}}
	The above equation can be solved to find the $\lambda_i$'s uniquely through following steps
	
	\vspace{-0.5cm}
	{\scriptsize\begin{eqnarray}
	a_N = \lambda_{N-1}+\lambda_{N}, \hspace{0.2cm}
	a_{N-1} = \lambda_{N-1}-\lambda_{N}, \hspace{0.5cm}    
	a_k = \lambda_k-\lambda_{k+1}, \hspace{0.2cm} k = 1,\cdots,N-2.  
	\end{eqnarray}}
which are inverted as:

    \vspace{-0.5cm}
	{\scriptsize\begin{eqnarray}\label{eq:soevenlambda-recursion}
	\lambda_{N} &=& \frac{1}{2}(a_{N}-a_{N-1}), \hspace{0.2cm}
	\lambda_{N-1} = \frac{1}{2}(a_{N}+a_{N-1}), \hspace{0.2cm}
	\lambda_{N-2} = a_{N-2}+\frac{1}{2}(a_{N}+a_{N-1}), \nonumber\\
	\lambda_k &=& a_k+a_{k+1}+\cdots+a_{N-2}+\frac{1}{2}(a_{N}+a_{N-1}), \hspace{0.5cm} k\leq N-2.
	\end{eqnarray}}
	The $r_i$'s are related to the $\lambda_i$'s through the following equation:
	
	\vspace{-0.5cm}
	{\scriptsize\begin{eqnarray}
		\vec{r} = \vec{\lambda}+\vec{\rho} \hspace{0.5cm}\text{where}\hspace{0.5cm}\rho_i = N-i, \hspace{0.5cm} i=1,2,..,N.
	\end{eqnarray}}
	Once $r_i$'s are noted down, we can compute the numerator using Eq.~\eqref{eq:soevenchar-numerator} and subsequently the full character.
	\\
	\\
	\textbf{Haar Measure}
	\\
	The Haar measure for $SO(2N)$ group can be written as \cite{Gray:2008yu,Hanany:2014dia,Henning:2017fpj}:
	
	\vspace{-0.6cm}
	{\scriptsize\begin{eqnarray}\label{eq:soevenhaar-measure}
			\int_{SO(2N)} d\mu_{SO(2N)} = \frac{1}{\left(2\pi i\right)^{N} 2^{N-1} N!}\oint_{|\epsilon_{l}|=1} \prod_{l=1}^{N}\frac{d\epsilon_l}{\epsilon_l}\Delta\left(\epsilon\right)\Delta\left(\epsilon^{-1}\right).
	\end{eqnarray}}
	$\Delta\left(\epsilon\right)$ is the denominator given in Eq.~\eqref{eq:soevenvandermonde}, $\Delta(\epsilon^{-1})$ can similarly be computed by substituting $\epsilon^{-1}_i$ in the place of $\epsilon_i$  in the expression of $\Delta\left(\epsilon\right)$. Finally, substituting them in Eq.~\eqref{eq:soevenhaar-measure} we can obtain the Haar measure.
	
	As examples, the characters for a few representations \cite{Slansky:1981yr,Yamatsu:2015npn} of $SO(8)$ and the  Haar measure are explicitly computed.
	
\begin{center}
	\rule{15cm}{1.25pt}	
\end{center}
\newpage
\noindent
\fbox{$SO(8)$}\label{examples:so8-char-haar}\\
\\
\textbf{Haar Measure}
\\
\\
The denominator for the character formula in this case is: 

\vspace{-0.6cm}
{\scriptsize\begin{eqnarray}\label{eq:vandermonde-so8}
	\Delta\left(\epsilon\right) = \begin{vmatrix}
	\epsilon_1^{3}+\epsilon_1^{-3}\hspace{0.1cm} \ \		\epsilon_1^{2}+\epsilon_1^{-2}\hspace{0.1cm} \ \ \epsilon_1^{1}+\epsilon_1^{-1}\hspace{0.1cm} \ \ 1\\
	\epsilon_2^{3}+\epsilon_2^{-3}\hspace{0.1cm} \ \		\epsilon_2^{2}+\epsilon_2^{-2}\hspace{0.1cm} \ \ \epsilon_2^{1}+\epsilon_2^{-1}\hspace{0.1cm} \ \ 1\\
	\epsilon_3^{3}+\epsilon_3^{-3}\hspace{0.1cm} \ \		\epsilon_3^{2}+\epsilon_3^{-2}\hspace{0.1cm} \ \ \epsilon_3^{1}+\epsilon_3^{-1}\hspace{0.1cm} \ \ 1\\
	\epsilon_4^{3}+\epsilon_4^{-3}\hspace{0.1cm} \ \		\epsilon_4^{2}+\epsilon_4^{-2}\hspace{0.1cm} \ \ \epsilon_4^{1}+\epsilon_4^{-1}\hspace{0.1cm} \ \ 1\\
	\end{vmatrix}.
	\end{eqnarray}}
$\Delta(\epsilon^{-1})$ is obtained by replacing $\epsilon_i$ by $\epsilon_i^{-1}$ in the above expression. Then using Eq.~\eqref{eq:soevenhaar-measure},

\vspace{-0.5cm}
{\scriptsize\begin{eqnarray}\label{eq:so8-haar-measure}
	d\mu_{SO(8)} &=& \frac{1}{4! 2^{3}\left(2\pi i\right)^{4}} \frac{d\epsilon_1}{\epsilon_1}\frac{d\epsilon_2}{\epsilon_2}\frac{d\epsilon_3}{\epsilon_3}\frac{d\epsilon_4}{\epsilon_4}\Delta\left(\epsilon\right)\Delta\left(\epsilon^{-1}\right).
	\end{eqnarray}}
\textbf{Characters}
\begin{itemize}
	\item The Fundamental Representation: $8_v \equiv (1,0,0,0)$ 
	
	Using Eq.~\eqref{eq:soevenlambda-recursion}, we get $\vec{\lambda}$ = $(1,0,0,0)$. Now, since $\vec{\rho}$ = $(3,2,1,0)$, therefore, $\vec{r} = \vec{\lambda}+\vec{\rho} = (4,2,1,0)$ and using Eq.~\eqref{eq:soeven-char-vec-formula} the character is obtained as:
	
	\vspace{-0.5cm}
	{\scriptsize\begin{eqnarray}\label{eq:so8-8v-char}
		\chi_{({SO(8)})_{8_v}}(\epsilon_1,\epsilon_2,\epsilon_3,\epsilon_4) &=& \frac{|\epsilon^{4}+\epsilon^{-4},\hspace{0.1cm}\epsilon^{2}+\epsilon^{-2},\hspace{0.1cm}\epsilon^{1}+\epsilon^{-1},\hspace{0.1cm}1|} {|\epsilon^{3}+\epsilon^{-3},\hspace{0.1cm}\epsilon^{2}+\epsilon^{-2},\hspace{0.1cm}\epsilon^{1}+\epsilon^{-1},\hspace{0.1cm}1|} 
		= \frac{
			\begin{vmatrix}
			\epsilon_1^{4}+\epsilon_1^{-4}\hspace{0.1cm} \ \		\epsilon_1^{2}+\epsilon_1^{-2}\hspace{0.1cm} \ \ \epsilon_1^{1}+\epsilon_1^{-1}\hspace{0.1cm} \ \ 1\\
			\epsilon_2^{4}+\epsilon_2^{-4}\hspace{0.1cm} \ \		\epsilon_2^{2}+\epsilon_2^{-2}\hspace{0.1cm} \ \ \epsilon_2^{1}+\epsilon_2^{-1}\hspace{0.1cm} \ \ 1\\
			\epsilon_3^{4}+\epsilon_3^{-4}\hspace{0.1cm} \ \		\epsilon_3^{2}+\epsilon_3^{-2}\hspace{0.1cm} \ \ \epsilon_3^{1}+\epsilon_3^{-1}\hspace{0.1cm} \ \ 1\\
			\epsilon_4^{4}+\epsilon_4^{-4}\hspace{0.1cm} \ \		\epsilon_4^{2}+\epsilon_4^{-2}\hspace{0.1cm} \ \ \epsilon_4^{1}+\epsilon_4^{-1}\hspace{0.1cm} \ \ 1\\
			\end{vmatrix}}{
			\begin{vmatrix}
			\epsilon_1^{3}+\epsilon_1^{-3}\hspace{0.1cm} \ \		\epsilon_1^{2}+\epsilon_1^{-2}\hspace{0.1cm} \ \ \epsilon_1^{1}+\epsilon_1^{-1}\hspace{0.1cm} \ \ 1\\
			\epsilon_2^{3}+\epsilon_2^{-3}\hspace{0.1cm} \ \		\epsilon_2^{2}+\epsilon_2^{-2}\hspace{0.1cm} \ \ \epsilon_2^{1}+\epsilon_2^{-1}\hspace{0.1cm} \ \ 1\\
			\epsilon_3^{3}+\epsilon_3^{-3}\hspace{0.1cm} \ \		\epsilon_3^{2}+\epsilon_3^{-2}\hspace{0.1cm} \ \ \epsilon_3^{1}+\epsilon_3^{-1}\hspace{0.1cm} \ \ 1\\
			\epsilon_4^{3}+\epsilon_4^{-3}\hspace{0.1cm} \ \		\epsilon_4^{2}+\epsilon_4^{-2}\hspace{0.1cm} \ \ \epsilon_4^{1}+\epsilon_4^{-1}\hspace{0.1cm} \ \ 1\\
			\end{vmatrix}}\nonumber\\
		&=& \epsilon_{1}+\frac{1}{\epsilon_{1}}+\epsilon_{2}+\frac{1}{\epsilon_{2}}+\epsilon_{3}+\frac{1}{\epsilon_{3}}+\epsilon_{4}+\frac{1}{\epsilon_{4}} .
		\end{eqnarray}} 
	
	\vspace{-0.7cm}
	\item The Spinor Representation: $8_s \equiv (0,0,0,1)$ 
	
	We obtain $\vec{\lambda}$ = $(\frac{1}{2},\frac{1}{2},\frac{1}{2},\frac{1}{2})$ and $\vec{r} = \vec{\lambda}+\vec{\rho} = (\frac{7}{2},\frac{5}{2},\frac{3}{2},\frac{1}{2})$ and using Eq.~\eqref{eq:soeven-CS-formula} we get:
	
	\vspace{-0.5cm}
	{\tiny\begin{eqnarray}
		\mathcal{C}_{({SO(8)})_{8_{s}}}(\epsilon_1,\epsilon_2,\epsilon_3,\epsilon_4) &=& \frac{|\epsilon^{\frac{7}{2}}+\epsilon^{-\frac{7}{2}},\hspace{0.1cm}\epsilon^{\frac{5}{2}}+\epsilon^{-\frac{5}{2}},\hspace{0.1cm}\epsilon^{\frac{3}{2}}+\epsilon^{-\frac{3}{2}},\hspace{0.1cm}\epsilon^{\frac{1}{2}}+\epsilon^{-\frac{1}{2}}|} {|\epsilon^{3}+\epsilon^{-3},\hspace{0.1cm}\epsilon^{2}+\epsilon^{-2},\hspace{0.1cm}\epsilon^{1}+\epsilon^{-1},\hspace{0.1cm} 1|}  
		=\frac{
			\begin{vmatrix}
			\epsilon_1^{\frac{7}{2}}+\epsilon_1^{-\frac{7}{2}}\hspace{0.1cm} \ \ \epsilon_1^{\frac{5}{2}}+\epsilon_1^{-\frac{5}{2}}\hspace{0.1cm} \ \ \epsilon_1^{\frac{3}{2}}+\epsilon_1^{-\frac{3}{2}}\hspace{0.1cm} \ \ \epsilon_1^{\frac{1}{2}}+\epsilon_1^{-\frac{1}{2}}\\
			\epsilon_2^{\frac{7}{2}}+\epsilon_2^{-\frac{7}{2}}\hspace{0.1cm} \ \ \epsilon_2^{\frac{5}{2}}+\epsilon_2^{-\frac{5}{2}}\hspace{0.1cm} \ \ \epsilon_2^{\frac{3}{2}}+\epsilon_2^{-\frac{3}{2}}\hspace{0.1cm} \ \ \epsilon_2^{\frac{1}{2}}+\epsilon_2^{-\frac{1}{2}}\\
			\epsilon_3^{\frac{7}{2}}+\epsilon_3^{-\frac{7}{2}}\hspace{0.1cm} \ \ \epsilon_3^{\frac{5}{2}}+\epsilon_3^{-\frac{5}{2}}\hspace{0.1cm} \ \ \epsilon_3^{\frac{3}{2}}+\epsilon_3^{-\frac{3}{2}}\hspace{0.1cm} \ \ \epsilon_3^{\frac{1}{2}}+\epsilon_3^{-\frac{1}{2}}\\
			\epsilon_4^{\frac{7}{2}}+\epsilon_4^{-\frac{7}{2}}\hspace{0.1cm} \ \ \epsilon_4^{\frac{5}{2}}+\epsilon_4^{-\frac{5}{2}}\hspace{0.1cm} \ \ \epsilon_4^{\frac{3}{2}}+\epsilon_4^{-\frac{3}{2}}\hspace{0.1cm} \ \ \epsilon_4^{\frac{1}{2}}+\epsilon_4^{-\frac{1}{2}}\\
			\end{vmatrix}}{
			\begin{vmatrix}
			\epsilon_1^{3}+\epsilon_1^{-3}\hspace{0.1cm} \ \		\epsilon_1^{2}+\epsilon_1^{-2}\hspace{0.1cm} \ \ \epsilon_1^{1}+\epsilon_1^{-1}\hspace{0.1cm} \ \ 1\\
			\epsilon_2^{3}+\epsilon_2^{-3}\hspace{0.1cm} \ \		\epsilon_2^{2}+\epsilon_2^{-2}\hspace{0.1cm} \ \ \epsilon_2^{1}+\epsilon_2^{-1}\hspace{0.1cm} \ \ 1\\
			\epsilon_3^{3}+\epsilon_3^{-3}\hspace{0.1cm} \ \		\epsilon_3^{2}+\epsilon_3^{-2}\hspace{0.1cm} \ \ \epsilon_3^{1}+\epsilon_3^{-1}\hspace{0.1cm} \ \ 1\\
			\epsilon_4^{3}+\epsilon_4^{-3}\hspace{0.1cm} \ \		\epsilon_4^{2}+\epsilon_4^{-2}\hspace{0.1cm} \ \ \epsilon_4^{1}+\epsilon_4^{-1}\hspace{0.1cm} \ \ 1\\
			\end{vmatrix}},\nonumber\\
		\mathcal{S}_{({SO(8)})_{8_{s}}}(\epsilon_1,\epsilon_2,\epsilon_3,\epsilon_4) &=& \frac{|\epsilon^{\frac{7}{2}}-\epsilon^{-\frac{7}{2}},\hspace{0.1cm}\epsilon^{\frac{5}{2}}-\epsilon^{-\frac{5}{2}},\hspace{0.1cm}\epsilon^{\frac{3}{2}}-\epsilon^{-\frac{3}{2}},\hspace{0.1cm}\epsilon^{\frac{1}{2}}-\epsilon^{-\frac{1}{2}}|} {|\epsilon^{3}+\epsilon^{-3},\hspace{0.1cm}\epsilon^{2}+\epsilon^{-2},\hspace{0.1cm}\epsilon^{1}+\epsilon^{-1},\hspace{0.1cm} 1|}  
		=\frac{
			\begin{vmatrix}
			\epsilon_1^{\frac{7}{2}}-\epsilon_1^{-\frac{7}{2}}\hspace{0.1cm} \ \ \epsilon_1^{\frac{5}{2}}-\epsilon_1^{-\frac{5}{2}}\hspace{0.1cm} \ \ \epsilon_1^{\frac{3}{2}}-\epsilon_1^{-\frac{3}{2}}\hspace{0.1cm} \ \ \epsilon_1^{\frac{1}{2}}-\epsilon_1^{-\frac{1}{2}}\\
			\epsilon_2^{\frac{7}{2}}-\epsilon_2^{-\frac{7}{2}}\hspace{0.1cm} \ \ \epsilon_2^{\frac{5}{2}}-\epsilon_2^{-\frac{5}{2}}\hspace{0.1cm} \ \ \epsilon_2^{\frac{3}{2}}-\epsilon_2^{-\frac{3}{2}}\hspace{0.1cm} \ \ \epsilon_2^{\frac{1}{2}}-\epsilon_2^{-\frac{1}{2}}\\
			\epsilon_3^{\frac{7}{2}}-\epsilon_3^{-\frac{7}{2}}\hspace{0.1cm} \ \ \epsilon_3^{\frac{5}{2}}-\epsilon_3^{-\frac{5}{2}}\hspace{0.1cm} \ \ \epsilon_3^{\frac{3}{2}}-\epsilon_3^{-\frac{3}{2}}\hspace{0.1cm} \ \ \epsilon_3^{\frac{1}{2}}-\epsilon_3^{-\frac{1}{2}}\\
			\epsilon_4^{\frac{7}{2}}-\epsilon_4^{-\frac{7}{2}}\hspace{0.1cm} \ \ \epsilon_4^{\frac{5}{2}}-\epsilon_4^{-\frac{5}{2}}\hspace{0.1cm} \ \ \epsilon_4^{\frac{3}{2}}-\epsilon_4^{-\frac{3}{2}}\hspace{0.1cm} \ \ \epsilon_4^{\frac{1}{2}}-\epsilon_4^{-\frac{1}{2}}\\
			\end{vmatrix}}{
			\begin{vmatrix}
			\epsilon_1^{3}+\epsilon_1^{-3}\hspace{0.1cm} \ \		\epsilon_1^{2}+\epsilon_1^{-2}\hspace{0.1cm} \ \ \epsilon_1^{1}+\epsilon_1^{-1}\hspace{0.1cm} \ \ 1\\
			\epsilon_2^{3}+\epsilon_2^{-3}\hspace{0.1cm} \ \		\epsilon_2^{2}+\epsilon_2^{-2}\hspace{0.1cm} \ \ \epsilon_2^{1}+\epsilon_2^{-1}\hspace{0.1cm} \ \ 1\\
			\epsilon_3^{3}+\epsilon_3^{-3}\hspace{0.1cm} \ \		\epsilon_3^{2}+\epsilon_3^{-2}\hspace{0.1cm} \ \ \epsilon_3^{1}+\epsilon_3^{-1}\hspace{0.1cm} \ \ 1\\
			\epsilon_4^{3}+\epsilon_4^{-3}\hspace{0.1cm} \ \		\epsilon_4^{2}+\epsilon_4^{-2}\hspace{0.1cm} \ \ \epsilon_4^{1}+\epsilon_4^{-1}\hspace{0.1cm} \ \ 1\\
			\end{vmatrix}}.\nonumber
		\end{eqnarray}}
	Using Eq.~\eqref{eq:soeven-char-spinor-formula} the character of this representation can be given as:
	
	\vspace{-0.5cm}
	{\scriptsize\begin{eqnarray}\label{eq:so8-8s-char}
		\chi_{(SO(8))_{8_s}}
		(\epsilon_1,\epsilon_2,\epsilon_3,\epsilon_4) &=& \frac{1}{2}(\mathcal{C}_{(SO(8))_{8_s}}
		(\epsilon_1,\epsilon_2,\epsilon_3,\epsilon_4)+\mathcal{S}_{(SO(8))_{8_s}}
		(\epsilon_1,\epsilon_2,\epsilon_3,\epsilon_4)) 
		 \\
		&=& \sqrt{\epsilon_{1} \epsilon_{2} \epsilon_{3}\epsilon_{4}}+\sqrt{\frac{\epsilon_{1}\epsilon_{2}}{\epsilon_{3} \epsilon_{4}}} +\sqrt{\frac{\epsilon_{1}\epsilon_{4}}{\epsilon_{3} \epsilon_{2}}}+\sqrt{\frac{\epsilon_{1}\epsilon_{3}}{\epsilon_{2} \epsilon_{4}}}+\sqrt{\frac{\epsilon_{3}\epsilon_{4}}{\epsilon_{1} \epsilon_{2}}}+\sqrt{\frac{\epsilon_{2}\epsilon_{3}}{\epsilon_{1} \epsilon_{4}}}+\sqrt{\frac{\epsilon_{2}\epsilon_{4}}{\epsilon_{1} \epsilon_{3}}} +\frac{1}{\sqrt{\epsilon_{1} \epsilon_{2} \epsilon_{3} \epsilon_{4}}}. \nonumber
		\end{eqnarray}} 
	
	\vspace{-0.7cm}
	\item The Conjugate Spinor Representation: $8_c \equiv (0,0,1,0)$ 
	
	We get $\vec{\lambda}$ = $(\frac{1}{2},\frac{1}{2},\frac{1}{2},-\frac{1}{2})$ and $\vec{r} = \vec{\lambda}+\vec{\rho} = (\frac{7}{2},\frac{5}{2},\frac{3}{2},-\frac{1}{2})$	and using Eq.~\eqref{eq:soeven-CS-formula} we get:
	
	\vspace{-0.5cm}
	{\tiny\begin{eqnarray}
		\mathcal{C}_{({SO(8)})_{8_{c}}}(\epsilon_1,\epsilon_2,\epsilon_3,\epsilon_4) &=& \frac{|\epsilon^{\frac{7}{2}}+\epsilon^{-\frac{7}{2}},\hspace{0.1cm}\epsilon^{\frac{5}{2}}+\epsilon^{-\frac{5}{2}},\hspace{0.1cm}\epsilon^{\frac{3}{2}}+\epsilon^{-\frac{3}{2}},\hspace{0.1cm}\epsilon^{\frac{1}{2}}+\epsilon^{-\frac{1}{2}}|} {|\epsilon^{3}+\epsilon^{-3},\hspace{0.1cm}\epsilon^{2}+\epsilon^{-2},\hspace{0.1cm}\epsilon^{1}+\epsilon^{-1},\hspace{0.1cm} 1|} = \frac{
			\begin{vmatrix}
			\epsilon_1^{\frac{7}{2}}+\epsilon_1^{-\frac{7}{2}}\hspace{0.1cm} \ \ \epsilon_1^{\frac{5}{2}}+\epsilon_1^{-\frac{5}{2}}\hspace{0.1cm} \ \ \epsilon_1^{\frac{3}{2}}+\epsilon_1^{-\frac{3}{2}}\hspace{0.1cm} \ \ \epsilon_1^{\frac{1}{2}}+\epsilon_1^{-\frac{1}{2}}\\
			\epsilon_2^{\frac{7}{2}}+\epsilon_2^{-\frac{7}{2}}\hspace{0.1cm} \ \ \epsilon_2^{\frac{5}{2}}+\epsilon_2^{-\frac{5}{2}}\hspace{0.1cm} \ \ \epsilon_2^{\frac{3}{2}}+\epsilon_2^{-\frac{3}{2}}\hspace{0.1cm} \ \ \epsilon_2^{\frac{1}{2}}+\epsilon_2^{-\frac{1}{2}}\\
			\epsilon_3^{\frac{7}{2}}+\epsilon_3^{-\frac{7}{2}}\hspace{0.1cm} \ \ \epsilon_3^{\frac{5}{2}}+\epsilon_3^{-\frac{5}{2}}\hspace{0.1cm} \ \ \epsilon_3^{\frac{3}{2}}+\epsilon_3^{-\frac{3}{2}}\hspace{0.1cm} \ \ \epsilon_3^{\frac{1}{2}}+\epsilon_3^{-\frac{1}{2}}\\
			\epsilon_4^{\frac{7}{2}}+\epsilon_4^{-\frac{7}{2}}\hspace{0.1cm} \ \ \epsilon_4^{\frac{5}{2}}+\epsilon_4^{-\frac{5}{2}}\hspace{0.1cm} \ \ \epsilon_4^{\frac{3}{2}}+\epsilon_4^{-\frac{3}{2}}\hspace{0.1cm} \ \ \epsilon_4^{\frac{1}{2}}+\epsilon_4^{-\frac{1}{2}}\\
			\end{vmatrix}}{
			\begin{vmatrix}
			\epsilon_1^{3}+\epsilon_1^{-3}\hspace{0.1cm} \ \		\epsilon_1^{2}+\epsilon_1^{-2}\hspace{0.1cm} \ \ \epsilon_1^{1}+\epsilon_1^{-1}\hspace{0.1cm} \ \ 1\\
			\epsilon_2^{3}+\epsilon_2^{-3}\hspace{0.1cm} \ \		\epsilon_2^{2}+\epsilon_2^{-2}\hspace{0.1cm} \ \ \epsilon_2^{1}+\epsilon_2^{-1}\hspace{0.1cm} \ \ 1\\
			\epsilon_3^{3}+\epsilon_3^{-3}\hspace{0.1cm} \ \		\epsilon_3^{2}+\epsilon_3^{-2}\hspace{0.1cm} \ \ \epsilon_3^{1}+\epsilon_3^{-1}\hspace{0.1cm} \ \ 1\\
			\epsilon_4^{3}+\epsilon_4^{-3}\hspace{0.1cm} \ \		\epsilon_4^{2}+\epsilon_4^{-2}\hspace{0.1cm} \ \ \epsilon_4^{1}+\epsilon_4^{-1}\hspace{0.1cm} \ \ 1\\
			\end{vmatrix}},\nonumber\\
		\mathcal{S}_{({SO(8)})_{8_{c}}}(\epsilon_1,\epsilon_2,\epsilon_3,\epsilon_4) &=& \frac{|\epsilon^{\frac{7}{2}}-\epsilon^{-\frac{7}{2}},\hspace{0.1cm}\epsilon^{\frac{5}{2}}-\epsilon^{-\frac{5}{2}},\hspace{0.1cm}\epsilon^{\frac{3}{2}}-\epsilon^{-\frac{3}{2}},\hspace{0.1cm}-(\epsilon^{\frac{1}{2}}-\epsilon^{-\frac{1}{2}})|} {|\epsilon^{3}+\epsilon^{-3},\hspace{0.1cm}\epsilon^{2}+\epsilon^{-2},\hspace{0.1cm}\epsilon^{1}+\epsilon^{-1},\hspace{0.1cm} 1|} = \frac{
			\begin{vmatrix}
			\epsilon_1^{\frac{7}{2}}-\epsilon_1^{-\frac{7}{2}}\hspace{0.1cm} \ \ \epsilon_1^{\frac{5}{2}}-\epsilon_1^{-\frac{5}{2}}\hspace{0.1cm} \ \ \epsilon_1^{\frac{3}{2}}-\epsilon_1^{-\frac{3}{2}}\hspace{0.1cm} \ \ -(\epsilon_1^{\frac{1}{2}}-\epsilon_1^{-\frac{1}{2}})\\
			\epsilon_2^{\frac{7}{2}}-\epsilon_2^{-\frac{7}{2}}\hspace{0.1cm} \ \ \epsilon_2^{\frac{5}{2}}-\epsilon_2^{-\frac{5}{2}}\hspace{0.1cm} \ \ \epsilon_2^{\frac{3}{2}}-\epsilon_2^{-\frac{3}{2}}\hspace{0.1cm} \ \ -(\epsilon_2^{\frac{1}{2}}-\epsilon_2^{-\frac{1}{2}})\\
			\epsilon_3^{\frac{7}{2}}-\epsilon_3^{-\frac{7}{2}}\hspace{0.1cm} \ \ \epsilon_3^{\frac{5}{2}}-\epsilon_3^{-\frac{5}{2}}\hspace{0.1cm} \ \ \epsilon_3^{\frac{3}{2}}-\epsilon_3^{-\frac{3}{2}}\hspace{0.1cm} \ \ -(\epsilon_3^{\frac{1}{2}}-\epsilon_3^{-\frac{1}{2}})\\
			\epsilon_4^{\frac{7}{2}}-\epsilon_4^{-\frac{7}{2}}\hspace{0.1cm} \ \ \epsilon_4^{\frac{5}{2}}-\epsilon_4^{-\frac{5}{2}}\hspace{0.1cm} \ \ \epsilon_4^{\frac{3}{2}}-\epsilon_4^{-\frac{3}{2}}\hspace{0.1cm} \ \ -(\epsilon_4^{\frac{1}{2}}-\epsilon_4^{-\frac{1}{2}})\\
			\end{vmatrix}}{
			\begin{vmatrix}
			\epsilon_1^{3}+\epsilon_1^{-3}\hspace{0.1cm} \ \		\epsilon_1^{2}+\epsilon_1^{-2}\hspace{0.1cm} \ \ \epsilon_1^{1}+\epsilon_1^{-1}\hspace{0.1cm} \ \ 1\\
			\epsilon_2^{3}+\epsilon_2^{-3}\hspace{0.1cm} \ \		\epsilon_2^{2}+\epsilon_2^{-2}\hspace{0.1cm} \ \ \epsilon_2^{1}+\epsilon_2^{-1}\hspace{0.1cm} \ \ 1\\
			\epsilon_3^{3}+\epsilon_3^{-3}\hspace{0.1cm} \ \		\epsilon_3^{2}+\epsilon_3^{-2}\hspace{0.1cm} \ \ \epsilon_3^{1}+\epsilon_3^{-1}\hspace{0.1cm} \ \ 1\\
			\epsilon_4^{3}+\epsilon_4^{-3}\hspace{0.1cm} \ \		\epsilon_4^{2}+\epsilon_4^{-2}\hspace{0.1cm} \ \ \epsilon_4^{1}+\epsilon_4^{-1}\hspace{0.1cm} \ \ 1\\
			\end{vmatrix}}.\nonumber
		\end{eqnarray}}
		Using Eq.~\eqref{eq:soeven-char-spinor-formula} the character of this representation can be given as:
		
		\vspace{-0.5cm}
	{\scriptsize\begin{eqnarray}\label{eq:so8-8c-char}
		\chi_{(SO(8))_{8_c}}
		(\epsilon_1,\epsilon_2,\epsilon_3,\epsilon_4) &=& \frac{1}{2}(\mathcal{C}_{(SO(8))_{8_c}}
		(\epsilon_1,\epsilon_2,\epsilon_3,\epsilon_4)+\mathcal{S}_{(SO(8))_{8_c}}
		(\epsilon_1,\epsilon_2,\epsilon_3,\epsilon_4)) \\
		&=& 
		\sqrt{\frac{\epsilon_{1}}{\epsilon_{2} \epsilon_{3} \epsilon_{4}}}+\sqrt{\frac{\epsilon_{2}}{\epsilon_{1} \epsilon_{3} \epsilon_{4}}}+\sqrt{\frac{\epsilon_{3}}{\epsilon_{1} \epsilon_{2} \epsilon_{4}}}+\sqrt{\frac{\epsilon_{4}}{\epsilon_{1} \epsilon_{2} \epsilon_{3}}}+\sqrt{\frac{\epsilon_{2} \epsilon_{3} \epsilon_{4}}{\epsilon_{1}}}+\sqrt{\frac{\epsilon_{1} \epsilon_{3} \epsilon_{4}}{\epsilon_{2}}}+\sqrt{\frac{\epsilon_{1} \epsilon_{2} \epsilon_{4}}{\epsilon_{3}}} +\sqrt{\frac{\epsilon_{1} \epsilon_{2} \epsilon_{3}}{\epsilon_{4}}}.\nonumber
		\end{eqnarray}} 
	
	\item The Adjoint Representation: $28 \equiv (0,1,0,0)$ 
	
	We obtain $\vec{\lambda}$ = $(1,1,0,0)$ and $\vec{r} = \vec{\lambda}+\vec{\rho}$ = $(4,3,1,0)$	and using Eq.~\eqref{eq:soeven-char-vec-formula} we get:
	
	\vspace{-0.5cm}
	{\scriptsize\begin{eqnarray}\label{eq:so8-28-char}
	\hspace{-0.5cm}\chi_{({SO(8)})_{28}}(\epsilon_1,\epsilon_2,\epsilon_3,\epsilon_4) &=& \frac{|\epsilon^{4}+\epsilon^{-4},\hspace{0.1cm}\epsilon^{3}+\epsilon^{-3},\hspace{0.1cm}\epsilon^{1}+\epsilon^{-1},\hspace{0.1cm} 1|} {|\epsilon^{3}+\epsilon^{-3},\hspace{0.1cm}\epsilon^{2}+\epsilon^{-2},\hspace{0.1cm}\epsilon^{1}+\epsilon^{-1},\hspace{0.1cm} 1|} = \frac{
	\begin{vmatrix}
	\epsilon_1^{4}+\epsilon_1^{-4}\hspace{0.1cm} \ \ \epsilon_1^{3}+\epsilon_1^{-3}\hspace{0.1cm} \ \ \epsilon_1^{1}+\epsilon_1^{-1}\hspace{0.1cm} \ \ 1\\
	\epsilon_2^{4}+\epsilon_2^{-4}\hspace{0.1cm} \ \ \epsilon_2^{3}+\epsilon_2^{-3}\hspace{0.1cm} \ \ \epsilon_2^{1}+\epsilon_2^{-1}\hspace{0.1cm} \ \ 1\\
	\epsilon_3^{4}+\epsilon_3^{-4}\hspace{0.1cm} \ \ \epsilon_3^{3}+\epsilon_3^{-3}\hspace{0.1cm} \ \ \epsilon_3^{1}+\epsilon_3^{-1}\hspace{0.1cm} \ \ 1\\
	\epsilon_4^{4}+\epsilon_4^{-4}\hspace{0.1cm} \ \ \epsilon_4^{3}+\epsilon_4^{-3}\hspace{0.1cm} \ \ \epsilon_4^{1}+\epsilon_4^{-1}\hspace{0.1cm} \ \ 1\\
	\end{vmatrix}}{
	\begin{vmatrix}
	\epsilon_1^{3}+\epsilon_1^{-3}\hspace{0.1cm} \ \		\epsilon_1^{2}+\epsilon_1^{-2}\hspace{0.1cm} \ \ \epsilon_1^{1}+\epsilon_1^{-1}\hspace{0.1cm} \ \ 1\\
	\epsilon_2^{3}+\epsilon_2^{-3}\hspace{0.1cm} \ \		\epsilon_2^{2}+\epsilon_2^{-2}\hspace{0.1cm} \ \ \epsilon_2^{1}+\epsilon_2^{-1}\hspace{0.1cm} \ \ 1\\
	\epsilon_3^{3}+\epsilon_3^{-3}\hspace{0.1cm} \ \		\epsilon_3^{2}+\epsilon_3^{-2}\hspace{0.1cm} \ \ \epsilon_3^{1}+\epsilon_3^{-1}\hspace{0.1cm} \ \ 1\\
	\epsilon_4^{3}+\epsilon_4^{-3}\hspace{0.1cm} \ \		\epsilon_4^{2}+\epsilon_4^{-2}\hspace{0.1cm} \ \ \epsilon_4^{1}+\epsilon_4^{-1}\hspace{0.1cm} \ \ 1\\
	\end{vmatrix}}\nonumber \\
	&=& 4+\epsilon_{1}\epsilon_{2}+\epsilon_{2}\epsilon_{3}+\epsilon_{1}\epsilon_{3}+\epsilon_{1}\epsilon_{4}+\epsilon_{2}\epsilon_{4}+\epsilon_{3}\epsilon_{4}+\frac{1}{\epsilon_1 \epsilon_2}+\frac{1}{\epsilon_2 \epsilon_3} +\frac{1}{\epsilon_1 \epsilon_3}+\frac{1}{\epsilon_1 \epsilon_4}+\frac{1}{\epsilon_2 \epsilon_4}+\frac{1}{\epsilon_4 \epsilon_3} \nonumber \\
	& & +\frac{\epsilon_1}{\epsilon_2}+\frac{\epsilon_2}{\epsilon_1}+ \frac{\epsilon_1}{\epsilon_3}+\frac{\epsilon_3}{\epsilon_1}+\frac{\epsilon_2}{\epsilon_3}+\frac{\epsilon_3}{\epsilon_2}+\frac{\epsilon_1}{\epsilon_4}+\frac{\epsilon_4}{\epsilon_1}+\frac{\epsilon_2}{\epsilon_4}+\frac{\epsilon_4}{\epsilon_2}+\frac{\epsilon_3}{\epsilon_4}+\frac{\epsilon_4}{\epsilon_3}.
	\end{eqnarray}} 
\end{itemize}
\noindent \rule{16cm}{1.25pt}

\subsection*{$4.~Sp(2N)$}\label{subsec:sp-haar-char}
\textbf{Characters}\\
The Weyl character formula for $Sp(2N)$ group is given as  \cite{Balantekin:2001id,koike1987,littlewood1977theory,rossmann2006lie}:

\vspace{-0.6cm}
{\scriptsize\begin{eqnarray}\label{eq:sp-char-formula}
\chi_{r_1,r_2,...,r_{N}}^{(M(\epsilon))} = \frac{|\epsilon^{r_1}-\epsilon^{-r_1},\epsilon^{r_2}-\epsilon^{-r_2},...,\epsilon^{r_N}-\epsilon^{-r_N}|}{|\epsilon^{N}-\epsilon^{-N},\epsilon^{N-1}-\epsilon^{-N+1},....,\epsilon^{1}-\epsilon^{-1}|},
\end{eqnarray}}
where $M(\epsilon)$ = diag$(\epsilon_1,\epsilon_2,...,\epsilon_N)$ represents a specific representation of $Sp(2N)$.  The $\epsilon_a$'s are determined using the matrix form of the maximal torus of $Sp(2N)$ \cite{Dieck}. The simple parametrization $\epsilon_{i} = e^{i\theta_i}$ yields the variables used in the definition of the character.

\vspace{-0.6cm}	
{\scriptsize\begin{eqnarray}
\mathbb{T}^N:
\begin{pmatrix}
e^{i\theta_1}& 0 & \cdots & 0\\
0 &e^{i\theta_2} & \cdots & 0\\
\vdots& \vdots & \ddots & \vdots\\
0 & 0 & \cdots & e^{i\theta_{N}}\\
\end{pmatrix}_{N\times N}\hspace{-0.7cm}.
\end{eqnarray}}
The numerator of Eq.~\eqref{eq:sp-char-formula} can be expressed as: 

\vspace{-0.6cm}	
{\scriptsize\begin{eqnarray}\label{eq:spchar-numerator}
|\epsilon^{r_1}-\epsilon^{-r_1},\epsilon^{r_2}-\epsilon^{-r_2},...,\epsilon^{r_N}-\epsilon^{-r_N}| = 
\begin{vmatrix}
\epsilon_1^{r_1}-\epsilon_1^{-r_1} & \epsilon_1^{r_2}-\epsilon_1^{-r_2} & \cdots & \epsilon_1^{r_N}-\epsilon_1^{-r_N} \\
\epsilon_2^{r_1}-\epsilon_2^{-r_1} & \epsilon_2^{r_2}-\epsilon_2^{-r_2} & \cdots & \epsilon_2^{r_N}-\epsilon_2^{-r_N} \\
\vdots & \vdots & \ddots & \vdots \\ 
\epsilon_N^{r_1}-\epsilon_N^{-r_1} & \epsilon_N^{r_2}-\epsilon_N^{-r_2} & \cdots & \epsilon_N^{r_N}-\epsilon_N^{-r_N} \\
\end{vmatrix},
\end{eqnarray}}
and the denominator is given as follows:

\vspace{-0.6cm}
{\scriptsize\begin{eqnarray}\label{eq:spvandermonde}
|\epsilon^{N}-\epsilon^{-N},\epsilon^{N-1}-\epsilon^{-N+1},....,\epsilon^{1}-\epsilon^{-1}| = \begin{vmatrix}
\epsilon_1^{N}-\epsilon_1^{-N} & \epsilon_1^{N-1}-\epsilon_1^{-N+1} & \cdots & \epsilon_1^{1}-\epsilon_1^{-1} \\
\epsilon_2^{N}-\epsilon_2^{-N} & \epsilon_2^{N-1}-\epsilon_2^{-N+1} & \cdots & \epsilon_2^{1}-\epsilon_2^{-1} \\
\vdots & \vdots & \ddots & \vdots \\ 
\epsilon_N^{N}-\epsilon_N^{-N} & \epsilon_N^{N-1}-\epsilon_N^{-N+1} &\ \cdots & \epsilon_N^{1}-\epsilon_N^{-1} \\
\end{vmatrix}.
\end{eqnarray}}

\vspace{-0.7cm}
\noindent \underline{\textbf{Calculating the $r_i$'s}}

\noindent The Cartan matrix for $Sp(2N)$ group is given as:

\vspace{-0.5cm}
{\scriptsize\begin{eqnarray}\label{eq:spcartan matrix}
\mathcal{A}_{Sp(2N)} = 
\begin{pmatrix}
- -\alpha_1- -\\
- -\alpha_2- -\\
\vline \\
- -\alpha_{N-1}- -\\
- -\alpha_{N}- -
\end{pmatrix} = 
\begin{pmatrix}
2 & \text{-}1 &  0 & \ \cdots & 0 & 0 & 0\\
\text{-}1 & 2 & \text{-}1 & \cdots & 0 & 0 & 0\\
\vdots & \vdots & \vdots & \ddots & \vdots & \vdots & \vdots \\
0 & 0 & 0 & \cdots & \text{-}1 & 2 & \text{-}1\\
0 & 0 & 0 & \cdots & 0 & \text{-}2 & 2\\
\end{pmatrix}_{N\times N}\hspace{-0.7cm}.
\end{eqnarray}}
Starting from the Dynkin label for the LDF representation of $Sp(2N)$ $\left(1,0,..,0\right)$ and successively subtracting the rows of the Cartan matrix, see Eq.~\eqref{eq:spcartan matrix}, the corresponding weight tree \cite{Foster:2016lectures} is computed as:

\vspace{-0.7cm}
{\scriptsize\begin{eqnarray}
L_1 &=& \underbrace{(1,0,0,...,0,0)}_{(\text{$N$-tuple})}, \nonumber\\
L_2 = L_1 - \alpha_1 &=& (-1,1,0,...,0,0),\nonumber\\
&\vdots&             \nonumber\\
L_k = L_{k-1}-\alpha_{k-1} &=& (0,..,-1,1,.,0),\nonumber\\
&\vdots&             \nonumber\\
L_{N-1} = L_{N-2}-\alpha_{N-2}&=& (0,0,...,-1,1,0),\nonumber\\
L_N = L_{N-1}-\alpha_{N-1}&=& (0,0,...,0,-1,1).
\end{eqnarray}}
Similar to the earlier cases, a particular representation of $Sp(2N)$ is uniquely identified by its Dynkin label $(a_1,a_2,.....,a_{N})$. Following the same trajectory, we can rewrite the $\lambda_i$'s in terms of the $a_j$'s using the weight tree of LDF. The successive paths adopted in this construction are as follows:

\vspace{-0.5cm}
{\scriptsize\begin{eqnarray}
(a_1,a_2,....,a_{N}) = \lambda_1 (1,0,...,0,0) + \lambda_2 (-1,1,0,...,0) + \cdots +\lambda_{N-1} (0,...,-1,1,0) + \lambda_{N}(0,0,...,-1,1). \nonumber
\end{eqnarray}}
Each entry of the Dynkin label can be recast as:

\vspace{-0.6cm}
{\scriptsize\begin{eqnarray}
a_N = \lambda_{N} \hspace{0.4cm} \text{and} \hspace{0.4cm} a_k &=& \lambda_k-\lambda_{k+1}, \hspace{0.7cm} k = 1,\cdots,N-1 \hspace{0.1cm}.  
\end{eqnarray}}

\vspace{-0.9cm}\noindent Then we can find unique $\lambda_i$'s in the following form

\vspace{-0.6cm}
{\scriptsize\begin{eqnarray}\label{eq:splambda-recursion}
\lambda_{N} = a_N, \hspace{0.2cm}
\lambda_{N-1} = a_{N-1}+a_N, \hspace{0.5cm}
\lambda_k = a_k+a_{k+1}+\cdots+a_{N}.
\end{eqnarray}}

\vspace{-0.7cm}\noindent The $r_i$'s are related to the $\lambda_i$'s through the following equation:

\vspace{-0.6cm}
{\scriptsize\begin{eqnarray}
\vec{r} = \vec{\lambda}+\vec{\rho} \hspace{0.5cm}\text{where}\hspace{0.5cm}\rho_i = N-i+1, \hspace{0.5cm} i=1,2,..,N.
\end{eqnarray}}

\vspace{-0.7cm}\noindent After computing the $r_i$'s, we can further simplify the numerator using Eq.~\eqref{eq:spchar-numerator} and compute the full character.
\\
\\
\noindent
\textbf{Haar Measure}
\\
Here, the Haar measure can be written as \cite{Gray:2008yu,Hanany:2014dia,Henning:2017fpj}:

\vspace{-0.6cm}
{\scriptsize\begin{eqnarray}\label{eq:sphaar-measure}
\int_{Sp(2N)} d\mu_{Sp(2N)} = \frac{1}{\left(2\pi i\right)^{N} 2^{N} N!}\oint_{|\epsilon_l|=1} \prod_{l=1}^{N}\frac{d\epsilon_l}{\epsilon_l}\Delta\left(\epsilon\right)\Delta\left(\epsilon^{-1}\right).
\end{eqnarray}}
$\Delta\left(\epsilon\right)$ is the denominator given in Eq.~\eqref{eq:spvandermonde}, $\Delta(\epsilon^{-1})$ can similarly be computed by substituting $\epsilon^{-1}_i$ in the place of $\epsilon_i$  in the expression of $\Delta\left(\epsilon\right)$. Then using these information we obtain the Haar measure for $Sp(2N)$ group.

Here, the characters for certain representations \cite{Slansky:1981yr,Yamatsu:2015npn} and the  Haar measures correspond to $Sp(4)$ and $Sp(6)$ groups  have been computed in detail.
\begin{center}
	\rule{15cm}{1.25pt}	
\end{center}
\noindent
\fbox{$Sp(4)$}\label{examples:sp4-char-haar}\\
\\
\textbf{Haar Measure}
\\
The denominator for the character formula in this case is: 

\vspace{-0.6cm}
{\scriptsize\begin{eqnarray}\label{eq:vandermonde-sp4}
	\Delta\left(\epsilon\right) = \begin{vmatrix}
	\epsilon_1^{2}-\epsilon_1^{-2}\hspace{0.1cm} \ \		\epsilon_1^{1}-\epsilon_1^{-1}\\
	\epsilon_2^{2}-\epsilon_2^{-2}\hspace{0.1cm} \ \		\epsilon_2^{1}-\epsilon_2^{-1}\\
	\end{vmatrix}.
	\end{eqnarray}}
$\Delta(\epsilon^{-1})$ is obtained by replacing $\epsilon_i$ by $\epsilon_i^{-1}$ in the above expression. Then using Eq.~\eqref{eq:sphaar-measure},

\vspace{-0.6cm}
{\scriptsize\begin{eqnarray}\label{eq:sp4-haar-measure}
	d\mu_{Sp(4)} &=& \frac{1}{2! 2^{2}\left(2\pi i\right)^{2}} \frac{d\epsilon_1}{\epsilon_1}\frac{d\epsilon_2}{\epsilon_2}\Delta\left(\epsilon\right)\Delta\left(\epsilon^{-1}\right) \nonumber \\
	&=& \frac{1}{8\left(2\pi i\right)^{2}} \frac{d\epsilon_1}{\epsilon_1^{5}}\frac{d\epsilon_2}{\epsilon_2^{5}}\left(-1+\epsilon_{1}^{2}\right)^{2}\left(-1+\epsilon_{2}^{2}\right)^{2}\left(\epsilon_{2}+\epsilon_{1}^{2} \epsilon_{2}-\epsilon_{1}\left(1+\epsilon_{2}^{2}\right)\right)^{2}.
	\end{eqnarray}}

\vspace{-0.7cm}\noindent \textbf{Characters}

\vspace{-0.4cm}
\begin{itemize}
	\item The Fundamental Representation: $4 \equiv (1,0)$ 
	
	Using Eq.~\eqref{eq:splambda-recursion}, we get $\vec{\lambda}$ = $(1,0)$. Now, since $\vec{\rho}$ = $(2,1)$ therefore, $\vec{r} = \vec{\lambda}+\vec{\rho} = (3,1)$ and the character is obtained as:
	
	\vspace{-1.1cm}
	{\scriptsize\begin{eqnarray}\label{eq:sp4-4-char}
		\chi_{({Sp(4)})_{4}}(\epsilon_1,\epsilon_2) &=& \frac{|\epsilon^{3}-\epsilon^{-3},\hspace{0.1cm}\epsilon^{1}-\epsilon^{-1}|} {|\epsilon^{2}-\epsilon^{-2},\hspace{0.1cm}\epsilon^{1}-\epsilon^{-1}|} = \frac{
			\begin{vmatrix}
			\epsilon_1^{3}-\epsilon_1^{-3}\hspace{0.1cm} \ \		\epsilon_1^{1}-\epsilon_1^{-1}\\
			\epsilon_2^{3}-\epsilon_2^{-3}\hspace{0.1cm} \ \		\epsilon_2^{1}-\epsilon_2^{-1}\\
			\end{vmatrix}}{
			\begin{vmatrix}
			\epsilon_1^{2}-\epsilon_1^{-2}\hspace{0.1cm} \ \		\epsilon_1^{1}-\epsilon_1^{-1}\\
			\epsilon_2^{2}-\epsilon_2^{-2}\hspace{0.1cm} \ \		\epsilon_2^{1}-\epsilon_2^{-1}\\
			\end{vmatrix}} = \epsilon_{1}+\frac{1}{\epsilon_{1}}+\epsilon_{2}+\frac{1}{\epsilon_{2}} .
		\end{eqnarray}} 
	
	\vspace{-0.5cm}
	\item The Quintuplet Representation: $5 \equiv (0,1)$ 
	
	We obtain $\vec{\lambda}$ = $(1,1)$ and $\vec{r} = \vec{\lambda}+\vec{\rho} = (3,2)$ and the character is obtained as:
	
	\vspace{-0.6cm}
	{\scriptsize\begin{eqnarray}\label{eq:sp4-5-char}
		\hspace{-1cm}\chi_{({Sp(4)})_{5}}(\epsilon_1,\epsilon_2) &=& \frac{|\epsilon^{3}-\epsilon^{-3},\hspace{0.1cm}\epsilon^{2}-\epsilon^{-2}|} {|\epsilon^{2}-\epsilon^{-2},\hspace{0.1cm}\epsilon^{1}-\epsilon^{-1}|} = \frac{
			\begin{vmatrix}
			\epsilon_1^{3}-\epsilon_1^{-3}\hspace{0.1cm} \ \		\epsilon_1^{2}-\epsilon_1^{-2}\\
			\epsilon_2^{3}-\epsilon_2^{-3}\hspace{0.1cm} \ \		\epsilon_2^{2}-\epsilon_2^{-2}\\
			\end{vmatrix}}{
			\begin{vmatrix}
			\epsilon_1^{2}-\epsilon_1^{-2}\hspace{0.1cm} \ \		\epsilon_1^{1}-\epsilon_1^{-1}\\
			\epsilon_2^{2}-\epsilon_2^{-2}\hspace{0.1cm} \ \		\epsilon_2^{1}-\epsilon_2^{-1}\\
			\end{vmatrix}} = 1+\frac{1}{\epsilon_{1} \epsilon_{2}}+\frac{\epsilon_{1}}{\epsilon_{2}}+\frac{\epsilon_{2}}{\epsilon_{1}}+\epsilon_{1} \epsilon_{2}.
		\end{eqnarray}}	
	
	\vspace{-0.5cm}
	\item The Adjoint Representation: $10 \equiv (2,0)$ 
	
	We obtain $\vec{\lambda}$ = $(2,0)$ and $\vec{r} = \vec{\lambda}+\vec{\rho}$ = $(4,1)$	and the character is obtained as:
	
	\vspace{-0.6cm}
	{\scriptsize\begin{eqnarray}\label{eq:sp4-10-char}
		\hspace{-1.2cm}\chi_{({Sp(4)})_{10}}(\epsilon_1,\epsilon_2) &=& \frac{|\epsilon^{4}-\epsilon^{-4},\hspace{0.1cm}\epsilon^{1}-\epsilon^{-1}|} {|\epsilon^{2}-\epsilon^{-2},\hspace{0.1cm}\epsilon^{1}-\epsilon^{-1}|} 
		= \frac{
			\begin{vmatrix}
			\epsilon_1^{4}-\epsilon_1^{-4}\hspace{0.1cm} \ \		\epsilon_1^{1}-\epsilon_1^{-1}\\
			\epsilon_2^{4}-\epsilon_2^{-4}\hspace{0.1cm} \ \		\epsilon_2^{1}-\epsilon_2^{-1}\\
			\end{vmatrix}}{
			\begin{vmatrix}
			\epsilon_1^{2}-\epsilon_1^{-2}\hspace{0.1cm} \ \		\epsilon_1^{1}-\epsilon_1^{-1}\\
			\epsilon_2^{2}-\epsilon_2^{-2}\hspace{0.1cm} \ \		\epsilon_2^{1}-\epsilon_2^{-1}\\
			\end{vmatrix}}\nonumber\\
		&=& 2+\frac{1}{\epsilon_{1}^{2}}+\epsilon_{1}^{2}+\frac{1}{\epsilon_{2}^{2}}+\frac{1}{\epsilon_{1} \epsilon_{2}}+\frac{\epsilon_{1}}{\epsilon_{2}}+\frac{\epsilon_{2}}{\epsilon_{1}}+\epsilon_{1} \epsilon_{2}+\epsilon_{2}^{2}.
		\end{eqnarray}} 
\end{itemize}
\noindent \rule{15.5cm}{1.25pt}
\\
\\
\noindent
\fbox{$Sp(6)$}\label{examples:sp6-char-haar}\\
\\
\textbf{Haar Measure}
\\
The denominator for the character formula in this case is: 

\vspace{-0.6cm}
{\scriptsize\begin{eqnarray}\label{eq:vandermonde-sp6}
	\Delta\left(\epsilon\right) = \begin{vmatrix}
	\epsilon_1^{3}-\epsilon_1^{-3}\hspace{0.1cm} \ \ \epsilon_1^{2}-\epsilon_1^{-2}\hspace{0.1cm} \ \		\epsilon_1^{1}-\epsilon_1^{-1}\\
	\epsilon_2^{3}-\epsilon_2^{-3}\hspace{0.1cm} \ \ \epsilon_2^{2}-\epsilon_2^{-2}\hspace{0.1cm} \ \		\epsilon_2^{1}-\epsilon_2^{-1}\\
	\epsilon_3^{3}-\epsilon_3^{-3}\hspace{0.1cm} \ \ \epsilon_3^{2}-\epsilon_3^{-2}\hspace{0.1cm} \ \		\epsilon_3^{1}-\epsilon_3^{-1}\\
	\end{vmatrix}.
	\end{eqnarray}}
$\Delta(\epsilon^{-1})$ is obtained by replacing $\epsilon_i$ by $\epsilon_i^{-1}$ in the above expression. Then using Eq.~\eqref{eq:sphaar-measure},

\vspace{-0.5cm}
{\scriptsize\begin{eqnarray}\label{eq:sp6-haar-measure}
	d\mu_{Sp(6)} &=& \frac{1}{3! 2^{3}\left(2\pi i\right)^{3}} \frac{d\epsilon_1}{\epsilon_1}\frac{d\epsilon_2}{\epsilon_2}\frac{d\epsilon_3}{\epsilon_3}\Delta\left(\epsilon\right)\Delta\left(\epsilon^{-1}\right) \nonumber \\
	&=& \frac{1}{48\left(2\pi i\right)^{3}} \frac{d\epsilon_1}{\epsilon_1^{7}}\frac{d\epsilon_2}{\epsilon_2^{7}}\frac{d\epsilon_3}{\epsilon_3^{7}}\left(-1+\epsilon_{1}^{2}\right)^{2}\left(-1+\epsilon_{2}^{2}\right)^{2}\left(\epsilon_{2}+\epsilon_{1}^{2} \epsilon_{2}-\epsilon_{1}\left(1+\epsilon_{2}^{2}\right)\right)^{2}\left(-1+\epsilon_{2}^{2}\right)^{2} \nonumber \\
	& & \left(\epsilon_{3}+\epsilon_{1}^{2} \epsilon_{3}-\epsilon_{1}\left(1+\epsilon_{3}^{2}\right)\right)^{2}\left(\epsilon_{3}+\epsilon_{2}^{2} \epsilon_{3}-\epsilon_{2}\left(1+\epsilon_{3}^{2}\right)\right)^{2}.
	\end{eqnarray}}

\vspace{-0.7cm}\noindent \textbf{Characters}

\vspace{-0.4cm}
\begin{itemize}
	\item The Fundamental Representation: $6 \equiv (1,0,0)$ 
	
	Using Eq.~\eqref{eq:splambda-recursion}, we get $\vec{\lambda}$ = $(1,0,0)$. Now, since $\vec{\rho}$ = $(3,2,1)$ therefore,\\	$\vec{r} = \vec{\lambda}+\vec{\rho} = (4,2,1)$ and the character is obtained as:
	
	\vspace{-0.8cm}
	{\scriptsize\begin{eqnarray}\label{eq:sp6-6-char}
		\chi_{({Sp(6)})_{6}}(\epsilon_1,\epsilon_2,\epsilon_3) &=& \frac{|\epsilon^{4}-\epsilon^{-4},\hspace{0.1cm}\epsilon^{2}-\epsilon^{-2},\hspace{0.1cm}\epsilon^{1}-\epsilon^{-1}|} {|\epsilon^{3}-\epsilon^{-3},\hspace{0.1cm}\epsilon^{2}-\epsilon^{-2},\hspace{0.1cm}\epsilon^{1}-\epsilon^{-1}|} 
		= \frac{
			\begin{vmatrix}
			\epsilon_1^{4}-\epsilon_1^{-4}\hspace{0.1cm} \ \ \epsilon_1^{2}-\epsilon_1^{-2}\hspace{0.1cm} \ \		\epsilon_1^{1}-\epsilon_1^{-1}\\
			\epsilon_2^{4}-\epsilon_2^{-4}\hspace{0.1cm} \ \ \epsilon_2^{2}-\epsilon_2^{-2}\hspace{0.1cm} \ \		\epsilon_2^{1}-\epsilon_2^{-1}\\
			\epsilon_3^{4}-\epsilon_3^{-4}\hspace{0.1cm} \ \ \epsilon_3^{2}-\epsilon_3^{-2}\hspace{0.1cm} \ \		\epsilon_3^{1}-\epsilon_3^{-1}\\
			\end{vmatrix}}{
			\begin{vmatrix}
			\epsilon_1^{3}-\epsilon_1^{-3}\hspace{0.1cm} \ \ \epsilon_1^{2}-\epsilon_1^{-2}\hspace{0.1cm} \ \		\epsilon_1^{1}-\epsilon_1^{-1}\\
			\epsilon_2^{3}-\epsilon_2^{-3}\hspace{0.1cm} \ \ \epsilon_2^{2}-\epsilon_2^{-2}\hspace{0.1cm} \ \		\epsilon_2^{1}-\epsilon_2^{-1}\\
			\epsilon_3^{3}-\epsilon_3^{-3}\hspace{0.1cm} \ \ \epsilon_3^{2}-\epsilon_3^{-2}\hspace{0.1cm} \ \		\epsilon_3^{1}-\epsilon_3^{-1}\\
			\end{vmatrix}} \nonumber \\
		&=& \epsilon_{1}+\frac{1}{\epsilon_{1}}+\epsilon_{2}+\frac{1}{\epsilon_{2}}+\epsilon_{3}+\frac{1}{\epsilon_{3}}.
		\end{eqnarray}} 
	
	\vspace{-0.6cm}
	\item The 14-dimensional Representation: $14 \equiv (0,1,0)$ 
	
	We obtain $\vec{\lambda}$ = $(1,1,0)$ and $\vec{r} = \vec{\lambda}+\vec{\rho}$ = $(4,3,1)$ and the character is obtained as:
	
	\vspace{-0.5cm}
	{\scriptsize\begin{eqnarray}\label{eq:sp6-14-char}
		\chi_{({Sp(6)})_{14}}(\epsilon_1,\epsilon_2,\epsilon_3) &=& \frac{|\epsilon^{4}-\epsilon^{-4},\hspace{0.1cm}\epsilon^{3}-\epsilon^{-3},\hspace{0.1cm}\epsilon^{1}-\epsilon^{-1}|} {|\epsilon^{3}-\epsilon^{-3},\hspace{0.1cm}\epsilon^{2}-\epsilon^{-2},\hspace{0.1cm}\epsilon^{1}-\epsilon^{-1}|}
		= \frac{
		\begin{vmatrix}
		\epsilon_1^{4}-\epsilon_1^{-4}\hspace{0.1cm} \ \ \epsilon_1^{3}-\epsilon_1^{-3}\hspace{0.1cm} \ \		\epsilon_1^{1}-\epsilon_1^{-1}\\
		\epsilon_2^{4}-\epsilon_2^{-4}\hspace{0.1cm} \ \ \epsilon_2^{3}-\epsilon_2^{-3}\hspace{0.1cm} \ \		\epsilon_2^{1}-\epsilon_2^{-1}\\
		\epsilon_3^{4}-\epsilon_3^{-4}\hspace{0.1cm} \ \ \epsilon_3^{3}-\epsilon_3^{-3}\hspace{0.1cm} \ \		\epsilon_3^{1}-\epsilon_3^{-1}\\
		\end{vmatrix}}{
		\begin{vmatrix}
		\epsilon_1^{3}-\epsilon_1^{-3}\hspace{0.1cm} \ \ \epsilon_1^{2}-\epsilon_1^{-2}\hspace{0.1cm} \ \		\epsilon_1^{1}-\epsilon_1^{-1}\\
		\epsilon_2^{3}-\epsilon_2^{-3}\hspace{0.1cm} \ \ \epsilon_2^{2}-\epsilon_2^{-2}\hspace{0.1cm} \ \		\epsilon_2^{1}-\epsilon_2^{-1}\\
		\epsilon_3^{3}-\epsilon_3^{-3}\hspace{0.1cm} \ \ \epsilon_3^{2}-\epsilon_3^{-2}\hspace{0.1cm} \ \		\epsilon_3^{1}-\epsilon_3^{-1}\\
		\end{vmatrix}} \\
		&=& 2+\frac{1}{\epsilon_{1} \epsilon_{2}}+\frac{\epsilon_{1}}{\epsilon_{2}}+\frac{\epsilon_{2}}{\epsilon_{1}}+\frac{1}{\epsilon_{1} \epsilon_{3}}+\frac{\epsilon_{1}}{\epsilon_{3}}+\frac{1}{\epsilon_{2} \epsilon_{3}}+\frac{\epsilon_{2}}{\epsilon_{3}}+\frac{\epsilon_{3}}{\epsilon_{1}}+\frac{\epsilon_{3}}{\epsilon_{2}}+\epsilon_{1} \epsilon_{2}+\epsilon_{1}\epsilon_{3}+\epsilon_{2} \epsilon_{3}.\nonumber
		\end{eqnarray}} 		
	\item The Adjoint Representation: $21 \equiv (2,0,0)$ 
	
	We obtain $\vec{\lambda}$ = $(2,0,0)$ and $\vec{r} = \vec{\lambda}+\vec{\rho}$ = $(5,2,1)$ and the character is obtained as:
	
	\vspace{-0.6cm}
	{\scriptsize\begin{eqnarray}\label{eq:sp6-21-char}
		\chi_{({Sp(6)})_{21}}(\epsilon_1,\epsilon_2,\epsilon_3) &=& \frac{|\epsilon^{5}-\epsilon^{-5},\hspace{0.1cm}\epsilon^{2}-\epsilon^{-2},\hspace{0.1cm}\epsilon^{1}-\epsilon^{-1}|} {|\epsilon^{3}-\epsilon^{-3},\hspace{0.1cm}\epsilon^{2}-\epsilon^{-2},\hspace{0.1cm}\epsilon^{1}-\epsilon^{-1}|} 
		= \frac{
			\begin{vmatrix}
			\epsilon_1^{5}-\epsilon_1^{-5}\hspace{0.1cm} \ \ \epsilon_1^{2}-\epsilon_1^{-2}\hspace{0.1cm} \ \		\epsilon_1^{1}-\epsilon_1^{-1}\\
			\epsilon_2^{5}-\epsilon_2^{-5}\hspace{0.1cm} \ \ \epsilon_2^{2}-\epsilon_2^{-2}\hspace{0.1cm} \ \		\epsilon_2^{1}-\epsilon_2^{-1}\\
			\epsilon_3^{5}-\epsilon_3^{-5}\hspace{0.1cm} \ \ \epsilon_3^{2}-\epsilon_3^{-2}\hspace{0.1cm} \ \		\epsilon_3^{1}-\epsilon_3^{-1}\\
			\end{vmatrix}}{
			\begin{vmatrix}
			\epsilon_1^{3}-\epsilon_1^{-3}\hspace{0.1cm} \ \ \epsilon_1^{2}-\epsilon_1^{-2}\hspace{0.1cm} \ \		\epsilon_1^{1}-\epsilon_1^{-1}\\
			\epsilon_2^{3}-\epsilon_2^{-3}\hspace{0.1cm} \ \ \epsilon_2^{2}-\epsilon_2^{-2}\hspace{0.1cm} \ \		\epsilon_2^{1}-\epsilon_2^{-1}\\
			\epsilon_3^{3}-\epsilon_3^{-3}\hspace{0.1cm} \ \ \epsilon_3^{2}-\epsilon_3^{-2}\hspace{0.1cm} \ \		\epsilon_3^{1}-\epsilon_3^{-1}\\
			\end{vmatrix}}\nonumber \\
		&=& 3+\frac{1}{\epsilon_{1}^{2}}+\epsilon_{1}^{2}+\frac{1}{\epsilon_{2}^{2}}+\frac{1}{\epsilon_{1} \epsilon_{2}}+\frac{\epsilon_{1}}{\epsilon_{2}}+\frac{\epsilon_{2}}{\epsilon_{1}}+\epsilon_{1} \epsilon_{2}+\epsilon_{2}^{2}+\frac{1}{\epsilon_{3}^{2}}+\frac{1}{\epsilon_{1} \epsilon_{3}} \nonumber \\
		& & +\frac{\epsilon_{1}}{\epsilon_{3}}+\frac{1}{\epsilon_{2} \epsilon_{3}}+\frac{\epsilon_{2}}{\epsilon_{3}}+\frac{\epsilon_{3}}{\epsilon_{1}}+\epsilon_{1} \epsilon_{3}+\frac{\epsilon_{3}}{\epsilon_{2}}+\epsilon_{2} \epsilon_{3}+\epsilon_{3}^{2}.
		\end{eqnarray}} 
\end{itemize}
\noindent \rule{15.5cm}{1.25pt}
\newpage
\subsection{The curious case of the non-compact Lorentz group}\label{subsec:hs-lorentz-group}

The quantum fields under consideration are dynamical in nature. Thus to perform a gauge invariant operator construction we need to include the  covariant derivative $\mathcal{D}_{\mu}$ in a consistent way. We have devoted this section to address that. We know that $\mathcal{D}_{\mu}$ transforms trivially under the internal symmetry groups while it has a non-trivial transformation property under the space-time transformations. Again, unlike the quantum fields we can not simply treat it as another degree of freedom because such an inclusion can introduce redundancies within the operator sets by virtue of integration by parts (IBP) and equation of motion of fields (EOM). It has been discussed in \cite{Grzadkowski:2010es,Henning:2015alf,Anisha:2019nzx} how some of the effective operator structures can be removed in favor of others by paying close attention to IBP and EOM. Thus  the incorporation of $\mathcal{D}_{\mu}$ is a highly involved task. The quantum fields carrying non-zero spins of any model as well as $\mathcal{D}_{\mu}$ transform non-trivially under the Lorentz group, i.e., the group of space-time transformations in $3+1$ dimensions. For the sake of the computation, we prefer to work with finite dimensional unitary representations. Thus, instead of working with non-compact Lorentz group $SO(3,1)$ we choose to work with the Euclidean conformal group $SO(4,\mathbb{C})$. It has been noted that in case of dimensions $d = p + q \geq 3$ dimensions, the conformal group is $SO(p + 1, q + 1)$. Thus for $d=3+1$ dimensions, the conformal group is $SO(4,2)$. While writing $SO(3,1)\cong SO(4,\mathbb{C})$ we recognize the fact that the presence of  2 extra dimensions increases the rank by one unit and this manifests itself as the scaling dimension ($\Delta_{\varphi}$) of the representation. We recall, here, that the Lorentz group is non-compact  and its unitary representations are infinite dimensional. Therefore, we will realize $SO(4,\mathbb{C})$ as $SU(2)\times SU(2)$ or more appropriately $SU(2)_l\times SU(2)_r$. In addition to that  we will be working in the Weyl(chiral) basis instead of the Dirac basis for the spinors. The similar treatment is extended to the field strength tensors where instead of $F_{\mu\nu}$ and  $\tilde{F}_{\mu\nu}$, we will work with $Fl$ and $Fr$:

\vspace{-0.5cm}
{\scriptsize\begin{eqnarray}
Fl = \frac{1}{2}\left(F^{\mu\nu}+i\tilde{F}^{\mu\nu}\right),\hspace{0.4cm}
Fr = \frac{1}{2}\left(F^{\mu\nu}-i\tilde{F}^{\mu\nu}\right).
\end{eqnarray}}
which transform under the $(1,0)$ and $(0,1)$ representations of $SU(2)_l\times SU(2)_r$ respectively.

The next step is the identification of our physical fields, i.e., scalars, spinors and vectors as Unitary Irreducible Representations (UIRs) of the conformal group. These representations can be categorized into long and short representations based on whether they satisfy certain unitarity bounds defined by the scaling dimension and the highest weight of the representation \cite{Ferrara:2000nu,Barabanschikov:2005ri,Dolan:2005wy,Siegel:1988gd,Gruber:1975sn,Dobrev:2004tk,Bourget:2017kik,Henning:2015alf,Henning:2015daa,Henning:2017fpj,Minwalla:1997ka}. Once a representation is identified as either long or short, the next step is to express the $SO(4,\mathbb{C})$ characters as a linear combination of $SU(2)\times SU(2)$ characters. The detailed procedure, as well as the precise meaning of each term, is given in \cite{Dolan:2005wy,Barabanschikov:2005ri,Minwalla:1997ka,Henning:2017fpj}. Instead of repeating the detailed computations,  we have provided the characters relevant for our analysis involving fields of spins-0, -1/2, and -1:

\vspace{-0.7cm}
{\scriptsize\begin{eqnarray}\label{eq:lorentz-char}
	\chi^{(4)}_{[1;(0,0)]}(\mathcal{D}, \alpha, \beta) &=&\mathcal{D} P^{(4)}(\mathcal{D}, \alpha, \beta) \times \Big[ 1 - \mathcal{D}^2 \Big],\nonumber\\
	\chi^{(4)}_{[\frac{3}{2},(\frac{1}{2},0)]}(\mathcal{D}, \alpha, \beta) &=& \mathcal{D}^{\frac{3}{2}} P^{(4)}(\mathcal{D}, \alpha, \beta) \times \Big[\alpha + \frac{1}{\alpha} -\mathcal{D}\Big(\beta + \frac{1}{\beta}\Big)\Big],\nonumber\\
	\chi^{(4)}_{[\frac{3}{2};(0,\frac{1}{2})]}(\mathcal{D}, \alpha, \beta) &=& \mathcal{D}^{\frac{3}{2}} P^{(4)}(\mathcal{D}, \alpha, \beta) \times \Big[\beta + \frac{1}{\beta} -\mathcal{D}\Big(\alpha + \frac{1}{\alpha}\Big)\Big],\nonumber\\
	\chi^{(4)}_{[2;(1,0)]}(\mathcal{D}, \alpha, \beta) &=& \mathcal{D}^2 P^{(4)}(\mathcal{D},\alpha, \beta) \times \Big[\alpha^2+\frac{1}{\alpha ^2}+1 - \mathcal{D}\Big(\alpha +\frac{1}{\alpha }\Big) \Big(\beta +\frac{1}{\beta }\Big) + \mathcal{D}^2\Big],\nonumber\\
	\chi^{(4)}_{[2;(0,1)]}(\mathcal{D}, \alpha, \beta) &=& \mathcal{D}^2 P^{(4)}(\mathcal{D},\alpha, \beta) \times \Big[ \beta^2+\frac{1}{\beta ^2}+1 - \mathcal{D}\Big(\alpha +\frac{1}{\alpha }\Big) \Big(\beta +\frac{1}{\beta }\Big) + \mathcal{D}^2\Big],    
	\end{eqnarray}}
where the subscripts on the LHS $[\Delta_{\varphi},(j_1,j_2)]$ contain information about the scaling dimension ($\Delta_{\varphi}$) and the representation under $SU(2)\times SU(2)$ as $(j_1, j_2)$. $P^{(4)}(\mathcal{D}, \alpha, \beta)$ is the momentum generating function which can be written as \cite{Dolan:2005wy,Henning:2015alf,Henning:2017fpj}:

\vspace{-0.5cm}
{\scriptsize\begin{eqnarray}\label{eq:momentum-gen-func}
	P^{(4)}(\mathcal{D}, \alpha, \beta) = \Bigg[\left(1-\mathcal{D}\alpha  \beta\right) \left(1-\frac{\mathcal{D}}{\alpha  \beta }\right) \left(1-\frac{\mathcal{D}\alpha}{\beta}\right) \left(1-\frac{\mathcal{D}\beta}{\alpha}\right)\Bigg]^{-1}.
	\end{eqnarray}}
The removal of redundancies in this construction due to the EOM and IBP are discussed in Refs. \cite{Henning:2017fpj,Henning:2015alf} in great detail. Next, the PE in Eq.~\eqref{eq:PE} also gets modified as \cite{Henning:2015alf}:

\vspace{-0.3cm}
{\scriptsize\begin{eqnarray}\label{eq:PE-modified}
	PE[\phi,\mathcal{D}, R] &=& \exp\left[\sum_{r=1}^{\infty}  \left(\frac{\phi}{\mathcal{D}^{\Delta_{\phi}} }\right)^r \frac{\chi_R(z^r_j,\alpha^r,\beta^r)}{r}\right],\\
	PE[\psi, \mathcal{D},R] &=& \exp\left[  \sum_{r=1}^{\infty}(-1)^{r+1}  \left(\frac{\psi}{\mathcal{D}^{\Delta_{\psi}} }\right)^r \frac{\chi_R(z^r_j,\alpha^r,\beta^r)}{ r}\right],
	\end{eqnarray}}

\vspace{-0.4cm}\noindent for bosons and fermions respectively.\\
\noindent
\\
\textbf{Haar Measure for the Lorentz Group}
\\
\\
Since the Lorentz group is non-compact the Haar measure is not defined for it. So, to enable the group space integration we transform Minkowskian $SO(3,1)$ to the Euclidean $SO(4,\mathbb{C})$ which is further decomposed into $SU(2)\times SU(2)$. Based on the knowledge of the Haar measure for $SU(2)$ Eq.~\eqref{eq:su2-haar-measure} the Haar measure for the Lorentz group is depicted as:

\vspace{-0.3cm}
{\scriptsize\begin{eqnarray}\label{eq:lorentz-haar-measure}
	\int d\mu_{Lorentz} &=& \int \frac{1}{P^{(4)}(\mathcal{D}, \alpha, \beta)} d\mu_{SU(2)}\times d\mu_{SU(2)}\\
	&=& \frac{1}{P^{(4)}(\mathcal{D}, \alpha, \beta)} \times \left[\frac{1}{2\left(2\pi i\right)} \frac{d\alpha}{\alpha}\left(1-\alpha^2\right)\left(1-\frac{1}{\alpha^2}\right)\right]\times\left[\frac{1}{2\left(2\pi i\right)} \frac{d\beta}{\beta}\left(1-\beta^2\right)\left(1-\frac{1}{\beta^2}\right)\right].\nonumber
	\end{eqnarray}}
The incorporation of the derivative operators and consequently the Lorentz group modifies the Hilbert Series in Eq.~\eqref{eq:HS} to the following form
\cite{Henning:2017fpj,Henning:2015alf,Henning:2015daa}:

\vspace{-0.3cm}
{\scriptsize\begin{eqnarray}\label{eq:HS-modified}
	\mathcal{H}[\varphi] = \int_{LG} \hspace{0.05cm} {d\mu_{SU(2)}\times d\mu_{SU(2)}} {\hspace{0.1cm}} \frac{1}{    P^{(4)}(\mathcal{D}, \alpha, \beta)} \prod^{n}_{j=1} \int_{\mathcal{G}_j} \hspace{0.1cm}{d\mu_j} \hspace{0.1cm}{PE[\varphi, \mathcal{D}, R]},
	\end{eqnarray}}
where, $\mathcal{D}$ is the spurion variable symbolizing the covariant derivative operator.

%% file: KnownLagrangians.tex
\section{Invariant Polynomial: Paving the Path to Lagrangian}\label{sec:hilbert-series-examples}

\subsection{Two Higgs Doublet Model (2HDM)}\label{subsec:2hdm}
The Two Higgs Doublet Model (2HDM) is a minimal extension of the Standard Model (SM) content through an additional $SU(2)$ complex doublet scalar \cite{Gunion:1989we,Gunion:2002zf,Branco:2011iw,Carena:2013ooa,Haber:2013mia,Chen:2013jvg,Mrazek:2011iu,Dev:2014yca,Bhattacharyya:2015nca,Crivellin:2015hha,Crivellin:2016ihg}. Here, we have considered a generic 2HDM scenario without imposing any $\mathbb{Z}_2$ symmetry. We start the discussion by first giving the full field content and their transformation properties under the gauge group $SU(3)_C\otimes SU(2)_L\otimes U(1)_Y$ and the Lorentz group in Table~\ref{table:2hdm-quantum-no}. Based on this information our aim is to the invariant polynomial, i.e., the Lagrangian for 2HDM.
\\
\begin{table}[h]
	\centering
	\renewcommand{\arraystretch}{1.8}
	{\tiny\begin{tabular}{|c|c|c|c|c|}
			\hline
			\textbf{2HDM Fields}&
			$SU(3)_C$&
			$SU(2)_L$&
			$U(1)_Y$&
			\textbf{Spin}\\
			\hline
			
			$\phi_1$&
			1&
			2&
			1/2&
			0\\
			
			$\phi_2$&
			1&
			2&
			1/2&
			0 \\
			
			$Q^p_L$&
			3&
			2&
			1/6&
			1/2 \\
			
			$u^p_R$&
			3&
			1&
			2/3&
			1/2\\
			
			$d^p_R$&
			3&
			1&
			-1/3&
			1/2\\
			
			$L^p_L$&
			1&
			2&
			-1/2&
			1/2 \\
			
			$e^p_R$&
			1&
			1&
			-1&
			1/2\\
			
			$B_{\mu\nu}$&
			1&
			1&
			0&
			1\\
			
			$W_{\mu\nu}^I$&
			1&
			3&
			0&
			1\\
			
			$G_{\mu\nu}^a$&
			8&
			1&
			0&
			1\\
			
			\hline
			\hline
			$\mathcal{D}_\mu$&
			\multicolumn{4}{c|}{\textbf{Covariant Derivative}}\\
			\hline
			
	\end{tabular}}
	\caption{\small 2HDM: Quantum numbers of fields under the gauge groups and their spins under the Lorentz group. $I$ = 1,2,3 and $a$ = 1,2,.....,8 correspond to $SU(2)$ and $SU(3)$ gauge indices respectively, and $p$ = 1,...,$N_f$ denotes the flavor index. The color and isospin indices have been suppressed. $L$ and $R$ denote the chirality, i.e., the left or right handedness of the field.}
	\label{table:2hdm-quantum-no}
\end{table} 

\noindent
\textbf{Gauge Group Characters and Haar Measure}
\\
The first step is to compute the characters corresponding to each field as well as its conjugate. The characters of the relevant representations of $SU(3)$, i.e., the $1$, $3$ and $8$-dimensional representations and $SU(2)$, i.e., the $1$, $2$ and $3$-dimensional representations have been computed in subsection \ref{examples:su2-char-haar}. Also, the $U(1)$ characters can be obtained using Eq.~\eqref{eq:u1-char}. Multiplying together the characters of representations under different groups one can obtain the total gauge group character for a field. One must also note that the Haar measures of these groups, which are important to carry out integration over the group space have been computed in Eqs.~\eqref{eq:u1-haar-measure}, \eqref{eq:su2-haar-measure} and \eqref{eq:su3-haar-measure}. \\
\\
\noindent
\textbf{Lorentz Characters and Haar Measure}
\\
Now, 2HDM contains only particles with spins-0, -1/2, and -1 particles. So, the relevant Lorentz characters are the ones given in Eq.~\eqref{eq:lorentz-char}. Multiplying together the gauge group characters and Lorentz characters we obtain the total character of each field. 
\\
\\
\textbf{Argument of the Plethystic Exponential}
\\
Having obtained the total character, the next step is to construct the Plethystic Exponential. The argument of the Plethystic Exponential is an infinite sum whose general term is the total character of a particular field weighted by a spurion denoting the field name. The variables parametrizing the characters as well as the spurion are raised to the power $r$, where $r$ runs from $1$ to $\infty$. Below we have enlisted the contributions to the argument of the Plethystic Exponential corresponding to each field, 

\vspace{-0.6cm}
{\scriptsize\begin{eqnarray}
	&Bl&\rightarrow\frac{1}{r} Bl^r \hspace{0.1cm} \chi^{(4)}_{\left(1,0\right)}\left(\mathcal{D}^r,\alpha^r,\beta^r\right), \hspace{2.85cm} Br\rightarrow \frac{1}{r} Br^r \hspace{0.1cm} \chi^{(4)}_{\left(0,1\right)} \left(\mathcal{D}^r,\alpha^r,\beta^r\right),  \nonumber\\
	&Wl&\rightarrow\frac{1}{r} Wl^r \hspace{0.1cm} \chi^{(4)}_{\left(1,0\right)}\left(\mathcal{D}^r,\alpha^r,\beta^r\right) \hspace{0.1cm} \chi_{({SU(2)})_{3}}(z_1^r), \hspace{0.8cm} Wr\rightarrow \frac{1}{r} Wr^r \hspace{0.1cm} \chi^{(4)}_{\left(0,1\right)} \left(\mathcal{D}^r,\alpha^r,\beta^r\right) \hspace{0.1cm} \chi_{({SU(2)})_{3}}(z_1^r),  \nonumber\\
	&Gl&\rightarrow\frac{1}{r} Gl^r \hspace{0.1cm} \chi^{(4)}_{\left(1,0\right)}\left(\mathcal{D}^r,\alpha^r,\beta^r\right) \hspace{0.1cm} \chi_{({SU(3)})_{8}}(z_1^r, z_2^r), \hspace{0.5cm} Gr\rightarrow \frac{1}{r} Gr^r \hspace{0.1cm} \chi^{(4)}_{\left(0,1\right)} \left(\mathcal{D}^r,\alpha^r,\beta^r\right) \hspace{0.1cm} \chi_{({SU(3)})_{8}}(z_1^r, z_2^r),  \nonumber\\
	&\phi_1&\rightarrow\frac{1}{r} \phi_1^r \hspace{0.1cm} \chi^{(4)}_{\left(0,0\right)}\left(\mathcal{D}^r,\alpha^r,\beta^r\right) \hspace{0.1cm} \chi_{({SU(2)})_{2}}(z_1^r) \hspace{0.1cm} z^{1/2},\hspace{0.45cm} \phi_1^{\dagger}\rightarrow \frac{1}{r} (\phi_1^{\dagger})^r \hspace{0.1cm} \chi^{(4)}_{\left(0,0\right)} \left(\mathcal{D}^r,\alpha^r,\beta^r\right) \hspace{0.1cm} \chi_{({SU(2)})_{\overline{2}}}(z_1^r) \hspace{0.1cm} z^{-r/2}, \nonumber\\
	&\phi_2&\rightarrow\frac{1}{r} \phi_2^r \hspace{0.1cm} \chi^{(4)}_{\left(0,0\right)}\left(\mathcal{D}^r,\alpha^r,\beta^r\right) \hspace{0.1cm} \chi_{({SU(2)})_{2}}(z_1^r) \hspace{0.1cm} z^{r/2},\hspace{0.45cm} \phi_2^{\dagger}\rightarrow \frac{1}{r} (\phi_2^{\dagger})^r \hspace{0.1cm} \chi^{(4)}_{\left(0,0\right)}\left(\mathcal{D}^r,\alpha^r,\beta^r\right) \hspace{0.1cm} \chi_{({SU(2)})_{\overline{2}}}(z_1^r)\hspace{0.1cm} z^{-r/2}, \nonumber\\
	&e&\rightarrow\frac{(-1)^{r+1}}{r} e^r \hspace{0.1cm} \chi^{(4)}_{\left(0,\frac{1}{2}\right)}\left(\mathcal{D}^r,\alpha^r,\beta^r\right) \hspace{0.1cm} z^{-r}, \hspace{1.4cm} e^{\dagger}\rightarrow \frac{(-1)^{r+1}}{r} (e^{\dagger})^r \hspace{0.1cm} \chi_{\left(\frac{1}{2},0\right)}\left(\mathcal{D}^r,\alpha^r,\beta^r\right) \hspace{0.1cm} z^{r}, \nonumber\\
	&u&\rightarrow\frac{(-1)^{r+1}}{r} u^r \hspace{0.1cm} \chi^{(4)}_{\left(0,\frac{1}{2}\right)}\left(\mathcal{D}^r,\alpha^r,\beta^r\right) \hspace{0.1cm} \chi^{(4)}_{({SU(3)})_{3}}(z_2^r, z_3^r) \hspace{0.1cm} z^{2r/3}, \nonumber\\
	&u^{\dagger}&\rightarrow \frac{(-1)^{r+1}}{r} (u^{\dagger})^r \hspace{0.1cm} \chi^{(4)}_{\left(\frac{1}{2},0\right)}\left(\mathcal{D}^r,\alpha^r,\beta^r\right) \hspace{0.1cm} \chi_{({SU(3)})_{\overline{3}}}(z_2^r, z_3^r) \hspace{0.1cm} z^{-2r/3}, \nonumber\\
	&d&\rightarrow\frac{(-1)^{r+1}}{r} d^r \hspace{0.1cm} \chi^{(4)}_{\left(0,\frac{1}{2}\right)}\left(\mathcal{D}^r,\alpha^r,\beta^r\right) \hspace{0.1cm} \chi_{({SU(3)})_{3}}(z_2^r, z_3^r) \hspace{0.1cm} z^{-r/3}, \nonumber\\
	&d^{\dagger}&\rightarrow \frac{(-1)^{r+1}}{r} (d^{\dagger})^r \hspace{0.1cm} \chi^{(4)}_{\left(\frac{1}{2},0\right)}\left(\mathcal{D}^r,\alpha^r,\beta^r\right) \hspace{0.1cm} \chi_{({SU(3)})_{\overline{3}}}(z_2^r, z_3^r) \hspace{0.1cm} z^{r/3}, \nonumber\\
	&L&\rightarrow\frac{(-1)^{r+1}}{r} L^r \hspace{0.1cm} \chi^{(4)}_{\left(\frac{1}{2},0\right)}\left(\mathcal{D}^r,\alpha^r,\beta^r\right) \hspace{0.1cm} \chi_{({SU(2)})_{2}}(z_1^r) \hspace{0.1cm} z^{-r/2},\nonumber\\ 
	&L^{\dagger}&\rightarrow \frac{(-1)^{r+1}}{r} (L^{\dagger})^r \hspace{0.1cm} \chi^{(4)}_{\left(0,\frac{1}{2}\right)}\left(\mathcal{D}^r,\alpha^r,\beta^r\right) \hspace{0.1cm} \chi_{({SU(2)})_{\overline{2}}}(z_1^r)\hspace{0.1cm} z^{r/2}, \nonumber\\
	&Q&\rightarrow\frac{(-1)^{r+1}}{r} Q^r \hspace{0.1cm} \chi^{(4)}_{\left(\frac{1}{2},0\right)}\left(\mathcal{D}^r,\alpha^r,\beta^r\right) \hspace{0.1cm} \chi_{({SU(3)})_{3}}(z_2^r, z_3^r) \hspace{0.1cm} \chi_{({SU(2)})_{2}}(z_1^r) \hspace{0.1cm} z^{r/6},\nonumber\\ 
	&Q^{\dagger}&\rightarrow \frac{(-1)^{r+1}}{r} (Q^{\dagger})^r \hspace{0.1cm} \chi^{(4)}_{\left(0,\frac{1}{2}\right)}\left(\mathcal{D}^r,\alpha^r,\beta^r\right) \hspace{0.1cm} \chi_{({SU(3)})_{\overline{3}}}(z_2^r, z_3^r) \hspace{0.1cm} \chi_{({SU(2)})_{\overline{2}}}(z_1^r)\hspace{0.1cm} z^{-r/6}.
	\end{eqnarray}}
Having thus constructed the Plethystic Exponential and with the knowledge of the momentum generating function Eq.~\eqref{eq:momentum-gen-func} and Haar measures of the gauge groups and the Lorentz group, we can compute the Hilbert Series using Eq.~\eqref{eq:HS-modified}.
From the full Hilbert Series we have filtered out the output based on the mass dimension of the fields and suitably categorized them. For operators up to mass dimension-4, we have provided the proper scheme to translate them into a covariant form in Table~\ref{table:2hdm-raw-to-covariant-transl} because the Hilbert Series construction is oblivious to parameters such as coupling constants as well as to the presence of invariant tensors such as $\gamma^{\mu}, \sigma_{\mu\nu}$ in the Lagrangian.
\begin{table}[h]
	\centering
	\renewcommand{\arraystretch}{1.8}
	{\tiny\begin{tabular}{|ccc|ccc|ccc|}
			\hline
			$Q$&
			$\rightarrow$&
			$Q^p_L$&
			$u$&
			$\rightarrow$&
			$u^p_R$&
			$d$&
			$\rightarrow$&
			$d^p_R$\\
			
			$Q^{\dagger}$&
			$\rightarrow$&
			$\overline{Q}^p_L$&
			$u^{\dagger}$&
			$\rightarrow$&
			$\overline{u}^p_R$&
			$d^{\dagger}$&
			$\rightarrow$&
			$\overline{d}^p_R$\\
			
			$L$&
			$\rightarrow$&
			$L^p_L$&
			$e$&
			$\rightarrow$&
			$e^p_R$&
			$\phi_{1,2}, \phi^{\dagger}_{1,2}$&
			$\rightarrow$&
			$\phi_{1,2}, \phi^{\dagger}_{1,2}$\\
			
			$L^{\dagger}$&
			$\rightarrow$&
			$\overline{L}^p_L$&
			$e^{\dagger}$&
			$\rightarrow$&
			$\overline{e}^p_R$&
			$\mathcal{D}$&
			$\rightarrow$&
			$\mathcal{D}_{\mu}$\\
			
			$(Wl, Wr)$&
			$\rightarrow$&
			$(W^I_{\mu\nu},\tilde{W}^I_{\mu\nu})$&
			$(Gl, Gr)$&
			$\rightarrow$&
			$(G^a_{\mu\nu},\tilde{G}^a_{\mu\nu})$&
			$(Bl, Br)$&
			$\rightarrow$&
			$(B_{\mu\nu},\tilde{B}_{\mu\nu})$\\
			\hline
	\end{tabular}}
	\caption{\small Dictionary for translation of operators from the Hilbert Series to their covariant forms.}
	\label{table:2hdm-raw-to-covariant-transl}
\end{table} 
In Table~\ref{table:2hdm-renorm-output} we have presented both the Hilbert Series output and the covariant form of operators up to mass dimension-4 for general $N_f$ (number of fermion flavours), side by side and we have categorized the operators based on whether they are constituted of purely SM fields, purely BSM fields, i.e., operators containing only the second doublet scalar or a mixture of the two. We have also catalogued higher dimensional effective operators up to dimension-6 in Table~\ref{table:2hdm-dim-5-6-op} and categorized the operators based on their composition in terms of scalars ($\phi$), fermions ($\psi$) and the field strength tensor of gauge bosons ($X$). Note, the covariant form of these operators (for $N_f=1$) were discussed in great detail in \cite{Anisha:2019nzx}. So, we do not repeat the same here. 
\begin{table}[h]
	\centering
	\renewcommand{\arraystretch}{2.0}
	{\tiny\begin{tabular}{|c|c|c|c|}
			\hline
			\multicolumn{4}{|c|}{\textbf{Mass Dimension-2}}\\
			\hline

			\multirow{2}{*}{\textbf{Operator Type}}&
			\multirow{2}{*}{\textbf{HS Output}}&
			\multirow{2}{*}{\textbf{Covariant Form}}&
			\multirow{2}{*}{\textbf{\textbf{No. of Operators}}}\\
			
			&
			&
			&
			\textbf{(including h.c.)}\\
			\hline
			\textbf{Pure SM}&
			$\phi_1^{\dagger}\phi_1$&
			$\phi_1^{\dagger}\phi_1$&
			1
			\\
			\hline
			
			\textbf{Pure BSM}&
			$\phi_2^{\dagger}\phi_2$&
			$\phi_2^{\dagger}\phi_2$&
			1
			\\
			\hline
			
			\textbf{Mixed}&
			$\color{blue}{\phi_1^{\dagger}\phi_2}$&
			$\color{blue}{\phi_1^{\dagger}\phi_2}$&
			2
			\\
			\hline
			\hline
			\multicolumn{4}{|c|}{\textbf{Mass Dimension-4}}\\
			\hline
			\multirow{2}{*}{\textbf{Operator Type}}&
			\multirow{2}{*}{\textbf{HS Output}}&
			\multirow{2}{*}{\textbf{Covariant Form}}&
			\multirow{2}{*}{\textbf{\textbf{No. of Operators}}}\\
			
			&
			&
			&
			\textbf{(including h.c.)}\\
			\hline
			\multirow{6}{*}{\textbf{Pure SM}}&
			$Bl^2 + Br^2,$&
			$B^{\mu\nu}B_{\mu\nu},\hspace{0.2cm} B^{\mu\nu}\tilde{B}_{\mu\nu},$&
			\multirow{6}{*}{$11N_f^2  +  8$}
			\\
			
			&
			$ Gl^2 + Gr^2,$&
			$G^{a\mu\nu}G^{a}_{\mu\nu},\hspace{0.2cm} G^{a\mu\nu}\tilde{G}^{a}_{\mu\nu},$&
			\\
			
			&
			$Wl^2 + Wr^2,$&
			$W^{I \mu\nu}W^{I}_{\mu\nu},\hspace{0.2cm} W^{I \mu\nu}\tilde{W}^{I}_{\mu\nu},$&
			\\
			
			&
			$\color{blue}{N_f^2L^{\dagger}\phi_1e},\hspace{0.1cm}\color{blue}{N_f^2Q^{\dagger}\phi_1d},\hspace{0.1cm}\color{blue}{N_f^2Q^{\dagger}\phi_1^{\dagger} u},$&
			$ \color{blue}{\overline{L}^r_L \phi_1e^s_R},\hspace{0.1cm} \color{blue}{\overline{Q}^r_L \phi_1d^s_R},\hspace{0.1cm} \color{blue}{\overline{Q}^r_L \tilde{\phi}_1 u^s_R},$&
			\\
			
			&
			${}^{\textcolor{black}{\clubsuit}}N_f^2Q^{\dagger}Q\mathcal{D},\hspace{0.1cm} N_f^2u^{\dagger}u\mathcal{D},\hspace{0.1cm} N_f^2d^{\dagger}d\mathcal{D},$&
			${}^{\textcolor{black}{\clubsuit}}\overline{Q}_L^{r}\slashed{\mathcal{D}}Q^{r}_L,\hspace{0.1cm}
			\overline{u}_R^{r}\slashed{\mathcal{D}}u^{r}_R,\hspace{0.1cm} \overline{d}_R^{r}\slashed{\mathcal{D}}d^{r}_R,$&
			\\
			
			&
			$N_f^2L^{\dagger}L\mathcal{D},\hspace{0.1cm} N_f^2e^{\dagger}e\mathcal{D},\hspace{0.1cm} \phi_1^{\dagger}\phi_1\mathcal{D}^2,\hspace{0.1cm}
			(\phi_1^{\dagger}\phi_1)^2$&
			$\overline{L}_L^{r}\slashed{\mathcal{D}}L^{r}_L,\hspace{0.1cm}
			\overline{e}_R^{r}\slashed{\mathcal{D}}e^{r}_R,\hspace{0.1cm} (\mathcal{D}_{\mu}\phi_1)^{\dagger}(\mathcal{D}^{\mu}\phi_1),\hspace{0.1cm}
			(\phi_1^{\dagger}\phi_1)^2$&
			\\
			\hline
			
			\textbf{Pure BSM}&
			$\phi_2^{\dagger}\phi_2 \mathcal{D}^2,\hspace{0.1cm} (\phi_2^{\dagger}\phi_2)^2$&
			$(\mathcal{D}_{\mu}\phi_2)^{\dagger}(\mathcal{D}^{\mu}\phi_2),\hspace{0.1cm} (\phi_2^{\dagger}\phi_2)^2$&
			2
			\\
			\hline
			
			\multirow{3}{*}{\textbf{Mixed}}&
			$\color{blue}{N_f^2L^{\dagger}\phi_2 e},\hspace{0.1cm} \color{blue}{N_f^2Q^{\dagger}\phi_2 d},\hspace{0.1cm} \color{blue}{N_f^2Q^{\dagger}\phi_2^{\dagger} u},$&
			$ \color{blue}{\overline{L}^r_L \phi_2 e^s_R},\hspace{0.1cm}  \color{blue}{\overline{Q}^r_L \phi_2 d^s_R},\hspace{0.1cm} \color{blue}{\overline{Q}^r_L \tilde{\phi}_2 u^s_R},$&
			\multirow{3}{*}{$6N_f^2  +  10$}\\
			
			&
			$\color{blue}{(\phi_1^{\dagger})^2(\phi_2)^2},\hspace{0.1cm} \color{blue}{(\phi_1^{\dagger})^2\phi_1\phi_2},\hspace{0.1cm} \color{blue}{(\phi_2^{\dagger})^2\phi_1\phi_2}$&
			$\color{blue}{(\phi_1^{\dagger}\phi_2)^2},\hspace{0.1cm} \color{blue}{(\phi_1^{\dagger}\phi_1)(\phi_1^{\dagger}\phi_2)},\hspace{0.1cm} \color{blue}{(\phi_2^{\dagger}\phi_1)(\phi_2^{\dagger}\phi_2)}$&\\

			&
			$2\phi_1^{\dagger}\phi_1\phi_2^{\dagger}\phi_2,\hspace{0.1cm}
			\hspace{0.1cm}\color{blue}{\phi_1^{\dagger}\phi_2 \mathcal{D}^2}$&
			$(\phi_1^{\dagger}\phi_1)(\phi_2^{\dagger}\phi_2), (\phi_1^{\dagger}\phi_2)(\phi_2^{\dagger}\phi_1),\hspace{0.1cm}
			\color{blue}{(\mathcal{D}_{\mu}\phi_1)^{\dagger}(\mathcal{D}^{\mu}\phi_2)}$& \\
			
			\hline
	\end{tabular}}
	\caption{\small 2HDM: Renormalizable operators as Hilbert Series output and their covariant form. Coefficients of each operator (which appear as functions of $N_f$) give the number of all possible operators with the same structure. The operators in blue have distinct hermitian conjugates which we have not written explicitly. Here, $I$ = 1,2,3 are $SU(2)$ indices; $a$ = 1,...,8 are $SU(3)$ indices and $r,s$ = $1,2,...,N_f$ are flavour indices which are summed over with the suitable coupling constants. ${\textcolor{black}{\clubsuit}}$ - In the Hilbert Series the fermion kinetic terms appear with a factor of $N_f^2$ but in the physical Lagrangian there is a flavour symmetry which forces the kinetic terms to be diagonal and the factor of $N_f^2$ is reduced to $N_f$.}
	\label{table:2hdm-renorm-output}
\end{table}
\newpage
\begin{table}[h]
	\centering
	\renewcommand{\arraystretch}{1.8}
	{\tiny\begin{tabular}{|c|l|}
			\hline
			\multicolumn{2}{|c|}{\textbf{Mass Dimension-5}}\\
			\hline
			\textbf{Operator Class}&
			\hspace{4cm}\textbf{Operators (in non-covariant form)}\\
			\hline
			$\Psi^2\Phi^2$&
			$\color{blue}{\frac{1}{2}(N_f^2 + N_f)\phi_1^2L^2,\hspace{0.3cm} \frac{1}{2}(N_f^2 + N_f)\phi_2^2L^2,\hspace{0.3cm}
				N_f^2\phi_1\phi_2L^2}$\\
			\hline
            \hline
            \multicolumn{2}{|c|}{\textbf{Mass Dimension-6}}\\
            \hline
            \textbf{Operator Class}&
            \hspace{4cm}\textbf{Operators (in non-covariant form)}\\
            \hline
			$X^3$&
			$ Wl^3,\hspace{0.2cm} Wr^3,\hspace{0.2cm} Gl^3,\hspace{0.2cm} Gr^3 \hspace{0.2cm}$\\
			\hline
			
			\multirow{2}{*}{$\Phi^6$}&
			$(\phi_1^{\dagger}\phi_1)^3, \hspace{0.2cm} (\phi_2^{\dagger}\phi_2)^3, \hspace{0.2cm}  2\phi_1^2(\phi_1^{\dagger})^2\phi_2\phi_2^{\dagger}, \hspace{0.2cm} 2\phi_1\phi_1^{\dagger}\phi_2^2(\phi_2^{\dagger})^2, \hspace{0.2cm}
			\color{blue}{(\phi_1^{\dagger})^3\phi_2^3, \hspace{0.2cm}
				\phi_1^2(\phi_1^{\dagger})^3\phi_2,\hspace{0.2cm} \phi_1(\phi_1^{\dagger})^3\phi_2^2,\hspace{0.2cm}}$\\
			
			&
			$\color{blue}{ (\phi_1^{\dagger})^2\phi_2^3\phi_2^{\dagger}, \hspace{0.2cm} \phi_1^{\dagger}\phi_2^3(\phi_2^{\dagger})^2, \hspace{0.2cm} 2\phi_1(\phi_1^{\dagger})^2\phi_2^2\phi_2^{\dagger} \hspace{0.2cm}} $\\
			\hline
			
			\multirow{2}{*}{$\Phi^2X^2$}&
			$\color{blue}{Gl^2\phi_1^{\dagger}\phi_1,\hspace{0.2cm}  Gl^2\phi_2^{\dagger}\phi_2,\hspace{0.2cm} Gl^2\phi_1^{\dagger}\phi_2,\hspace{0.2cm} Gl^2\phi_1\phi_2^{\dagger},\hspace{0.2cm} Wl^2\phi_1^{\dagger}\phi_1,\hspace{0.2cm} Wl^2\phi_2^{\dagger}\phi_2,\hspace{0.2cm} Wl^2\phi_1^{\dagger}\phi_2,\hspace{0.2cm} Wl^2\phi_1\phi_2^{\dagger},\hspace{0.2cm} }$\\
			
			&
			$\color{blue}{Bl^2\phi_1^{\dagger}\phi_1,\hspace{0.2cm} Bl^2\phi_2^{\dagger}\phi_2,\hspace{0.2cm} Bl^2\phi_1^{\dagger}\phi_2,\hspace{0.2cm}
			Bl^2\phi_1\phi_2^{\dagger},\hspace{0.2cm} BlWl\phi_1^{\dagger}\phi_1,\hspace{0.2cm} BlWl\phi_2^{\dagger}\phi_2,\hspace{0.2cm} BlWl\phi_1^{\dagger}\phi_2,\hspace{0.2cm} BlWl\phi_1\phi_2^{\dagger}\hspace{0.2cm}} $\\
			\hline
			
			\multirow{3}{*}{$\Psi^2\Phi X$}&
			$\color{blue}{   (N_f^2)Gl\phi_1^{\dagger}d^{\dagger}Q,\hspace{0.2cm} (N_f^2)Gl\phi_2^{\dagger}d^{\dagger}Q,\hspace{0.2cm}
			(N_f^2)Gl\phi_1u^{\dagger}Q,\hspace{0.2cm} (N_f^2)Gl\phi_2u^{\dagger}Q,\hspace{0.2cm} (N_f^2)Wl\phi_1^{\dagger}d^{\dagger}Q,\hspace{0.2cm} (N_f^2)Wl\phi_2^{\dagger}d^{\dagger}Q,\hspace{0.2cm}}$\\
			
			&
			$\color{blue}{(N_f^2)Wl\phi_1u^{\dagger}Q,\hspace{0.2cm} (N_f^2)Wl\phi_2u^{\dagger}Q,\hspace{0.2cm} (N_f^2)Wl\phi_1^{\dagger}e^{\dagger}L,\hspace{0.2cm} (N_f^2)Wl\phi_2^{\dagger}e^{\dagger}L,\hspace{0.2cm}
			(N_f^2)Bl\phi_1^{\dagger}d^{\dagger}Q,\hspace{0.2cm} (N_f^2)Bl\phi_2^{\dagger}d^{\dagger}Q,\hspace{0.2cm}}$\\
			
			&
			$\color{blue}{(N_f^2)Bl\phi_1u^{\dagger}Q,\hspace{0.2cm} (N_f^2)Bl\phi_2u^{\dagger}Q,\hspace{0.2cm} (N_f^2)Bl\phi_1^{\dagger}e^{\dagger}L,\hspace{0.2cm} (N_f^2)Bl\phi_2^{\dagger}e^{\dagger}L \hspace{0.2cm}}$\\
			\hline
			
			\multirow{4}{*}{$\Psi^2\Phi^2 \mathcal{D}$}&
			$(N_f^2)dd^{\dagger}\phi_1\phi_1^{\dagger}\mathcal{D},\hspace{0.2cm}(N_f^2)dd^{\dagger}\phi_2\phi_2^{\dagger}\mathcal{D},\hspace{0.2cm}
			(N_f^2)ee^{\dagger}\phi_1\phi_1^{\dagger}\mathcal{D},\hspace{0.2cm} (N_f^2)ee^{\dagger}\phi_2\phi_2^{\dagger}\mathcal{D},\hspace{0.2cm} (2N_f^2)LL^{\dagger}\phi_1\phi_1^{\dagger}\mathcal{D},\hspace{0.2cm}$\\
			
			&
			$(2N_f^2)LL^{\dagger}\phi_2\phi_2^{\dagger}\mathcal{D}, \hspace{0.2cm}
			(2N_f^2)QQ^{\dagger}\phi_1\phi_1^{\dagger}\mathcal{D}, \hspace{0.2cm} (2N_f^2)QQ^{\dagger}\phi_2\phi_2^{\dagger}\mathcal{D}, \hspace{0.2cm} (N_f^2)uu^{\dagger}\phi_1\phi_1^{\dagger}\mathcal{D},\hspace{0.2cm} (N_f^2)uu^{\dagger}\phi_2\phi_2^{\dagger}\mathcal{D},\hspace{0.2cm} $\\
			
			&
			$\color{blue}{(N_f^2)dd^{\dagger}\phi_1^{\dagger}\phi_2\mathcal{D}, \hspace{0.2cm} (N_f^2)ee^{\dagger}\phi_1^{\dagger}\phi_2\mathcal{D}, \hspace{0.2cm}
				(N_f^2)uu^{\dagger}\phi_1^{\dagger}\phi_2\mathcal{D}, \hspace{0.2cm} 
				(2N_f^2)LL^{\dagger}\phi_1^{\dagger}\phi_2\mathcal{D}, \hspace{0.2cm} (2N_f^2)QQ^{\dagger}\phi_1^{\dagger}\phi_2\mathcal{D}, \hspace{0.2cm}}$\\
			
			&
			$\color{blue}{ (N_f^2)ud^{\dagger}(\phi_1^{\dagger})^2\mathcal{D},\hspace{0.2cm} (N_f^2)ud^{\dagger}(\phi_2^{\dagger})^2\mathcal{D},\hspace{0.2cm} (N_f^2)ud^{\dagger}(\phi_1^{\dagger})(\phi_2^{\dagger})\mathcal{D} \hspace{0.2cm}}$\\
			\hline
			
			\multirow{4}{*}{$\Psi^2\Phi^3$}&
			$\color{blue}{ (N_f^2)\phi_1(\phi_1^{\dagger})^2e^{\dagger}L, \hspace{0.2cm} (N_f^2)\phi_2(\phi_2^{\dagger})^2e^{\dagger}L, \hspace{0.2cm}
			(N_f^2)\phi_1(\phi_2^{\dagger})^2e^{\dagger}L, \hspace{0.2cm}  (N_f^2)(\phi_1^{\dagger})^2\phi_2e^{\dagger}L, \hspace{0.2cm}  (2N_f^2)\phi_1\phi_1^{\dagger}\phi_2^{\dagger}e^{\dagger}L, \hspace{0.2cm}}$\\
			
			&
			$\color{blue}{ (2N_f^2)\phi_1^{\dagger}\phi_2\phi_2^{\dagger}e^{\dagger}L,\hspace{0.2cm} (N_f^2)\phi_1(\phi_1^{\dagger})^2d^{\dagger}Q, \hspace{0.2cm} (N_f^2)\phi_2(\phi_2^{\dagger})^2d^{\dagger}Q, \hspace{0.2cm}
				(N_f^2)\phi_1(\phi_2^{\dagger})^2d^{\dagger}Q, \hspace{0.2cm}  (N_f^2)(\phi_1^{\dagger})^2\phi_2d^{\dagger}Q, \hspace{0.2cm}}$\\
			
			&
			$\color{blue}{	(2N_f^2)\phi_1\phi_1^{\dagger}\phi_2^{\dagger}d^{\dagger}Q, \hspace{0.2cm} (2N_f^2)\phi_1^{\dagger}\phi_2\phi_2^{\dagger}d^{\dagger}Q, \hspace{0.2cm} (N_f^2)\phi_1^2\phi_1^{\dagger}u^{\dagger}Q,\hspace{0.2cm}(N_f^2)\phi_2^2\phi_2^{\dagger}u^{\dagger}Q,\hspace{0.2cm}
				(N_f^2)\phi_1^2\phi_2^{\dagger}u^{\dagger}Q,\hspace{0.2cm}}$\\
			
			&
			$\color{blue}{ (N_f^2)\phi_1^{\dagger}\phi_2^2u^{\dagger}Q,\hspace{0.2cm} (2N_f^2)\phi_1\phi_1^{\dagger}\phi_2u^{\dagger}Q, \hspace{0.2cm} (2N_f^2)\phi_1\phi_2\phi_2^{\dagger}u^{\dagger}Q \hspace{0.2cm}}$\\
			\hline
			
			$\Phi^4 \mathcal{D}^2$&
			$ 2\phi_1^2(\phi_1^{\dagger})^2\mathcal{D}^2,\hspace{0.2cm} 2\phi_2^2(\phi_2^{\dagger})^2\mathcal{D}^2,\hspace{0.2cm} 4\phi_1\phi_2\phi_1^{\dagger}\phi_2^{\dagger}\mathcal{D}^2, \hspace{0.2cm}
			\color{blue}{2\phi_1^2(\phi_2^{\dagger})^2\mathcal{D}^2, \hspace{0.2cm} 2\phi_1\phi_2(\phi_2^{\dagger})^2\mathcal{D}^2, \hspace{0.2cm} 2\phi_1(\phi_1^{\dagger})^2\phi_2\mathcal{D}^2 \hspace{0.2cm}} $\\
			\hline
			
			\multirow{5}{*}{$\Psi^4$}&
			$(N_f^4)ee^{\dagger}uu^{\dagger},\hspace{0.2cm}
			(\frac{1}{2}N_f^2+\frac{1}{2}N_f^4)L^2(L^{\dagger})^2, \hspace{0.2cm} 
			(N_f^2+N_f^4)Q^2(Q^{\dagger})^2,\hspace{0.2cm} (2N_f^4)LL^{\dagger}QQ^{\dagger},\hspace{0.2cm}$\\
			
			&
			$(N_f^4)ee^{\dagger}dd^{\dagger},\hspace{0.2cm}
			(\frac{1}{4}N_f^2+\frac{1}{2}N_f^3+\frac{1}{4}N_f^4)e^2(e^{\dagger})^2,\hspace{0.2cm}
			(\frac{1}{2}N_f^2+\frac{1}{2}N_f^4)d^2(d^{\dagger})^2,\hspace{0.2cm} (\frac{1}{2}N_f^2+\frac{1}{2}N_f^4)u^2(u^{\dagger})^2,\hspace{0.2cm}$\\
			
			&
			$(2N_f^4)dd^{\dagger}uu^{\dagger},\hspace{0.2cm}
			(N_f^4)ee^{\dagger}LL^{\dagger},\hspace{0.2cm} (N_f^4)uu^{\dagger}LL^{\dagger},\hspace{0.2cm} (N_f^4)dd^{\dagger}LL^{\dagger},\hspace{0.2cm}
			(N_f^4)ee^{\dagger}QQ^{\dagger},\hspace{0.2cm} (2N_f^4)uu^{\dagger}QQ^{\dagger},\hspace{0.2cm}$\\
			
			&
			$(2N_f^4)dd^{\dagger}QQ^{\dagger},\hspace{0.2cm}
			\color{blue}{  (N_f^4)eL^{\dagger}d^{\dagger}Q,\hspace{0.2cm} (2N_f^4)ud(Q^{\dagger})^2,\hspace{0.2cm} (2N_f^4)eL^{\dagger}uQ^{\dagger},\hspace{0.2cm}
				(\frac{1}{3}N_f^2+\frac{2}{3}N_f^4)LQ^3,\hspace{0.2cm} 
			}$\\
			
			&
			$\color{blue}{(\frac{1}{2}N_f^3+\frac{1}{2}N_f^4)e\,u\,Q^2,\hspace{0.2cm}
				(N_f^4)eu^2d,\hspace{0.2cm} (N_f^4)L\,u\,d\,Q} $\\
			\hline		    
	\end{tabular}}
	\caption{\small 2HDM: Operators of mass dimensions-5 and -6.}
	\label{table:2hdm-dim-5-6-op}
\end{table} 

\vspace{-0.6cm}
\subsection{The Pati-Salam Model} \label{subsec:patisalam}
The Pati-Salam model can be thought of as a partially unified scenario where $SU(3)_C$ and $U(1)_{B-L}$ are embedded to form $SU(4)_C$. This leads to quark-lepton unification and the lepton is considered to be the fourth color \cite{Pati:1973rp,Pati:1973uk,Pati:1974yy}. The underlying gauge symmetry is $SU(4)_C\otimes SU(2)_L\otimes SU(2)_R$. The Pati-Salam model also has a rich scalar structure to facilitate symmetry breaking. We consider the most general form of Pati-Salam \cite{Saad:2017pqj} where scalar fields transform as $(1,1,1)$, $(1,2,2)$, $(15,2,2)$, $(10,3,1)$ and $(10,1,3)$. On top of that, a global Pecci-Quinn $U(1)_{PQ}$ symmetry is imposed. The field content and their transformation properties under the gauge groups and the Lorentz group, as well as their $U(1)_{PQ}$ charges, are provided in Table~\ref{table:patisalam-quantum-no}. The necessary information related to characters and Haar measures has already been discussed in earlier sections. Based on that we have computed the Hilbert Series. It must be kept in mind that since the model consists of two distinct $SU(2)$s, their characters must be parametrized using distinct variables. Unlike 2HDM, for this particular case, we have limited ourselves up to dimension-5 operators. Due to the rich scalar structure, we have emphasized the scalar potential and Yukawa terms by explicitly giving their covariant forms in Tables~\ref{table:patisalam-renorm-output} and \ref{table:patisalam-renorm-output-2}. The dimension-5 result is collected in Table~\ref{table:patisalam-dim-5-op}. On comparing with \cite{Saad:2017pqj}, we have found some discrepancies. We observe that the operator structure \textcolor{red}{$\Phi\Sigma(\Sigma^{\dagger})^2$} is absent in \cite{Saad:2017pqj}, whereas the Hilbert Series output contains \textcolor{red}{4} operators having this structure as well as their hermitian conjugates. We have shown the explicit forms of those operators in Table~\ref{table:patisalam-renorm-output-2}. We also observe an under-counting w.r.t. the operator structure \textcolor{red}{$\Sigma\Sigma^{\dagger}\Delta_R\Delta_L$}, while \cite{Saad:2017pqj} contains only 1 such operator (and its hermitian conjugate), Hilbert Series output contains \textcolor{red}{2} such operators (and their hermitian conjugates). 
\begin{table}[h]
	\centering
	\renewcommand{\arraystretch}{1.6}
	{\tiny\begin{tabular}{|c|c|c|c|c|c|c|}
			\hline
			\textbf{Fields}&
			$SU(4)_C$&
			$SU(2)_L$&
			$SU(2)_R$&
			$U(1)_{PQ}$&
			\textbf{Spin}&
			\textbf{Gauge Group Characters}\\
			\hline
			
			$\mathcal{S}$&
			1&
			1&
			1&
			4&
			0&
			1\\
			
			$\Phi$&
			1&
			2&
			2&
			2&
			0&
			$\chi_{({SU(2)_L})_{2}}\cdot\chi_{({SU(2)_R})_{2}}$\\
			
			$\Sigma$&
			15&
			2&
			2&
			2&
			0&
			$\chi_{({SU(4)})_{15}}\cdot\chi_{({SU(2)_L})_{2}}\cdot\chi_{({SU(2)_R})_{2}}$\\
			
			$\Delta_L$&
			10&
			3&
			1&
			2&
			0&
			$\chi_{({SU(4)})_{10}}\cdot\chi_{({SU(2)_L})_{3}}$\\
			
			$\Delta_R$&
			10&
			1&
			3&
			-2&
			0&
			$\chi_{({SU(4)})_{10}}\cdot\chi_{({SU(2)_R})_{3}}$\\
			
			$\psi_L$&
			4&
			2&
			1&
			1&
			1/2&
			$\chi_{({SU(4)})_{4}}\cdot\chi_{({SU(2)_L})_{2}}$\\
			
			$\psi_R$&
			4&
			1&
			2&
			-1&
			1/2&
			$\chi_{({SU(4)})_{4}}\cdot\chi_{({SU(2)_R})_{2}}$\\
			
			$W_{L\mu\nu}$&
			1&
			3&
			1&
			0&
			1&
			$\chi_{({SU(2)_L})_{3}}$\\
			
			$W_{R\mu\nu}$&
			1&
			1&
			3&
			0&
			1&
			$\chi_{({SU(2)_R})_{3}}$\\
			
			$V_{\mu\nu}$&
			15&
			1&
			1&
			0&
			1&
			$\chi_{({SU(4)})_{15}}$\\
			
			\hline
			\hline
			
			$\mathcal{D}_\mu$&
			\multicolumn{6}{c|}{\textbf{Covariant Derivative}}\\
			\hline
	\end{tabular}}
	\caption{\small Pati-Salam Model: Quantum numbers of fields under the gauge groups, their spins and gauge group characters. Since the model only contains fields having spins-0, -1/2, and -1, the relevant Lorentz characters are the ones given in Eq.~\eqref{eq:lorentz-char}. Here, $U(1)_{PQ}$ appears as a global symmetry hence we do not assign any character to the fields corresponding to it. Internal symmetry indices have been suppressed. $L$ and $R$ denote whether the fields transform non-trivially under $SU(2)_L$ and $SU(2)_R$ respectively.}
	\label{table:patisalam-quantum-no}
\end{table}  

\begin{table}[h]
	\centering
	\renewcommand{\arraystretch}{2.0}
	{\tiny\begin{tabular}{||c|c||c|c||}
			\hline
			\multicolumn{4}{|c|}{\textbf{Mass Dimension-2}}\\
			\hline
			
			\textbf{HS Output}&
			\textbf{Covariant Form}&
			\textbf{HS Output}&
			\textbf{Covariant Form}\\
			\hline
			
			$\Phi^{\dagger}\Phi$&
			$\Phi^{\dot{\alpha}}_{\alpha}
			\Phi^{\dagger\alpha}_{\dot{\alpha}}$&
			$\Sigma^{\dagger}\Sigma$&
			$\Sigma^{\nu\dot{\alpha}}_{\mu\alpha}
			\Sigma^{\dagger\mu\alpha}_{\nu\dot{\alpha}}$\\
			\cline{1-4}
			
			$\Delta_L^{\dagger}\Delta_L$&
			$\Delta^{\beta}_{L\mu\nu\alpha}
			\Delta^{\dagger\mu\nu\alpha}_{L\beta}$&
			$\Delta_R^{\dagger}\Delta_R$&
			$\Delta_{R\mu\nu\dot{\alpha}}^{\dot{\beta}}
			\Delta^{\dagger\mu\nu\dot{\alpha}}_{R\dot{\beta}}$
			\\
			\cline{1-4}
			
			$\mathcal{S}^{\dagger}\mathcal{S}$&
			$\mathcal{S}^{\dagger}\mathcal{S}$&
			&
			\\
			\hline
			\hline
			
			\multicolumn{4}{|c|}{\textbf{Mass Dimension-3}}\\
			\hline
			\textbf{HS Output}&
			\textbf{Covariant Form}&
			\textbf{HS Output}&
			\textbf{Covariant Form}\\
			\hline
			
			$\color{blue}{\Phi\Phi\mathcal{S}^{\dagger}}$&
			$\color{blue}{\epsilon_{\dot{\alpha}\dot{\beta}} \epsilon^{\alpha\beta} \Phi^{\dot{\alpha}}_{\alpha} 
			\Phi^{\dot{\beta}}_{\beta} \mathcal{S}^{\dagger}}$&
			$\color{blue}{\Sigma\Sigma\mathcal{S}^{\dagger}}$&
			$\color{blue}{\epsilon_{\dot{\alpha}\dot{\beta}} \epsilon^{\alpha\beta} \Sigma^{\nu\dot{\alpha}}_{\mu\alpha}
			\Sigma^{\mu\dot{\beta}}_{\nu\beta}\mathcal{S}^{\dagger}}$\\
			\hline\hline
			\multicolumn{4}{|c|}{\textbf{Mass Dimension-4}}\\
			\hline
			\textbf{HS Output}&
			\textbf{Covariant Form}&
			\textbf{HS Output}&
			\textbf{Covariant Form}\\
			\hline

			$\color{blue}{\psi_L^{\dagger}\psi_R\Phi}$&
			$\color{blue}{\overline{\psi}_L\Phi\psi_R}$&
			$\color{blue}{\psi_L^2\Delta_L^{\dagger}}$&
			$\color{blue}{\psi^T_LC\Delta_L^{\dagger}\psi_L}$\\
			\cline{1-4}

			$\color{blue}{\psi_L^{\dagger}\psi_R\Sigma}$&
			$\color{blue}{\overline{\psi}_L\Sigma\psi_R}$&
			$\color{blue}{\psi_R^2\Delta_R^{\dagger}}$&
			$\color{blue}{\psi^T_RC\Delta_R^{\dagger}\psi_R}$
			\\
			\cline{1-4}
			$(\mathcal{S}^{\dagger})^2\mathcal{S}^2$&
			$(\mathcal{S}^{\dagger}\mathcal{S})^2$&
			$\mathcal{S}^{\dagger}\mathcal{S}\Phi^{\dagger}\Phi$&
			$\left(\mathcal{S}^{\dagger}\mathcal{S}\right) \left(\Phi^{\dot{\alpha}}_{\alpha}
			\Phi^{\dagger\alpha}_{\dot{\alpha}}\right)$
			\\
			\cline{1-4}
			$\mathcal{S}^{\dagger}\mathcal{S}\Delta_L^{\dagger}\Delta_L$&
			$\left(\mathcal{S}^{\dagger}\mathcal{S}\right) \left(\Delta^{\beta}_{L\mu\nu\alpha}
			\Delta^{\dagger\mu\nu\alpha}_{L\beta}\right)$&
			$\mathcal{S}^{\dagger}\mathcal{S}\Sigma^{\dagger}\Sigma$&
			$\left(\mathcal{S}^{\dagger}\mathcal{S}\right) \left(\Sigma^{\nu\dot{\alpha}}_{\mu\alpha}
			\Sigma^{\dagger\mu\alpha}_{\nu\dot{\alpha}}\right)$
			\\
			\hline
			
			$\color{blue}{\Delta_L\Delta_L\Delta_R\Delta_R}$&
			$\color{blue}{\epsilon^{\mu\rho\lambda\zeta}\epsilon^{\nu\tau\xi\omega}\left(\Delta^{\dot{\beta}}_{R\mu\nu\dot{\alpha}}\Delta^{\dot{\alpha}}_{R\rho\tau\dot{\beta}}\Delta^{\beta}_{L\lambda\xi\alpha}\Delta^{\alpha}_{L\zeta\omega\beta}\right)}$&
			$\color{blue}{\Phi\Phi\Delta_R\Delta_L^{\dagger}}$&
			$\color{blue}{\epsilon_{\alpha\kappa}\epsilon^{\dot{\alpha}\dot{\kappa}}\left(\Phi^{\dot{\alpha}}_{\alpha}\Phi^{\dot{\beta}}_{\beta}\Delta^{\mu\nu\dot{\beta}}_{R\dot{\kappa}}\Delta_{L\mu\nu\kappa}^{\dagger\beta}\right)}$
			\\
			\hline		
	\end{tabular}}
	\caption{\small Pati-Salam Model: Renormalizable operators as Hilbert Series output and their covariant form. The operators in blue have distinct hermitian conjugates which we have not written explicitly continued. We have suppressed the indices in the Yukawa terms. Notation for the indices is same as in \cite{Saad:2017pqj}, i.e., $\alpha, \beta, \gamma, \kappa = 1,2$ are $SU(2)_L$ indices, $\dot{\alpha}, \dot{\beta}, \dot{\gamma}, \dot{\kappa} = 1,2$ are $SU(2)_R$ indices and $\mu, \nu, \rho, \tau, \lambda, \xi, \zeta, \omega = 1,2,3,4$  are $SU(4)_C$ indices.}
	\label{table:patisalam-renorm-output}
\end{table}
\newpage
\begin{table}[h]
	\centering
	\renewcommand{\arraystretch}{2.0}
	{\tiny\begin{tabular}{||c|c||c|c||}
			\hline
			\multicolumn{4}{|c|}{\textbf{Mass Dimension-4}}\\
			\hline
			\textbf{HS Output}&
			\textbf{Covariant Form}&
			\textbf{HS Output}&
			\textbf{Covariant Form}\\
			\hline			
			$\mathcal{S}^{\dagger}\mathcal{S}\Delta_R^{\dagger}\Delta_R$&
			$\left(\mathcal{S}^{\dagger}\mathcal{S}\right)\left(\Delta_{R\mu\nu\dot{\alpha}}^{\dot{\beta}}
			\Delta^{\dagger\mu\nu\dot{\alpha}}_{R\dot{\beta}}\right)$&
			$\color{blue}{\Phi\Sigma\Delta_R\Delta_L^{\dagger}}$&
			$\color{blue}{\epsilon^{\alpha\kappa}\epsilon_{\dot{\alpha}\dot{\kappa}}\left(\Phi^{\dot{\alpha}}_{\alpha}\Sigma^{\nu\dot{\beta}}_{\mu\beta}\Delta_{R\nu\tau\dot{\beta}}^{\dot{\kappa}}\Delta_{L\kappa}^{\dagger\tau\mu\beta}\right)}$
			\\
			\cline{1-4}
			\multirow{2}{*}{$2(\Phi^{\dagger})^2\Phi^2 $}&
			$\left(\Phi^{\dot{\alpha}}_{\alpha}
			\Phi^{\dagger\alpha}_{\dot{\alpha}}\right) \left(\Phi^{\dot{\beta}}_{\beta}
			\Phi^{\dagger\beta}_{\dot{\beta}}\right)$&
			\multirow{4}{*}{$4\Phi^{\dagger}\Phi\Sigma^{\dagger}\Sigma $}&
			$\Phi^{\dot{\alpha}}_{\alpha}\Phi^{\dagger\alpha}_{\dot{\alpha}}\Sigma^{\nu\dot{\beta}}_{\mu\beta}\Sigma^{\dagger\mu\beta}_{\nu\dot{\beta}}$
			\\
			
			&
			$\left(\Phi^{\dot{\alpha}}_{\alpha}
			\Phi^{\dagger\alpha}_{\dot{\beta}} \Phi^{\dot{\beta}}_{\beta}
			\Phi^{\dagger\beta}_{\dot{\alpha}}\right)$&
			&
			$\Phi^{\dot{\alpha}}_{\alpha}\Phi^{\dagger\alpha}_{\dot{\beta}}\Sigma^{\nu\dot{\beta}}_{\mu\beta}\Sigma^{\dagger\mu\beta}_{\nu\dot{\alpha}}$
			\\
			\cline{1-2}
			
			\multirow{14}{*}{$14(\Sigma^{\dagger})^2\Sigma^2 $}&
			$\left(\Sigma_{\mu\alpha}^{\nu\dot{\alpha}}\Sigma^{\dagger\mu\alpha}_{\nu\dot{\alpha}}\right)\left(\Sigma_{\rho\beta}^{\tau\dot{\beta}}\Sigma^{\dagger\rho\beta}_{\tau\dot{\beta}}\right)$&
			&
			$\Phi^{\dot{\alpha}}_{\alpha}\Phi^{\dagger\beta}_{\dot{\alpha}}\Sigma^{\nu\dot{\beta}}_{\mu\beta}\Sigma^{\dagger\mu\alpha}_{\nu\dot{\beta}}$
			\\
			
			&
			$\Sigma_{\mu\alpha}^{\nu\dot{\alpha}}\Sigma^{\dagger\rho\alpha}_{\tau\dot{\alpha}}\Sigma_{\nu\beta}^{\mu\dot{\beta}}\Sigma^{\dagger\rho\beta}_{\tau\dot{\beta}}$&
			&
			$\Phi^{\dot{\alpha}}_{\alpha}\Phi^{\dagger\beta}_{\dot{\beta}}\Sigma^{\nu\dot{\beta}}_{\mu\beta}\Sigma^{\dagger\mu\alpha}_{\nu\dot{\alpha}}$
			\\
			\cline{3-4}
			
			&
			$\Sigma_{\mu\alpha}^{\nu\dot{\alpha}}\Sigma^{\dagger\tau\alpha}_{\rho\dot{\alpha}}\Sigma_{\tau\beta}^{\rho\dot{\beta}}\Sigma^{\dagger\mu\beta}_{\nu\dot{\beta}}$&
			\multirow{2}{*}{$\color{blue}{2\Phi\Phi\Sigma^{\dagger}\Sigma^{\dagger}} $}&
			$\color{blue}{\Phi^{\dot{\alpha}}_{\alpha}\Phi^{\dot{\beta}}_{\beta}\Sigma^{\dagger\nu\alpha}_{\mu\dot{\alpha}}\Sigma^{\dagger\mu\beta}_{\nu\dot{\beta}}}$
			\\

			&
			$\Sigma_{\mu\alpha}^{\nu\dot{\alpha}}\Sigma^{\dagger\rho\alpha}_{\nu\dot{\alpha}}\Sigma_{\rho\beta}^{\tau\dot{\beta}}\Sigma^{\dagger\mu\beta}_{\tau\dot{\beta}}$&
			&
			$\color{blue}{\Phi^{\dot{\alpha}}_{\alpha}\Phi^{\dot{\beta}}_{\beta}\Sigma^{\dagger\nu\alpha}_{\mu\dot{\beta}}\Sigma^{\dagger\mu\beta}_{\nu\dot{\alpha}}}$
			\\
			\cline{3-4}
			
			&
			$\Sigma_{\mu\alpha}^{\nu\dot{\alpha}}\Sigma^{\dagger\mu\alpha}_{\tau\dot{\alpha}}\Sigma_{\nu\beta}^{\rho\dot{\beta}}\Sigma^{\dagger\tau\beta}_{\rho\dot{\beta}}$&
			\multirow{5}{*}{$5(\Delta_L^{\dagger})^2\Delta_L^2 $}&
			$\Delta_{L\mu\nu\alpha}^{\beta}\Delta_{L\beta}^{\dagger\mu\nu\alpha}\Delta_{L\rho\tau\gamma}^{\kappa}\Delta_{L\kappa}^{\dagger\rho\tau\gamma}$
			\\

			&
			$\Sigma_{\mu\alpha}^{\nu\dot{\alpha}}\Sigma^{\dagger\tau\alpha}_{\rho\dot{\alpha}}\Sigma_{\tau\beta}^{\mu\dot{\beta}}\Sigma^{\dagger\rho\beta}_{\nu\dot{\beta}}$&
			&
			$\Delta_{L\mu\nu\alpha}^{\beta}\Delta_{L\kappa}^{\dagger\mu\nu\gamma}\Delta_{L\rho\tau\beta}^{\alpha}\Delta_{L\gamma}^{\dagger\rho\tau\kappa}$
			\\

			&
			$\Sigma_{\mu\alpha}^{\nu\dot{\alpha}}\Sigma^{\dagger\mu\alpha}_{\nu\dot{\beta}}\Sigma_{\rho\beta}^{\tau\dot{\beta}}\Sigma^{\dagger\rho\beta}_{\tau\dot{\alpha}}$&
			&
			$\Delta_{L\mu\nu\alpha}^{\beta}\Delta_{L\gamma}^{\dagger\mu\nu\kappa}\Delta_{L\rho\tau\kappa}^{\gamma}\Delta_{L\beta}^{\dagger\rho\tau\alpha}$
			\\

			&
			$\Sigma_{\mu\alpha}^{\nu\dot{\alpha}}\Sigma^{\dagger\rho\alpha}_{\tau\dot{\beta}}\Sigma_{\nu\beta}^{\mu\dot{\beta}}\Sigma^{\dagger\tau\beta}_{\rho\dot{\alpha}}$&
			&
			$\Delta_{L\mu\nu\alpha}^{\beta}\Delta_{L\beta}^{\dagger\nu\rho\alpha}\Delta_{L\rho\tau\gamma}^{\kappa}\Delta_{L\kappa}^{\dagger\tau\mu\gamma}$
			\\

			&
			$\Sigma_{\mu\alpha}^{\nu\dot{\alpha}}\Sigma^{\dagger\rho\alpha}_{\nu\dot{\beta}}\Sigma_{\rho\beta}^{\tau\dot{\beta}}\Sigma^{\dagger\mu\beta}_{\tau\dot{\alpha}}$&
			&
			$\Delta_{L\mu\nu\alpha}^{\beta}\Delta_{L\kappa}^{\dagger\nu\rho\gamma}\Delta_{L\rho\tau\beta}^{\alpha}\Delta_{L\gamma}^{\dagger\tau\mu\kappa}$
			\\
			\cline{3-4}
			
			&
			$\Sigma_{\mu\alpha}^{\nu\dot{\alpha}}\Sigma^{\dagger\mu\alpha}_{\tau\dot{\beta}}\Sigma_{\nu\beta}^{\rho\dot{\beta}}\Sigma^{\dagger\tau\beta}_{\rho\dot{\alpha}}$&
			\multirow{5}{*}{$5(\Delta_R^{\dagger})^2\Delta_R^2 $}&
			$\Delta_{R\mu\nu\dot{\alpha}}^{\dot{\beta}}\Delta_{R\dot{\beta}}^{\dagger\mu\nu\dot{\alpha}}\Delta_{R\rho\tau\dot{\gamma}}^{\dot{\kappa}}\Delta_{R\dot{\kappa}}^{\dagger\rho\tau\dot{\gamma}}$
			\\

			&
			$\epsilon^{\alpha\gamma}\epsilon_{\dot{\alpha}\dot{\gamma}}\epsilon_{\beta\kappa}\epsilon^{\dot{\beta}\dot{\kappa}}\left(\Sigma_{\mu\alpha}^{\nu\dot{\alpha}}\Sigma^{\mu\dot{\gamma}}_{\nu\gamma}\Sigma^{\dagger\tau\beta}_{\rho\dot{\beta}}\Sigma^{\dagger\rho\kappa}_{\tau\dot{\kappa}}\right)$&
			&
			$\Delta_{R\mu\nu\dot{\alpha}}^{\dot{\beta}}\Delta_{R\dot{\kappa}}^{\dagger\mu\nu\dot{\gamma}}\Delta_{R\rho\tau\dot{\beta}}^{\dot{\alpha}}\Delta_{R\dot{\gamma}}^{\dagger\rho\tau\dot{\kappa}}$
			\\

			&
			$\epsilon^{\alpha\gamma}\epsilon_{\dot{\alpha}\dot{\gamma}}\epsilon_{\beta\kappa}\epsilon^{\dot{\beta}\dot{\kappa}}\left(\Sigma_{\mu\alpha}^{\nu\dot{\alpha}}\Sigma^{\rho\dot{\gamma}}_{\tau\gamma}\Sigma^{\dagger\mu\beta}_{\nu\dot{\beta}}\Sigma^{\dagger\tau\kappa}_{\rho\dot{\kappa}}\right)$&
			&
			$\Delta_{R\mu\nu\dot{\alpha}}^{\dot{\beta}}\Delta_{R\dot{\gamma}}^{\dagger\mu\nu\dot{\kappa}}\Delta_{R\rho\tau\dot{\kappa}}^{\dot{\gamma}}\Delta_{R\dot{\beta}}^{\dagger\rho\tau\dot{\alpha}}$
			\\

			&
			$\epsilon^{\alpha\gamma}\epsilon_{\dot{\alpha}\dot{\gamma}}\epsilon_{\beta\kappa}\epsilon^{\dot{\beta}\dot{\kappa}}\left(\Sigma_{\mu\alpha}^{\nu\dot{\alpha}}\Sigma^{\rho\dot{\gamma}}_{\nu\gamma}\Sigma^{\dagger\tau\beta}_{\rho\dot{\beta}}\Sigma^{\dagger\mu\kappa}_{\tau\dot{\kappa}}\right)$&
			&
			$\Delta_{R\mu\nu\dot{\alpha}}^{\dot{\beta}}\Delta_{R\dot{\beta}}^{\dagger\nu\rho\dot{\alpha}}\Delta_{R\rho\tau\dot{\gamma}}^{\dot{\kappa}}\Delta_{R\dot{\kappa}}^{\dagger\tau\mu\dot{\gamma}}$
			\\

			&
			$\epsilon^{\alpha\gamma}\epsilon_{\dot{\alpha}\dot{\gamma}}\epsilon_{\beta\kappa}\epsilon^{\dot{\beta}\dot{\kappa}}\left(\Sigma_{\mu\alpha}^{\nu\dot{\alpha}}\Sigma^{\tau\dot{\gamma}}_{\rho\gamma}\Sigma^{\dagger\mu\beta}_{\tau\dot{\beta}}\Sigma^{\dagger\rho\kappa}_{\nu\dot{\kappa}}\right)$&
			&
			$\Delta_{R\mu\nu\dot{\alpha}}^{\dot{\beta}}\Delta_{R\dot{\kappa}}^{\dagger\nu\rho\dot{\gamma}}\Delta_{R\rho\tau\dot{\beta}}^{\dot{\alpha}}\Delta_{R\dot{\gamma}}^{\dagger\tau\mu\dot{\kappa}}$
			\\
			\hline
			\multirow{2}{*}{$2\Phi^{\dagger}\Phi\Delta_R^{\dagger}\Delta_R $}&
			$\Phi^{\dot{\alpha}}_{\alpha}\Phi^{\dagger\alpha}_{\dot{\alpha}}\Delta_{R\mu\nu\dot{\beta}}^{\dot{\gamma}}\Delta_{R\dot{\gamma}}^{\dagger\mu\nu\dot{\beta}}$&
			\multirow{4}{*}{$\color{red}{4\Phi\Sigma(\Sigma^{\dagger})^2}$}&
			$\color{blue}{\Phi^{\dot{\alpha}}_{\alpha}\Sigma_{\mu\beta}^{\nu\dot{\beta}}\Sigma_{\rho\dot{\beta}}^{\dagger\mu\beta}\Sigma_{\nu\dot{\alpha}}^{\dagger\rho\alpha}}$
			\\
			
			&
			$\Phi^{\dot{\alpha}}_{\alpha}\Phi^{\dagger\alpha}_{\dot{\beta}}\Delta_{R\mu\nu\dot{\alpha}}^{\dot{\gamma}}\Delta_{R\dot{\gamma}}^{\dagger\mu\nu\dot{\beta}}$&
			&
			$\color{blue}{\Phi^{\dot{\alpha}}_{\alpha}\Sigma_{\mu\beta}^{\nu\dot{\beta}}\Sigma_{\rho\dot{\alpha}}^{\dagger\mu\beta}\Sigma_{\nu\dot{\beta}}^{\dagger\rho\alpha}}$
			\\
			\cline{1-2}
			
			\multirow{2}{*}{$2\Phi^{\dagger}\Phi\Delta_L^{\dagger}\Delta_L $}&
			$\Phi^{\dot{\alpha}}_{\alpha}\Phi^{\dagger\alpha}_{\dot{\alpha}}\Delta_{L\mu\nu\beta}^{\gamma}\Delta_{L\gamma}^{\dagger\mu\nu\beta}$&
			&
			$\color{blue}{\epsilon_{\dot{\alpha}\dot{\kappa}}\epsilon^{\dot{\beta}\dot{\gamma}}\left(\Phi^{\dot{\alpha}}_{\alpha}\Sigma_{\mu\kappa}^{\nu\dot{\kappa}}\Sigma_{\rho\dot{\beta}}^{\dagger\mu\kappa}\Sigma_{\nu\dot{\gamma}}^{\dagger\rho\alpha}\right)}$
			\\
			
			&
			$\Phi^{\dot{\alpha}}_{\alpha}\Phi^{\dagger\gamma}_{\dot{\alpha}}\Delta_{L\mu\nu\beta}^{\alpha}\Delta_{L\gamma}^{\dagger\mu\nu\beta}$&
			&
			$\color{blue}{\epsilon_{\dot{\alpha}\dot{\kappa}}\epsilon^{\alpha\kappa}\epsilon^{\dot{\beta}\dot{\gamma}}\epsilon_{\beta\gamma}\left(\Phi^{\dot{\alpha}}_{\alpha}\Sigma_{\mu\kappa}^{\nu\dot{\kappa}}\Sigma_{\rho\dot{\beta}}^{\dagger\mu\beta}\Sigma_{\nu\dot{\gamma}}^{\dagger\rho\gamma}\right)}$
			\\
			\cline{1-4}
			
			\multirow{8}{*}{$8\Sigma^{\dagger}\Sigma\Delta_L^{\dagger}\Delta_L $}&
			$\Sigma_{\rho\alpha}^{\tau\dot{\alpha}}\Sigma^{\dagger\rho\alpha}_{\tau\dot{\alpha}}\Delta_{L\mu\nu\beta}^{\gamma}\Delta_{L\gamma}^{\dagger\mu\nu\beta}$&
			\multirow{8}{*}{$8\Sigma^{\dagger}\Sigma\Delta_R^{\dagger}\Delta_R $}&
			$\Sigma_{\rho\alpha}^{\tau\dot{\alpha}}\Sigma^{\dagger\rho\alpha}_{\tau\dot{\alpha}}\Delta_{R\mu\nu\dot{\beta}}^{\dot{\gamma}}\Delta_{R\dot{\gamma}}^{\dagger\mu\nu\dot{\beta}}$
			\\

			&
			$\Sigma_{\rho\alpha}^{\tau\dot{\alpha}}\Sigma^{\dagger\mu\alpha}_{\tau\dot{\alpha}}\Delta_{L\mu\nu\beta}^{\gamma}\Delta_{L\gamma}^{\dagger\nu\rho\beta}$&
			&
			$\Sigma_{\rho\alpha}^{\tau\dot{\alpha}}\Sigma^{\dagger\mu\alpha}_{\tau\dot{\alpha}}\Delta_{R\mu\nu\dot{\beta}}^{\dot{\gamma}}\Delta_{R\dot{\gamma}}^{\dagger\nu\rho\dot{\beta}}$
			\\

			&
			$\Sigma_{\rho\alpha}^{\tau\dot{\alpha}}\Sigma^{\dagger\rho\alpha}_{\mu\dot{\alpha}}\Delta_{L\tau\nu\beta}^{\gamma}\Delta_{L\gamma}^{\dagger\nu\mu\beta}$&
			&
			$\Sigma_{\rho\alpha}^{\tau\dot{\alpha}}\Sigma^{\dagger\rho\alpha}_{\mu\dot{\alpha}}\Delta_{R\tau\nu\dot{\beta}}^{\dot{\gamma}}\Delta_{R\dot{\gamma}}^{\dagger\nu\mu\dot{\beta}}$
			\\

			&
			$\Sigma_{\rho\alpha}^{\tau\dot{\alpha}}\Sigma^{\dagger\nu\alpha}_{\mu\dot{\alpha}}\Delta_{L\tau\nu\beta}^{\gamma}\Delta_{L\gamma}^{\dagger\rho\mu\beta}$&
			&
			$\Sigma_{\rho\alpha}^{\tau\dot{\alpha}}\Sigma^{\dagger\nu\alpha}_{\mu\dot{\alpha}}\Delta_{R\tau\nu\dot{\beta}}^{\dot{\gamma}}\Delta_{R\dot{\gamma}}^{\dagger\rho\mu\dot{\beta}}$
			\\

			&
			$\Sigma_{\rho\alpha}^{\tau\dot{\alpha}}\Sigma^{\dagger\rho\gamma}_{\tau\dot{\alpha}}\Delta_{L\mu\nu\beta}^{\alpha}\Delta_{L\gamma}^{\dagger\mu\nu\beta}$&
			&
			$\Sigma_{\rho\alpha}^{\tau\dot{\alpha}}\Sigma^{\dagger\rho\alpha}_{\tau\dot{\beta}}\Delta_{R\mu\nu\dot{\alpha}}^{\dot{\gamma}}\Delta_{R\dot{\gamma}}^{\dagger\mu\nu\dot{\beta}}$
			\\
			
			&
			$\Sigma_{\rho\alpha}^{\tau\dot{\alpha}}\Sigma^{\dagger\mu\gamma}_{\tau\dot{\alpha}}\Delta_{L\mu\nu\beta}^{\alpha}\Delta_{L\gamma}^{\dagger\nu\rho\beta}$&
			&
			$\Sigma_{\rho\alpha}^{\tau\dot{\alpha}}\Sigma^{\dagger\mu\alpha}_{\tau\dot{\beta}}\Delta_{R\mu\nu\dot{\alpha}}^{\dot{\gamma}}\Delta_{R\dot{\gamma}}^{\dagger\nu\rho\dot{\beta}}$
			\\

			&
			$\Sigma_{\rho\alpha}^{\tau\dot{\alpha}}\Sigma^{\dagger\rho\gamma}_{\mu\dot{\alpha}}\Delta_{L\tau\nu\beta}^{\alpha}\Delta_{L\gamma}^{\dagger\nu\mu\beta}$&
			&
			$\Sigma_{\rho\alpha}^{\tau\dot{\alpha}}\Sigma^{\dagger\rho\alpha}_{\mu\dot{\beta}}\Delta_{R\tau\nu\dot{\alpha}}^{\dot{\gamma}}\Delta_{R\dot{\gamma}}^{\dagger\nu\mu\dot{\beta}}$
			\\
			
			&
			$\Sigma_{\rho\alpha}^{\tau\dot{\alpha}}\Sigma^{\dagger\nu\gamma}_{\mu\dot{\alpha}}\Delta_{L\tau\nu\beta}^{\alpha}\Delta_{L\gamma}^{\dagger\rho\mu\beta}$&
			&
			$\Sigma_{\rho\alpha}^{\tau\dot{\alpha}}\Sigma^{\dagger\nu\alpha}_{\mu\dot{\beta}}\Delta_{R\tau\nu\dot{\alpha}}^{\dot{\gamma}}\Delta_{R\dot{\gamma}}^{\dagger\rho\mu\dot{\beta}}$
			\\
			\cline{1-4}
			
			\multirow{2}{*}{$\color{blue}{2\Sigma\Sigma\Delta_L^{\dagger}\Delta_L^{\dagger}} $}&
			$\color{blue}{\epsilon^{\alpha\beta}\epsilon_{\dot{\alpha}\dot{\beta}}\left(\Sigma^{\nu\dot{\alpha}}_{\mu\alpha}\Sigma^{\tau\dot{\beta}}_{\rho\beta}\Delta_{L\kappa}^{\dagger\mu\lambda\gamma}\Delta_{L\gamma}^{\dagger\rho\xi\kappa}\right)\epsilon_{\nu\tau\lambda\xi}}$&
			\multirow{2}{*}{$\color{blue}{2\Sigma\Sigma\Delta_R\Delta_R} $}&
			$\color{blue}{\epsilon^{\alpha\beta}\epsilon_{\dot{\alpha}\dot{\beta}}\left(\Sigma^{\nu\dot{\alpha}}_{\mu\alpha}\Sigma^{\tau\dot{\beta}}_{\rho\beta}\Delta_{R\nu\lambda\dot{\kappa}}^{\dot{\gamma}}\Delta_{R\tau\xi\dot{\gamma}}^{\dot{\kappa}}\right)\epsilon^{\mu\rho\lambda\xi}}$
			\\
			
			&
			$\color{blue}{\epsilon^{\alpha\kappa}\epsilon_{\dot{\alpha}\dot{\beta}}\left(\Sigma^{\nu\dot{\alpha}}_{\mu\alpha}\Sigma^{\tau\dot{\beta}}_{\rho\beta}\Delta_{L\kappa}^{\dagger\mu\lambda\gamma}\Delta_{L\gamma}^{\dagger\rho\xi\beta}\right)\epsilon_{\nu\tau\lambda\xi}}$&
			&
			$\color{blue}{\epsilon^{\alpha\beta}\epsilon_{\dot{\alpha}\dot{\kappa}}\left(\Sigma^{\nu\dot{\alpha}}_{\mu\alpha}\Sigma^{\tau\dot{\beta}}_{\rho\beta}\Delta_{R\nu\lambda\dot{\beta}}^{\dot{\gamma}}\Delta_{R\tau\xi\dot{\gamma}}^{\dot{\kappa}}\right)\epsilon^{\mu\rho\lambda\xi}}$
			\\
			\cline{1-4}
			
			\multirow{3}{*}{$3\Delta_L^{\dagger}\Delta_L\Delta_R^{\dagger}\Delta_R $}&
			$\Delta_{R\mu\nu\dot{\alpha}}^{\dot{\beta}}\Delta_{R\dot{\beta}}^{\dagger\mu\nu\dot{\alpha}}\Delta_{L\rho\tau\alpha}^{\beta}\Delta_{L\beta}^{\dagger\rho\tau\alpha}$&
			\multirow{2}{*}{$\color{blue}{2\Phi^{\dagger}\Sigma\Delta_L^{\dagger}\Delta_L} $}&
			$\color{blue}{\Phi_{\dot{\alpha}}^{\dagger\alpha}\Sigma^{\nu\dot{\alpha}}_{\mu\alpha}\Delta_{L\nu\rho\beta}^{\gamma}\Delta_{L\gamma}^{\dagger\rho\mu\beta}}$
			\\
			
			&
			$\Delta_{R\mu\nu\dot{\alpha}}^{\dot{\beta}}\Delta_{R\dot{\beta}}^{\dagger\nu\rho\dot{\alpha}}\Delta_{L\rho\tau\alpha}^{\beta}\Delta_{L\beta}^{\dagger\tau\mu\alpha}$&
			&
			$\color{blue}{\Phi_{\dot{\alpha}}^{\dagger\beta}\Sigma^{\nu\dot{\alpha}}_{\mu\alpha}\Delta_{L\nu\rho\beta}^{\gamma}\Delta_{L\gamma}^{\dagger\rho\mu\alpha}}$
			\\
			\cline{3-4}
			&
			$\Delta_{R\mu\nu\dot{\alpha}}^{\dot{\beta}}\Delta_{R\dot{\beta}}^{\dagger\rho\tau\dot{\alpha}}\Delta_{L\rho\tau\alpha}^{\beta}\Delta_{L\beta}^{\dagger\mu\nu\alpha}$&
			\multirow{2}{*}{$\color{blue}{2\Phi^{\dagger}\Sigma\Delta_R^{\dagger}\Delta_R} $}&
			$\color{blue}{\Phi_{\dot{\alpha}}^{\dagger\alpha}\Sigma^{\nu\dot{\alpha}}_{\mu\alpha}\Delta_{R\nu\rho\dot{\beta}}^{\dot{\gamma}}\Delta_{R\dot{\gamma}}^{\dagger\rho\mu\dot{\beta}}}$
			\\
			\cline{1-2}

			\multirow{3}{*}{$\color{blue}{3\Sigma\Sigma\Delta_L^{\dagger}\Delta_R} $}&
			$\color{blue}{\epsilon^{\alpha\kappa}\epsilon_{\dot{\alpha}\dot{\kappa}}\left(\Sigma^{\nu\dot{\alpha}}_{\mu\alpha}\Sigma^{\mu\dot{\beta}}_{\nu\beta}\Delta_{R\lambda\xi\dot{\beta}}^{\dot{\kappa}}\Delta_{L\kappa}^{\dagger\lambda\xi\beta}\right)}$&
			&
			$\color{blue}{\Phi_{\dot{\alpha}}^{\dagger\alpha}\Sigma^{\nu\dot{\beta}}_{\mu\alpha}\Delta_{R\nu\rho\dot{\beta}}^{\dot{\gamma}}\Delta_{R\dot{\gamma}}^{\dagger\rho\mu\dot{\alpha}}}$
			\\
			\cline{3-4}
			&
			$\color{blue}{\epsilon^{\alpha\kappa}\epsilon_{\dot{\alpha}\dot{\kappa}}\left(\Sigma^{\nu\dot{\alpha}}_{\mu\alpha}\Sigma^{\mu\dot{\beta}}_{\rho\beta}\Delta_{R\nu\tau\dot{\beta}}^{\dot{\kappa}}\Delta_{L\kappa}^{\dagger\tau\rho\beta}\right)}$&
			\multirow{2}{*}{$\color{red}{2\Sigma\Sigma^{\dagger}\Delta_R\Delta_L}$}&
			$\color{blue}{\epsilon^{\mu\rho\lambda\xi}\left(\Sigma_{\mu\alpha}^{\nu\dot{\alpha}}\Sigma_{\rho\dot{\beta}}^{\dagger\tau\beta}\Delta_{R\nu\lambda\dot{\alpha}}^{\dot{\beta}}\Delta_{L\tau\xi\beta}^{\alpha}\right)}$
			\\
			
			&
			$\color{blue}{\epsilon^{\alpha\kappa}\epsilon_{\dot{\alpha}\dot{\kappa}}\left(\Sigma^{\nu\dot{\alpha}}_{\mu\alpha}\Sigma^{\tau\dot{\beta}}_{\rho\beta}\Delta_{R\nu\tau\dot{\beta}}^{\dot{\kappa}}\Delta_{L\kappa}^{\dagger\mu\rho\beta}\right)}$&
			&
			$\color{blue}{\epsilon^{\mu\rho\tau\xi}\left(\Sigma_{\mu\alpha}^{\nu\dot{\alpha}}\Sigma_{\rho\dot{\beta}}^{\dagger\lambda\beta}\Delta_{R\nu\lambda\dot{\alpha}}^{\dot{\beta}}\Delta_{L\tau\xi\beta}^{\alpha}\right)}$
			\\
			\hline
	\end{tabular}}
	\caption{\small Table~\ref{table:patisalam-renorm-output} continued. The operator structures highlighted in red show discrepancies when compared with the results of \cite{Saad:2017pqj}. }
	\label{table:patisalam-renorm-output-2}
\end{table}

\begin{table}[h]
	\centering
	\renewcommand{\arraystretch}{1.8}
	{\tiny\begin{tabular}{|c|l|}
			\hline
			\multicolumn{2}{|c|}{\textbf{Mass Dimension-5}}\\
			\hline
			\textbf{Operator Class}&
			\hspace{2.5cm}\textbf{Operators (in non-covariant form)}\\
			\hline
			$\Psi^2\Phi^2$&
			$\color{blue}{\mathcal{S}\Sigma^{\dagger} \psi_L^{\dagger}\psi_R},\hspace{0.1cm} \mathcal{S}\Phi^{\dagger} \psi_L^{\dagger}\psi_R$\\
			\hline
			\multirow{4}{*}{$\Phi^5$}&
			$\color{blue}{\mathcal{S}(\mathcal{S}^{\dagger})^2\Sigma^2,\hspace{0.1cm} 6\mathcal{S}^{\dagger}\Sigma^3\Sigma^{\dagger},\hspace{0.1cm} 2\mathcal{S}(\mathcal{S}^{\dagger})^2\Phi^2,\hspace{0.1cm} \mathcal{S}^{\dagger}\Phi^3\Phi^{\dagger},\hspace{0.1cm} 2\mathcal{S}(\Sigma^{\dagger})^3\Phi,\hspace{0.1cm} 2\mathcal{S}^{\dagger}\Sigma^2\Phi\Phi^{\dagger},\hspace{0.1cm} 2\mathcal{S}^{\dagger}\Sigma\Sigma^{\dagger}\Phi^2,}$\\
			
			&
			$\color{blue}{4\mathcal{S}^{\dagger}\Sigma^2\Sigma^{\dagger}\Phi,\hspace{0.1cm} \mathcal{S}\Delta_L\Delta_R(\Sigma^{\dagger})^2,\hspace{0.1cm} \mathcal{S}^{\dagger}\Delta_L\Delta_R\Sigma^2,\hspace{0.1cm} 2\mathcal{S}^{\dagger}\Delta_L^2\Sigma\Sigma^{\dagger},\hspace{0.1cm} 2\mathcal{S}\Delta_R^2\Sigma\Sigma^{\dagger},\hspace{0.1cm} 4\mathcal{S}\Delta_R\Delta_L^{\dagger}\Sigma\Sigma^{\dagger},}$\\

			&
			$\color{blue}{4\mathcal{S}^{\dagger}\Delta_L\Delta_L^{\dagger}(\Sigma)^2,\hspace{0.1cm} \mathcal{S}\Delta_L^{\dagger}\Delta_R\Phi\Phi^{\dagger}, \hspace{0.1cm} \mathcal{S}^{\dagger}\Delta_L^{\dagger}\Delta_L\Phi^2,\hspace{0.1cm} \mathcal{S}^{\dagger}\Delta_R^{\dagger}\Delta_R\Phi^2,\hspace{0.1cm}
			\mathcal{S}\Delta_L^{\dagger}\Delta_R\Phi\Sigma^{\dagger}, \hspace{0.1cm} }$
			\\
			
			&
			$\color{blue}{4\mathcal{S}^{\dagger}\Delta_R\Delta_R^{\dagger}(\Sigma)^2,\hspace{0.1cm} \mathcal{S}^{\dagger}\Delta_L\Delta_R^{\dagger}\Phi\Sigma^{\dagger}, \hspace{0.1cm} \mathcal{S}^{\dagger}\Delta_R^{\dagger}\Delta_R\Phi^2,\hspace{0.1cm} 2\mathcal{S}^{\dagger}\Delta_L\Delta_L^{\dagger}\Phi\Sigma, \hspace{0.1cm} 2\mathcal{S}^{\dagger}\Delta_R\Delta_R^{\dagger}\Phi\Sigma}$\\
			\hline
	\end{tabular}}
	\caption{\small Pati-Salam Model: Operators of mass dimension-5.}
	\label{table:patisalam-dim-5-op}
\end{table} 

\section{Bridging the Theory and the Program}
In the previous sections we have delineated the mathematical ideas in an algorithmic way and we have shown how they can be used to write down the Lagrangian as a polynomial constituted of quantum fields. We have accentuated the procedure through an in-depth study of two distinct non-supersymmetric models. The same guiding principles can be employed to construct group invariant polynomials of superfields. Based on these ideas and to automatize the process a $\text{\texttt{Mathematica}}^{\tiny\textregistered}$ package has been developed named ``\textbf{GrIP}". This program asks for minimal information from the user about the model. One needs to provide only the particle content and their transformation properties under connected compact internal symmetry groups and the Lorentz group to generate the a complete and independent set of operators at any (canonical)mass dimension.  The remaining sections depict the anatomy of \textbf{GrIP} and illustrate its utility in the context of supersymmetric as well as non-supersymmetric model building.

%% file: GrIPCodeDescription.tex
\section{GrIP: Automatizing Invariant Polynomial Computation}\label{sec:grip-description}
\begin{figure}[h!]
	\centering
	{
		\includegraphics[trim = 0 80 0 0,scale=0.6]{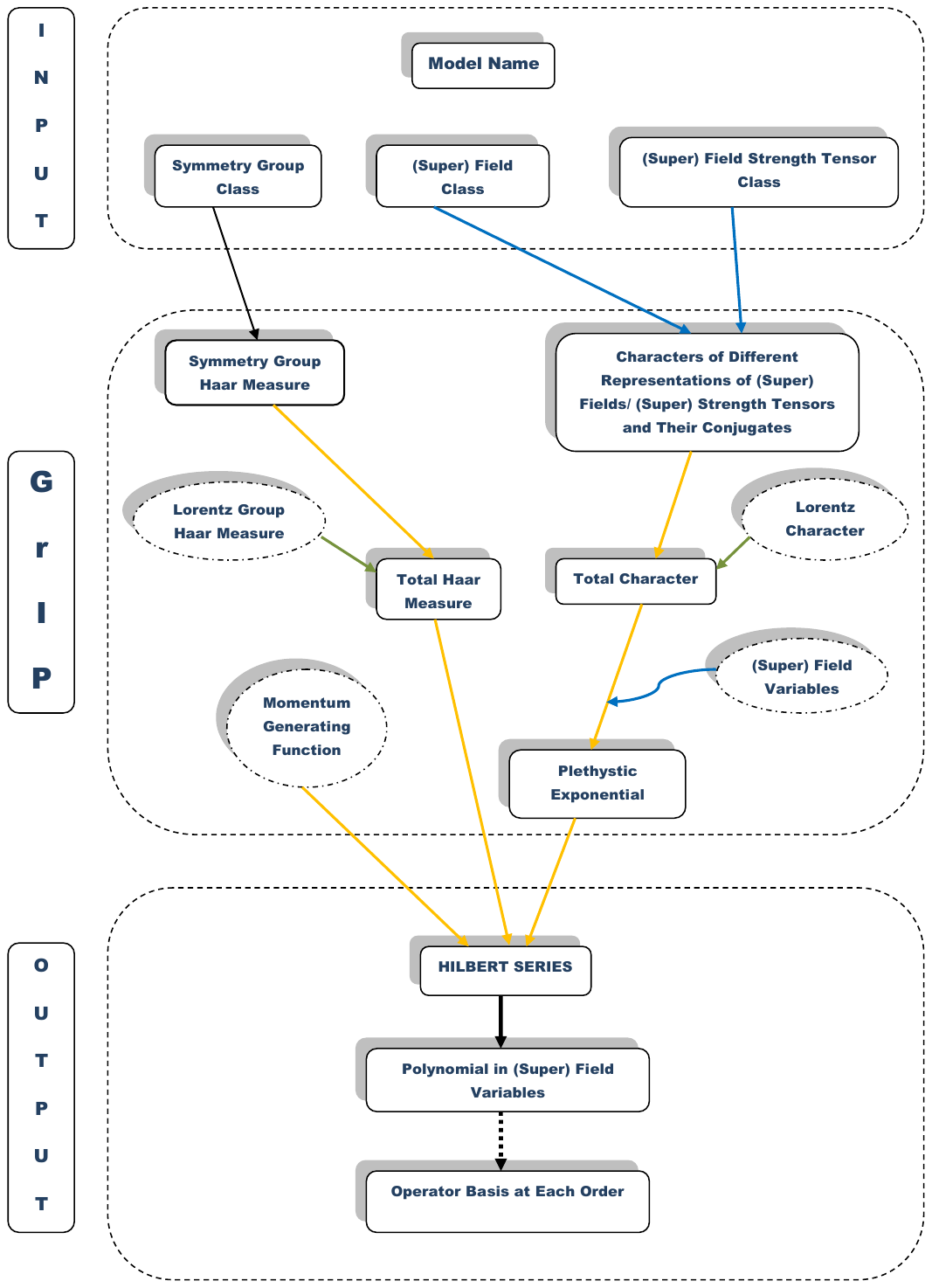}
	}
	\caption{A flow chart depicting the major components of \textbf{GrIP} and the algorithm on which it is based.}
	\label{fig:flowchart-code}
\end{figure}

\textbf{GrIP} is a \texttt{Mathematica}${}^{\tiny\textregistered}$\footnote{The current version of this code is compatible with \texttt{Mathematica}${}^{\tiny\textregistered}$\,V.11 and higher.} based scientific package that automatizes the computation of group invariant polynomials following the algorithm sketched in Fig.~\ref{fig:flowchart-code}.

\subsection{Introducing \textbf{GrIP}}\label{subsec:grip-download}
The complete package can be downloaded from \url{https://TeamGrIP.github.io/GrIP/} in \texttt{.zip} as well as \texttt{.tar.gz} formats. The downloaded file contains :\

\begin{enumerate}
	\item `GrIP.m' $\to$ This is the main program where all the relevant functions have been defined, which enables the user to carry out all of the computations. 
	\item `GrpInfo.m' $\to$ This file contains information about dimensions of the representations of various $SU(N)$ groups with $N\leq5$ and their corresponding Dynkin labels.
	\item `MODEL' $\to$ This folder contains examples for model input files. Two model files are provided one in terms of the Dynkin label (\_Dyn is appended at the end) another in terms of the representation (\_Rep is appended at the end), for each of the following scenarios.\footnote{The effective operators corresponding to the models depicted in italics are  discussed in this paper. The respective input files for all scenarios, given below, can be found in the downloaded package.}
	\begin{enumerate}
		\item SM and its Extensions: \textit{Standard Model} and its extension by the other lighter degrees of freedoms: \textit{Singly Charged Scalar}; \textit{Doubly Charged Scalar}; \textit{Complex Triplet Scalar}; \textit{$SU(2)$ Quadruplet Scalar}; \textit{$SU(2)$ Quintuplet Scalar}; \textit{Left-Handed Triplet Fermion}; \textit{Right-Handed Singlet Fermion}; \textit{Scalar Lepto-Quarks}; $SU(2)$ Doublet Scalar (with different hypercharge);  Real Triplet Scalar; Color Triplet Scalars and Sterile Neutrino;  $SU(2)$ Triplet and Quadruplet Fermions, 
		\item Supersymmetric scenarios: \textit{MSSM}, \textit{NMSSM}, \textit{Supersymmetric Pati-Salam}, \textit{Minimal Supersymmetric Left-Right} models.
		\item UV models: \textit{Two Higgs doublet}, \textit{Minimal Left-Right Symmetric}, \textit{Pati-Salam}, and \textit{$SU(5)$ Grand Unified} models.
		\item Models below electroweak scale: $SU(3)_C\otimes U(1)_{em}$ and its extension by additional:  \textit{Scalar Dark Matter}; \textit{Vector-like Fermion Dark Matter}.
	\end{enumerate}
	\item `Example\_SM.nb', `Example\_MSSM.nb' $\to$ Two $\text{\texttt{Mathematica}}^{\tiny\textregistered}$ Notebook files, for non-supersymmetric and supersymmetric input models are provided. The results there are generated using \texttt{"SM\_Rep.m"} and \texttt{"MSSM\_Rep.m"} input files from the `MODEL' folder respectively. 
	\item `CHaar.m' $\to$ This sub-program calculates character for a particular representation  and  Haar measure for a given connected compact group when the corresponding Dynkin label is provided. This program can be used without preparing any input file.  
	\item `Example\_CHaar.nb' $\to$ A $\text{\texttt{Mathematica}}^{\tiny\textregistered}$ Notebook file that contains illustrative examples showing how the functions of \textbf{CHaar} work. 
\end{enumerate}

\subsection{User Input: Model description}\label{subsec:grip-input}
%%%%%%%%%%%%%%%%%%%%%%%%%%%%%%%%%%%%%%%%%%%%%%%%%%%%%%%%%%%%%%%%%%%%%%%%
The user is required to prepare an input file to feed information into the main program. This file should contain information about the symmetry groups, particles and their representations and(or) charges under the given symmetries. The conjugate fields need not to be mentioned separately in the input file. For a given field, the respective conjugate will be created by the program.

\noindent
There are four main classes within the input file:

\subsubsection*{\underline{Model Name}}

There is a provision to save the results in a folder for each input model file. The user must provide a suitable name for the model as a string as shown below:
\begin{alltt}	
ModelName= "Name of the model".
\end{alltt}	
%%%%%%%%%%%%%%%%%%%%%%%%%%%%%%%%%%%%%%%%%%%%%%%%%%%%%%%%%%%%%%%%%%%%%%%%%
\subsubsection*{\underline{Group Class}}\label{subsubsec:grp_cls}
%%%%%%%%%%%%%%%%%%%%%%%%%%%%%%%%%%%%%%%%%%%%%%%%%%%%%%%%%%%%%%%%%%%%%%%%%
The user must provide information about the gauge and(or) global symmetry groups within the  \texttt{"SymmetryGroupClass"}. These groups must be connected and compact, but can represent global or gauge symmetries. The user needs to provide the name of the groups in the following format, i.e., one should write \texttt{"SUN"} for $SU(N)$ and \texttt{"U1"} for $U(1)$ groups. So far this program can handle symmetries of the form of $U(1)$ and $SU(N)$. A point to be noted is that for multiple occurrence of the same group one must give different names as shown in Table~\ref{table:gaugegroup}.

\begin{alltt}
SymmetryGroupClass =\{

Group[1] = \{
"GroupName" \(\to\) "\(\langle\)name of the group\(\rangle\)",
"N" \(\to\) \(\langle N \)for \(SU(N)\), 1 for \(U(1)\rangle\) 
\},
\(\vdots\)
\};
\end{alltt}

\begin{table}[t]
	\centering 
	\renewcommand{\arraystretch}{1.6}
	{\scriptsize\begin{tabular}{|lcl|}
		\hline
		\bf{Keys} & \bf{Values}  & \bf{Details} \\ 
		\hline
		\textcolor{mmaUndefined}{\texttt{GroupName}} & \texttt{"SUN","U1",} & Name of the groups.\\ 
		& \texttt{"SUNL","SUNR",} & There is no restriction on the\\ 
		& \texttt{"U1Y","U1X".} &  number and ordering of the groups. \\ 
		& &   \\
		\textcolor{mmaUndefined}{\texttt{N}} &  \texttt{N} for \texttt{"SUN",} & Degree of the  group.   \\ 
		&  \texttt{1} for \texttt{"U1".}& No distinction for different $SU(N)$'s. \\
		\hline
	\end{tabular}}
	\caption{\small Information related to symmetry groups that must be specified within the \texttt{"SymmetryGroupClass"} in the input file along with the possible values.}\label{table:gaugegroup}
\end{table}
%%%%%%%%%%%%%%%%%%%%%%%%%%%%%%%%%%%%%%%%%%%%%%%%%%%%%%%%%%%%%%%%%%%%%%%%%
\subsubsection*{\underline{(Super)Field Class}}\label{subsubsec:field_class}
%%%%%%%%%%%%%%%%%%%%%%%%%%%%%%%%%%%%%%%%%%%%%%%%%%%%%%%%%%%%%%%%%%%%%%%%%

(Super)Field content of the model should be given in this class. Within this class following information must be provided:\\ \texttt{\{ <"FieldName">,  <"Self-Conjugate">, <"Lorentz Behaviour">, <"Chirality">,\\
<"Baryon Number">, <"Lepton Number">, <"G1Rep/G1Dyn">, $\cdots$\}}. If there are multiple symmetry groups, then the quantum numbers or representations of a particle under each group must be provided in the same order in which the groups are mentioned in the  \texttt{"SymmetryGroupClass"}. If the model does not contain any (Super)Field, the user should leave this part empty. The possible values of these attributes are described in Table~\ref{table:particlelist}.

\begin{table}[h]
	\centering
	\renewcommand{\arraystretch}{1.5}
	{\scriptsize\begin{tabular}{|lcl|}
			\hline
			\bf{Keys} & \bf{Values}  & \bf{Details} \\
			\hline
			\textcolor{mmaUndefined}{\texttt{FieldName}} & \texttt{Alphabets, characters} & It represents the particle (superfield). Operators\\ & \texttt{or symbols.} & will be obtained in terms of these variables.  \\
			& & \\
			\textcolor{mmaUndefined}{\texttt{Self-Conjugate}} & \texttt{True} or \texttt{False.} & It decides whether to prepare the conjugate \\ & & field of this particular (super)field or not. \\
			& & The option should be selected considering the\\
			& &  gauge and Lorentz symmetry behavior.\\
			& & \\
			\textcolor{mmaUndefined}{\texttt{Lorentz Behaviour}} & \texttt{"SCALAR",} & These values describe the spin of the particle:\\
			& \texttt{"FERMION",} &  0 (scalar), 1/2 (fermion), 1 (vector).\\
			& \texttt{"VECTOR".}& In the case of superfields we attach the prefix \\
			& & ``\texttt{SUPER}" to the respective values. Superfields\\
			& &  behave like scalars under the Lorentz group.\\
			& & \\
			
			\textcolor{mmaUndefined}{\texttt{Chirality}} & \texttt{"l",} & \texttt{"l"} denotes left-handed(LH) field or conjugate of \\
			& \texttt{"r",} &  right-handed(RH) one, \texttt{"r"} denotes RH field or\\
			& \texttt{"NA".} &  conjugate of LH one. \texttt{"NA"} is for scalar field only.\\
			& & \\
			\textcolor{mmaUndefined}{\texttt{Baryon Number}} & \texttt{Numerical Value.} & Assignment should be consistent with the\\ & ($\in \mathbb{Z}$) & SM hypercharge. \\
			& & \\
			\textcolor{mmaUndefined}{\texttt{Lepton Number}} & \texttt{Numerical Value.} & Assignment should be consistent with the\\ & ($\in \mathbb{Z}$) & SM hypercharge. \\
			& & \\
			\textcolor{mmaUndefined}{\texttt{G1Rep}} & \texttt{+ve integer.} & Dimensionality of the  representation of the\\
			& ($\in \mathbb{Z^+}$) &  field under group G1. \\ 
			& & \\
			\textcolor{mmaUndefined}{\texttt{G1Dyn}} & \texttt{n-tuple - \{ \}}&  Dynkin label of the representation. \\
			\hline
	\end{tabular}}
	\caption{\small Characteristics of the particles that must be specified within \texttt{"(Super)FieldClass"} and \texttt{"(Super)FieldTensorClass"} in the input file along with the possible values.}\label{table:particlelist}
\end{table}
\noindent
The characteristics of the particle (or superfield) under group \texttt{"G1"} can be incorporated in two ways: \\
either \texttt{"G1Rep"-> "<dimension of the representation under Group[1]>"},\\
or \texttt{"G1Dyn"-> "<Dynkin label of the representation under Group[1]>"}.\\
\\
\noindent 
The \texttt{"(Super)FieldClass"} when representations are provided in terms of dimension is: 
\begin{alltt}
	(Super)FieldClass=\{
	
	(Super)Field[1]=\{
	"FieldName"-> < >,	
	"Self-Conjugate"-> < >,	
	"Lorentz Behaviour"-> <" ">,	
	"Chirality"-> <" ">,
	"Baryon Number"-> < # >, 
	"Lepton Number"-> < # >,
	"G1Rep"-> <dimension of the representation under Group[1]>,
\(\vdots\)
\},\(\cdots\)
\};
\end{alltt}

\noindent
The \texttt{"(Super)FieldClass"} when representations are provided in terms of Dynkin label is: 
\begin{alltt}
(Super)FieldClass=\{
(Super)Field[1]=\{
"FieldName"-> < >, 
"Self-Conjugate"-> < >,
"Lorentz Behaviour"-> <" ">,
"Chirality"-> <" ">,
"Baryon Number"-> < # >,
"Lepton Number"-> < # >,
"G1Dyn"-> <Dynkin label of the representation under Group[1]>,
\(\vdots\)
\},\(\cdots\)
\};
\end{alltt}
%%%%%%%%%%%%%%%%%%%%%%%%%%%%%%%%%%%%%%%%%%%%%%%%%%%%%%%%%%%%%%%%%%%%%%%%%
\subsubsection*{\underline{(Super)Field Strength Tensor Class}}\label{subsubsec:field_strength}
%%%%%%%%%%%%%%%%%%%%%%%%%%%%%%%%%%%%%%%%%%%%%%%%%%%%%%%%%%%%%%%%%%%%%%%%%
A gauge invariant Lagrangian does not explicitly contain the gauge fields, instead, it contains the corresponding field strength tensors. In the \texttt{"(Super)FieldTensorClass"} the user must provide the information about these (super)field strength tensors. In absence of any gauge symmetry, this part should be left empty. We must emphasize that within the \texttt{"SymmetryGroupClass"} all the symmetry groups, irrespective of gauge or global symmetries, must be enlisted. But field strength tensors correspond to only the gauge symmetries and not the global symmetries. The user must keep this in mind while framing this sector.
\begin{alltt}
(Super)FieldTensorClass=\{

(Super)TensorField[1]=\{
"FieldName"-> < >,
"Self-Conjugate"-> < >,
"Lorentz Behaviour"-> <" ">,
"Chirality"-> <" ">,
"Baryon Number"-> < # >,
"Lepton Number"-> < # >,
"G1Rep"-> representation under Group[1],
\(\vdots\)
\},\(\cdots\)
\};
\end{alltt}

%%%%%%%%%%%%%%%%%%%%%%%%%%%%%%%%%%%%%%%%%%%
\subsection{Running the code}\label{subsec:grip-run}
The user must perform the following tasks to run the code successfully:
\begin{enumerate}
	
\item \textbf{Step-1:} The user should prepare a `Model' file, if it is not already inside the `MODEL' folder and keep it inside the `MODEL' folder which already contains many example model input files. 
	
\item \textbf{Step-2:} Then load the model file using \texttt{"Get[]"} function in \texttt{Mathematica}${}^{\tiny\textregistered}$ by\\ providing proper \texttt{\$Path} to access the model file:

{\footnotesize\textcolor{mmaLabel}{\fontfamily{phv}\selectfont In[1]:=}}
\textcolor{mmaUndefined}{\texttt{SetDirectory}}\texttt{[\textcolor{mmaString}{"<provide address of the MODEL folder>"}]}

{\footnotesize\textcolor{mmaLabel}{\fontfamily{phv}\selectfont In[2]:=}}
\textcolor{mmaUndefined}{\texttt{Get}}\texttt{[\textcolor{mmaString}{"MODEL/Model.m"}]}
	
\item \textbf{Step-3:} Now one has to install the main program. There are two ways to do that:
	\begin{enumerate}
	\item If the user keeps `GrIP.m' in a local folder, to load one has to use \texttt{"Get[]"} with proper address of the main program file as shown in \textbf{Step-2}:

	\vspace{0.15cm}	
	{\footnotesize\textcolor{mmaLabel}{\fontfamily{phv}\selectfont In[3]:=}}
	\textcolor{mmaUndefined}{\texttt{SetDirectory}}\texttt{[\textcolor{mmaString}{"<provide address of the package>"}]}
	
	{\footnotesize\textcolor{mmaLabel}{\fontfamily{phv}\selectfont In[4]:=}}
	\textcolor{mmaUndefined}{\texttt{Get}}\texttt{[\textcolor{mmaString}{"GrIP.m"}]}
	\vspace{0.15cm}	
	
	\item If the user keeps it in the  `Applications' folder in \texttt{\$UserBaseDirectory}, it can be loaded using \texttt{"Needs[]"}:

	\vspace{0.15cm}
	{\footnotesize\textcolor{mmaLabel}{\fontfamily{phv}\selectfont In[3]:=}}
	\textcolor{mmaUndefined}{\texttt{Needs}}\texttt{[\textcolor{mmaString}{"GrIP\,$\grave{}$"}]}
	\end{enumerate}
\end{enumerate}
If the program is loaded correctly, a text cell will appear in the Notebook file to notify the user.

\subsection{Saving private output}\label{subsec:programme_output}
Once the program is loaded, a folder named \texttt{"Name of the model"} will be created in the working directory. All the results generated through specific commands will be saved in this folder in TeXForm. 

The  output functions available in \textbf{GrIP} can be grouped into the following categories based on their specific purpose. Details on these have also been provided in Tables~\ref{table:grip_output}-\ref{table:susy_function_details}.

%%%%%%%%%%%%%%%%%%%%%%%%%%%%%%%%%%%%%%%%%%%%%%%%%%%%%%%%%%%%%%%%%%%%%%%%%
\subsubsection*{\underline{Input verification}}\label{subsubsec:Grp_input_table}
%%%%%%%%%%%%%%%%%%%%%%%%%%%%%%%%%%%%%%%%%%%%%%%%%%%%%%%%%%%%%%%%%%%%%%%%%

The input information provided by the user can be verified using:

\vspace{0.15cm}
{\footnotesize\textcolor{mmaLabel}{\fontfamily{phv}\selectfont In[1]:=}}
\textcolor{mmaUndefined}{\texttt{DisplayUserInputTable}}

\vspace{0.15cm}
\noindent
We also need the conjugate fields to construct valid Lagrangian. These conjugate fields will be automatically generated internally based on the information entered by the user in the model input file, and the full list of particles can be verified through:

\vspace{0.15cm}
{\footnotesize\textcolor{mmaLabel}{\fontfamily{phv}\selectfont In[2]:=}}
\textcolor{mmaUndefined}{\texttt{DisplayWorkingInputTable}}

\subsubsection*{\underline{Group theoretic output}}\label{subsubsec:Grp_output}
\noindent
The characters of the representations of each field and their respective conjugate fields can be obtained through:

\vspace{0.15cm}
{\footnotesize\textcolor{mmaLabel}{\fontfamily{phv}\selectfont In[1]:=}}
\textcolor{mmaUndefined}{\texttt{DisplayCharacterTable}}

\vspace{0.15cm}
\noindent
Similarly, the associated Haar measure of the underlying symmetry groups are displayed using:

\vspace{0.15cm}
{\footnotesize\textcolor{mmaLabel}{\fontfamily{phv}\selectfont In[2]:=}}
\textcolor{mmaUndefined}{\texttt{DisplayHaarMeasure}}

\begin{table}[t]
		\centering
		\renewcommand{\arraystretch}{1.6}
		{\scriptsize\begin{tabular}{|ll|}
			\hline
			\textbf{Functions} & \textbf{Details of the output} \\
			\hline
			\textcolor{mmaUndefined}{\texttt{DisplayUserInputTable}} & A list of particles  and their properties\\ 
			&  provided by the user.\\
			&  \\
			\textcolor{mmaUndefined}{\texttt{DisplayWorkingInputTable}} & A  list of particles \& the respective conjugate   \\ 
			& fields and their properties based on the  \\ 
			& information provided by the user.\\
			&  \\
			\textcolor{mmaUndefined}{\texttt{DisplayCharacterVariables}} & The list of independent variables for\\ 
			&  different symmetry groups. \\
			&  The no. of variables = Rank of the group.\\
			&  The characters of the representations\\
			&  are functions of these variables.\\
			
			& \\
			\textcolor{mmaUndefined}{\texttt{DisplayCharacterTable}}  & Explicit structures of the character  function \\
			&  correspond to the particle representation\\
			&  in terms of the character variables.\\
			&  \\
			\textcolor{mmaUndefined}{\texttt{DisplayHaarMeasure}} & The Haar measure of the symmetry groups\\
			& in terms of the character variables. \\
			\hline
		\end{tabular}}
	\caption{\small Different output functions of \textbf{GrIP} and their details.}\label{table:grip_output}
\end{table}

%%%%%%%%%%%%%%%%%%%%%%%%%%%%%%%%%%%%%%%%%%%%%%%%%%%%%%%%%%%%%%%%%%%%%%%%%
\subsubsection*{\underline{Hilbert Series output: polynomial of fields}}\label{subsubsec:grip-lag-output}
%%%%%%%%%%%%%%%%%%%%%%%%%%%%%%%%%%%%%%%%%%%%%%%%%%%%%%%%%%%%%%%%%%%%%%%%%
This program aims to compute the independent group invariant polynomials that can be expressed as the monomial basis of different mass dimensions. We can further classify the operators for a given mass dimension as a polynomial whose order is determined by specific baryon number and lepton number violation. Here, we have provided two options to get the operators: (i) for a specific mass dimension setting \texttt{"OnlyMassDimOutput"-> True} and (ii) up to a specific mass dimension setting \texttt{"OnlyMassDimOutput"-> False}.  One can be more specific while selecting the operator set by probing specific values (\texttt{\#}) to 
\texttt{"$\Delta$B"->\#} and \texttt{"$\Delta$L"->\#}.

\vspace{0.15cm}
{\footnotesize\textcolor{mmaLabel}{\fontfamily{phv}\selectfont In[1]:=}}
\textcolor{mmaUndefined}{\texttt{DisplayHSOutput}}\texttt{[\textcolor{mmaString}{"MassDim"}$\to$\textcolor{mmaLocal}{\#}, \textcolor{mmaString}{"OnlyMassDimOutput"}$\to$\textcolor{mmaLocal}{\$},\\ \hspace*{5cm}\textcolor{mmaString}{"$\Delta$B"}$\to$\textcolor{mmaLocal}{ \#}, \textcolor{mmaString}{"$\Delta$L"}$\to$\textcolor{mmaLocal}{\#}, \textcolor{mmaString}{"Flavours"}$\to$\textcolor{mmaLocal}{\#}]}

\vspace{0.15cm}
\noindent
Note that for the case of supersymmetric models we have the following function which serves the same purpose:

\vspace{0.15cm}
{\footnotesize\textcolor{mmaLabel}{\fontfamily{phv}\selectfont In[1]:=}}
\textcolor{mmaUndefined}{\texttt{DisplaySHSOutput}}\texttt{[\textcolor{mmaString}{"CanonicalDim"}$\to$\textcolor{mmaLocal}{\#}, \textcolor{mmaString}{"OnlyCanonicalDimOutput"}$\to$\textcolor{mmaLocal}{\$},\\ \hspace*{5cm}\textcolor{mmaString}{"$\Delta$B"}$\to$\textcolor{mmaLocal}{ \#}, \textcolor{mmaString}{"$\Delta$L"}$\to$\textcolor{mmaLocal}{\#}, \textcolor{mmaString}{"Flavours"}$\to$\textcolor{mmaLocal}{\#}]}

\vspace{0.15cm}
\noindent
One can also look for operators violating baryon number (\texttt{B}) and lepton number (\texttt{L}) by specific amounts using the following command:\\

\vspace{0.15cm}
{\footnotesize\textcolor{mmaLabel}{\fontfamily{phv}\selectfont In[1]:=}}
\textcolor{mmaUndefined}{\texttt{DisplayBLviolatingOperators}}\texttt{[\textcolor{mmaString}{"HighestMassDim"}$\to$\textcolor{mmaLocal}{\#}, \textcolor{mmaString}{"$\Delta$B"}$\to$\textcolor{mmaLocal}{\#},\\ 
\hspace*{7cm} \textcolor{mmaString}{"$\Delta$L"}$\to$\textcolor{mmaLocal}{\#}, \textcolor{mmaString}{"Flavours"}$\to$\textcolor{mmaLocal}{\#}]}

\vspace{0.15cm}
\noindent
The argument \texttt{"HighestMassDim"} sets the upper limit of search process. The function always returns the lowest dimensional operator amounting the required \texttt{B} and \texttt{L} violations if any within the given \texttt{"HighestMassDim"}.

If the user wants to obtain a suggestive form of the Lagrangian, it is also possible to collect all the terms of the Hilbert Series polynomial and give them the schematic form of a Lagrangian through suitable incorporation of the dimension full parameters. This can be obtained using the following function:

\vspace{0.15cm}
{\footnotesize\textcolor{mmaLabel}{\fontfamily{phv}\selectfont In[1]:=}}
\textcolor{mmaUndefined}{\texttt{DisplayLagOutput}}\texttt{[\textcolor{mmaString}{"MassDim"}$\to$\textcolor{mmaLocal}{\#}, \textcolor{mmaString}{"OnlyMassDimOutput"}$\to$\textcolor{mmaLocal}{\$},\\ \hspace*{5cm}\textcolor{mmaString}{"$\Delta$B"}$\to$\textcolor{mmaLocal}{ \#}, \textcolor{mmaString}{"$\Delta$L"}$\to$\textcolor{mmaLocal}{\#}, \textcolor{mmaString}{"Flavours"}$\to$\textcolor{mmaLocal}{\#}]}

\vspace{0.15cm}
\noindent
There is also the provision of storing the output (i.e., the operator set) under a variable to allow for further manipulations for both non-supersymmetric and supersymmetric models. For the non-supersymmetric case, this job is performed using:

\vspace{0.15cm}
{\footnotesize\textcolor{mmaLabel}{\fontfamily{phv}\selectfont In[1]:=}}
\textcolor{mmaLocal}{\texttt{Poly}} = \textcolor{mmaUndefined}{\texttt{SaveHSOutput}}\texttt{[\textcolor{mmaString}{"MassDim"}$\to$\textcolor{mmaLocal}{\#}, \textcolor{mmaString}{"$\Delta$B"}$\to$\textcolor{mmaLocal}{ \#}, \textcolor{mmaString}{"$\Delta$L"}$\to$\textcolor{mmaLocal}{\#}, \textcolor{mmaString}{"Flavours"}$\to$\textcolor{mmaLocal}{\#}]}

\vspace{0.15cm}
\noindent
Same can be achieved for the supersymmetric case using:

\vspace{0.15cm}
{\footnotesize\textcolor{mmaLabel}{\fontfamily{phv}\selectfont In[1]:=}}
\textcolor{mmaLocal}{\texttt{Poly}} = \textcolor{mmaUndefined}{\texttt{SaveSHSOutput}}\texttt{[\textcolor{mmaString}{"CanonicalDim"}$\to$\textcolor{mmaLocal}{\#}, \textcolor{mmaString}{"$\Delta$B"}$\to$\textcolor{mmaLocal}{ \#}, \textcolor{mmaString}{"$\Delta$L"}$\to$\textcolor{mmaLocal}{\#},\\
\hspace*{5.5cm} \textcolor{mmaString}{"Flavours"}$\to$\textcolor{mmaLocal}{\#}]}

\vspace{0.15cm}
\noindent
To count the number of terms in any polynomial one can use the function \texttt{\textcolor{mmaUndefined}{OpCounter}}. If the polynomial is composed of chiral and vector superfields then the corresponding task is carried out by \texttt{\textcolor{mmaUndefined}{SusyOpCounter}}:

\vspace{0.15cm}
{\footnotesize\textcolor{mmaLabel}{\fontfamily{phv}\selectfont In[2]:=}}
\textcolor{mmaUndefined}{\texttt{OpCounter}}\texttt{[\textcolor{mmaLocal}{Poly}]}

\vspace{0.15cm}
{\footnotesize\textcolor{mmaLabel}{\fontfamily{phv}\selectfont In[2]:=}}
\textcolor{mmaUndefined}{\texttt{SusyOpCounter}}\texttt{[\textcolor{mmaLocal}{Poly}]}

\vspace{0.15cm}
\noindent
Given a polynomial in the input fields, one can also impose external global symmetries which were not defined in \texttt{"SymmetryGroupClass"} to sort out specific operators. The function \texttt{\textcolor{mmaUndefined}{ReOutput}} enables the user to do that:

\vspace{0.15cm}
{\footnotesize\textcolor{mmaLabel}{\fontfamily{phv}\selectfont In[3]:=}}
\textcolor{mmaUndefined}{\texttt{ReOutput}}\texttt{[\textcolor{mmaString}{"NameOfPoly"}$\to$\textcolor{mmaLocal}{Poly}, \textcolor{mmaString}{"SymmetryName"}$\to$\{"QN1","QN2",$\cdots$\},\\ 
\hspace*{4cm} \textcolor{mmaString}{"Qno"}$\to$\{\{FieldName$\to$\textcolor{mmaLocal}{\#}\},$\cdots$\},
\textcolor{mmaString}{"$\Delta$sym"}$\to$\{\textcolor{mmaLocal}{\$},$\cdots$\}]}

\begin{table}[htb!]
	\centering
	\renewcommand{\arraystretch}{1.4}
		{\scriptsize\begin{tabular}{|lcl|}
			\hline	
			\textbf{Functions} & \textbf{Options} & \textbf{Details of the output} \\
			\hline
			\textcolor{mmaUndefined}{\texttt{DisplayHSOutput}}	& 
			\textcolor{mmaLocal}{\texttt{MassDim}} $\to$ \texttt{positive integer},&
			Operators at a given mass dimension\\
			
			\texttt{[\textcolor{mmaLocal}{MassDim}, \textcolor{mmaLocal}{OnlyMassDimOutput},}&
			\textcolor{mmaLocal}{\texttt{OnlyMassDimOutput}}$\to$ \texttt{True}, &   
			irrespective of B \& L number violation\\
			
			\texttt{\textcolor{mmaLocal}{$\Delta$B}, \textcolor{mmaLocal}{$\Delta$L}, \textcolor{mmaLocal}{Flavours}]}& 
			\texttt{\textcolor{mmaLocal}{$\Delta$B}} $\to$ \texttt{"NA"},
			\texttt{\textcolor{mmaLocal}{$\Delta$L}} $\to$ \texttt{"NA"},  & 
			with specific no. of fermionic flavours. \\

			& 	
			\texttt{\textcolor{mmaLocal}{Flavours}} $\to$ \texttt{1, $N_f$}.& 
			\\
			
			&   & \\

			& \texttt{\textcolor{mmaLocal}{OnlyMassDimOutput}}
			$\to$ \texttt{True}, & Operators at a given mass dimension  \\
			&  \texttt{\textcolor{mmaLocal}{$\Delta$B}} $\to$ \texttt{integer},
			\texttt{\textcolor{mmaLocal}{$\Delta$L}} $\to$ \texttt{integer}. & posing specific B \& L number violation.\\ 
			& & \\
			
			& \texttt{\textcolor{mmaLocal}{OnlyMassDimOutput}}
			$\to$ \texttt{False}, & Operators up to a  given mass dimension\\
			&  \texttt{\textcolor{mmaLocal}{$\Delta$B}} $\to$ \texttt{"NA"},
			\texttt{\textcolor{mmaLocal}{$\Delta$L}} $\to$ \texttt{"NA"}. &  irrespective of B \& L number violation. \\
			& & \\
			
			& \texttt{\textcolor{mmaLocal}{OnlyMassDimOutput}}
			$\to$ \texttt{False}, & Operators up to a  given mass dimension \\
			& \texttt{\textcolor{mmaLocal}{$\Delta$B}} $\to$ \texttt{integer},
			\texttt{\textcolor{mmaLocal}{$\Delta$L}} $\to$ \texttt{integer}. & posing specific B \& L number violation. \\
			& & \\
			
			\texttt{\textcolor{mmaUndefined}{DisplayLagOutput}} & Same as 	\texttt{\textcolor{mmaUndefined}{DisplayHSOutput}}. & Similar as \texttt{\textcolor{mmaUndefined}{DisplayHSOutput}}, where all\\
			
			\texttt{[\textcolor{mmaLocal}{MassDim}, \textcolor{mmaLocal}{OnlyMassDimOutput},}
			& 	  &  the operators are collected and multiplied \\
			
			\texttt{\textcolor{mmaLocal}{$\Delta$B}, \textcolor{mmaLocal}{$\Delta$L}, \textcolor{mmaLocal}{Flavours}]}
			&   & by suitable mass-dimensional couplings. \\ 
			& & \\
			
			\texttt{\textcolor{mmaUndefined}{DisplayBLviolatingOperators}} & \texttt{\textcolor{mmaLocal}{$\Delta$B}} $\to$ \texttt{integer},
			\texttt{\textcolor{mmaLocal}{$\Delta$L}} $\to$ \texttt{integer},  & The lowest mass dimensional operator \\
			
			\texttt{[\textcolor{mmaLocal}{HighestMassDim}, \textcolor{mmaLocal}{$\Delta$B}, \textcolor{mmaLocal}{$\Delta$L},}
			& 	\texttt{\textcolor{mmaLocal}{HighestMassDim}} $\to$ \texttt{\#}, & that posses the mentioned amount of  \\
			\texttt{\textcolor{mmaLocal}{Flavours}]}
			& 	\texttt{\textcolor{mmaLocal}{Flavours}} $\to$ 1, $N_f$. & B \& L violation within the specified \texttt{\#}. \\
			& & \\
			
			\texttt{\textcolor{mmaLocal}{Poly}} = \texttt{\textcolor{mmaUndefined}{SaveHSOutput}} & \texttt{\textcolor{mmaLocal}{Poly}} $\to$
			\texttt{any variable}. & Saves the operators as a polynomial \\			
			
			\texttt{[\textcolor{mmaLocal}{MassDim}, \textcolor{mmaLocal}{$\Delta$B}, \textcolor{mmaLocal}{$\Delta$L}, \textcolor{mmaLocal}{Flavours}]} & & named \texttt{\textcolor{mmaLocal}{Poly}}.\\ 
			&  & \\		
			
			\texttt{\textcolor{mmaUndefined}{OpCounter}[\textcolor{mmaLocal}{Poly}]} & \texttt{\textcolor{mmaLocal}{Poly}} $\to$ \texttt{any variable}.  & Number of independent invariant \\
			& & operators that constitute the polynomial.\\	
			& & \\
			
			\texttt{\textcolor{mmaUndefined}{ReOutput}} & \texttt{\textcolor{mmaLocal}{NameOfPoly}}$\to$ \texttt{Poly}, & Assigns qunatum no. \texttt{\#} to the \\
			
			\texttt{[\textcolor{mmaLocal}{NameOfPoly},}  &\texttt{\textcolor{mmaLocal}{SymmetryName}}$\to$\texttt{\{"QN1","QN2",...\}}, & particle and returns those operators\\
			
			\texttt{\textcolor{mmaLocal}{SymmetryName}, \textcolor{mmaLocal}{Qno}, \textcolor{mmaLocal}{$\Delta$sym}}\texttt{]} &\texttt{\textcolor{mmaLocal}{Qno}}$\to$\texttt{\{\{FieldName $\to$ \#\}}, & which violates the quantum no by \texttt{\$} unit. \\
			
			&\{\texttt{FieldName} $\to$ \#\},...\}, & \\
			
			& \texttt{\textcolor{mmaLocal}{$\Delta$sym}} $\to$ \texttt{\{\$,\$...\}}. & \\
			
			&  & \\		
			
			\hline
		\end{tabular}}
\caption{\small \textbf{GrIP} functions and their working principles.}\label{table:function_details}
\end{table}
\clearpage
\begin{table}[htb!]
	\centering
	\renewcommand{\arraystretch}{1.4}
	{\scriptsize\begin{tabular}{|lcl|}
			\hline	
			\textbf{Functions} & \textbf{Options} & \textbf{Details of the Output} \\
			\hline
			\texttt{\textcolor{mmaUndefined}{DisplaySHSOutput}}	&
			\texttt{\textcolor{mmaLocal}{CanonicalDim}} $\to$ \texttt{positive integer}, &
			Operators at a given canonical dimension \\
			
			\texttt{[\textcolor{mmaLocal}{CanonicalDim}, \textcolor{mmaLocal}{OnlyCanonicalDimOutput},} &
			\texttt{\textcolor{mmaLocal}{OnlyCanonicalDimOutput}}$\to$ \texttt{True}, &   
			irrespective of B \& L number violation\\
						
			\texttt{\textcolor{mmaLocal}{$\Delta$B}, \textcolor{mmaLocal}{$\Delta$L}, \textcolor{mmaLocal}{Flavours}]} & 
			\texttt{\textcolor{mmaLocal}{$\Delta$B}} $\to$ \texttt{"NA"},
			\texttt{\textcolor{mmaLocal}{$\Delta$L}} $\to$ \texttt{"NA"},  & 
			with specific no. of fermionic flavours.\\
			
			& 	
			\texttt{\textcolor{mmaLocal}{Flavours}} $\to$ 1, $N_f$.	& 
			\\
			
			&   & \\

			& \texttt{\textcolor{mmaLocal}{OnlyCanonicalDimOutput}}
			$\to$ \texttt{True}, & Operators at a given canonical dimension  \\
			&  \texttt{\textcolor{mmaLocal}{$\Delta$B}} $\to$ \texttt{integer},
			\texttt{\textcolor{mmaLocal}{$\Delta$L}} $\to$ \texttt{integer}. & posing specific B \& L number violation.\\ 
			& & \\
			
			& \texttt{\textcolor{mmaLocal}{OnlyCanonicalDimOutput}}
			$\to$ \texttt{False}, & Operators up to a  given canonical dimension\\
			&  \texttt{\textcolor{mmaLocal}{$\Delta$B}} $\to$ \texttt{"NA"},
			\texttt{\textcolor{mmaLocal}{$\Delta$L}} $\to$ \texttt{"NA"}. &  irrespective of B \& L number violation. \\
			& & \\
			
			& \texttt{\textcolor{mmaLocal}{OnlyCanonicalDimOutput}}
			$\to$ \texttt{False}, & Operators up to a  given mass dimension \\
			&  \texttt{\textcolor{mmaLocal}{$\Delta$B}} $\to$ \texttt{integer},
			\texttt{\textcolor{mmaLocal}{$\Delta$L}} $\to$ \texttt{integer}. & posing specific B \& L number violation. \\
			& & \\

			\texttt{\textcolor{mmaLocal}{Poly}} = \texttt{\textcolor{mmaUndefined}{SaveSHSOutput}} & \texttt{\textcolor{mmaLocal}{Poly}} $\to$
			\texttt{any variable}. & Saves the supersymmetric operators as a \\
			
			\texttt{[\textcolor{mmaLocal}{CanonicalDim}, \textcolor{mmaLocal}{$\Delta$B}, \textcolor{mmaLocal}{$\Delta$L}, \textcolor{mmaLocal}{Flavours}]}& & polynomial\\
			
			& & \\
			
			\texttt{\textcolor{mmaUndefined}{SusyOpCounter}[\textcolor{mmaLocal}{Poly}]} & \texttt{\textcolor{mmaLocal}{Poly}} $\to$ \texttt{any variable}.   & Number of independent invariant \\
			& & operators that constitute the polynomial\\	
			& & in supersymmetric case.\\
			\hline
		\end{tabular}}
	\caption{\small \textbf{GrIP} SUSY functions and their working principles.}\label{table:susy_function_details}
\end{table}

\subsection{\textbf{CHaar} and its working principle}\label{subsubsec:Chaar_output}

In addition to \textbf{GrIP}, the package also includes a separate sub-program \textbf{CHaar} which would enable the user to obtain Haar measures and characters for any given irreducible representations of connected compact groups: $SU(N), SO(2N+1), SO(2N)$ and $Sp(2N)$. Note that unlike \textbf{GrIP}, it does not rely on any input file and it operates independently. The process of loading `CHaar.m' in the Notebook file is the same as what has been followed for `GrIP.m'. Again one can use both  \texttt{"Get[]"} or \texttt{"Needs[]"} to load the program after setting up the package directory \texttt{\$Path} using \texttt{"SetDirectory[]"}.  
If the package is installed in a local folder, `CHaar.m' can be launched in the following manner:

\vspace{0.15cm}
{\footnotesize\textcolor{mmaLabel}{\fontfamily{phv}\selectfont In[1]:=}}
\textcolor{mmaUndefined}{\texttt{SetDirectory}}\texttt{[\textcolor{mmaString}{"<provide address of the package>"}]}

\vspace{0.15cm}
{\footnotesize\textcolor{mmaLabel}{\fontfamily{phv}\selectfont In[2]:=}}
\textcolor{mmaUndefined}{\texttt{Get}}\texttt{[\textcolor{mmaString}{"CHaar.m"}]}

\vspace{0.15cm}
\noindent
If the package is installed in the `Applications' folder in \texttt{\$UserBaseDirectory} one can use \texttt{"Needs[]"} to load the program. In this case, there is no need to set up the \texttt{\$Path} of the folder it is installed in.

\vspace{0.15cm}
{\footnotesize\textcolor{mmaLabel}{\fontfamily{phv}\selectfont In[1]:=}}
\textcolor{mmaUndefined}{\texttt{Needs}}\texttt{[\textcolor{mmaString}{"CHaar\,$\grave{}$"}]}

\vspace{0.15cm}
\noindent
We illustrate the functions of \textbf{CHaar} and their working principles in Table~\ref{table:CHaar_output}. 

%%%%%%%%%%%%%%%%%%%%%%%%%%%%%%%%%%%%%%%%%%%%%%%%%%%%%%%%%%%%%%%%%%%%%%%%%
\begin{table}[htb!]
	\centering
	\renewcommand{\arraystretch}{1.6}
	{\scriptsize\begin{tabular}{|lcl|}
			\hline
			\textbf{Functions} & \textbf{Options} & \textbf{Details of the output} \\
			\hline
			\texttt{\textcolor{mmaUndefined}{HaarMeasure}} & \texttt{\textcolor{mmaLocal}{Group}}$\to$ \texttt{String},
			& Haar measure for a group for a \\
			\texttt{[\textcolor{mmaLocal}{Group}, \textcolor{mmaLocal}{Argument}]}			& \texttt{\textcolor{mmaLocal}{Argument}}$\to$ \texttt{\#}. &  given degree \texttt{\#}. The user must\\
			&  & provide this as string.\\

			\texttt{\textcolor{mmaUndefined}{CharacterFunction}} & \texttt{\textcolor{mmaLocal}{Dynkin}}$\to$ \text{Dynkin label} 
			& The character function for a particular \\
			\texttt{[\textcolor{mmaLocal}{Group}, \textcolor{mmaLocal}{Argument},}			& of the representation. & representation of a group of given\\
			\texttt{\textcolor{mmaLocal}{Dynkin}]}	& &  degree \texttt{\#} and Dynkin label of the representation.\\
			
			\hline
	\end{tabular}}
	\caption{\small Output functions of \textbf{CHaar} and their working principles.}\label{table:CHaar_output}
\end{table}

\newpage
\noindent
Examples of output provided by the functions of \textbf{CHaar} are shown below:
\begin{table}[h]
\hspace*{0.25cm}{\footnotesize\textcolor{mmaLabel}{\fontfamily{phv}\selectfont In[1]:=}}
\textcolor{mmaUndefined}{\texttt{CharacterFunction}}\texttt{[\textcolor{mmaString}{"Group"}$\to$"SO", \textcolor{mmaString}{"Argument"}$\to$7, \textcolor{mmaString}{"Dynkin"}$\to$\{0,0,1\}]}

\vspace{0.2cm}
{\footnotesize\textcolor{mmaLabel}{\fontfamily{phv}\selectfont Out[1]:=}}

\vspace{-1.15cm}
{\scriptsize\begin{eqnarray}
\hspace{0.2cm}& &\frac{1}{\sqrt{z_{1}} \sqrt{z_{2}} \sqrt{z_{3}}}+\frac{\sqrt{z_{1}}}{\sqrt{z_{2}} \sqrt{z_{3}}}+\frac{\sqrt{z_{2}}}{\sqrt{z_{1}} \sqrt{z_{3}}}+\frac{\sqrt{z_{1}} \sqrt{z_{2}}}{\sqrt{z_{3}}}+\frac{\sqrt{z_{3}}}{\sqrt{z_{1}} \sqrt{z_{2}}}+\frac{\sqrt{z_{1}} \sqrt{z_{3}}}{\sqrt{z_{2}}}+\frac{\sqrt{z_{2}} \sqrt{z_{3}}}{\sqrt{z_{1}}}+\sqrt{z_{1}} \sqrt{z_{2}} \sqrt{z_{3}} \nonumber
\end{eqnarray}}

\hspace*{0.25cm}{\footnotesize\textcolor{mmaLabel}{\fontfamily{phv}\selectfont In[2]:=}}
\textcolor{mmaUndefined}{\texttt{HaarMeasure}}\texttt{[\textcolor{mmaString}{"Group"}$\to$"Sp", \textcolor{mmaString}{"Argument"}$\to$4]}

\vspace{0.25cm}
{\footnotesize\textcolor{mmaLabel}{\fontfamily{phv}\selectfont Out[2]:=}}

\vspace{-1.25cm}
{\scriptsize\begin{eqnarray}
\hspace{-4.5cm} -\frac{dz_1 dz_2\left(-1+z_{1}^{2}\right)^{2}\left(-1+z_{2}^{2}\right)^{2}\left(z_{2}+z_{1}^{2} z_{2}-z_{1}\left(1+z_{2}^{2}\right)\right)^{2}}{32 \pi^{2} z_{1}^{5} z_{2}^{5}}\nonumber
\end{eqnarray}}
\end{table}

%% file: GrIPExamples.tex
\section{GrIPping: Illustrating the action of GrIP}\label{sec:grip-examples}
\subsection{Operator construction for a non-supersymmetric model using \textbf{GrIP}}\label{subsec:smeft}

This subsection illustrates how \textbf{GrIP} works for a non-supersymmetric model. The input file and the results from \textbf{GrIP} are shown explicitly with the proper functions. The Standard Model (SM) is used as an example.

\subsubsection*{\underline{The fields and their transformation properties}}
The field content and the transformation properties of those fields under the gauge groups $SU(3)_C\otimes SU(2)_L\otimes U(1)_Y$ and the Lorentz group are enlisted in Table~\ref{table:sm-quantum-no}.

\begin{table}[h]
	\centering
	\renewcommand{\arraystretch}{1.8}
	{\tiny\begin{tabular}{|c|c|c|c|c c|}
			\hline
			\textbf{SM Fields}&
			$SU(3)_C$&
			$SU(2)_L$&
			$U(1)_Y$&
			\multicolumn{2}{c|}{Lorentz Group $(SU(2)_l\otimes SU(2)_r)$}\\
			\hline
			
			$H$&
			1&
			2&
			1/2&
			\hspace{0.5cm}Scalar& (0,0)\\
			
			$Q^p_l$&
			3&
			2&
			1/6&
			\hspace{0.5cm}Spinor& (1/2,0)\\
			
			$u^p_r$&
			3&
			1&
			2/3&
			\hspace{0.5cm}Spinor& (0,1/2)\\
			
			$d^p_r$&
			3&
			1&
			-1/3&
			\hspace{0.5cm}Spinor& (0,1/2)\\
			
			$L^p_l$&
			1&
			2&
			-1/2&
			\hspace{0.5cm}Spinor& (1/2,0)\\
			
			$e^p_r$&
			1&
			1&
			-1&
			\hspace{0.5cm}Spinor& (0,1/2)\\
			
			$Bl$&
			1&
			1&
			0&
			\hspace{0.5cm}Vector& (1,0)\\
			
			$Wl$&
			1&
			3&
			0&
			\hspace{0.5cm}Vector& (1,0)\\
			
			$Gl$&
			8&
			1&
			0&
			\hspace{0.5cm}Vector& (1,0)\\
			\hline
			\hline
			\multicolumn{4}{|l|}{\textbf{Covariant Derivative}\hspace{0.5cm} $\mathcal{D}_\mu$}&
			\hspace{.2cm}Bi-spinor& (1/2,1/2)\\
			\hline
	\end{tabular}}
	\caption{\small Quantum numbers of fields under the SM gauge groups and Lorentz group. $I$ = 1,2,3; $a$ = 1,2,.....,8; $p$ = 1,2,3 denotes the flavor index. The color and isospin indices have been suppressed. $l$ and $r$ denote the chirality, i.e., the left or right handedness of the field.}
	\label{table:sm-quantum-no}
\end{table} 
\noindent

\subsubsection*{\underline{The \textbf{GrIP} Input file}}
The structure of the model input file \texttt{"SM\_Rep.m"} is shown in detail below:

{\scriptsize\begin{alltt}
ModelName="StandardModel"
\end{alltt}}

{\scriptsize\begin{alltt}
SymmetryGroupClass =\{
Group[1]=\{"GroupName"->"SU3", "N"->3\},
Group[2]=\{"GroupName"->"SU2", "N"->2\},
Group[3]=\{"GroupName"->"U1",  "N"->1\} \};
\end{alltt}}

{\scriptsize\begin{alltt}
FieldClass=\{
Field[1]=\{
"FieldName"-> \(H\),  "Self-Conjugate"-> False,  "Lorentz Behaviour"-> "SCALAR",  "Chirality"-> "NA",
"Baryon Number"-> 0,  "Lepton Number"-> 0,  "SU3Rep"-> "1",  "SU2Rep"-> "2",  "U1Rep"-> 1/2\},

Field[2]=\{
"FieldName"-> \(Q\),  "Self-Conjugate"-> False,  "Lorentz Behaviour"-> "FERMION",  "Chirality"-> "l",
"Baryon Number"-> 1/3,  "Lepton Number"-> 0,  "SU3Rep"-> "3",  "SU2Rep"-> "2",  "U1Rep"-> 1/6\},

Field[3]=\{
"FieldName"-> \(u\),  "Self-Conjugate"-> False,  "Lorentz Behaviour"-> "FERMION",  "Chirality"-> "r",
"Baryon Number"-> 1/3,  "Lepton Number"-> 0,  "SU3Rep"-> "3",  "SU2Rep"-> "1",  "U1Rep"-> 2/3\},

Field[4]=\{
"FieldName"-> \(d\),  "Self-Conjugate"-> False,  "Lorentz Behaviour"-> "FERMION",  "Chirality"-> "r",  
"Baryon Number"-> 1/3,  "Lepton Number"-> 0,  "SU3Rep"-> "3",  "SU2Rep"-> "1",  "U1Rep"-> -1/3\},

Field[5]=\{
"FieldName"-> \(L\),  "Self-Conjugate"-> False,  "Lorentz Behaviour"-> "FERMION",  "Chirality"-> "l",
"Baryon Number"-> 0,  "Lepton Number"-> -1,  "SU3Rep"-> "1",  "SU2Rep"-> "2",  "U1Rep"-> -1/2\},

Field[6]=\{
"FieldName"-> \(el\),
"Self-Conjugate"-> False,  "Lorentz Behaviour"-> "FERMION",  "Chirality"-> "r",
"Baryon Number"-> 0,  "Lepton Number"-> -1,  "SU3Rep"-> "1",  "SU2Rep"-> "1",  "U1Rep"-> -1\}\};
\end{alltt}}

{\scriptsize\begin{alltt}
FieldTensorClass=\{
TensorField[1]=\{
"FieldName"-> \(Bl\),  "Self-Conjugate"-> False,  "Lorentz Behaviour"-> "VECTOR",  "Chirality"-> "l",
"Baryon Number"-> 0,  "Lepton Number"-> 0,  "SU3Rep"-> "1",  "SU2Rep"-> "1",  "U1Rep"-> 0\},

TensorField[2]=\{
"FieldName"-> \(Wl\),  "Self-Conjugate"-> False,  "Lorentz Behaviour"-> "VECTOR",  "Chirality"-> "l",
"Baryon Number"-> 0,  "Lepton Number"-> 0,  "SU3Rep"-> "1",  "SU2Rep"-> "3",  "U1Rep"-> 0\},

TensorField[3]=\{
"FieldName"-> \(Gl\),  "Self-Conjugate"-> False,  "Lorentz Behaviour"-> "VECTOR",  "Chirality"-> "l",
"Baryon Number"-> 0,  "Lepton Number"-> 0,  "SU3Rep"-> "8",  "SU2Rep"-> "1",  "U1Rep"-> 0\} \};
\end{alltt}}
%%%%%%%%%%%%%%%%%%%%%%%%%%%%%%%%%%%%%%%%%%%%%%%%%%%%%%%%
\noindent Alternate provision: Providing Dynkin labels instead of dimension of the representation. 

{\scriptsize\begin{alltt}
Field[2]=\{
"FieldName"-> \(Q\),  "Self-Conjugate"-> False,  "Lorentz Behaviour"-> "FERMION",  "Chirality"-> "l",
"Baryon Number"-> 1/3,  "Lepton Number"-> 0,  "SU3Dyn"-> \{1,0\},  "SU2Dyn"-> \{1\},  "U1Dyn"-> 1/6\},
\end{alltt}}

\input{GrIP_Err.tex}

\subsubsection*{\underline{Details of the user interface for the Standard Model}}
Here, we provide an illustration of how to run \textbf{GrIP} and utilize its various commands to obtain specific outputs based on the Standard Model and how to further modify those results.

\vspace{0.25cm}
{\footnotesize\textcolor{mmaLabel}{\fontfamily{phv}\selectfont In[1]:=}}
\textcolor{mmaUndefined}{\texttt{SetDirectory}}\texttt{[\textcolor{mmaString}{"\,$\tilde{}$/home"}]}

\vspace{0.25cm}
{\footnotesize\textcolor{mmaLabel}{\fontfamily{phv}\selectfont In[2]:=}}
\textcolor{mmaUndefined}{\texttt{Get}}\texttt{[\textcolor{mmaString}{"MODEL/SM\_Rep.m"}]}
\vspace{0.25cm}

\begin{Verbatim}[frame=single]
Model Name: Standard Model

Authors: Upalaparna Banerjee, Joydeep Chakrabortty, Suraj Prakash, Shakeel 
Ur Rahaman 

Institutes: Indian Institute of Technology Kanpur, India 

Emails: upalab, joydeep, surajprk, shakel@iitk.ac.in
\end{Verbatim}

{\footnotesize\textcolor{mmaLabel}{\fontfamily{phv}\selectfont In[3]:=}}
\textcolor{mmaUndefined}{\texttt{Get}}\texttt{[\textcolor{mmaString}{"GrIP.m"}]}
\vspace{0.15cm}	

\begin{Verbatim}[frame=single]
GrIP-V.1.0.0
\end{Verbatim}

\begin{Verbatim}[frame=single]
Authors: Upalaparna Banerjee, Joydeep Chakrabortty, Suraj Prakash, 
Shakeel Ur Rahaman
Indian Institute of Technology Kanpur, India
\end{Verbatim}

\begin{Verbatim}[frame=single]
"GrIP is successfully loaded and ready to compute! 
A folder, named [StandardModel] has been created in your working 
directory and all the output will be saved in that folder.
Thank You!!"
\end{Verbatim}

\begin{table}[h]
		\centering
		\renewcommand{\arraystretch}{1.8}
\hspace*{-7.05cm}{\footnotesize\textcolor{mmaLabel}{\fontfamily{phv}\selectfont In[4]:=}}
\textcolor{mmaUndefined}{\texttt{DisplayUserInputTable}}

\vspace{0.25cm}
\hspace*{-11.7cm}{\footnotesize\textcolor{mmaLabel}{\fontfamily{phv}\selectfont Out[4]:=}}

\vspace{0.35cm}
	{\tiny\begin{tabular}{|c|c|c|c|c|c|c|c|c|}
			\hline
	\textbf{Field} & \textbf{Self} & \textbf{Lorentz} & \textbf{Chirality} & \textbf{Baryon} & \textbf{Lepton} & \textbf{SU3Rep} & \textbf{SU2Rep} & \textbf{U1Rep} \\
	\textbf{Name} &  \textbf{Conjugate} &  \textbf{Behaviour}  &    &  \textbf{Number}    &  \textbf{Number}    &   &    &  \\
	\hline
	$H$  & False & SCALAR  & NA & 0   &  0   &  1   & 2 & 1/2 \\
	$Q$  & False & FERMION & l  & 1/3 &  0   &  3   & 2 & 1/6 \\
	$u$  & False & FERMION & r  & 1/3 &  0   &  3   & 1 & 2/3 \\
	$d$  & False & FERMION & r  & 1/3 &  0   &  3   & 1 & -1/3 \\
	$L$  & False & FERMION & l  & 0   &  -1  &  1   & 2 & -1/2 \\
	$el$ & False & FERMION & r  & 0   &  -1  &  1   & 1 & -1 \\
	$Bl$ & False & VECTOR  & l  & 0   &  0   &  1   & 1 & 0 \\
	$Wl$ & False & VECTOR  & l  & 0   &  0   &  1   & 3 & 0 \\
	$Gl$ & False & VECTOR  & l  & 0   &  0   &  8   & 1 & 0 \\
	\hline
	\end{tabular}}
\end{table}

\begin{table}[h]
	\centering
	\renewcommand{\arraystretch}{1.8}
\hspace*{-6.5cm}{\footnotesize\textcolor{mmaLabel}{\fontfamily{phv}\selectfont In[5]:=}}
\textcolor{mmaUndefined}{\texttt{DisplayWorkingInputTable}}

\vspace{0.25cm}
\hspace*{-11.7cm}{\footnotesize\textcolor{mmaLabel}{\fontfamily{phv}\selectfont Out[5]:=}}

\vspace{0.35cm}
	{\tiny\begin{tabular}{|c|c|c|c|c|c|c|c|c|}
			\hline
			\textbf{Field} & \textbf{Self} & \textbf{Lorentz} & \textbf{Chirality} & \textbf{Baryon} & \textbf{Lepton} & \textbf{SU3Dyn} & \textbf{SU2Dyn} & \textbf{U1Dyn} \\
			\textbf{Name}  &  \textbf{Conjugate} &  \textbf{Behaviour}  &    &  \textbf{Number}    &  \textbf{Number}    &   &    &   \\
			\hline
			$H$ & False & SCALAR  & NA & 0   & 0  & \{0,0\} & \{1\} & 1/2 \\
			$Q$ & False & FERMION & l  & 1/3 & 0  & \{1,0\} & \{1\} & 1/6 \\
			$u$ & False & FERMION & r  & 1/3 & 0  & \{1,0\} & \{0\} & 2/3 \\
			$d$ & False & FERMION & r  & 1/3 & 0  & \{1,0\} & \{0\} & -1/3 \\
			$L$ & False & FERMION & l  & 0   &-1  & \{0,0\} & \{1\} & -1/2 \\
			$el$ & False & FERMION & r & 0   &-1  & \{0,0\} & \{0\} & -1 \\
			$H^{\dagger}$ & False & SCALAR   & NA & 0 & 0 & \{0,0\} & \{1\}  & -1/2 \\
			$Q^{\dagger}$ & False & FERMION & r & -1/3 & 0 & \{0,1\} & \{1\} & -1/6 \\
			$u^{\dagger}$ & False & FERMION & l & -1/3 & 0 & \{0,1\} & \{0\} & -2/3 \\
			$d^{\dagger}$ & False & FERMION & l & -1/3 & 0 & \{0,1\} & \{0\} & 1/3 \\
			$L^{\dagger}$ & False & FERMION & r & 0 & 1 & \{0,0\} & \{1\} & 1/2 \\
			$el^{\dagger}$ & False & FERMION & l & 0 & 1 & \{0,0\} & \{0\} & 1 \\
			$Bl$ & False & VECTOR & l & 0 & 0 & \{0,0\} & \{0\} & 0 \\
			$Wl$ & False & VECTOR & l & 0 & 0 & \{0,0\} & \{2\} & 0 \\
			$Gl$ & False & VECTOR & l & 0 & 0 & \{1,1\} & \{0\} & 0 \\
			$Br$ & False & VECTOR & r & 0 & 0 & \{0,0\} & \{0\} & 0 \\
			$Wr$ & False & VECTOR & r & 0 & 0 & \{0,0\} & \{2\} & 0 \\
			$Gr$ & False & VECTOR & r & 0 & 0 & \{1,1\} & \{0\} & 0 \\
			\hline
			
	\end{tabular}}
\end{table} 
\clearpage

\begin{table}[h]
	\centering
	\renewcommand{\arraystretch}{2.2}
\hspace*{-7.2cm}{\footnotesize\textcolor{mmaLabel}{\fontfamily{phv}\selectfont In[6]:=}}
\textcolor{mmaUndefined}{\texttt{DisplayCharacterTable}}

\vspace{0.25cm}
\hspace*{-11.8cm}{\footnotesize\textcolor{mmaLabel}{\fontfamily{phv}\selectfont Out[6]:=}}

\vspace{0.35cm}	
	{\tiny\begin{tabular}{|c|c|c|c|}
			\hline
			\textbf{Dyn} & \textbf{SU3} & \textbf{SU2} & \textbf{U1} \\
			\hline
			\{\{0,0\},\{1\},1/2\} & $1$ & $\cfrac{1}{G2z_1}+G2z_1 $& $\sqrt{G3z}$ \\
			\hline
			\{\{1,0\},\{1\},1/6\} & $G1z_1+\cfrac{1}{G1z_2}+\frac{G1z_2}{G1z_1}$ & $\frac{1}{G2z_1}+G2z_1$
			& $G3z^{1/6}$ \\
			\hline
			\{\{1,0\},\{0\},2/3\} & $G1z_1+\cfrac{1}{G1z_2}+\cfrac{G1z_2}{G1z_1}$ & 1 & $G3z^{2/3}$ \\
			\hline
			\{\{1,0\},\{0\},-1/3\} & $G1z_1+\cfrac{1}{G1z_2}+\cfrac{G1z_2}{G1z_1}$ & $1$ & $\frac{1}{G3z^{1/3}}$
			\\
			\hline
			\{\{0,0\},\{1\},-1/2\} & 1 & $\frac{1}{G2z_1}+G2z_1$ & $\frac{1}{\sqrt{G3z}}$ \\
			\hline
			\{\{0,0\},\{0\},-1\} & 1 & 1 & $\cfrac{1}{G3z}$ \\
			\hline
			\{\{0,0\},\{1\},-1/2\} & $1$ & $\cfrac{1}{G2z_1}+G2z_1$ & $\cfrac{1}{\sqrt{G3z}}$ \\
			\hline
			\{\{0,1\},\{1\},-1/6\} & $\cfrac{1}{{G1z}_1}+\cfrac{{G1z}_1}{{G1z}_2}+G1z_2$ & $\cfrac{1}{{G2z}_1}+G2z_1$
			& $\cfrac{1}{{G3z}^{1/6}}$ \\
			\hline
			\{\{0,1\},\{0\},-2/3\} & $\cfrac{1}{{G1z}_1}+\cfrac{{G1z}_1}{{G1z}_2}+{G1z}_2$ & 1 & $\cfrac{1}{{G3z}^{2/3}}$
			\\
			\hline
			\{\{0,1\},\{0\},1/3\} & $\cfrac{1}{{G1z}_1}+\cfrac{{G1z}_1}{{G1z}_2}+G1z_2$ & 1 & ${G3z}^{1/3}$ \\
			\hline
			\{\{0,0\},\{1\},1/2\} & 1 & $\cfrac{1}{{G2z}_1}+G2z_1$ & $\sqrt{G3z}$ \\
			\hline
			\{\{0,0\},\{0\},1\} & 1 & 1 & $G3z$ \\
			\hline
			\{\{0,0\},\{0\},0\} & 1 & 1 & 1 \\
			\hline
			\{\{0,0\},\{2\},0\} & 1 & $1+\cfrac{1}{{G2z}_1^2}+{G2z}_1^2$ & 1 \\
			\hline
			\{\{1,1\},\{0\},0\} & $2+\cfrac{{G1z}_1}{{G1z}_2^2}+\cfrac{1}{{G1z}_1 {G1z}_2}+\cfrac{{G1z}_1^2}{{G1z}_2}+\cfrac{{G1z}_2}{{G1z}_1^2}+{G1z}_1
			{G1z}_2+\cfrac{{G1z}_2^2}{{G1z}_1}$ & 1 & 1 \\
			\hline
			\{\{0,0\},\{0\},0\} & 1 & 1 & 1 \\
			\hline
			\{\{0,0\},\{2\},0\} & 1 & $1+\cfrac{1}{{G2z}_1^2}+G2z_1^2$ & 1 \\
			\hline
			\{\{1,1\},\{0\},0\} & $2+\cfrac{{G1z}_1}{{G1z}_2^2}+\cfrac{1}{{G1z}_1 {G1z}_2}+\cfrac{{G1z}_1^2}{{G1z}_2}+\cfrac{{G1z}_2}{{G1z}_1^2}+{G1z}_1
			{G1z}_2+\cfrac{{G1z}_2^2}{{G1z}_1}$ & 1 & 1 \\
			\hline
			
	\end{tabular}}
\end{table} 

\begin{table}[h]
	\centering
	\renewcommand{\arraystretch}{2}
\hspace*{-7.8cm}{\footnotesize\textcolor{mmaLabel}{\fontfamily{phv}\selectfont In[7]:=}}
\textcolor{mmaUndefined}{\texttt{DisplayHaarMeasure}}

\vspace{0.25cm}
\hspace*{-11.8cm}{\footnotesize\textcolor{mmaLabel}{\fontfamily{phv}\selectfont Out[7]:=}}

\vspace{0.35cm}	
{\tiny\begin{tabular}{|c|c|c|}
		\hline
		\textbf{SU3} & \textbf{SU2} & \textbf{U1} \\
		\hline
		$-\cfrac{\left(\text{-}{G1z}_1^4 {G1z}_2+{G1z}_2^3+{G1z}_1^3 \left(1+{G1z}_2^3\right)-{G1z}_1 \left({G1z}_2+{G1z}_2^4\right)\right)^2}{6{G1z}_1^5 {G1z}_2^5} $ & $-\cfrac{\left(\text{-}1+{G2z}_1^2\right)^2}{2 {G2z}_1^3}$ & $\cfrac{1}{G3z}$ \\
		\hline
\end{tabular}}
\end{table}

\begin{table}[h]
\hspace*{1cm}{\footnotesize\textcolor{mmaLabel}{\fontfamily{phv}\selectfont In[8]:=}}
\textcolor{mmaUndefined}{\texttt{DisplayHSOutput}}\texttt{[\textcolor{mmaString}{"MassDim"}$\to$4, \textcolor{mmaString}{"OnlyMassDimOutput"}$\to$True,\\ \hspace*{5cm}\textcolor{mmaString}{"$\Delta$B"}$\to$0, \textcolor{mmaString}{"$\Delta$L"}$\to$0, \textcolor{mmaString}{"Flavours"}$\to$1]}

\vspace{0.25cm}
\hspace*{0.8cm}{\footnotesize\textcolor{mmaLabel}{\fontfamily{phv}\selectfont Out[8]:=}}

\vspace{-1.cm}
{\scriptsize\begin{eqnarray}
& &\hspace{1.0cm}-Bl^2-Br^2-Gl^2-Gr^2-Wl^2-Wr^2-Bl\; \mathcal{D}^2-Br\; \mathcal{D}^2+d\; \mathcal{D}\; d^{\dagger}+el\; \mathcal{D}\;el^{\dagger}+H\;\mathcal{D}^2\ H^{\dagger} \nonumber\\ 
& &\hspace{1.0cm}-Q\;d^{\dagger}\;H^{\dagger}- L\; el^{\dagger}\; H^{\dagger}-H^2\;\left(H^{\dagger}\right)^2-el\; H\; L^{\dagger}+L\;\mathcal{D}\; L^{\dagger}-d\; H\; Q^{\dagger}+Q\; \mathcal{D}\; Q^{\dagger}-u\; H^{\dagger}\; Q^{\dagger}\;\nonumber\\
& &\hspace{1.0cm}-H\; Q\; u^{\dagger}\;+u\;  \mathcal{D}\; u^{\dagger}-\mathcal{D}^4\; \nonumber
\end{eqnarray}}
	
\end{table}

\begin{table}[h]
\hspace*{1cm}{\footnotesize\textcolor{mmaLabel}{\fontfamily{phv}\selectfont In[9]:=}}
\textcolor{mmaUndefined}{\texttt{DisplayBLviolatingOperators}}\texttt{[\textcolor{mmaString}{"HighestMassDim"}$\to$10, \textcolor{mmaString}{"$\Delta$B"}$\to$ $+1(-1)$,\\ 
\hspace*{6.5cm} \textcolor{mmaString}{"$\Delta$L"}$\to$ $-1(+1)$, \textcolor{mmaString}{"Flavours"}$\to$1]}

\vspace{0.25cm}
\hspace*{0.75cm}{\footnotesize\textcolor{mmaLabel}{\fontfamily{phv}\selectfont Out[9]:=}}

\vspace{-1.08cm}
{\scriptsize\begin{eqnarray}
\hspace{-1.0cm}L\;Q^3+d\;L\;Q\;u+el\;Q^2\;u+d\;el\;u^2 \ \ \text{(corresponding hermitian conjugates)} \nonumber
\end{eqnarray}}
	
\end{table}

\begin{table}[h]
\hspace*{1cm}{\footnotesize\textcolor{mmaLabel}{\fontfamily{phv}\selectfont In[10]:=}}
\textcolor{mmaUndefined}{\texttt{DisplayLagOutput}}\texttt{[\textcolor{mmaString}{"MassDim"}$\to$4, \textcolor{mmaString}{"OnlyMassDimOutput"}$\to$False,\\ \hspace*{5cm}\textcolor{mmaString}{"$\Delta$B"}$\to$"NA", \textcolor{mmaString}{"$\Delta$L"}$\to$"NA", \textcolor{mmaString}{"Flavours"}$\to$ $N_f$]}

\vspace{0.25cm}
\hspace*{0.75cm}{\footnotesize\textcolor{mmaLabel}{\fontfamily{phv}\selectfont Out[10]:=}}

\vspace{-1.05cm}
{\scriptsize\begin{eqnarray}
& & \hspace{1.9cm} -Bl^2-Br^2-Gl^2-Gr^2-Wl^2-Wr^2+d\;N_f\;\mathcal{D}\; d^{\dagger}+el\; N_f\; \mathcal{D}\;el^{\dagger} +H\; \mathcal{D}^2 H^{\dagger}-N_f^2\; Q \mathit{y}_1\; d^{\dagger}\; H^{\dagger}\nonumber\\
& &\hspace{1.9cm} -H^2\;\lambda\;\left(H^{\dagger}\right)^2+L\; N_f\;\mathcal{D}\;L^{\dagger}- el\; H\;N_f^2\;\mathit{y}_2 L^{\dagger} +N_f\;Q\; \mathcal{D}
Q^{\dagger}-d\; H\;N_f^2\; \mathit{y}_3\; Q^{\dagger}-N_f^2\;u\;\mathit{y}_4 \;H^{\dagger}\; Q^{\dagger}+\nonumber\\
& &\hspace{1.9cm}N_f\; u\;\mathcal{D}\; u^{\dagger} -H\; N_f^2\; Q\; \mathit{y}_5\; u^{\dagger} - L\;N_f^2 \mathit{y}_6\; el^{\dagger} H^{\dagger}\nonumber
\end{eqnarray}}
\end{table}

\begin{table}[h]
\hspace*{1cm}{\footnotesize\textcolor{mmaLabel}{\fontfamily{phv}\selectfont In[11]:=}}
\textcolor{mmaUndefined}{\texttt{DisplayLagOutput}}\texttt{[\textcolor{mmaString}{"MassDim"}$\to$4, \textcolor{mmaString}{"OnlyMassDimOutput"}$\to$False,\\ \hspace*{5cm}\textcolor{mmaString}{"$\Delta$B"}$\to$0, \textcolor{mmaString}{"$\Delta$L"}$\to$0, \textcolor{mmaString}{"Flavours"}$\to$ 1]}
	
\vspace{0.25cm}
\hspace*{0.75cm}{\footnotesize\textcolor{mmaLabel}{\fontfamily{phv}\selectfont Out[11]:=}}
	
\vspace{-1.05cm}
{\scriptsize\begin{eqnarray}
& &\hspace{1.6cm} -Bl^2-Br^2+d\; \mathcal{D}\; d^{\dagger}-Q\;\mathit{y}_1\; d^{\dagger}\; H^{\dagger}-d\; H\; \mathit{y}_4\; Q^{\dagger}+\mathcal{D}\; el\; el^{\dagger}-L\; \mathit{y}_2\;  el^{\dagger}\;	H^{\dagger}-el\; H\;  \mathit{y}_3\; L^{\dagger}\nonumber\\ 
& &\hspace{1.6cm} +\mathcal{D}^2  H \; H^{\dagger}-H  \mathit{m}^2 H^{\dagger}-u\;  \mathit{y}_5\; H^{\dagger} Q^{\dagger}-H^2\; \lambda  \left(H^{\dagger}\right)^2-H\; Q\; \mathit{y}_6\; u^{\dagger}+\mathcal{D}\; L\; L^{\dagger}+\mathcal{D}\; Q\; Q^{\dagger}+\mathcal{D}\; u\; u^{\dagger} \nonumber\\
& & \hspace{1.6cm} -Gl^2-Gr^2  -Wl^2-Wr^2\nonumber
\end{eqnarray}}
\end{table}

\begin{table}[h]
\hspace*{1cm}{\footnotesize\textcolor{mmaLabel}{\fontfamily{phv}\selectfont In[12]:=}}
\textcolor{mmaUndefined}{\texttt{PolyA}}\texttt{=}\textcolor{mmaUndefined}{\texttt{SaveHSOutput}}\texttt{[\textcolor{mmaString}{"MassDim"}$\to$4, \textcolor{mmaString}{"$\Delta$B"}$\to$0, \textcolor{mmaString}{"$\Delta$L"}$\to$0,\\ \hspace*{7cm}\textcolor{mmaString}{"Flavours"}$\to$1];}
	
\vspace{0.35cm}
\hspace*{1cm}{\footnotesize\textcolor{mmaLabel}{\fontfamily{phv}\selectfont In[13]:=}}
\texttt{\textcolor{mmaUndefined}{OpCounter}[PolyA]}
\vspace{0.25cm}
	
\hspace*{0.75cm}{\footnotesize\textcolor{mmaLabel}{\fontfamily{phv}\selectfont Out[13]:=}}\hspace*{0.4cm}\texttt{22}
\end{table}

\begin{table}[h]
\hspace*{1cm}{\footnotesize\textcolor{mmaLabel}{\fontfamily{phv}\selectfont In[14]:=}}
\textcolor{mmaUndefined}{\texttt{PolyB}}\texttt{=}\textcolor{mmaUndefined}{\texttt{SaveHSOutput}}\texttt{[\textcolor{mmaString}{"MassDim"}$\to$4, \textcolor{mmaString}{"$\Delta$B"}$\to$"NA", \textcolor{mmaString}{"$\Delta$L"}$\to$"NA",\\ \hspace*{7cm}\textcolor{mmaString}{"Flavours"}$\to$ $N_f$];}

\vspace{0.35cm}
\hspace*{1cm}{\footnotesize\textcolor{mmaLabel}{\fontfamily{phv}\selectfont In[15]:=}}
\texttt{\textcolor{mmaUndefined}{OpCounter}[PolyB]}
\vspace{0.25cm}

\hspace*{0.75cm}{\footnotesize\textcolor{mmaLabel}{\fontfamily{phv}\selectfont Out[15]:=}}\hspace*{0.4cm}\texttt{11$N_f$+11$N_f^2$}
\end{table}

\clearpage
\subsection{Operator construction for a supersymmetric model using \textbf{GrIP}}\label{subsec:mssm}

Our prescription is not restricted to non-supersymmetric models. The program \textbf{GrIP} enables one to construct the polynomial in terms of chiral and vector superfields. In our construction, we have taken care of the transformation of vector superfield (V) under the Wess-Zumino \cite{Wess:1992cp,Bailin:1994qt} gauge as 
$V^{\prime} = V + i(\Lambda-\Lambda^{\dagger})$. The chiral superfield $\Phi$ transforms as $\Phi \rightarrow e^{-iq\Lambda}\Phi$ and its conjugate as $\Phi^{\dagger} \rightarrow e^{iq\Lambda^{\dagger}}\Phi^{\dagger}$. This leaves $\Phi^{\dagger}e^{qV}\Phi $ invariant.

For multiple chiral superfields ($\Phi$) and gauge symmetries, the respective $V_{\Phi}$ can be written as linear combinations of vector superfields corresponding to the individual gauge groups suitably accompanied by the gauge charges of $\Phi$.
We have enlisted the transformation properties of the superfields of the Minimal Supersymmetric Standard Model \cite{Baer:2006rs,Martin:1997ns} under the gauge group $SU(3)_C\otimes SU(2)_L\otimes U(1)_Y$ in Table~\ref{table:mssm-quantum-no}. The $V_{\Phi}$ for this particular scenario is summarized in Table~\ref{table:mssm-vector-superfield}.

\begin{table}[h]
	\centering
	\renewcommand{\arraystretch}{1.8}
	{\tiny\begin{tabular}{|c|c|c|c|}
			\hline
			\textbf{Superfields}&
			$SU(3)_C$&
			$SU(2)_L$&
			$U(1)_Y$\\
			\hline
			
			$H_u$&
			1&
			2&
			1/2\\
			
			$H_d$&
			1&
			2&
			-1/2\\
			
			$Q^i$&
			3&
			2&
			1/6\\
			
			$U^i$&
			$\overline{3}$&
			1&
			-2/3\\
			
			$D^i$&
			$\overline{3}$&
			1&
			1/3\\
			
			$L^i$&
			1&
			2&
			-1/2\\
			
			$E^i$&
			1&
			1&
			1\\
			
			$B$&
			1&
			1&
			0\\
			
			$W$&
			1&
			3&
			0\\
			
			$G$&
			8&
			1&
			0\\
			\hline
	\end{tabular}}
	\caption{\small MSSM: Quantum numbers of superfields under the gauge groups. Internal symmetry indices have been suppressed. $i$=1,2,...,$N_f$ is the flavor index.}
	\label{table:mssm-quantum-no}
\end{table}

\begin{table}[h]
	\centering
	\renewcommand{\arraystretch}{2.0}
	{\tiny\begin{tabular}{|r l|c|r l|c|}
			\hline
			$V_{H_u}$&
			$\rightarrow$&
			$q^{H_u}_{SU(2)_L} W + q^{H_u}_{U(1)_Y} B $&
			$V_{H_d}$&
			$\rightarrow$&
			$q^{H_d}_{SU(2)_L} W + q^{H_d}_{U(1)_Y} B $\\
			
			$V_{Q}$&
			$\rightarrow$&
			$q^{Q}_{SU(3)_C} G + q^{Q}_{SU(2)_L} W + q^{Q}_{U(1)_Y} B $&
			$V_{L}$&
			$\rightarrow$&
			$q^{L}_{SU(2)_L} W + q^{L}_{U(1)_Y} B $\\
			
			$V_{U}$&
			$\rightarrow$&
			$q^{U}_{SU(3)_C} G + q^{U}_{U(1)_Y} B $&
			$V_{E}$&
			$\rightarrow$&
			$q^{E}_{U(1)_Y} B $\\
			
			$V_{D}$&
			$\rightarrow$&
			$q^{D}_{SU(3)_C} G + q^{D}_{U(1)_Y} B $&
			&
			&
			\\
			\hline
	\end{tabular}}
	\caption{\small Vector superfields corresponding to each of the given chiral superfields expressed as a linear combination of the vector superfields corresponding to the gauge groups of the model weighted by the appropriate charges of the chiral superfields. ($q_\mathcal{G}^F $ denotes the charge of the fields $F$ under the gauge group $\mathcal{G}$.) }
	\label{table:mssm-vector-superfield}	
\end{table}

\subsubsection*{\underline{ \textbf{GrIP} Input File for MSSM}}

The input file is prepared following the similar rules prescribed for the SM case:
{\scriptsize\begin{alltt}
ModelName="MSSM"
\end{alltt}}

{\scriptsize\begin{alltt}
SymmetryGroupClass = \{
Group[1] = \{"GroupName" -> "SU3", "N" -> 3\},
Group[2] = \{"GroupName" -> "SU2", "N" -> 2\},
Group[3] = \{"GroupName" -> "U1",  "N" -> 1\} \};
\end{alltt}}

{\scriptsize\begin{alltt}
SuperFieldClass=\{

SuperField[1]=\{
"FieldName"-> \(H\sb{u}\),  "Self-Conjugate"-> False,  "Lorentz Behaviour"-> "SUPERSCALAR",  "Chirality"-> "NA",
"Baryon Number"-> 0,  "Lepton Number"-> 0,  "SU3Rep"-> "1",  "SU2Rep"-> "2",  "U1Rep"-> 1/2\},

SuperField[2]=\{
"FieldName"-> \(H\sb{d}\),  "Self-Conjugate"->False,  "Lorentz Behaviour"-> "SUPERSCALAR",  "Chirality"-> "NA",
"Baryon Number"-> 0,  "Lepton Number"-> 0,  "SU3Rep"-> "1",  "SU2Rep"-> "2",  "U1Rep"-> -1/2\},

SuperField[3]=\{
"FieldName"-> \(Q\),  "Self-Conjugate"-> False,  "Lorentz Behaviour"-> "SUPERFERMION",  "Chirality"-> "NA",
"Baryon Number"-> 1/3,  "Lepton Number"-> 0,  "SU3Rep"-> "3",  "SU2Rep"-> "2",  "U1Rep"-> 1/6\},

SuperField[4]=\{
"FieldName"-> \(U\),  "Self-Conjugate"-> False,  "Lorentz Behaviour"-> "SUPERFERMION",  "Chirality"-> "NA",
"Baryon Number"-> 1/3,  "Lepton Number"-> 0,  "SU3Rep"-> "3 bar",  "SU2Rep"-> "1",  "U1Rep"-> -2/3\},

SuperField[5]=\{
"FieldName"-> \(D\),  "Self-Conjugate"-> False,  "Lorentz Behaviour"-> "SUPERFERMION",  "Chirality"-> "NA",
"Baryon Number"-> 1/3,  "Lepton Number"-> 0,  "SU3Rep"-> "3 bar",  "SU2Rep"-> "1",  "U1Rep"-> 1/3\},

SuperField[6]=\{
"FieldName"-> \(L\),  "Self-Conjugate"-> False,  "Lorentz Behaviour"-> "SUPERFERMION",  "Chirality"-> "NA",
"Baryon Number"-> 0,  "Lepton Number"-> -1,  "SU3Rep"-> "1",  "SU2Rep"-> "2",  "U1Rep"-> -1/2\},

SuperField[7]=\{
"FieldName"-> \(E\),  "Self-Conjugate"-> False,  "Lorentz Behaviour"-> "SUPERFERMION",  "Chirality"-> "NA",
"Baryon Number"-> 0,  "Lepton Number"-> -1,  "SU3Rep"-> "1",  "SU2Rep"-> "1",  "U1Rep"-> 1\}
\};
\end{alltt}}

{\scriptsize\begin{alltt}
SuperFieldTensorClass=\{

TensorSuperField[1]=\{
"FieldName"-> \(Bl\),  "Self-Conjugate"-> False,  "Lorentz Behaviour"-> "SUPERVECTOR",  "Chirality"-> "l",
"Baryon Number"-> 0,  "Lepton Number"-> 0,  "SU3Rep"-> "1",  "SU2Rep"-> "1",  "U1Rep"-> 0\},

TensorSuperField[2]=\{
"FieldName"-> \(Wl\),  "Self-Conjugate"-> False,  "Lorentz Behaviour"-> "SUPERVECTOR",  "Chirality"-> "l",
"Baryon Number"-> 0,  "Lepton Number"-> 0, "SU3Rep"-> "1",  "SU2Rep"-> "3",  "U1Rep"-> 0\},

TensorSuperField[3]=\{
"FieldName"-> \(Gl\),  "Self-Conjugate"-> False,  "Lorentz Behaviour"-> "SUPERVECTOR",  "Chirality"-> "l",
"Baryon Number"-> 0,  "Lepton Number"-> 0,  "SU3Rep"-> "8",  "SU2Rep"-> "1",  "U1Rep"-> 0\}
\};
\end{alltt}}

\subsubsection*{\underline{User Interface for MSSM}}

\hspace*{0.6cm}
{\footnotesize\textcolor{mmaLabel}{\fontfamily{phv}\selectfont In[1]:=}}
\textcolor{mmaUndefined}{\texttt{SetDirectory}}\texttt{[\textcolor{mmaString}{"\,$\tilde{}$\,/home"}]}

\vspace{0.25cm}
{\footnotesize\textcolor{mmaLabel}{\fontfamily{phv}\selectfont In[2]:=}}
\textcolor{mmaUndefined}{\texttt{Get}}\texttt{[\textcolor{mmaString}{"MODEL/MSSM\_Rep.m"}]}

\vspace{0.25cm}
{\footnotesize\textcolor{mmaLabel}{\fontfamily{phv}\selectfont In[3]:=}}
\textcolor{mmaUndefined}{\texttt{Get}}\texttt{[\textcolor{mmaString}{"GrIP.m"}]}

\vspace{0.25cm}
\noindent
The above commands display similar output as shown in the non-supersymmetric case. 

\clearpage

\begin{table}[h]
	\centering
	\renewcommand{\arraystretch}{2.0}
\hspace*{-7.25cm}{\footnotesize\textcolor{mmaLabel}{\fontfamily{phv}\selectfont In[4]:=}}
\textcolor{mmaUndefined}{\texttt{DisplayUserInputTable}}

\vspace{0.25cm}
\hspace*{-11.9cm}{\footnotesize\textcolor{mmaLabel}{\fontfamily{phv}\selectfont Out[4]:=}}

\vspace{0.35cm}
	{\tiny\begin{tabular}{|c|c|c|c|c|c|c|c|c|}
	\hline
\textbf{Super} & \textbf{Self-} & \textbf{Lorentz} & \textbf{Chirality} & \textbf{Baryon} & \textbf{Lepton} & \textbf{SU3Rep} & \textbf{SU2Rep} & \textbf{U1Rep} \\
\textbf{Field} & \textbf{Conjugate} & \textbf{Behaviour} & & \textbf{Number} & \textbf{Number}
&  &  &  \\
\hline
$H_u$ & False & SUPERSCALAR & NA & 0 & 0 & 1 & 2 & 1/2 \\
$H_d$ & False & SUPERSCALAR & NA & 0 & 0 & 1 & 2 & -1/2 \\
$Q$ & False & SUPERFERMION & NA & 1/3 & 0 & 3 & 2 & 1/6 \\
$U$ & False & SUPERFERMION & NA & 1/3 & 0 & 3 bar & 1 & -2/3 \\
$D$ & False & SUPERFERMION & NA & 1/3 & 0 & 3 bar & 1 & 1/3 \\
$L$ & False & SUPERFERMION & NA & 0 & -1 & 1 & 2 & -1/2 \\
$E$ & False & SUPERFERMION & NA & 0 & -1 & 1 & 1 & 1 \\
$Bl$ & False & SUPERVECTOR & l & 0 & 0 & 1 & 1 & 0 \\
$Wl$ & False & SUPERVECTOR & l & 0 & 0 & 1 & 3 & 0 \\
$Gl$ & False & SUPERVECTOR & l & 0 & 0 & 8 & 1 & 0 \\
\hline
\end{tabular}}
	
\end{table}

\begin{table}[h]
	\centering
	\renewcommand{\arraystretch}{2.0}
\hspace*{-6.7cm}{\footnotesize\textcolor{mmaLabel}{\fontfamily{phv}\selectfont In[5]:=}}
\textcolor{mmaUndefined}{\texttt{DisplayWorkingInputTable}}

\vspace{0.25cm}
\hspace*{-11.9cm}{\footnotesize\textcolor{mmaLabel}{\fontfamily{phv}\selectfont Out[5]:=}}

\vspace{0.35cm}
	{\tiny\begin{tabular}{|c|c|c|c|c|c|c|c|c|}
			\hline
			\textbf{Super} & \textbf{Self-} & \textbf{Lorentz} & \textbf{Chirality} & \textbf{Baryon} & \textbf{Lepton} & \textbf{SU3Dyn} & \textbf{SU2Dyn} & \textbf{U1Dyn} \\
			\textbf{Field} & \textbf{Conjugate} & \textbf{Behaviour} & & \textbf{Number} & \textbf{Number}
			&  &  &  \\
			\hline
$H_{u}$ & False & SUPERSCALAR & NA & 0 & 0 & \{0,0\} & \{1\} & 1/2
\\
${H}_{d}$ & False & SUPERSCALAR & NA & 0 & 0 & \{0,0\} & \{1\} & -1/2
\\
$Q$ & False & SUPERFERMION & NA & 1/3 & 0 & \{1,0\} & \{1\} & 1/6 \\
$U$ & False & SUPERFERMION & NA & 1/3 & 0 & \{0,1\} & \{0\} & -2/3 \\
$D$ & False & SUPERFERMION & NA & 1/3 & 0 & \{0,1\} & \{0\} & 1/3 \\
$L$ & False & SUPERFERMION & NA & 0 & -1 & \{0,0\} & \{1\} & -1/2 \\
$E$ & False & SUPERFERMION & NA & 0 & -1 & \{0,0\} & \{0\} & 1 \\
$\left({H}_{u}\right)^{\dagger} e^{V_{H_u}}$ & False & SUPERSCALAR & NA & 0 & 0 & \{0,0\} & \{1\} & -1/2 \\
$\left(H_d\right)^{\dagger} e^{V_{H_d}}$ & False & SUPERSCALAR & NA & 0 & 0 & \{0,0\} & \{1\} & 1/2 \\
$Q^{\dagger} e^{{V}_{Q}}$ & False & SUPERFERMION & NA & -1/3 & 0 & \{0,1\} & \{1\} & -1/6 \\
$U^{\dagger} e^{{V}_{U}}$ & False & SUPERFERMION & NA & -1/3 & 0 & \{1,0\} & \{0\} & 2/3 \\
$D^{\dagger} e^{{V}_{D}}$ & False & SUPERFERMION & NA & -1/3 & 0 & \{1,0\} & \{0\} & -1/3 \\
$L^{\dagger} e^{{V}_{L}}$ & False & SUPERFERMION & NA & 0 & 1 & \{0,0\} & \{1\} &  1/2 \\
$E^{\dagger} e^{{V}_{E}}$ & False & SUPERFERMION & NA & 0 & 1 & \{0,0\} & \{0\} & -1 \\
$Bl$ & False & SUPERVECTOR & l & 0 & 0 & \{0,0\} & \{0\} & 0 \\
$Wl$ & False & SUPERVECTOR & l & 0 & 0 & \{0,0\} & \{2\} & 0 \\
$Gl$ & False & SUPERVECTOR & l & 0 & 0 & \{1,1\} & \{0\} & 0 \\
$Br$ & False & SUPERVECTOR & r & 0 & 0 & \{0,0\} & \{0\} & 0 \\
$Wr$ & False & SUPERVECTOR & r & 0 & 0 & \{0,0\} & \{2\} & 0 \\
$Gr$ & False & SUPERVECTOR & r & 0 & 0 & \{1,1\} & \{0\} & 0 \\
\hline
\end{tabular}}

\end{table}
\newpage
\noindent
The functions - \texttt{\textcolor{mmaUndefined}{DisplayCharacterTable}} and  \texttt{\textcolor{mmaUndefined}{DisplayHaarMeasure}} generate similar output as in the case of SM since the gauge groups and and the transformation properties of the particles are similar.

\begin{table}[h]
\hspace*{0.25cm}{\footnotesize\textcolor{mmaLabel}{\fontfamily{phv}\selectfont In[6]:=}}
\textcolor{mmaUndefined}{\texttt{DisplaySHSOutput}}\texttt{[\textcolor{mmaString}{"CanonicalDim"}$\to$3, \textcolor{mmaString}{"OnlyCanonicalDimOutput"}$\to$False,\\ \hspace*{5cm}\textcolor{mmaString}{"$\Delta$B"}$\to$"NA", \textcolor{mmaString}{"$\Delta$L"}$\to$"NA", \textcolor{mmaString}{"Flavours"}$\to$ $N_f$]}

\vspace{0.25cm}
{\footnotesize\textcolor{mmaLabel}{\fontfamily{phv}\selectfont Out[6]:=}}

\vspace{-0.5cm}
\hspace*{1.4cm}\texttt{Total number\,of independent operators at  dimension 1 is 0},\\
\hspace*{1.4cm}\texttt{Operators: 0}\\

\hspace*{1.4cm}\texttt{Total number\,of independent operators at  dimension 2 is {\small$4+4 N_f+5 N_f^2$}}\\
\hspace*{1.4cm}\texttt{Operators:}
{\tiny\begin{eqnarray}
& &\hspace*{1.2cm} L\; N_f\; H_u+H_d\; H_u+D\;N_f^2\; D^{\dagger}\; e^{V_D}+E\; N_f^2\; E^{\dagger}\; e^{V_E}+L\; N_f^2\; L^{\dagger}\; e^{V_L}+N_f\; H_d\; L^{\dagger}\; e^{V_L}+ N_f^2\; Q\; Q^{\dagger}\; e^{V_Q}+ N_f^2\; U U^{\dagger}\; e^{V_U}+ \nonumber \\
& &\hspace*{1.2cm} L\;N_f\; \left(H_d \right)^{\dagger}\; e^{V_{H_d}}+H_d\; \left(H_d \right)^{\dagger}\; e^{V_{H_d}}+H_u\;\left(H_u \right)^{\dagger} e^{V_{H_u}}+N_f\; L^{\dagger}\; \left(H_u \right)^{\dagger}\; e^{V_L} e^{V_{H_u}}+\left(H_d \right)^{\dagger}\; \left(H_u \right)^{\dagger}\; e^{V_{H_d}}\; e^{V_{H_u}} \nonumber
\end{eqnarray}}

\noindent
\hspace*{1.4cm}\texttt{Total number\,of independent operators at  dimension 3 is \\
\hspace*{1.4cm}	{\small$3+2N_f+11N_f^2+9N_f^3$}} \\
\\
\hspace*{1.4cm}\texttt{Operators:}
{\tiny\begin{eqnarray}
& &\hspace*{1.2cm} \frac{Bl^2}{2}+\frac{Br^2}{2}+\frac{Gl^2}{2}+\frac{Gr^2}{2}-\frac{1}{2} E\; L^2\; N_f^2+\frac{1}{2} E\; L^2\; N_f^3+D\; L\; N_f^3\; Q-\frac{1}{2} D^2\; N_f^2\; U+\frac{1}{2} D^2\; N_f^3\;U+ \frac{Wl^2}{2}+ \frac{Wr^2}{2}+ E\; L\; N_f^2\; H_d+ \nonumber \\
& &\hspace*{1.2cm}  D\; N_f^2\; Q\; H_d+N_f^2\; Q\; U\; H_u+\frac{1}{2} N_f^2\; Q^2\; D^{\dagger}\; e^{V_D}+\frac{1}{2}\;N_f^3\; Q^2 D^{\dagger}\; e^{V_D}+E\; N_f^3\; U\; D^{\dagger}\; e^{V_D}+N_f^3\; Q\; U\; L^{\dagger}\; e^{V_L}+ N_f^2\; H_u E^{\dagger}\; L^{\dagger}\; e^{V_E}\; e^{V_L}- \nonumber \\
& &\hspace*{1.2cm} \frac{1}{2} N_f^2\; E^{\dagger}\; \left(L^{\dagger}\right)^2\; e^{V_E} \left(e^{V_L}\right)^2+\frac{1}{2}\; N_f^3\; E^{\dagger} \left(L^{\dagger}\right)^2\; e^{V_E} \left(e^{V_L}\right)^2+ N_f^2\; H_u\; D^{\dagger}\; Q^{\dagger}\; e^{V_D}\; e^{V_Q}+N_f^3\; D^{\dagger}\; L^{\dagger}\; Q^{\dagger}\; e^{V_D}\; e^{V_L}\; e^{V_Q}\;+ \nonumber \\
& &\hspace*{1.2cm} \frac{1}{2} D\; N_f^2\; \left(Q^{\dagger}\right)^2\; \left(e^{V_Q}\right)^2+\frac{1}{2} D\; N_f^3\; \left(Q^{\dagger}\right)^2\;
\left(e^{V_Q}\right)^2\;- \frac{1}{2} N_f^2\; \left({D}^{\dagger}\right)^2\; U^{\dagger}\; \left(e^{V_D}\right)^2\;
e^{V_U}+\frac{1}{2} N_f^3\; \left(D^{\dagger}\right)^2\; U^{\dagger}\; \left(e^{V_D}\right)^2\; e^{V_U}+\nonumber \\
& &\hspace*{1.2cm} D\; N_f^3\; E^{\dagger}\; U^{\dagger}\; e^{V_E}\; e^{V_U}+ 
L\; N_f^3\; Q^{\dagger}\; U^{\dagger}\; e^{V_Q}\; e^{V_U}+N_f^2\; H_d\; Q^{\dagger}\; U^{\dagger}\; e^{V_Q}\; e^{V_U}\;+N_f^2\; Q\; U\; \left(H_d \right)^{\dagger}\; e^{V_{H_d}}+N_f\;H_u\; E^{\dagger}\; \left(H_d \right)^{\dagger}\; e^{V_E}\; e^{V_{H_d}}+ \nonumber \\
& &\hspace*{1.2cm} N_f^2\; E^{\dagger}\; L^{\dagger}\; \left(H_d \right)^{\dagger}\; e^{V_E}\; e^{V_L}\;e^{V_{H_d}}\;+ N_f^2\;D^{\dagger}\; Q^{\dagger}\;\left(H_d \right)^{\dagger}\; e^{V_D}\; e^{V_Q}\;e^{V_{H_d}}+E\; L\; N_f^2\; \left(H_u \right)^{\dagger}\; e^{V_{H_u}}+\nonumber \\
& &\hspace*{1.2cm} D\;N_f^2\; Q\; \left(H_u \right)^{\dagger}\; e^{V_{H_u}}+E\; N_f\; H_d\; \left(H_u \right)^{\dagger}\; e^{V_{H_u}}+N_f^2\; Q^{\dagger}\; U^{\dagger}\; \left(H_u \right)^{\dagger}\; e^{V_Q}\; e^{V_U}\; e^{V_{H_u}} \nonumber
\end{eqnarray}}
\end{table}

\begin{table}[h]
\hspace*{0.25cm}{\footnotesize\textcolor{mmaLabel}{\fontfamily{phv}\selectfont In[7]:=}}
\textcolor{mmaUndefined}{\texttt{DisplaySHSOutput}}\texttt{[\textcolor{mmaString}{"CanonicalDim"}$\to$4, \textcolor{mmaString}{"OnlyCanonicalDimOutput"}$\to$True,\\ \hspace*{5cm}\textcolor{mmaString}{"$\Delta$B"}$\to$"NA", \textcolor{mmaString}{"$\Delta$L"}$\to$"NA", \textcolor{mmaString}{"Flavours"}$\to$ $N_f$]}

\vspace{0.25cm}
{\footnotesize\textcolor{mmaLabel}{\fontfamily{phv}\selectfont Out[7]:=}}

\vspace{-1cm}
{\tiny\begin{eqnarray}
& &\hspace{1.3cm} -\frac{1}{3} L N_f^2 Q^3+\frac{1}{3} L N_f^4 Q^3+ E L N_f^4 Q U+D N_f^4 Q^2 U-\frac{1}{2} D E N_f^3 U^2+\frac{1}{2} D E N_f^4 U^2-\frac{1}{3} N_f Q^3 H_d+\frac{1}{3} N_f^3 Q^3 H_d+E N_f^3 Q U H_d+ \nonumber \\
& & L N_f H_d H_u^2+H_d^2 H_u^2+ D L N_f^3 H_u D^{\dagger} e^{V_D}+D N_f^2 H_d H_u D^{\dagger} e^{V_D}+\frac{1}{2} D^2 N_f^2 \left(D^{\dagger}\right)^2 \left(e^{V_D}\right)^2+\frac{1}{2} D^2 N_f^4 \left(D^{\dagger}\right)^2 \left(e^{V_D}\right)^2-\frac{1}{2} N_f^2 Q H_u \left(D^{\dagger}\right)^2 \left(e^{V_D}\right)^2+\nonumber \\ 
& & \frac{1}{2} N_f^3 Q H_u \left(D^{\dagger}\right)^2 \left(e^{V_D}\right)^2+\frac{1}{3} E N_f^2 \left(D^{\dagger}\right)^3 \left(e^{V_D}\right)^3-\frac{1}{2} E N_f^3 \left(D^{\dagger}\right)^3 \left(e^{V_D}\right)^3+\frac{1}{6}E N_f^4 \left(D^{\dagger}\right)^3 \left(e^{V_D}\right)^3+\frac{1}{3} D^3 N_f^2 E^{\dagger} e^{V_E}-\nonumber \\
& & \frac{1}{2} D^3 N_f^3 E^{\dagger} e^{V_E}+\frac{1}{6} D^3 N_f^4 E^{\dagger} e^{V_E}+ E L N_f^3 H_u E^{\dagger} e^{V_E}+D N_f^3 Q H_u E^{\dagger} e^{V_E}+E N_f^2 H_d H_u E^{\dagger} e^{V_E}+D E N_f^4 D^{\dagger} E^{\dagger} e^{V_D} e^{V_E}+\frac{1}{4} E^2 N_f^2 \left(E^{\dagger}\right)^2 \left(e^{V_E}\right)^2+\nonumber \\
& & \frac{1}{2} E^2 N_f^3 \left(E^{\dagger}\right)^2 \left(e^{V_E}\right)^2+\frac{1}{4} E^2 N_f^4 \left(E^{\dagger}\right)^2 \left(e^{V_E}\right)^2+L^2 N_f^3 H_u L^{\dagger} e^{V_L}+2 L N_f^2 H_d H_u L^{\dagger} e^{V_L}+N_f H_d^2 H_u L^{\dagger} e^{V_L}+D L N_f^4 D^{\dagger} L^{\dagger} e^{V_D} e^{V_L}+\nonumber \\
& &  D N_f^3 H_d D^{\dagger} L^{\dagger} e^{V_D} e^{V_L}+N_f^3 U H_u D^{\dagger} L^{\dagger}e^{V_D} e^{V_L}-\frac{1}{2} N_f^3 Q \left(D^{\dagger}\right)^2 L^{\dagger} \left(e^{V_D}\right)^2 e^{V_L}+ \frac{1}{2} N_f^4 Q \left(D^{\dagger}\right)^2 L^{\dagger} \left(e^{V_D}\right)^2 e^{V_L}+E L N_f^4 E^{\dagger} L^{\dagger} e^{V_E} e^{V_L}+\nonumber \\
& &D N_f^4 Q E^{\dagger} L^{\dagger} e^{V_E}
e^{V_L}+E N_f^3 H_d E^{\dagger} L^{\dagger} e^{V_E} e^{V_L}+\frac{1}{2} L^2 N_f^2\left(L^{\dagger}\right)^2 \left(e^{V_L}\right)^2+\frac{1}{2} L^2 N_f^4 \left(L^{\dagger}\right)^2 \left(e^{V_L}\right)^2+L N_f^3 H_d \left(L^{\dagger}\right)^2 \left(e^{V_L}\right)^2+\nonumber \\ 
& & \frac{1}{2} N_f H_d^2 \left(L^{\dagger}\right)^2 \left(e^{V_L}\right)^2+\frac{1}{2}N_f^2 H_d^2 \left(L^{\dagger}\right)^2 \left(e^{V_L}\right)^2-\frac{1}{2} N_f^3 U D^{\dagger} \left(L^{\dagger}\right)^2 e^{V_D}\left(e^{V_L}\right)^2+\frac{1}{2} N_f^4 U D^{\dagger} \left(L^{\dagger}\right)^2e^{V_D}\left(e^{V_L}\right)^2-\frac{1}{2} D^2 L N_f^3 Q^{\dagger} e^{V_Q}+\nonumber \\
& &\frac{1}{2} D^2L N_f^4 Q^{\dagger} e^{V_Q}-\frac{1}{2} D^2 N_f^2 H_d Q^{\dagger} e^{V_Q}+\frac{1}{2} D^2 N_f^3 H_d Q^{\dagger} e^{V_Q}+2 L N_f^3 Q H_u Q^{\dagger} e^{V_Q}+D N_f^3 U H_u Q^{\dagger} e^{V_Q}+2N_f^2 Q H_d H_u Q^{\dagger} e^{V_Q}+E L N_f^4 D^{\dagger} Q^{\dagger} e^{V_D} e^{V_Q}+\nonumber \\ 
& & 2D N_f^4 Q D^{\dagger} Q^{\dagger} e^{V_D} e^{V_Q}+E N_f^3 H_d D^{\dagger} Q^{\dagger} e^{V_D} e^{V_Q}+E N_f^4 Q E^{\dagger} Q^{\dagger} e^{V_E}e^{V_Q}+2 L N_f^4 Q L^{\dagger} Q^{\dagger} e^{V_L} e^{V_Q}+D N_f^4 U L^{\dagger} Q^{\dagger}e^{V_L} e^{V_Q}+\nonumber \\
& & 2 N_f^3 Q H_d L^{\dagger} Q^{\dagger} e^{V_L} e^{V_Q}+N_f^2 Q^2 \left(Q^{\dagger}\right)^2 \left(e^{V_Q}\right)^2+N_f^4 Q^2 \left(Q^{\dagger}\right)^2\left(e^{V_Q}\right)^2+\frac{1}{2} EN_f^3 U \left(Q^{\dagger}\right)^2 \left(e^{V_Q}\right)^2+\frac{1}{2} E N_f^4 U \left(Q^{\dagger}\right)^2 \left(e^{V_Q}\right)^2-\nonumber \\
& & \frac{1}{3}N_f H_u \left(Q^{\dagger}\right)^3 \left(e^{V_Q}\right)^3+ \frac{1}{3} N_f^3 H_u \left(Q^{\dagger}\right)^3 \left(e^{V_Q}\right)^3-\frac{1}{3}N_f^2 L^{\dagger} \left(Q^{\dagger}\right)^3 e^{V_L}\left(e^{V_Q}\right)^3+\frac{1}{3} N_f^4 L^{\dagger} \left(Q^{\dagger}\right)^3 e^{V_L} \left(e^{V_Q}\right)^3-\frac{1}{2} D L^2 N_f^3 U^{\dagger} e^{V_U}+\nonumber \\
& & \frac{1}{2} DL^2 N_f^4 U^{\dagger} e^{V_U}+D L N_f^3 H_d U^{\dagger} e^{V_U}+L N_f^3 U H_u U^{\dagger} e^{V_U}+N_f^2U H_d H_u U^{\dagger} e^{V_U}+L N_f^4 Q D^{\dagger} U^{\dagger} e^{V_D} e^{V_U}+2 DN_f^4 U D^{\dagger} U^{\dagger} e^{V_D} e^{V_U}+\nonumber \\
& & N_f^3 Q H_d D^{\dagger} U^{\dagger} e^{V_D} e^{V_U}+\frac{1}{2} N_f^3 Q^2 E^{\dagger} U^{\dagger} e^{V_E} e^{V_U}+\frac{1}{2}N_f^4 Q^2 E^{\dagger} U^{\dagger} e^{V_E} e^{V_U}+E N_f^4 U E^{\dagger} U^{\dagger} e^{V_E} e^{V_U}+L N_f^4 U L^{\dagger} U^{\dagger} e^{V_L} e^{V_U}+\nonumber \\ 
& & N_f^3U H_d L^{\dagger} U^{\dagger} e^{V_L} e^{V_U}+2 N_f^4 Q U Q^{\dagger} U^{\dagger} e^{V_Q} e^{V_U}+N_f^3H_u E^{\dagger} Q^{\dagger} U^{\dagger} e^{V_E} e^{V_Q} e^{V_U}+N_f^4 E^{\dagger}L^{\dagger} Q^{\dagger} U^{\dagger} e^{V_E} e^{V_L} e^{V_Q} e^{V_U}+N_f^4 D^{\dagger} \left(Q^{\dagger}\right)^2 U^{\dagger} e^{V_D} \left(e^{V_Q}\right)^2 e^{V_U}+\nonumber \\
& & \frac{1}{2} N_f^2U^2 \left(U^{\dagger}\right)^2 \left(e^{V_U}\right)^2+\frac{1}{2} N_f^4 U^2 \left(U^{\dagger}\right)^2 \left(e^{V_U}\right)^2- \frac{1}{2}N_f^3 D^{\dagger} E^{\dagger}\left(U^{\dagger}\right)^2 e^{V_D} e^{V_E} \left(e^{V_U}\right)^2+\frac{1}{2}N_f^4 D^{\dagger} E^{\dagger} \left(U^{\dagger}\right)^2 e^{V_D} e^{V_E} \left(e^{V_U}\right)^2+L^2N_f^2 H_u \left(H_d \right)^{\dagger} e^{V_{H_d}}+ \nonumber \\
& & 2 L N_f H_d H_u \left(H_d \right)^{\dagger} e^{V_{H_d}}+ H_d^2H_u \left(H_d \right)^{\dagger} e^{V_{H_d}}+D L N_f^3 D^{\dagger} \left(H_d \right)^{\dagger} e^{V_D}e^{V_{H_d}}+D N_f^2 H_d D^{\dagger} \left(H_d \right)^{\dagger} e^{V_D} e^{V_{H_d}}+ N_f^2U H_u D^{\dagger} \left(H_d \right)^{\dagger} e^{V_D} e^{V_{H_d}}-\nonumber \\
& &\frac{1}{2} N_f^2 Q \left(D^{\dagger}\right)^2 \left(H_d \right)^{\dagger} \left(e^{V_D}\right)^2e^{V_{H_d}}+\frac{1}{2} N_f^3 Q \left(D^{\dagger}\right)^2 \left(H_d \right)^{\dagger} \left(e^{V_D}\right)^2 e^{V_{H_d}}+ E L N_f^3 E^{\dagger} \left(H_d \right)^{\dagger} e^{V_E} e^{V_{H_d}}+D N_f^3 Q E^{\dagger} \left(H_d \right)^{\dagger} e^{V_E} e^{V_{H_d}}+\nonumber \\
& &E N_f^2 H_d E^{\dagger} \left(H_d\right)^{\dagger} e^{V_E}e^{V_{H_d}}+L^2 N_f^3 L^{\dagger} \left(H_d \right)^{\dagger} e^{V_L} e^{V_{H_d}}+2 L N_f^2 H_d L^{\dagger} \left(H_d \right)^{\dagger} e^{V_L} e^{V_{H_d}}+N_f H_d^2 L^{\dagger} \left(H_d \right)^{\dagger} e^{V_L}e^{V_{H_d}}+N_f^3 U D^{\dagger} L^{\dagger} \left(H_d\right)^{\dagger} e^{V_D} e^{V_L}e^{V_{H_d}}+\nonumber \\
& & 2 L N_f^3 Q Q^{\dagger} \left(H_d \right)^{\dagger} e^{V_Q} e^{V_{H_d}}+D N_f^3 UQ^{\dagger} \left(H_d \right)^{\dagger} e^{V_Q} e^{V_{H_d}}+2 N_f^2 Q H_d Q^{\dagger} \left(H_d\right)^{\dagger} e^{V_Q}e^{V_{H_d}}-\frac{1}{3} N_f \left(Q^{\dagger}\right)^3 \left(H_d \right)^{\dagger} \left(e^{V_Q}\right)^3 e^{V_{H_d}}+\nonumber \\
& & L N_f^3 U U^{\dagger} \left(H_d \right)^{\dagger} e^{V_U} e^{V_{H_d}}+N_f^2 U H_d U^{\dagger} \left(H_d \right)^{\dagger} e^{V_U}e^{V_{H_d}}+N_f^3 E^{\dagger} Q^{\dagger} U^{\dagger} \left(H_d \right)^{\dagger} e^{V_E} e^{V_Q}e^{V_U} e^{V_{H_d}}+ \frac{1}{2} L^2 N_f \left(\left(H_d \right)^{\dagger}\right)^2 \left(e^{V_{H_d}}\right)^2+\nonumber \\
& & \frac{1}{2}L^2 N_f^2 \left(\left(H_d \right)^{\dagger}\right){}^2 \left(e^{V_{H_d}}\right)^2+L N_f H_d \left(\left(H_d \right)^{\dagger}\right)^2 \left(e^{V_{H_d}}\right)^2+H_d^2\left(\left(H_d \right)^{\dagger}\right)^2 \left(e^{V_{H_d}}\right)^2-\frac{1}{3}N_f Q^3 \left(H_u \right)^{\dagger} e^{V_{H_u}}+\frac{1}{3} N_f^3 Q^3 \left(H_u \right)^{\dagger} e^{V_{H_u}}+\nonumber \\
& & EN_f^3 Q U \left(H_u \right)^{\dagger} e^{V_{H_u}}+L N_f H_u^2 \left(H_u \right)^{\dagger} e^{V_{H_u}}+H_d H_u^2\left(H_u \right)^{\dagger} e^{V_{H_u}}+D N_f^2 H_u D^{\dagger} \left(H_u \right)^{\dagger} e^{V_D}e^{V_{H_u}}+E N_f^2 H_u E^{\dagger} \left(H_u \right)^{\dagger} e^{V_E} e^{V_{H_u}}+\nonumber \\
& & 2L N_f^2 H_u L^{\dagger} \left(H_u \right)^{\dagger} e^{V_L} e^{V_{H_u}}+  2 N_f H_d H_u L^{\dagger} \left(H_u \right)^{\dagger} e^{V_L} e^{V_{H_u}}+D N_f^3 D^{\dagger} L^{\dagger} \left(H_u \right)^{\dagger} e^{V_D}e^{V_L} e^{V_{H_u}}+E N_f^3 E^{\dagger} L^{\dagger} \left(H_u \right){}^{\dagger} e^{V_E}e^{V_L} e^{V_{H_u}}+\nonumber \\
& & L N_f^3 \left(L^{\dagger}\right)^2 \left(H_u \right)^{\dagger} \left(e^{V_L}\right)^2 e^{V_{H_u}}+N_f^2H_d \left(L^{\dagger}\right)^2 \left(H_u \right)^{\dagger} \left(e^{V_L}\right){}^2 e^{V_{H_u}}-\frac{1}{2} D^2 N_f^2Q^{\dagger} \left(H_u \right)^{\dagger} e^{V_Q} e^{V_{H_u}}+\frac{1}{2} D^2 N_f^3 Q^{\dagger} \left(H_u \right)^{\dagger} e^{V_Q} e^{V_{H_u}}+\nonumber \\
& & 2 N_f^2 Q H_u Q^{\dagger} \left(H_u \right)^{\dagger} e^{V_Q} e^{V_{H_u}}+E N_f^3 D^{\dagger} Q^{\dagger}\left(H_u \right)^{\dagger} e^{V_D} e^{V_Q} e^{V_{H_u}}+2N_f^3 Q L^{\dagger} Q^{\dagger} \left(H_u \right)^{\dagger} e^{V_L} e^{V_Q} e^{V_{H_u}}+D L N_f^3U^{\dagger} \left(H_u \right)^{\dagger} e^{V_U} e^{V_{H_u}}+\nonumber \\
& & D N_f^2 H_d U^{\dagger} \left(H_u \right)^{\dagger} e^{V_U} e^{V_{H_u}}+N_f^2 U H_u U^{\dagger} \left(H_u \right)^{\dagger} e^{V_U} e^{V_{H_u}}+N_f^3Q D^{\dagger} U^{\dagger} \left(H_u \right)^{\dagger} e^{V_D} e^{V_U} e^{V_{H_u}}+N_f^3U L^{\dagger} U^{\dagger} \left(H_u \right)^{\dagger} e^{V_L} e^{V_U} e^{V_{H_u}}+\nonumber \\
& & 2 L N_f H_u\left(H_d \right)^{\dagger} \left(H_u \right)^{\dagger} e^{V_{H_d}} e^{V_{H_u}}+2 H_d H_u \left(H_d \right)^{\dagger} \left(H_u \right)^{\dagger} e^{V_{H_d}}e^{V_{H_u}}+D N_f^2 D^{\dagger} \left(H_d \right)^{\dagger} \left(H_u \right)^{\dagger} e^{V_D}e^{V_{H_d}} e^{V_{H_u}}+\frac{1}{2} L^2 N_f H_u^2+\frac{1}{2} L^2 N_f^2 H_u^2+\nonumber \\
& & E N_f^2 E^{\dagger} \left(H_d \right)^{\dagger} \left(H_u \right)^{\dagger} e^{V_E} e^{V_{H_d}} e^{V_{H_u}}+ 2 L N_f^2 L^{\dagger} \left(H_d \right)^{\dagger} \left(H_u \right)^{\dagger} e^{V_L} e^{V_{H_d}} e^{V_{H_u}}+2 N_f H_d L^{\dagger} \left(H_d \right){}^{\dagger} \left(H_u \right)^{\dagger}e^{V_L} e^{V_{H_d}} e^{V_{H_u}}+\nonumber \\
& & 2 N_f^2 Q Q^{\dagger} \left(H_d \right)^{\dagger} \left(H_u \right)^{\dagger} e^{V_Q} e^{V_{H_d}} e^{V_{H_u}}+N_f^2 U U^{\dagger} \left(H_d \right)^{\dagger} \left(H_u \right)^{\dagger} e^{V_U} e^{V_{H_d}} e^{V_{H_u}}+H_u^2 \left(\left(H_u \right)^{\dagger}\right)^2 \left(e^{V_{H_u}}\right)^2+L N_f \left(\left(H_d \right)^{\dagger}\right)^2 \left(H_u \right)^{\dagger} \left(e^{V_{H_d}}\right)^2 e^{V_{H_u}}+\nonumber \\
& & H_d \left(\left(H_d \right)^{\dagger}\right)^2 \left(H_u\right)^{\dagger} \left(e^{V_{H_d}}\right)^2e^{V_{H_u}}+ N_f H_u L^{\dagger} \left(\left(H_u \right)^{\dagger}\right)^2 e^{V_L} \left(e^{V_{H_u}}\right)^2+\frac{1}{2} N_f \left(L^{\dagger}\right)^2 \left(\left(H_u \right)^{\dagger}\right)^2 \left(e^{V_L}\right)^2 \left(e^{V_{H_u}}\right)^2+\frac{1}{2} N_f^2 \left(L^{\dagger}\right)^2 \left(\left(H_u \right)^{\dagger}\right)^2 \left(e^{V_L}\right)^2 \left(e^{V_{H_u}}\right)^2\nonumber \\
& & + H_u \left(H_d \right)^{\dagger} \left(\left(H_u\right)^{\dagger}\right)^2 e^{V_{H_d}} \left(e^{V_{H_u}}\right){}^2+ N_f L^{\dagger} \left(H_d \right)^{\dagger} \left(\left(H_u \right)^{\dagger}\right)^2 e^{V_L} e^{V_{H_d}} \left(e^{V_{H_u}}\right)^2+\left(\left(H_d\right)^{\dagger}\right)^2\left(\left(H_u\right)^{\dagger}\right)^2 \left(e^{V_{H_d}}\right)^2 \left(e^{V_{H_u}}\right)^2 \nonumber	
\end{eqnarray}}
\end{table}
\clearpage

%% file: GrIP_Err.tex
\subsubsection*{\underline{Possible sources of errors}}\label{subsubsec:error}

Here, we have highlighted possible errors one can make while preparing the input file for a new model. The prime focus should be on the keys: \texttt{"FieldName"}, \texttt{"Self-Conjugate"}, \texttt{"Lorentz Behaviour"}, \texttt{"Chirality"}, \texttt{"Baryon Number"} and \texttt{"Lepton Number"}. Their sequence must be unaltered and none of the keys should be omitted.
 
The entries within the \texttt{"SymmetryGroupClass"} must be incorporated systematically, respecting the following thumb-rules:

$\bullet$ 
The sequence of group information in \texttt{"SymmetryGroupClass"} and  \texttt{"Field[i]"} must be the same. 
 
$\bullet$
Different names to the same group must be assigned in case of their repetitive appearance, e.g, two $SU(2)$ groups should be named as $``SU2L"$ and $``SU2R"$ to distinguish them from each other.
\begin{figure}[h]
	\centering
	\includegraphics[scale=0.6]{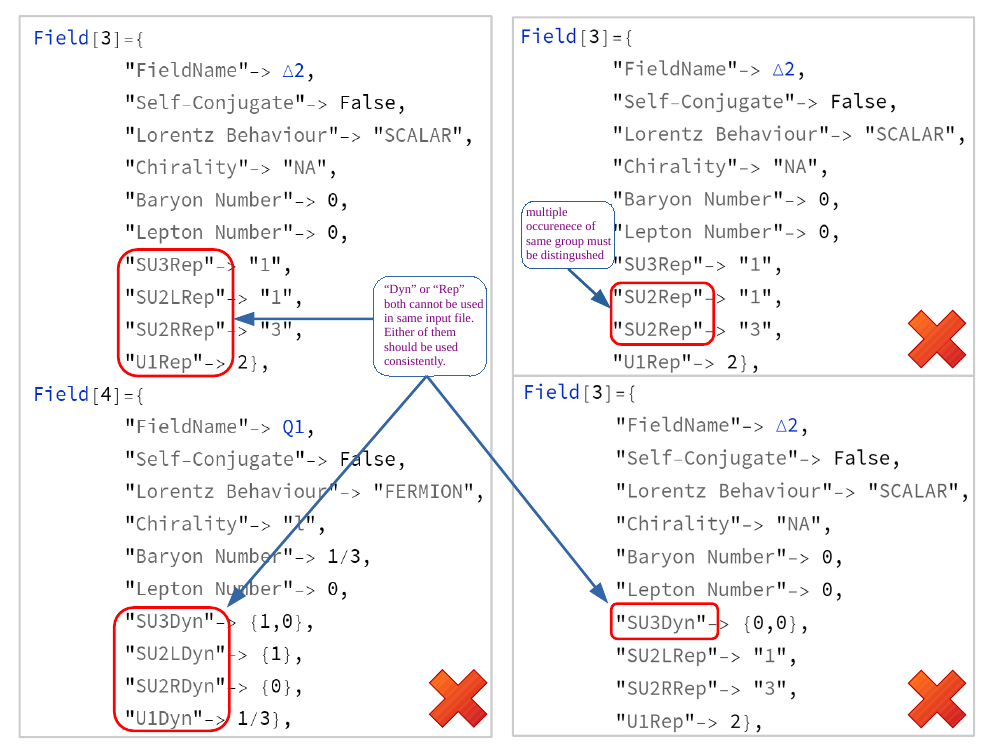}
	\caption{\small This figure highlights how one \emph{should not} name the groups in describing the particle transformation property.}
	\label{fig:grp_wrong}
\end{figure}

\begin{figure}[h]
	\centering
	\includegraphics[scale=0.6]{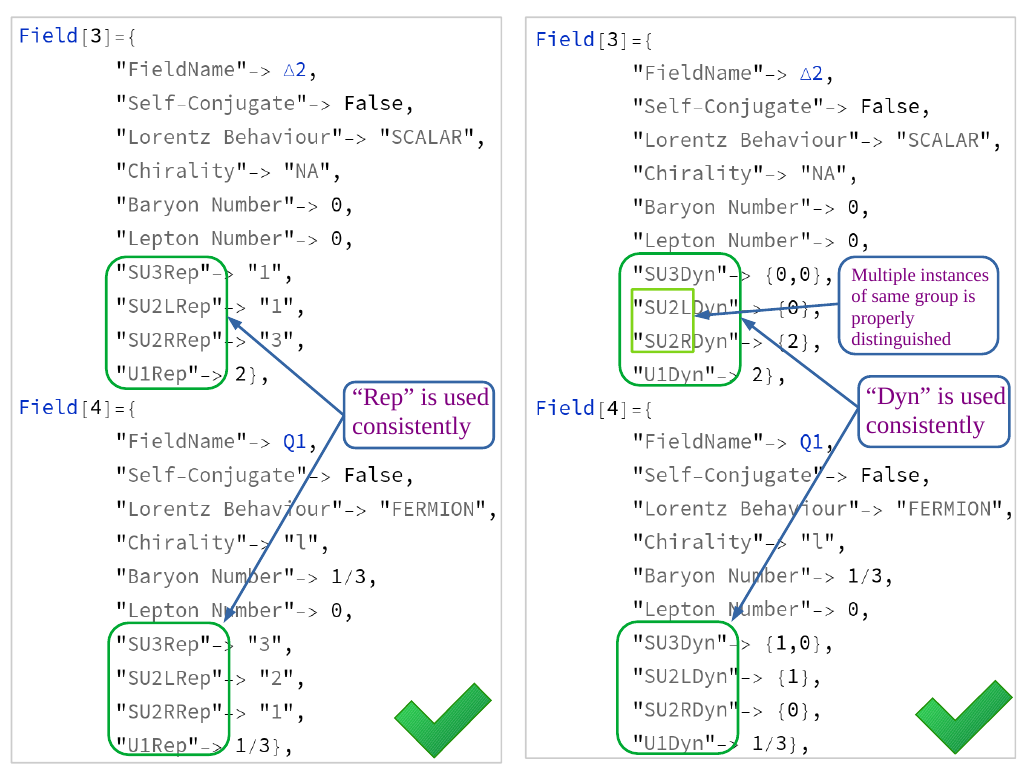}
	\caption{\small The correct way of providing details about the gauge groups.}
	\label{fig:grp_ok}
\end{figure}

$\bullet$
The transformation property under a particular group must be entered in the form of the dimension of the representation (``Rep") or Dynkin label corresponding to that representation (``Dyn"). These options should not be mixed up and must be used uniformly for the whole input file, see Figs.~\ref{fig:grp_wrong} and \ref{fig:grp_ok}.

\begin{figure}[h]
	\centering
	\includegraphics[scale=0.5]{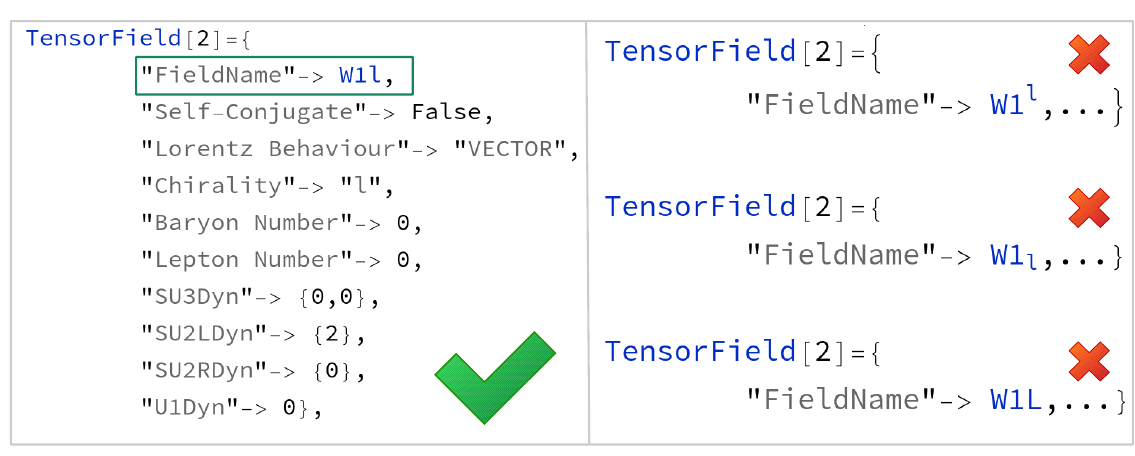}
	\caption{\small Correct and incorrect ways of encoding the details of the chiral nature of the field strength tensors in the \texttt{"FieldName"}.}
	\label{fig:ft_ok}
\end{figure}

$\bullet$
The chiral nature of the field strength tensors must be reflected at the end of \texttt{"FieldName"} by appending ``$l$" or ``$r$", e.g.,``$Bl$" or ``$Br$", see Fig.~\ref{fig:ft_ok}. \\

$\bullet$
The dimensions of the representations must be provided keeping the following points in mind:
\begin{enumerate}
	\item For non-abelian groups the dimension of the representation must be written as a string, see Fig.~\ref{fig:grp_ok}. Also, for abelian groups such as $U(1)$, the charge must be entered as a number.  
	\item The conjugate of any representation must contain \texttt{"bar"}, e.g.: $\overline{3} \, \to$ \texttt{"3 bar"}.
	\item One must distinguish different representations of the same dimension: $8^{\prime} \, \to$ \texttt{"8 p"}, similarly, $8^{\prime\prime} \, \to$ \texttt{"8 pp"}, see Table~\ref{table:rep_input}.
\end{enumerate}

\begin{table}[t]
\centering
\renewcommand{\arraystretch}{1.8}	
{\scriptsize\begin{tabular}[h]{|c|c|}
		\hline
		\textbf{Dimension} & \textbf{Input Form} \\
		\hline
		20 & \texttt{"20"} \\
		\hline
		$\overline{20}$ & \texttt{"20 bar"}\\
		\hline
		$20^{\prime}$ & \texttt{"20 p"} \\
		\hline
		$20^{\prime\prime}$ & \texttt{"20 pp"}\\
		\hline
		$\overline{20^{\prime\prime}}$  & \texttt{"20 ppbar"}\\
		\hline
	\end{tabular}}
\caption{\small Input form of the dimension.}
\label{table:rep_input}
\end{table}

%% file: OperatorCategorization.tex
\subsection{Categorizing the operators and writing them in a covariant form}
\subsubsection*{\underline{Covariant form of the SM Lagrangian}}
To convert the \textbf{GrIP} output into a covariant form, we employ the translation between symbols as outlined in Table~\ref{table:sm-raw-to-covariant-transl}. The correspondence between the Hilbert Series output generated by \textbf{GrIP} and their covariant forms is shown in Table~\ref{table:sm-renorm-output}. 
\begin{table}[h]
	\centering
	\renewcommand{\arraystretch}{1.8}
	{\tiny\begin{tabular}{|ccc|ccc|ccc|}
			\hline
			$Q$&
			$\rightarrow$&
			$Q^p_l$&
			$u$&
			$\rightarrow$&
			$u^p_r$&
			$d$&
			$\rightarrow$&
			$d^p_r$\\
			
			$Q^{\dagger}$&
			$\rightarrow$&
			$\overline{Q}^p_l$&
			$u^{\dagger}$&
			$\rightarrow$&
			$\overline{u}^p_r$&
			$d^{\dagger}$&
			$\rightarrow$&
			$\overline{d}^p_r$\\
			
			$L$&
			$\rightarrow$&
			$L^p_l$&
			$el$&
			$\rightarrow$&
			$e^p_r$&
			$H, H^{\dagger}$&
			$\rightarrow$&
			$H, H^{\dagger}$\\
			
			$L^{\dagger}$&
			$\rightarrow$&
			$\overline{L}^p_l$&
			$el^{\dagger}$&
			$\rightarrow$&
			$\overline{e}^p_r$&
			$\mathcal{D}$&
			$\rightarrow$&
			$\mathcal{D}_{\mu}$\\
			
			$(Wl, Wr)$&
			$\rightarrow$&
			$(W^I_{\mu\nu},\tilde{W}^I_{\mu\nu})$&
			$(Gl, Gr)$&
			$\rightarrow$&
			$(G^a_{\mu\nu},\tilde{G}^a_{\mu\nu})$&
			$(Bl, Br)$&
			$\rightarrow$&
			$(B_{\mu\nu},\tilde{B}_{\mu\nu})$\\
			\hline
	\end{tabular}}
	\caption{\small Dictionary for translation of operators from the Hilbert Series output to their covariant forms.}
	\label{table:sm-raw-to-covariant-transl}
\end{table} 

\begin{table}[h]
	\centering
	\renewcommand{\arraystretch}{2.0}
	{\scriptsize\begin{tabular}{||c|c||c|}
			\hline
			\multicolumn{3}{|c|}{\textbf{Mass Dimension-2}}\\
			\hline
			\multirow{2}{*}{\textbf{HS Output}}&
			\multirow{2}{*}{\textbf{Covariant Form}}&
			\multirow{2}{*}{\textbf{\textbf{No. of Operators}}}\\

			&
			&
			\textbf{(including h.c.)}\\
			\hline
			$H^{\dagger}H$&
			$H^{\dagger}H$&
			1
			\\
			\hline
			\hline
			\multicolumn{3}{|c|}{\textbf{Mass Dimension-4}}\\
			\hline
			\multirow{2}{*}{\textbf{HS Output}}&
			\multirow{2}{*}{\textbf{Covariant Form}}&
			\multirow{2}{*}{\textbf{\textbf{No. of Operators}}}\\
			
			&
			&
			\textbf{(including h.c.)}\\
			\hline
			$Bl^2 + Br^2,$&
			$B^{\mu\nu}B_{\mu\nu},\hspace{0.2cm} B^{\mu\nu}\tilde{B}_{\mu\nu},$&
			\multirow{6}{*}{$11N_f^2  +  8$}
			\\
			
			$ Gl^2 + Gr^2,$&
			$G^{a\mu\nu}G^{a}_{\mu\nu},\hspace{0.2cm} G^{a\mu\nu}\tilde{G}^{a}_{\mu\nu},$&
			\\
			
			$Wl^2 + Wr^2, $&
			$W^{I \mu\nu}W^{I}_{\mu\nu},\hspace{0.2cm} W^{I \mu\nu}\tilde{W}^{I}_{\mu\nu}, $&
			\\
			
			$H^{\dagger}H \mathcal{D}^2,\hspace{0.2cm}(H^{\dagger}H)^2,$&
			$(\mathcal{D}_{\mu}H)^{\dagger}(\mathcal{D}^{\mu}H),\hspace{0.2cm}(H^{\dagger}H)^2,$&
			\\
			
			$N_f^{2} L^{\dagger}L\mathcal{D},\hspace{0.2cm} N_f^{2} el^{\dagger}el\mathcal{D}, $&
			$\overline{L}_l^{q}\slashed{\mathcal{D}}L^{q}_l,\hspace{0.2cm}
			\overline{e}_r^{q}\slashed{\mathcal{D}}e^{q}_r$,&
			\\
			
			${}^{\textcolor{black}{\clubsuit}}N_f^{2} Q^{\dagger}Q\mathcal{D},\hspace{0.2cm} N_f^{2} u^{\dagger}u\mathcal{D},\hspace{0.2cm} N_f^{2} d^{\dagger}d\mathcal{D},$&
			${}^{\textcolor{black}{\clubsuit}}\overline{Q}_l^{q}\slashed{\mathcal{D}}Q^{q}_l,\hspace{0.2cm}
			\overline{u}_r^{q}\slashed{\mathcal{D}}u^{q}_r,\hspace{0.2cm}
			\overline{d}_r^{q}\slashed{\mathcal{D}}d^{q}_r,$&
			\\
			
			$ \color{blue}{N_f^2L^{\dagger}H el},\hspace{0.2cm}
			\color{blue}{N_f^2Q^{\dagger}H d},\hspace{0.2cm} \color{blue}{N_f^2Q^{\dagger}H^{\dagger} u},$&
			$\color{blue}{\overline{L}^q_l  H e^s_r},\hspace{0.2cm}  \color{blue}{\overline{Q}^q_l  H d^s_r},\hspace{0.2cm} \color{blue}{\overline{Q}^q_l \tilde{ H} u^s_r}$&
			\\
			
			\hline
	\end{tabular}}
	\caption{\small SM: Renormalizable operators as Hilbert Series output and their covariant form. Coefficients of each operator (which appear as functions of $N_f$) tell us the number of all possible operators with the same structure. The operators in blue have distinct hermitian conjugates which we have not written explicitly. Here, $I$ = 1,2,3 are $SU(2)$ indices; $a$ = 1,...,8 are $SU(3)$ indices and $q,s$ = $1,2,...,N_f$ are flavour indices which are summed over with the suitable coupling constants. ${\textcolor{black}{\clubsuit}}$ - In the Hilbert Series the fermion kinetic terms appear with a factor of $N_f^2$ but in the physical Lagrangian there is a flavour symmetry which forces the kinetic terms to be diagonal and the factor of $N_f^2$ is reduced to $N_f$.}
	\label{table:sm-renorm-output}
\end{table}

\subsubsection*{\underline{Higher Dimension Effective Operators of the SM}}  
We have enlisted the higher dimensional effective operators upto mass dimension 6 in Table~\ref{table:sm-dim-5-6-op}. These operators and their implications in the context of SMEFT have been heavily studied \cite{Jenkins:2013zja,Jenkins:2013wua,Alonso:2013hga,Brivio:2017vri,Berthier:2015oma,Berthier:2015gja,Bjorn:2016zlr,Falkowski:2017pss,Gomez-Ambrosio:2018pnl,Vryonidou:2018eyv,Dedes:2018seb,Dawson:2018jlg,Dawson:2018liq,Barzinji:2018xvu,deVries:2017ncy}.
\vspace{0.5cm}

\begin{table}[t]
	\centering
	\renewcommand{\arraystretch}{1.8}
	{\scriptsize\begin{tabular}{|c|l|}
			\hline
			\multicolumn{2}{|c|}{\textbf{Mass Dimension-5}}\\
			\hline
			\textbf{Operator Class}&
			\hspace{4cm}\textbf{Operators (in non-covariant form)}\\
			\hline
			$\Psi^2\Phi^2$&
			$\color{blue}{\frac{1}{2}(N_f^2 + N_f)H^2L^2}$\\
			\hline
			\hline
			\multicolumn{2}{|c|}{\textbf{Mass Dimension-6}}\\
			\hline
			\textbf{Operator Class}&
			\hspace{4cm}\textbf{Operators  (in non-covariant form)}\\
			\hline
			$X^3$&
			$ Wl^3,\hspace{0.2cm} Wr^3,\hspace{0.2cm} Gl^3,\hspace{0.2cm} Gr^3 \hspace{0.2cm}$\\
			\hline
			
			$\Phi^6$&
			$ (H^\dagger H)^3\hspace{0.2cm}$\\
			\hline
			
			$\Phi^2X^2$&
			$\color{blue}{Gl^2H^{\dagger}H,\hspace{0.2cm}  Wl^2H^{\dagger}H,\hspace{0.2cm} Bl^2H^{\dagger}H,\hspace{0.2cm}  BlWlH^{\dagger}H\hspace{0.2cm}}$\\
			\hline
			
			\multirow{2}{*}{$\Psi^2\Phi X$}&
			$\color{blue}{   (N_f^2)GlH^{\dagger}d^{\dagger}Q,\hspace{0.2cm} 
			(N_f^2)GlH u^{\dagger}Q,\hspace{0.2cm}  (N_f^2)WlH^{\dagger}d^{\dagger}Q,\hspace{0.2cm}
			(N_f^2)WlH u^{\dagger}Q,\hspace{0.2cm}
			(N_f^2)Wl H^{\dagger}el^{\dagger}L,\hspace{0.2cm}}$\\
			
			&
			$\color{blue}{(N_f^2)Bl H^{\dagger}d^{\dagger}Q,\hspace{0.2cm}	
			(N_f^2)Bl H u^{\dagger}Q,\hspace{0.2cm}  (N_f^2)Bl H^{\dagger}el^{\dagger}L\hspace{0.2cm}}$\\
			\hline
			
			\multirow{2}{*}{$\Psi^2\Phi^2 \mathcal{D}$}&
			$(N_f^2)dd^{\dagger} H H^{\dagger}\mathcal{D},\hspace{0.2cm} 
			(N_f^2)elel^{\dagger} H H^{\dagger}\mathcal{D},\hspace{0.2cm}  (2N_f^2)LL^{\dagger} H H^{\dagger}\mathcal{D},\hspace{0.2cm} 
			(2N_f^2)QQ^{\dagger} H H^{\dagger}\mathcal{D},\hspace{0.2cm}
			(N_f^2)uu^{\dagger} H H^{\dagger}\mathcal{D},\hspace{0.2cm}$\\
			
			&
			$\color{blue}{(N_f^2)ud^{\dagger}( H^{\dagger})^2\mathcal{D}}$\\
			\hline
			
			$\Psi^2\Phi^3$&
			$\color{blue}{ (N_f^2)H(H^{\dagger})^2el^{\dagger}L,\hspace{0.2cm}
				(N_f^2)H (H^{\dagger})^2d^{\dagger}Q,\hspace{0.2cm} (N_f^2)\phi^2H^{\dagger}u^{\dagger}Q\hspace{0.2cm} }$\\
			\hline
			
			$\Phi^4\mathcal{D}^2$&
			$2H^2(H^{\dagger})^2\mathcal{D}^2\hspace{0.2cm}$\\
			\hline
			
			\multirow{5}{*}{$\Psi^4$}&
			$(N_f^4)elel^{\dagger}uu^{\dagger},\hspace{0.2cm}
			\frac{1}{2}(N_f^2+N_f^4)L^2(L^{\dagger})^2, \hspace{0.2cm} 
			(N_f^2+N_f^4)Q^2(Q^{\dagger})^2,\hspace{0.2cm} (2N_f^4)LL^{\dagger}QQ^{\dagger},\hspace{0.2cm}$\\
			
			&
			$(N_f^4)elel^{\dagger}dd^{\dagger},\hspace{0.2cm}
			\frac{1}{4}(N_f^2+2N_f^3+N_f^4)el^2(el^{\dagger})^2,\hspace{0.2cm}
			\frac{1}{2}(N_f^2+N_f^4)d^2(d^{\dagger})^2,\hspace{0.2cm} \frac{1}{2}(N_f^2+N_f^4)u^2(u^{\dagger})^2,\hspace{0.2cm}$\\
			
			&
			$(2N_f^4)dd^{\dagger}uu^{\dagger},\hspace{0.2cm}
			(N_f^4)elel^{\dagger}LL^{\dagger},\hspace{0.2cm} (N_f^4)uu^{\dagger}LL^{\dagger},\hspace{0.2cm} (N_f^4)dd^{\dagger}LL^{\dagger},\hspace{0.2cm}
			(N_f^4)elel^{\dagger}QQ^{\dagger},\hspace{0.2cm} (2N_f^4)uu^{\dagger}QQ^{\dagger},\hspace{0.2cm}$\\
			
			&
			$(2N_f^4)dd^{\dagger}QQ^{\dagger},\hspace{0.2cm}
			\color{blue}{  (N_f^4)elL^{\dagger}d^{\dagger}Q,\hspace{0.2cm} (2N_f^4)ud(Q^{\dagger})^2,\hspace{0.2cm} (2N_f^4)elL^{\dagger}uQ^{\dagger},\hspace{0.2cm}
			\frac{1}{3}(N_f^2+2N_f^4)LQ^3,\hspace{0.2cm} 
			}$\\
			
			&
			$\color{blue}{\frac{1}{2}(N_f^3+N_f^4)eluQ^2,\hspace{0.2cm}	(N_f^4)elu^2d,\hspace{0.2cm} (N_f^4)LudQ} $\\
			\hline		    
	\end{tabular}}
	\caption{\small SM: Operators of mass dimensions-5 and -6.}
	\label{table:sm-dim-5-6-op}
\end{table}

\subsubsection*{\underline{Categorizing the operators of the MSSM based on canonical dimension}}

For supersymmetric scenarios, the operators are constructed out of superfields. Unlike the usual quantum fields, e.g. non-supersymmetric models, mass dimension is not a suitable index for superfields and it cannot be used as the order parameter of the polynomial. Instead, the canonical dimension does the required job\footnote{For chiral superfields the canonical dimension is 1, while for vector superfields it is 0. The covariant derivatives have canonical dimension 1/2, hence the field strength tensors which are composed of a vector superfield and 3 covariant derivatives have canonical dimension 3/2.}. 
At the end of the previous subsection, we showed a representative MSSM Lagrangian comprised of operators having canonical dimension 4. We can similarly obtain operators having different canonical dimensions. We have tabulated those operators in Tables\,\ref{table:mssm-chiral-only-output} and \ref{table:mssm-chiral-vector-output} and also classified them based on whether they only contain chiral superfields or a mixture of chiral as well as vector superfields.
\newpage
\noindent While comparing our results with those of \cite{Piriz:1997id}, one must keep in mind that while we have categorized the operators based on their canonical dimension, in \cite{Piriz:1997id} the categorization is based on the mass dimension of the F-term (for operators composed solely of chiral superfields) or the D-term (for operators composed of chiral as well as vector superfields) obtained after expanding the superfields in terms of the quantum fields. Also, the mass dimension of the F-term is one more than the canonical dimension of the original operator while that of the D-term is two more.
\begin{table}[h]
	\centering
	\renewcommand{\arraystretch}{2.3}
	{\tiny\begin{tabular}{|c|c|c|}
			\hline
			
			\textbf{Canonical Dim.}&
			\textbf{Operators}&
			\textbf{No. of Operators}\\
			\hline
			2&
			$H_d H_u,\hspace{0.1cm} N_fH_uL$&
			$N_f+1$\\
			\hline
			
			\multirow{2}{*}{3}&
			$N_f^2LEH_d,\hspace{0.1cm}N_f^2QDH_d,\hspace{0.1cm}N_f^2QUH_u,$&
			\multirow{2}{*}{$2N_f^3+2N_f^2$}\\
			
			&
			$N_f^3LQD,\hspace{0.1cm} \boxed{\frac{1}{2}\left(N_f^3-N_f^2\right)L^2E},\hspace{0.1cm} \boxed{\frac{1}{2}\left(N_f^3-N_f^2\right)D^2U}$&
			\\
			
			\hline
			
			\multirow{2}{*}{4}&
			$(H_dH_u)^2,\hspace{0.1cm} N_fH_dH_u^2L,\hspace{0.1cm} \frac{1}{2}\left(N_f^2+N_f\right)(H_uL)^2,\hspace{0.1cm}
			N_f^4Q^2DU,\hspace{0.1cm}
			N_f^4QULE,$&
			\multirow{2}{*}{$\frac{17}{6}N_f^4 + \frac{5}{6}N_f^3 + \frac{1}{6}N_f^2$}
			\\
			
			&
			${N_f}^3QUEH_d,\hspace{0.1cm} \boxed{\frac{1}{2}\left(N_f^4-N_f^3\right)U^2DE},\hspace{0.1cm} \boxed{\frac{1}{3}\left(N_f^3-N_f\right)H_dQ^3},\hspace{0.1cm} \boxed{\frac{1}{3}\left(N_f^4-N_f^2\right)LQ^3}$&
			$ + \frac{7}{6}N_f +1$\\
			\hline

			\multirow{6}{*}{5}&
			$N_f^2H_uH_d^2LE,\hspace{0.1cm} N_f^3H_uH_dL^2E,\hspace{0.1cm} \boxed{\frac{1}{3}\left(N_f^4-N_f^2\right)H_uL^3E},\hspace{0.1cm} 2N_f^3H_uH_dLDQ,$&
			\multirow{6}{*}{$\frac{11}{12}N_f^5+\frac{3}{2}N_f^4+\frac{37}{12}N_f^3$}\\
			
			&
			$N_f^4H_uL^2DQ,\hspace{0.1cm} N_f^2H_uH_d^2DQ,\hspace{0.1cm} \boxed{\frac{1}{2}\left(N_f^4-N_f^3\right)H_uLD^2U},\hspace{0.1cm} \boxed{\frac{1}{2}\left(N_f^3-N_f^2\right)H_uH_dD^2U},$&
			\\
			
			&
			$N_f^3H_u^2LQU,\hspace{0.1cm} N_f^2H_u^2H_dQU,\hspace{0.1cm} \boxed{\frac{1}{2}\left(N_f^5-N_f^3\right)Q^2U^2E},\hspace{0.1cm} \boxed{\frac{1}{4}\left(N_f^5-N_f^3\right)Q^4U},$&
			\\
			
			&
			$\boxed{\frac{1}{6}\left(N_f^4-3N_f^3+2N_f^2\right)H_dLD^3},\hspace{0.1cm} \boxed{\frac{1}{12}\left(N_f^5-4N_f^4+5N_f^3-2N_f^2\right)L^2D^3},\hspace{0.1cm}$&
			$+\frac{5}{2}N_f^2+4N_f+4 $\\
			
			&
			$\boxed{\frac{1}{12}\left(N_f^5-2N_f^4-N_f^3+2N_f^2\right)E^2U^3},\hspace{0.1cm} X_B^2H_uH_d,\hspace{0.1cm} N_fX_B^2H_uL,\hspace{0.1cm}
			X_G^2H_uH_d,$&\\
			
			&
			$N_fX_G^2H_uL,\hspace{0.1cm}
			X_W^2H_uH_d,\hspace{0.1cm} N_fX_W^2H_uL,\hspace{0.1cm}
			X_BX_WH_uH_d,\hspace{0.1cm} N_fX_BX_WH_uL$&
			\\
			\hline
	\end{tabular}}
	\caption{\small MSSM: Operators composed solely of chiral superfields. $X_B, X_W, X_G$ are field strength tensors corresponding to the vector superfields $B,W $ and $G$. Boxed operators vanish for $N_f=1$.}
	\label{table:mssm-chiral-only-output}
\end{table}
\\
\noindent
Here, we have elaborated on the detailed structures of the operators and their flavour dependence:
\begin{enumerate}
	\item We have noted that the operators $\mathcal{O}_D^{(8)}\equiv E^{\dagger}e^{V_E}H_d^2$,\hspace{0.3cm} $\mathcal{O}_D^{(53)}\equiv U^{\dagger}e^{V_U}DH_u^2$,\hspace{0.3cm} $\mathcal{O}_D^{(54)}\equiv D^{\dagger}e^{V_D}UH_d^2$,\hspace{0.3cm} $\mathcal{O}_F^{(19)}\equiv U^2DH_d^2$,\hspace{0.3cm}  $\mathcal{O}_F^{(20)}\equiv D^3H_u^2$, \hspace{0.3cm}  $\mathcal{O}_F^{(27)}\equiv EH_u^3H_d$ \hspace{0.2cm} which appear in \cite{Piriz:1997id}, are absent from our set. We can justify this as follows. Here, we have $N_f$ fermion-like superfields while only one flavour of the Higgs-like superfields $H_u$ and $H_d$. So, each of the structures $\mathcal{O}_D^{(8)},\hspace{0.2cm} \mathcal{O}_D^{(53)},\hspace{0.2cm} \mathcal{O}_D^{(54)},\hspace{0.2cm} \mathcal{O}_F^{(19)},\hspace{0.2cm} \mathcal{O}_F^{(20)},\hspace{0.2cm} \mathcal{O}_F^{(27)}$ would appear if $H_u, H_d$ were to have more than one flavour.
	
\begin{table}[h]
	\centering
	\renewcommand{\arraystretch}{2.0}
	{\tiny\begin{tabular}{|c|c|}
			\hline
			\multicolumn{2}{|c|}{\textbf{Operators of Canonical Dimension-2}}\\
			\hline
			
			\multirow{2}{*}{\textbf{Operators}}&
			\multirow{2}{*}{\textbf{\textbf{No. of Operators}}}\\
			
			&
			\textbf{(including h.c.)}\\
			\hline
			
			$H_d^{\dagger}e^{V_{H_d}}H_d,\hspace{0.1cm} H_u^{\dagger}e^{V_{H_u}}H_u,\hspace{0.1cm} 
			N_f^2Q^{\dagger}e^{V_{Q}}Q,\hspace{0.1cm} \color{blue}{N_fH_d^{\dagger}e^{V_{H_d}}L}$&
			\multirow{2}{*}{$5N_f^2+2N_f+2$}\\
			
			$N_f^2L^{\dagger}e^{V_{L}}L,\hspace{0.1cm} N_f^2E^{\dagger}e^{V_{E}}E,\hspace{0.1cm} N_f^2D^{\dagger}e^{V_{D}}D,\hspace{0.1cm} N_f^2U^{\dagger}e^{V_{U}}U$&
			\\
			\hline
			\hline
			\multicolumn{2}{|c|}{\textbf{Operators of Canonical Dimension-3}}\\
			\hline
			
			\multirow{2}{*}{\textbf{Operators}}&
			\multirow{2}{*}{\textbf{\textbf{No. of Operators}}}\\
			
			&
			\textbf{(including h.c.)}\\
			\hline
			$\color{blue}{\frac{1}{2}X_B^2,\hspace{0.1cm} \frac{1}{2}X_W^2,\hspace{0.1cm} \frac{1}{2}X_G^2,\hspace{0.1cm} N_fH_u^{\dagger}e^{V_{H_u}}H_dE,\hspace{0.1cm} N_f^2H_u^{\dagger}e^{V_{H_u}}LE,\hspace{0.1cm} N_f^2H_u^{\dagger}e^{V_{H_u}}QD} $&
			\multirow{2}{*}{$5N_f^3+7N_f^2+2N_f+3$}\\
			
			$\color{blue}{N_f^2H_d^{\dagger}e^{V_{H_d}}QU,\hspace{0.1cm} N_f^3L^{\dagger}e^{V_{L}}QU,\hspace{0.1cm} N_f^3D^{\dagger}e^{V_{D}}UE,\hspace{0.1cm} 
			\frac{1}{2}\left(N_f^3+N_f^2\right)D^{\dagger}e^{V_{D}}Q^2}$&
			\\
			\hline
			\hline
			\multicolumn{2}{|c|}{\textbf{Operators of Canonical Dimension-4}}\\
			\hline
			\multirow{2}{*}{\textbf{Operators}}&
			\multirow{2}{*}{\textbf{\textbf{No. of Operators}}}\\
			
			&
			\textbf{(including h.c.)}\\
			\hline
			
			$(H_d^{\dagger}e^{V_{H_d}}H_d)^2,\hspace{0.1cm} (H_u^{\dagger}e^{V_{H_u}}H_u)^2,\hspace{0.1cm} 2(H_d^{\dagger}e^{V_{H_d}}H_d)(H_u^{\dagger}e^{V_{H_u}}H_u),\hspace{0.1cm} \frac{1}{2}\left(N_f^4+N_f^2\right)(U^{\dagger}e^{V_{U}}U)^2,$&
			\multirow{21}{*}{$\frac{289}{12}N_f^4+\frac{235}{6}N_f^3+\frac{413}{12}N_f^2+\frac{43}{3}N_f+8$}\\
			
			$\frac{1}{2}\left(N_f^4+N_f^2\right)(D^{\dagger}e^{V_{D}}D)^2,\hspace{0.1cm} \frac{1}{4}\left(N_f^4+2N_f^3+N_f^2\right)(E^{\dagger}e^{V_{E}}E)^2,\hspace{0.1cm} \left(N_f^4+N_f^2\right)(Q^{\dagger}e^{V_{Q}}Q)^2,$&
			\\
			
			$\frac{1}{2}\left(N_f^4+N_f^2\right)(L^{\dagger}e^{V_{L}}L)^2,\hspace{0.1cm} N_f^2(D^{\dagger}e^{V_{D}}DH_d^{\dagger}e^{V_{H_d}}H_d),\hspace{0.1cm}
			N_f^2(D^{\dagger}e^{V_{D}}DH_u^{\dagger}e^{V_{H_u}}H_u),$&
			\\
			
			$N_f^2(E^{\dagger}e^{V_{E}}EH_d^{\dagger}e^{V_{H_d}}H_d),\hspace{0.1cm} N_f^2(U^{\dagger}e^{V_{U}}UH_d^{\dagger}e^{V_{H_d}}H_d),\hspace{0.1cm} N_f^2(U^{\dagger}e^{V_{U}}UH_u^{\dagger}e^{V_{H_u}}H_u),$&
			\\
			
			$N_f^2(E^{\dagger}e^{V_{E}}EH_u^{\dagger}e^{V_{H_u}}H_u),\hspace{0.1cm} 2N_f^2(L^{\dagger}e^{V_{L}}LH_d^{\dagger}e^{V_{H_d}}H_d),\hspace{0.1cm} 2N_f^2(L^{\dagger}e^{V_{L}}LH_u^{\dagger}e^{V_{H_u}}H_u),$&
			\\
			
			$2N_f^2(Q^{\dagger}e^{V_{Q}}QH_d^{\dagger}e^{V_{H_d}}H_d),\hspace{0.1cm} 2N_f^2(Q^{\dagger}e^{V_{Q}}QH_u^{\dagger}e^{V_{H_u}}H_u),\hspace{0.1cm} 2N_f^4(U^{\dagger}e^{V_{U}}UD^{\dagger}e^{V_{D}}D),$&
			\\
			
			$2N_f^4(Q^{\dagger}e^{V_{Q}}QL^{\dagger}e^{V_{L}}L),\hspace{0.1cm} 2N_f^4(Q^{\dagger}e^{V_{Q}}QD^{\dagger}e^{V_{D}}D),\hspace{0.1cm} 2N_f^4(Q^{\dagger}e^{V_{Q}}QU^{\dagger}e^{V_{U}}U),$&
			\\
			
			$N_f^4(Q^{\dagger}e^{V_{Q}}QE^{\dagger}e^{V_{E}}E),\hspace{0.1cm} N_f^4(L^{\dagger}e^{V_{L}}LE^{\dagger}e^{V_{E}}E),\hspace{0.1cm} N_f^4(L^{\dagger}e^{V_{L}}LU^{\dagger}e^{V_{U}}U),$&
			\\
			
			$N_f^4(L^{\dagger}e^{V_{L}}LD^{\dagger}e^{V_{D}}D),\hspace{0.1cm} N_f^4(E^{\dagger}e^{V_{E}}ED^{\dagger}e^{V_{D}}D),\hspace{0.1cm} 
			N_f^4(E^{\dagger}e^{V_{E}}EU^{\dagger}e^{V_{U}}U),$&
			\\
			
			$\color{blue}{N_f(H_d^{\dagger}e^{V_{H_d}}H_dH_d^{\dagger}e^{V_{H_d}}L),\hspace{0.1cm} 2N_f(H_u^{\dagger}e^{V_{H_u}}H_uH_d^{\dagger}e^{V_{H_d}}L),\hspace{0.1cm} \frac{1}{2}(N_f^2+N_f)(H_d^{\dagger}e^{V_{H_d}}L)^2,}$&
			\\
			
			$\color{blue}{2N_f^3(Q^{\dagger}e^{V_{Q}}QH_d^{\dagger}e^{V_{H_d}}L),\hspace{0.1cm}
				N_f^3(L^{\dagger}e^{V_{L}}LH_d^{\dagger}e^{V_{H_d}}L),\hspace{0.1cm} N_f^3(E^{\dagger}e^{V_{E}}EH_d^{\dagger}e^{V_{H_d}}L),}$&
			\\
			
			$\color{blue}{ N_f^3(D^{\dagger}e^{V_{D}}DH_d^{\dagger}e^{V_{H_d}}L),\hspace{0.1cm} N_f^3(U^{\dagger}e^{V_{U}}UH_d^{\dagger}e^{V_{H_d}}L),\hspace{0.1cm} 
				\boxed{\frac{1}{2}\left(N_f^3-N_f^2\right)Q^{\dagger}e^{V_{Q}}H_u^{\dagger}e^{V_{H_u}}D^2,}}$&
			\\
			
			$\color{blue}{\frac{1}{2}\left(N_f^4+N_f^3\right)U^{\dagger}e^{V_{U}}E^{\dagger}e^{V_{E}}Q^2,\hspace{0.1cm} 
				N_f^3(E^{\dagger}e^{V_{E}}H_d^{\dagger}e^{V_{H_d}}DQ),\hspace{0.1cm} 
				N_f^3(Q^{\dagger}e^{V_{Q}}H_d^{\dagger}e^{V_{H_d}}DU),}$&
			\\
			
			$\color{blue}{N_f^2(D^{\dagger}e^{V_{D}}H_d^{\dagger}e^{V_{H_d}}H_uU),\hspace{0.1cm} 
				N_f^3(D^{\dagger}e^{V_{D}}L^{\dagger}e^{V_{L}}H_uU),\hspace{0.1cm} 
				N_f^4(Q^{\dagger}e^{V_{Q}}L^{\dagger}e^{V_{L}}UD),}$&
			\\
			
			$\color{blue}{N_f^4(E^{\dagger}e^{V_{E}}L^{\dagger}e^{V_{L}}DQ),\hspace{0.1cm} \boxed{\frac{1}{2}\left(N_f^4-N_f^3\right)U^{\dagger}e^{V_{U}}DL^2},\hspace{0.1cm}
				\boxed{\frac{1}{2}\left(N_f^4-N_f^3\right)Q^{\dagger}e^{V_{Q}}D^2L},}$&
			\\
			
			$\color{blue}{N_f^3U^{\dagger}e^{V_{U}}DLH_d,\hspace{0.1cm} N_f^2U^{\dagger}e^{V_{U}}UH_uH_d,\hspace{0.1cm} N_f^3U^{\dagger}e^{V_{U}}UH_uL,\hspace{0.1cm} N_f^3H_u^{\dagger}e^{V_{H_u}}QUE,}$&
			\\
			
			$\color{blue}{N_f^3Q^{\dagger}e^{V_{Q}}DUH_u,\hspace{0.1cm} 2N_f^2Q^{\dagger}e^{V_{Q}}QH_uH_d,\hspace{0.1cm} 2N_f^3Q^{\dagger}e^{V_{Q}}QH_uL,\hspace{0.1cm} N_f^2H_d^{\dagger}e^{V_{H_d}}H_uL^2}$&
			\\
			
			$\color{blue}{
				\boxed{\frac{1}{3}\left(N_f^3-N_f\right)H_u^{\dagger}e^{V_{H_u}}Q^3},\hspace{0.1cm} \boxed{\frac{1}{2}\left(N_f^3-N_f^2\right)Q^{\dagger}e^{V_{Q}}D^2H_d},\hspace{0.1cm}
				N_fL^{\dagger}e^{V_{L}}H_d^2H_u, }$&
			\\
			
			$\color{blue}{2N_f^2L^{\dagger}e^{V_{L}}LH_uH_d,\hspace{0.1cm} N_f^3L^{\dagger}e^{V_{L}}H_uL^2,\hspace{0.1cm}
				\boxed{\frac{1}{6}\left(N_f^4-3N_f^3+2N_f^2\right)E^{\dagger}e^{V_{E}}D^3} }$&
			\\
			
			$\color{blue}{N_f^2E^{\dagger}e^{V_{E}}EH_uH_d,\hspace{0.1cm} N_f^3E^{\dagger}e^{V_{E}}EH_uL,\hspace{0.1cm}
				N_f^2D^{\dagger}e^{V_{D}}DH_uH_d,\hspace{0.1cm} N_f^3D^{\dagger}e^{V_{D}}DH_uL,}$&
			\\
			
			$\color{blue}{N_f^3E^{\dagger}e^{V_{E}}DQH_u,\hspace{0.1cm} 2N_fH_d^{\dagger}e^{V_{H_d}}H_dH_uL,\hspace{0.1cm}
				N_fH_u^{\dagger}e^{V_{H_u}}H_u^2L,\hspace{0.1cm}
				H_d^{\dagger}e^{V_{H_d}}H_d^2H_u,\hspace{0.1cm}
				H_u^{\dagger}e^{V_{H_u}}H_u^2H_d}$&
			\\
			\hline
	\end{tabular}}
	\caption{\small MSSM: Operators composed of chiral as well as vector superfields. The operators in blue have distinct hermitian conjugates which we have not written explicitly. Boxed operators vanish for $N_f=1$.}
	\label{table:mssm-chiral-vector-output}
\end{table}
	
	\clearpage
	
	\item Operators enclosed in boxes vanish for certain values of $N_f$. The following operator structures vanish for $N_f = 1$:
	
	\vspace{-0.5cm}
	{\scriptsize\begin{eqnarray}
		& &\frac{1}{2}\left(N_f^3-N_f^2\right)L^2E, \hspace{0.1cm} \frac{1}{2}\left(N_f^3-N_f^2\right)D^2U, \hspace{0.1cm}
		\frac{1}{2}\left(N_f^4-N_f^3\right)U^2DE, \hspace{0.1cm}
		\frac{1}{3}\left(N_f^3-N_f\right)H_dQ^3, \hspace{0.1cm}
		\frac{1}{3}\left(N_f^4-N_f^2\right)LQ^3, \nonumber\\
		& &\frac{1}{3}\left(N_f^4-N_f^2\right)H_uL^3E, \hspace{0.1cm}
		\frac{1}{2}\left(N_f^4-N_f^3\right)H_uLD^2U, \hspace{0.1cm}
		\frac{1}{2}\left(N_f^3-N_f^2\right)H_uH_dD^2U, \hspace{0.1cm}
		\frac{1}{2}\left(N_f^5-N_f^3\right)Q^2U^2E, \nonumber\\
		& & \frac{1}{4}\left(N_f^5-N_f^3\right)Q^4U, \hspace{0.1cm}
		\color{blue}{\frac{1}{2}\left(N_f^3-N_f^2\right)Q^{\dagger}e^{V_{Q}}H_u^{\dagger}e^{V_{H_u}}D^2, \hspace{0.1cm} \frac{1}{2}\left(N_f^4-N_f^3\right)U^{\dagger}e^{V_{U}}DL^2, \hspace{0.1cm} \frac{1}{2}\left(N_f^4-N_f^3\right)Q^{\dagger}e^{V_{Q}}D^2L,}\nonumber \\
		& & \color{blue}{ \frac{1}{3}\left(N_f^3-N_f\right)H_u^{\dagger}e^{V_{H_u}}Q^3, \hspace{0.1cm} \frac{1}{2}\left(N_f^3-N_f^2\right)Q^{\dagger}e^{V_{Q}}D^2H_d}. \nonumber
		\end{eqnarray}}
	Another set of operators vanish for $N_f = 1,2$
	{\scriptsize\begin{eqnarray}
		& &\frac{1}{6}\left(N_f^4-3N_f^3+2N_f^2\right)H_dLD^3, \hspace{0.1cm} \frac{1}{12}\left(N_f^5-4N_f^4+5N_f^3-2N_f^2\right)L^2D^3,  \nonumber\\
		& &
		\frac{1}{12}\left(N_f^5-2N_f^4-N_f^3+2N_f^2\right)E^2U^3, \hspace{0.1cm} \color{blue}{\frac{1}{6}\left(N_f^4-3N_f^3+2N_f^2\right)E^{\dagger}e^{V_{E}}D^3}. \nonumber 
		\end{eqnarray}} 
	It is easy to understand this vanishing behaviour. Each of these operators contains 3 similar superfields which transform under the fundamental representation of $SU(3)$ ($D, U$). The explicit index contraction can be written as, for example, for $H_{d}LD^3$:
	
	\vspace{-0.5cm}
	{\scriptsize\begin{eqnarray}
		H_{d}LD^3 \equiv \epsilon_{ab}\epsilon_{\alpha\,\beta\,\gamma} H^a_{d}L^b D^{\alpha} D^{\beta} D^{\gamma} \nonumber
		\end{eqnarray}}
	where $a,b$ are $SU(2)$ indices and $\alpha,\beta,\gamma$ are $SU(3)$ indices. Here, $\epsilon_{\alpha\,\beta\,\gamma}$ is fully anti-symmetric whereas $D^{\alpha} D^{\beta} D^{\gamma}$ is fully symmetric. Now, if we have 2 flavors of $D$, then the expression is still symmetric in 2 of the indices. Therefore, this operator and other similar operators are non-vanishing only if $N_f>2$.	 
\end{enumerate}

%% file: GrIPpheno.tex
\subsection{Extracting phenomenologically relevant operators using \textbf{GrIP}}

\subsubsection*{\underline{Baryon and Lepton Number violating rare processes in the SM}}\label{subsec:bviol-lviol-sm}

The renormalizable SM Lagrangian has accidental symmetries in the form of baryon and lepton number conservation. But these do get violated at higher mass dimensions \cite{Pontecorvo:1967fh,Fritzsch:1974nn,Weinberg:1979sa,Chikashige:1980ui,Schechter:1981cv}. Operators displaying such violations have attracted a lot of attention due to a variety of reasons. The most popular being contributions to neutrino mass \cite{Weinberg:1979sa,Hambye:2013jsa,Anamiati:2018cuq,Mohapatra:1979ia,Mohapatra:1980yp} as well as predictions of exotic processes, for instance, neutrinoless double beta decay \cite{KlapdorKleingrothaus:2001ke,Doi:1985dx,Dev:2013vxa,Fonseca:2016jbm}, and the decay of proton and other nucleons into leptons \cite{Hambye:2017qix,Fonseca:2018ehk}. As we look for higher and higher mass dimensional operators the number of possible combinations of the fields increases. Thus it becomes difficult to filter out operators that violate baryon and lepton numbers by some specific amount. Keeping this in mind, we have defined a function \texttt{"\textcolor{mmaUndefined}{DisplayBLviolatingOperators}"} within \textbf{GrIP}, which enables one to obtain the lowest dimensional operators for a particular combination of $\Delta B$ and $\Delta L$. We have collected the results in Table~\ref{table:sm-bno-lno-viol-operator} and also described the action of this particular function.

\begin{table}[h]
	\centering
	\renewcommand{\arraystretch}{1.8}
	{\tiny\begin{tabular}{|c|c|c|c|c|c|}
			\hline
			\multirow{2}{*}{$\Delta$\textbf{B}}&
			\multirow{2}{*}{$\Delta$\textbf{L}}&
			\textbf{Lowest}&
			\multirow{2}{*}{\textbf{Operators}}&
			\multirow{2}{*}{\textbf{Remarks}}\\
						
			&
			&
			\textbf{Dimension}&
			&
			\\
			\hline 
			
			\multirow{2}{*}{$0$}&
			\multirow{2}{*}{$-2$}&
			\multirow{2}{*}{$5$}&
			\multirow{2}{*}{$H^2L^2$}&
			Contributes to Neutrino Mass \\
						
			&
			&
			&
			&
			\cite{Weinberg:1979sa,Anamiati:2018cuq,Hambye:2013jsa}\\
			\hline
			
			\multirow{2}{*}{$+1$}&
			\multirow{2}{*}{$-1$}&
			\multirow{2}{*}{$6$}&
			\multirow{2}{*}{$LQ^3,\hspace{0.1cm} LQdu,\hspace{0.1cm} eluQ^2,\hspace{0.1cm} elu^2d$}&
			\multirow{2}{*}{These contribute to proton decay.}\\
			
			&
			&
			&
			&
			\\
			\hline
			
			\multirow{2}{*}{$1$}&
			\multirow{2}{*}{$1$}&
			\multirow{2}{*}{$7$}&
			\multirow{2}{*}{$el^{\dagger}d^3\mathcal{D},\hspace{0.1cm} L^{\dagger}Qd^2\mathcal{D},\hspace{0.1cm} H^{\dagger}L^{\dagger}dQ^2,\hspace{0.1cm} H^{\dagger}L^{\dagger}ud^2$}&
			These violate Baryon and Lepton numbers \\
			
			&
			&
			&
			&
			by equal amounts thus preserving $B-L$ \\
			\hline
			
			\multirow{2}{*}{$+2$}&
			\multirow{2}{*}{$0$}&
			\multirow{2}{*}{$9$}&
			\multirow{2}{*}{$d^2Q^4,\hspace{0.1cm} d^3uQ^2,\hspace{0.1cm} 2u^2d^4$}&
			\multirow{2}{*}{Leads to nucleon-nucleon scattering}\\
			
			&
			&
			&
			&
			\\
			\hline

			\multirow{2}{*}{$+1$}&
			\multirow{2}{*}{$-3$}&
			\multirow{2}{*}{$9$}&
			\multirow{2}{*}{$\color{purple}{L^3Qu^2,\hspace{0.1cm} L^2elu^3}$}&
			Leads to decay of nucleons to \\
			
			&
			&
			&
			&
			3 charged leptons \cite{Hambye:2017qix}\\
			\hline
			
			\multirow{2}{*}{$+1$}&
			\multirow{2}{*}{$+3$}&
			\multirow{2}{*}{$10$}&
			\multirow{2}{*}{$\color{purple}{H^{\dagger}(L^{\dagger})^3d^3}$}&
			Leads to decay of nucleons to charged \\
			
			&
			&
			&
			&
			leptons mediated through a scalar \cite{Hambye:2017qix,Fonseca:2018ehk}\\
			\hline

			\multirow{2}{*}{$0$}&
			\multirow{2}{*}{$-4$}&
			\multirow{2}{*}{$10$}&
			\multirow{2}{*}{$\color{purple}{H^4L^4}$}&
			Suggests the possibility of a neutrinoless \\

			&
			&
			&
			&
			quadruple beta decay \cite{Fonseca:2018aav,Heeck:2013rpa}\\
			\hline
\end{tabular}}
\caption{\small Baryon and Lepton number violating SM operators. The operators in red vanish for $N_f=1$. The hermitian conjugates of each of these operators will have $\Delta B\rightarrow -\Delta B$ and $\Delta L\rightarrow -\Delta L$.}
\label{table:sm-bno-lno-viol-operator}
\end{table}

The baryon and lepton number violations are also signatures of various extended SM scenarios. It is important to note that operators with the same amounts of $\Delta B$ and $\Delta L$ appear at different mass dimensions for different models. If in the near future such rare process(es) are discovered one can perform a comparative analysis across a variety of models.

\begin{table}[h]
\hspace*{0.25cm}{\footnotesize\textcolor{mmaLabel}{\fontfamily{phv}\selectfont In[1]:=}}
\textcolor{mmaUndefined}{\texttt{DisplayBLviolatingOperators}}\texttt{[\textcolor{mmaString}{"HighestMassDim"}$\to$10, \textcolor{mmaString}{"$\Delta$B"}$\to$ 0,\\ 
\hspace*{6.5cm} \textcolor{mmaString}{"$\Delta$L"}$\to$ -2, \textcolor{mmaString}{"Flavours"}$\to$ $N_f$]}

\vspace{0.25cm}
{\footnotesize\textcolor{mmaLabel}{\fontfamily{phv}\selectfont Out[1]:=}}

\vspace{-0.5cm}
\hspace*{1.4cm}\texttt{First instance of $\Delta$B=0 and $\Delta$L=-2 occurs at mass dimension 5,}\\ 

\vspace{-0.3cm}
\hspace*{1.4cm}\texttt{Operators:} {\scriptsize$\frac{1}{2} H^2 L^2N_f^2+\frac{1}{2} H^2 L^2 N_f$}\\

\vspace{-0.2cm}
\hspace*{0.25cm}{\footnotesize\textcolor{mmaLabel}{\fontfamily{phv}\selectfont In[2]:=}}
\textcolor{mmaUndefined}{\texttt{DisplayBLviolatingOperators}}\texttt{[\textcolor{mmaString}{"HighestMassDim"}$\to$10, \textcolor{mmaString}{"$\Delta$B"}$\to$ 1,\\ 
\hspace*{6.5cm} \textcolor{mmaString}{"$\Delta$L"}$\to$ -1, \textcolor{mmaString}{"Flavours"}$\to$ $N_f$]}

\vspace{0.25cm}
{\footnotesize\textcolor{mmaLabel}{\fontfamily{phv}\selectfont Out[2]:=}}

\vspace{-0.5cm}
\hspace*{1.4cm}\texttt{First instance of $\Delta$B=1 and $\Delta$L=-1 occurs at mass dimension 6,}\\ 

\vspace{-0.3cm}
\hspace*{1.4cm}\texttt{Operators:} {\scriptsize$ d \;el\; N_f^4\; u^2+d\; L\; N_f^4\; Q\; u+\frac{1}{2}el\; N_f^4\; Q^2\; u+\frac{1}{2}  el\;N_f^3\; Q^2 u+\frac{2}{3}\; L N_f^4\; Q^3+\frac{1}{3} L\; N_f^2\; Q^3$}
\end{table}
\vspace*{-0.5cm}
\subsubsection*{\underline{External $U(1)_{R}$ Global Symmetry and extended MSSM}}

We have kept the provision to assign quantum numbers under some external global symmetry, which is not introduced in the input file, to each particle. Based on this new assignment, one can identify operators that either conserve this symmetry or violate it by a specific amount. This can be achieved through by judiciously using the functions \texttt{"\textcolor{mmaUndefined}{SaveSHSOutput}"}\footnote{For non-supersymmetric cases, one must use \texttt{"\textcolor{mmaUndefined}{SaveHSOutput}"} instead of \texttt{"\textcolor{mmaUndefined}{SaveHSOutput}"}.} and \texttt{"\textcolor{mmaUndefined}{ReOutput}"} in tandem. As an illustration, we have worked with the model given in \cite{Chakraborty:2017dfg}. We have listed the superfield content and their transformation properties under the gauge group $SU(3)_C\otimes SU(2)_L\otimes U(1)_Y$ as well as under the global $U(1)_R$ in Table~\ref{table:susy+globalU1-quantum-no}. One can obtain the superpotential up to canonical dimension 4 with the help of \textbf{GrIP} in the same way in which we obtained the superpotential for MSSM. Then, using the function \texttt{"\textcolor{mmaUndefined}{ReOutput}"}, see Table~\ref{table:function_details}, we can assign the $U(1)_R$ charges and segregate between operators having different values of the overall $R$-charge. In Table~\ref{table:susy+globalU1-renorm-output}, we have collected the segregated superpotential for $N_f = 1$. 
\newpage
\begin{table}[h]
	\centering
	\renewcommand{\arraystretch}{1.8}
	{\tiny\begin{tabular}{|c|c|c|c|c|c|}
			\hline
			\textbf{Superfields}&
			$SU(3)_C$&
			$SU(2)_L$&
			$U(1)_Y$&
			$U(1)_R$\\
			\hline
			
			$H_u$&
			1&
			2&
			1/2&
			0\\
			
			$H_d$&
			1&
			2&
			-1/2&
			0\\
			
			$Q^i$&
			3&
			2&
			1/6&
			1\\
			
			$U^i$&
			$\overline{3}$&
			1&
			-2/3&
			1\\
			
			$D^i$&
			$\overline{3}$&
			1&
			1/3&
			1\\
			
			$L^i$&
			1&
			2&
			-1/2&
			2\\
			
			$E^i$&
			1&
			1&
			1&
			0\\
			
			$R_u$&
			1&
			2&
			1/2&
			2\\
			
			$R_d$&
			1&
			2&
			-1/2&
			2\\
			
			$\mathcal{S}$&
			1&
			1&
			0&
			0\\
			
			$\mathcal{T}$&
			1&
			3&
			0&
			0\\
			
			$\mathcal{O}$&
			8&
			1&
			0&
			0\\
			\hline
	\end{tabular}}
	\caption{\small MSSM + Global $U(1)_R$: Quantum numbers of superfields under the gauge groups. Internal symmetry indices have been suppressed. $i=1,2,..,N_f$ is the flavour index.}
	\label{table:susy+globalU1-quantum-no}
\end{table} 

\hspace*{0.25cm}{\footnotesize\textcolor{mmaLabel}{\fontfamily{phv}\selectfont In[1]:=}}
\textcolor{mmaUndefined}{\texttt{dim1}} = \textcolor{mmaUndefined}{\texttt{SaveSHSOutput}}\texttt{[\textcolor{mmaString}{"CanonicalDim"}$\to$1, \textcolor{mmaString}{"$\Delta$B"}$\to$"NA", \textcolor{mmaString}{"$\Delta$L"}$\to$"NA", \textcolor{mmaString}{"Flavours"}$\to$1];}

\vspace{0.15cm}
\hspace*{0.25cm}{\footnotesize\textcolor{mmaLabel}{\fontfamily{phv}\selectfont In[2]:=}}
\textcolor{mmaUndefined}{\texttt{ReOutput}}\texttt{[\textcolor{mmaString}{"NameOfPoly"}$\to$dim1, \textcolor{mmaString}{"SymmetryName"}$\to$\{"U(1)${}_{R}$"\},\\ 
\hspace*{4cm} \textcolor{mmaString}{"Qno"}$\to$\{Q$\to$1,U$\to$1,D$\to$1,L$\to$2,R${}_{u}$ $\to$2,R${}_{d}$ $\to$2\},
\textcolor{mmaString}{"$\Delta$sym"}$\to$\{0\}]}

\vspace{0.25cm}
{\footnotesize\textcolor{mmaLabel}{\fontfamily{phv}\selectfont Out[2]:=}}
\vspace{-0.5cm}
\texttt{Total No.\,of Operators: 1}\\

\hspace*{1.2cm} \texttt{Operators:} {\scriptsize$\mathcal{S}$}\\

\vspace{-0.3cm}
\hspace*{0.25cm}{\footnotesize\textcolor{mmaLabel}{\fontfamily{phv}\selectfont In[3]:=}}
\textcolor{mmaUndefined}{\texttt{dim2}} = \textcolor{mmaUndefined}{\texttt{SaveSHSOutput}}\texttt{[\textcolor{mmaString}{"CanonicalDim"}$\to$2, \textcolor{mmaString}{"$\Delta$B"}$\to$"NA", \textcolor{mmaString}{"$\Delta$L"}$\to$"NA", \textcolor{mmaString}{"Flavours"}$\to$1]}

\vspace{0.25cm}
{\footnotesize\textcolor{mmaLabel}{\fontfamily{phv}\selectfont Out[3]:=}}
\vspace{-0.75cm}
{\scriptsize$
R_d\;H_u+H_d\;R_u+H_d\; H_u+R_d\; R_u+L\; H_u+L\; R_u+\mathcal{O}^2+\mathcal{S}^2+\mathcal{T}^2$}\\

\vspace{0.5cm}
\hspace*{0.25cm}{\footnotesize\textcolor{mmaLabel}{\fontfamily{phv}\selectfont In[4]:=}}
\textcolor{mmaUndefined}{\texttt{ReOutput}}\texttt{[\textcolor{mmaString}{"NameOfPoly"}$\to$dim2, \textcolor{mmaString}{"SymmetryName"}$\to$\{"U(1)${}_{R}$"\},\\ 
	\hspace*{4cm} \textcolor{mmaString}{"Qno"}$\to$\{Q$\to$1,U$\to$1,D$\to$1,L$\to$2,R${}_{u}$ $\to$2,R${}_{d}$ $\to$2\},
	\textcolor{mmaString}{"$\Delta$sym"}$\to$\{0\}]}

\vspace{0.25cm}
{\footnotesize\textcolor{mmaLabel}{\fontfamily{phv}\selectfont Out[4]:=}}
\vspace{-0.5cm}
\texttt{Total No.\,of Operators: 4}\\

\hspace*{1.2cm} \texttt{Operators:} {\scriptsize$H_d H_u+\mathcal{O}^2+S^2+T^2$}\\

\vspace{-0.3cm}
\hspace*{0.25cm}{\footnotesize\textcolor{mmaLabel}{\fontfamily{phv}\selectfont In[5]:=}}
\textcolor{mmaUndefined}{\texttt{ReOutput}}\texttt{[\textcolor{mmaString}{"NameOfPoly"}$\to$dim2, \textcolor{mmaString}{"SymmetryName"}$\to$\{"U(1)${}_{R}$"\},\\ 
	\hspace*{4cm} \textcolor{mmaString}{"Qno"}$\to$\{Q$\to$1,U$\to$1,D$\to$1,L$\to$2,R${}_{u}$ $\to$2,R${}_{d}$ $\to$2\},
	\textcolor{mmaString}{"$\Delta$sym"}$\to$\{2\}]}

\vspace{0.25cm}
{\footnotesize\textcolor{mmaLabel}{\fontfamily{phv}\selectfont Out[5]:=}}
\vspace{-0.5cm}
\texttt{Total No.\,of Operators: 3}\\

\hspace*{1.2cm} \texttt{Operators:} {\scriptsize$R_d H_u+H_d R_u+L H_u$}\\

\vspace{-0.3cm}
\hspace*{0.25cm}{\footnotesize\textcolor{mmaLabel}{\fontfamily{phv}\selectfont In[6]:=}}
\textcolor{mmaUndefined}{\texttt{ReOutput}}\texttt{[\textcolor{mmaString}{"NameOfPoly"}$\to$dim2, \textcolor{mmaString}{"SymmetryName"}$\to$\{"U(1)${}_{R}$"\},\\ 
	\hspace*{4cm} \textcolor{mmaString}{"Qno"}$\to$\{Q$\to$1,U$\to$1,D$\to$1,L$\to$2,R${}_{u}$ $\to$2,R${}_{d}$ $\to$2\},
	\textcolor{mmaString}{"$\Delta$sym"}$\to$\{4\}]}

\vspace{0.25cm}
{\footnotesize\textcolor{mmaLabel}{\fontfamily{phv}\selectfont Out[6]:=}}
\vspace{-0.5cm}
\texttt{Total No.\,of Operators: 2}\\

\hspace*{1.2cm} \texttt{Operators:} {\scriptsize$R_d R_u+L R_u$}\\

\vspace{-0.3cm}	
\hspace*{0.25cm}{\footnotesize\textcolor{mmaLabel}{\fontfamily{phv}\selectfont In[7]:=}}
\textcolor{mmaUndefined}{\texttt{dim3}} = \textcolor{mmaUndefined}{\texttt{SaveSHSOutput}}\texttt{[\textcolor{mmaString}{"CanonicalDim"}$\to$3, \textcolor{mmaString}{"$\Delta$B"}$\to$"NA", \textcolor{mmaString}{"$\Delta$L"}$\to$"NA", \textcolor{mmaString}{"Flavours"}$\to$1]}

\vspace{0.25cm}
{\footnotesize\textcolor{mmaLabel}{\fontfamily{phv}\selectfont Out[7]:=}}

\vspace{-1.2cm}
{\scriptsize\begin{eqnarray}
& &\hspace{1.65cm} D\; Q\; H_d+D\; Q\; R_d+E\; L\; H_d\;+E\; H_d\; R_d +E\; L\; R_d+\mathcal{S}\; R_d\; H_u+\mathcal{S}\; H_d\; R_u+\mathcal{T}\; R_d H_u+\mathcal{T}\; H_d\; R_u+\mathcal{S}\; H_d\; H_u+ \nonumber\\
& &\hspace{1.65cm} \mathcal{T}\; H_d\; H_u +\mathcal{S}\; R_d\; R_u\;+\mathcal{T}\; R_d\;  {R_u}\; +D\;  L\;  Q +L\; \mathcal{S}\; H_u +L\; \mathcal{T}\; H_u+Q\; U\;  H_u\; +L\; \mathcal{S}\; {R_u} +L\; \mathcal{T}\; {R_u} +\mathcal{O}^3\;+\nonumber\\
& &\hspace{1.65cm} \mathcal{O}^2\; \mathcal{S}\;+Q\; U\; {R_u} +\mathcal{S}^3+\mathcal{S}\; \mathcal{T}^2 \nonumber
\end{eqnarray}}

\hspace*{0.25cm}{\footnotesize\textcolor{mmaLabel}{\fontfamily{phv}\selectfont In[8]:=}}
\textcolor{mmaUndefined}{\texttt{ReOutput}}\texttt{[\textcolor{mmaString}{"NameOfPoly"}$\to$dim3, \textcolor{mmaString}{"SymmetryName"}$\to$\{"U(1)${}_{R}$"\},\\ 
\hspace*{4cm} \textcolor{mmaString}{"Qno"}$\to$\{Q$\to$1,U$\to$1,D$\to$1,L$\to$2,R${}_{u}$ $\to$2,R${}_{d}$ $\to$2\},
\textcolor{mmaString}{"$\Delta$sym"}$\to$\{0\}]}

\vspace{0.25cm}
{\footnotesize\textcolor{mmaLabel}{\fontfamily{phv}\selectfont Out[8]:=}}
\vspace{-0.5cm}
\texttt{Total No.\,of Operators: 6}\\

\hspace*{1.15cm} \texttt{Operators:} {\scriptsize$\mathcal{S}\; H_d\; H_u +\mathcal{T}\; H_d\; H_u +\mathcal{O}^3+\mathcal{O}^2\; \mathcal{S}+\mathcal{S}^3+\mathcal{S}\; \mathcal{T}^2$}\\

\vspace{-0.3cm}
\hspace*{0.25cm}{\footnotesize\textcolor{mmaLabel}{\fontfamily{phv}\selectfont In[9]:=}}
\textcolor{mmaUndefined}{\texttt{ReOutput}}\texttt{[\textcolor{mmaString}{"NameOfPoly"}$\to$dim3, \textcolor{mmaString}{"SymmetryName"}$\to$\{"U(1)${}_{R}$"\},\\ 
\hspace*{4cm} \textcolor{mmaString}{"Qno"}$\to$\{Q$\to$1,U$\to$1,D$\to$1,L$\to$2,R${}_{u}$ $\to$2,R${}_{d}$ $\to$2\},
\textcolor{mmaString}{"$\Delta$sym"}$\to$\{2\}]}

\vspace{0.25cm}
{\footnotesize\textcolor{mmaLabel}{\fontfamily{phv}\selectfont Out[9]:=}}
\vspace{-0.5cm}
\texttt{Total No.\,of Operators: 10}\\

\hspace*{1.15cm} \texttt{Operators:} {\scriptsize$D\; Q\; H_d +E\; L\; H_d\;+E\; H_d\; R_d + \mathcal{S} H_d\; R_u +\mathcal{S}\; R_d\; H_u\;+\mathcal{T}\; H_d\; R_u+\mathcal{T}\; R_d\; H_u +$\\
\hspace*{4.1cm} $L\; \mathcal{S}\; H_u +L\; \mathcal{T}\; H_u +Q\; U\; H_u$}\\

\vspace{-0.3cm}
\hspace*{0.25cm}{\footnotesize\textcolor{mmaLabel}{\fontfamily{phv}\selectfont In[10]:=}}
\textcolor{mmaUndefined}{\texttt{ReOutput}}\texttt{[\textcolor{mmaString}{"NameOfPoly"}$\to$dim3, \textcolor{mmaString}{"SymmetryName"}$\to$\{"U(1)${}_{R}$"\},\\ 
\hspace*{4cm} \textcolor{mmaString}{"Qno"}$\to$\{Q$\to$1,U$\to$1,D$\to$1,L$\to$2,R${}_{u}$ $\to$2,R${}_{d}$ $\to$2\},
\textcolor{mmaString}{"$\Delta$sym"}$\to$\{4\}]}

\vspace{0.25cm}
{\footnotesize\textcolor{mmaLabel}{\fontfamily{phv}\selectfont Out[10]:=}}
\vspace{-0.5cm}
\texttt{Total No.\,of Operators: 8}\\

\hspace*{1.3cm} \texttt{Operators:} {\scriptsize$D\; Q\; R_d+E\; L\; R_d+\mathcal{S}\; R_d\; R_u +\mathcal{T}\; R_d\; R_u+D\; L\; Q+L\; \mathcal{S}\; R_u+L\; \mathcal{T}\; R_u+Q\; U\; R_u$}

\begin{table}[h]
	\centering
	\renewcommand{\arraystretch}{2.0}
	{\tiny\begin{tabular}{|c|c|c|}
			\hline
			\multicolumn{3}{|c|}{\textbf{ Superpotential  with $U(1)_R$ charge $=0$}}\\
			\hline
			
			\textbf{Canonical Dim.}&
			\textbf{Operators}&
			\textbf{No. of Operators}\\
			\hline
			
			1&
			$\mathcal{S}$&
			1\\
			\hline
			
			2&
			$\mathcal{S}^2,\hspace{0.1cm} \mathcal{T}^2,\hspace{0.1cm} \mathcal{O}^2,\hspace{0.1cm} H_dH_u$&
			4\\
			\hline
			
			3&
			$\mathcal{S}^3,\hspace{0.1cm} \mathcal{S}\mathcal{T}^2,\hspace{0.1cm} \mathcal{S}\mathcal{O}^2,\hspace{0.1cm} \mathcal{O}^3,\hspace{0.1cm} H_dH_u\mathcal{S},\hspace{0.1cm} H_dH_u\mathcal{T}$&
			6\\
			\hline
			\hline
			\multicolumn{3}{|c|}{\textbf{ Superpotential  with $U(1)_R$ charge $=2$}}\\
			\hline
			
			\textbf{Canonical Dim.}&
			\textbf{Operators}&
			\textbf{No. of Operators}\\
			\hline
			
			2&
			$\color{purple}{H_u L,\hspace{0.1cm} H_u R_d,\hspace{0.1cm} H_d R_u}$&
			3\\
			\hline
			
			3&
			$\color{purple}{LEH_d,\hspace{0.1cm} QDH_d,\hspace{0.1cm} QUH_u,\hspace{0.1cm} EH_dR_d,\hspace{0.1cm} H_uR_d\mathcal{S}, H_uL\mathcal{S},\hspace{0.1cm} H_dR_u\mathcal{S},\hspace{0.1cm} H_uL\mathcal{T},\hspace{0.1cm} H_uR_d\mathcal{T},\hspace{0.1cm} H_dR_u\mathcal{T}}$&
			10\\
			
			\hline
			\hline
			\multicolumn{3}{|c|}{\textbf{ Superpotential  with $U(1)_R$ charge $=4$}}\\
			\hline
			
			\textbf{Canonical Dim.}&
			\textbf{Operators}&
			\textbf{No. of Operators}\\
			\hline
			
			2&
			$R_dR_u,\hspace{0.1cm}R_u L$&
			2\\
			\hline
			
			3&
			$QLD,\hspace{0.1cm} LER_d,\hspace{0.1cm} QDR_d,\hspace{0.1cm} QUR_u, LR_u\mathcal{T},\hspace{0.1cm} R_dR_u\mathcal{T},\hspace{0.1cm} LR_u\mathcal{S},\hspace{0.1cm} R_dR_u\mathcal{S}$&
			8\\
			\hline
	\end{tabular}}
	\caption{\small MSSM + Global $U(1)_R$: Superpotential terms (for $N_f = 1$). Terms in red correspond to the operators given in \cite{Chakraborty:2017dfg}.}
	\label{table:susy+globalU1-renorm-output}
\end{table}
\clearpage

%% file: BSMeftBasis.tex
\section{Operator bases of SM extended by Infrared degrees of freedom}\label{sec:bsmeft}
\begin{figure}[h]
	\centering
	{
		\includegraphics[scale=0.5]{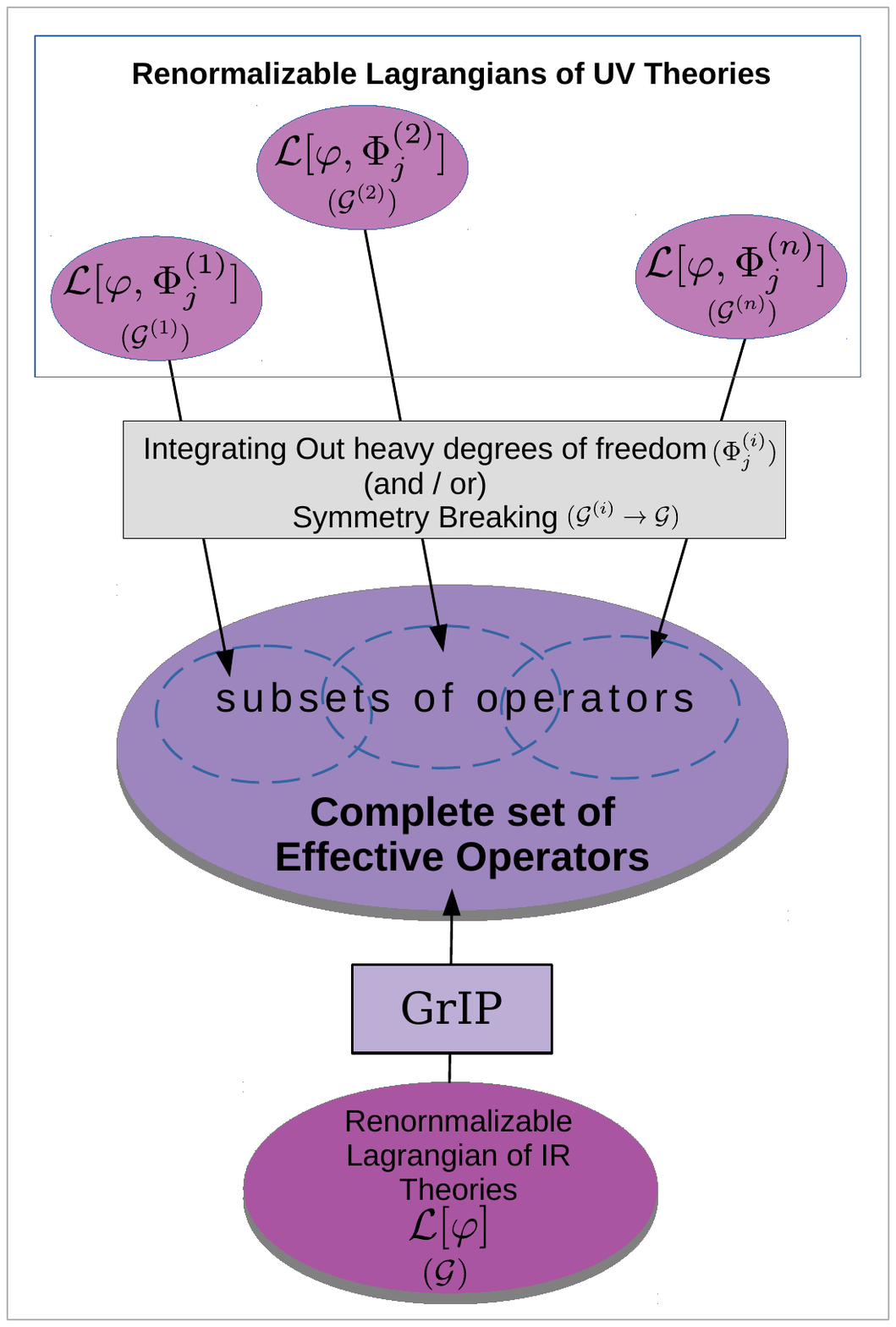}
	}
	\caption{A schematic depiction of the two approaches to EFT and their interplay. Here, $\varphi$ and $\mathcal{G}$ represent the degrees of freedom and gauge group of the IR theory while $\Phi^{(i)}_j$ represent the heavy fields of the $i$-th UV theory and $\mathcal{G}^{(i)}$ represents the corresponding gauge groups. \textbf{GrIP} generates a superset of effective operators for a given low energy theory.}
	\label{fig:flowchart}
\end{figure}

The shortcomings of the Standard Model are attempted to be cured by extending the gauge symmetry and (or) adding new particles. Some of the additional particles, i.e., DOFs, are expected to be lying around the electroweak scale (within the TeV scale). These particles can be remnants of a complete UV theory where the other heavier modes are beyond the reach of present days experiments. In that case, the light enough BSM particles or multiplets need to be taken into consideration as IR DOF along with the SM ones.  Thus we need to include the higher dimensional effective operators involving these new IR DOFs which extend the SM-EFT operator basis.  This is the key idea behind this section where we have provided the complete set of dimension-6 operators for a few popular choices of beyond SM infrared degrees of freedom. We have tabulated the results for a few more scenarios in the appendix. A pertinent detail that must be kept in mind is that different UV theories may lead to the same set of IR DOFs, after integrating out suitable heavier modes for respective theories. Our program \textbf{GrIP} allows one to construct the complete basis of effective operators which is always a superset of the operator sets achieved through the top-down (integrating out) method \cite{Bakshi:2018ics,Chiang:2015ura,Huo:2015exa,Huo:2015nka,Fuentes-Martin:2016uol,Kramer:2019fwz,Alonso:2019mok,CriadoAlamo:2019kbt}. We have illustrated the complementary nature of these constructions in Fig.~\ref{fig:flowchart}. This construction also opens up the possibility to use the EFT method in the light of present and future experiments to address the ``inverse problem". This work is a significant step towards that final goal of identifying the experimentally favoured BSM physics.

We start by constructing operator sets for minimal, single particle extensions of the Standard Model. These extra particles are kept in the same footing as the SM ones and are considered to be IR DOFs for further operator construction. Here, we will not repeat the SM interactions, which are neatly categorized in Table~\ref{table:sm-dim-5-6-op}. Rather, our focus will be on the interactions of the non-SM particles among themselves and with the SM ones.
We have considered a variety of exotic scalars and fermions that transform differently under the SM gauge symmetry. These extra fields and their transformation properties under the gauge groups $SU(3)_C\otimes SU(2)_L\otimes U(1)_Y$ and their spins are enlisted in Table~\ref{table:models-quantum-no}.
\begin{table}[h]
	\centering
	\renewcommand{\arraystretch}{1.8}
	{\tiny\begin{tabular}{|c|c|c|c|c|c|}
			\hline
			\textbf{Model No.}&
			\textbf{Extra Particle}&
			$SU(3)_C$&
			$SU(2)_L$&
			$U(1)_Y$&
			\textbf{Spin}\\
			\hline
			
			1&
			$\delta^{ + }$&
			1&
			1&
			1&
			0\\
			\hline
			
			2&
			$\delta^{++}$&
			1&
			1&
			2&
			0\\
			\hline
			
			3&
			$\Delta$&
			1&
			3&
			1&
			0\\
			\hline
			
			4&
			$\Theta$&
			1&
			4&
			3/2&
			0\\
			\hline
			
			5&
			$\Omega$&
			1&
			5&
			0&
			0\\
			\hline
			
			6&
			$\Sigma$&
			1&
			3&
			0&
			1/2\\
			\hline
			
			7&
			$\mathcal{N}$&
			1&
			1&
			0&
			1/2\\
			\hline
	\end{tabular}}
	\caption{\small Quantum numbers of various BSM fields under the SM gauge groups and their spins.}
	\label{table:models-quantum-no}
\end{table} 

\subsubsection*{\underline{SM + Singly Charged Scalar}}

Our first minimal non-trivial extension, the trivial case being that of a real gauge singlet scalar \cite{Adhikari:2020vqo}, is the inclusion of a singly charged color and isospin singlet scalar. Now, such a scalar could exist outside of any multiplet or it could be a part of some $n$-plet whose other components acquire larger masses. As those heavy modes are integrated out, their footprints can be captured through the effective operators composed of SM fields and a light singly charged scalar. We have collected the operator sets at mass dimensions-5 and -6 in Table~\ref{table:model1-operators}.   

\begin{table}[h]
	\centering
	\renewcommand{\arraystretch}{2.0}
	{\tiny\begin{tabular}{|c|c|c|}
			\hline
			\multicolumn{3}{|c|}{\textbf{ Mass Dimension-5}}\\
			\hline
			\textbf{Operator Class}&
			\multicolumn{2}{c|}{\textbf{Operators (in non-covariant form)}}\\
			\hline
			
			$\Psi^2\Phi^2$&
			\multicolumn{2}{l|}{$\color{blue}{
			\frac{1}{2}(N_f^2 + N_f)el^2\delta^2,\hspace{0.2cm}
			(N_f^2)el\delta H^{\dagger}L^{\dagger},\hspace{0.2cm}
			(N_f^2)d\delta H^{\dagger}Q^{\dagger},\hspace{0.2cm}
			(N_f^2)u^{\dagger}\delta H^{\dagger}Q}$}\\
			\hline
			\hline
			\multicolumn{3}{|c|}{\textbf{Mass Dimension-6}}\\
			\hline
			\textbf{Operator Class}&
			\multicolumn{2}{c|}{\textbf{Operators (in non-covariant form)}}\\
			\hline
	
			$\Phi^6$&
			\multicolumn{2}{l|}{$\delta^3(\delta^{\dagger})^3,\hspace{0.2cm} H H^{\dagger}\delta^2(\delta^{\dagger})^2,\hspace{0.2cm} 
			\delta\delta^{\dagger} H^2( H^{\dagger})^2$}\\
			\hline
			
			$\Phi^2X^2$&
			\multicolumn{2}{l|}{$\color{blue}{Bl^2\delta\delta^{\dagger},\hspace{0.2cm} Gl^2\delta\delta^{\dagger},\hspace{0.2cm}
			Wl^2\delta\delta^{\dagger}}$}\\
			\hline
			
			$\Psi^2\Phi X$&
			\multicolumn{2}{l|}{$\color{blue}{\frac{1}{2}\left(N_f^2+N_f\right)Bl L^2\delta,\hspace{0.2cm}
			\boxed{\frac{1}{2}\left(N_f^2-N_f\right)Wl L^2\delta}}$}\\
			\hline
						
			$\Psi^2\Phi^2 \mathcal{D}$&
			\multicolumn{2}{l|}{$(N_f^2)QQ^{\dagger}\delta\delta^{\dagger}\mathcal{D},\hspace{0.2cm}
			(N_f^2)LL^{\dagger}\delta\delta^{\dagger}\mathcal{D},\hspace{0.2cm} 
			(N_f^2)uu^{\dagger}\delta\delta^{\dagger}\mathcal{D},\hspace{0.2cm}
			(N_f^2)dd^{\dagger}\delta\delta^{\dagger}\mathcal{D},\hspace{0.2cm} 
			(N_f^2)elel^{\dagger}\delta\delta^{\dagger}\mathcal{D},\hspace{0.2cm} 
			\color{blue}{(N_f^2) H\delta Lel\mathcal{D}}$}\\
			\hline
			$\Psi^2\Phi^3$&
			\multicolumn{2}{l|}{$\color{blue}{ (N_f^2) H H^{\dagger}L^2\delta,\hspace{0.2cm}
			(N_f^2) H\delta\delta^{\dagger}L^{\dagger}el, \hspace{0.2cm} 
			(N_f^2) H\delta\delta^{\dagger}dQ^{\dagger}, \hspace{0.2cm}
			(N_f^2) H^{\dagger}\delta\delta^{\dagger}uQ^{\dagger}, \hspace{0.2cm}
			\boxed{\frac{1}{2}\left(N_f^2-N_f\right)L^2\delta^2\delta^{\dagger}}}$}\\
			\hline
			
			$\Phi^4\mathcal{D}^2$&
			\multicolumn{2}{l|}{$\delta^2(\delta^{\dagger})^2\mathcal{D}^2,\hspace{0.2cm}
			2 H H^{\dagger}\delta\delta^{\dagger}\mathcal{D}^2 $}\\
			\hline		    
	\end{tabular}}
	\caption{\small SM + Singly Charged Scalar\,($\delta^{+}$): Operators of mass dimensions-5 and -6 excluding pure SM operators. The operators in blue have distinct hermitian conjugates which we have not written explicitly. Boxed operators vanish for $N_f = 1$. Here, $\delta\rightarrow\delta^+$ and $\delta^{\dagger}\rightarrow\delta^-$.}
	\label{table:model1-operators}
\end{table} 

\subsubsection*{\underline{SM + Doubly Charged Scalar}}

Similar to the earlier case, the doubly charged scalar can be the IR DOF and several theories can lead to this scenario \cite{Chakrabortty:2015zpm}. The additional effective operators in the presence of $\delta^{++}$ are shown in Table~\ref{table:model2-operators}. 
\begin{table}[h]
	\centering
	\renewcommand{\arraystretch}{1.8}
	{\tiny\begin{tabular}{|c|c|c|}
			\hline
			\multicolumn{3}{|c|}{\textbf{Mass Dimension-6}}\\
			\hline
			
			\textbf{Operator Class}&
			\multicolumn{2}{c|}{\textbf{Operators (in non-covariant form)}}\\
			\hline
			
			$\Phi^6$&
			\multicolumn{2}{l|}{$ H H^{\dagger}\delta^2(\delta^{\dagger})^2,\hspace{0.1cm}  H^2( H^{\dagger})^2\delta\delta^{\dagger},\hspace{0.1cm} \delta^3(\delta^{\dagger})^3$}\\
			\hline
			
			$\Phi^2X^2$&
			\multicolumn{2}{l|}{$\color{blue}{Bl^2\delta\delta^{\dagger},\hspace{0.1cm} Gl^2\delta\delta^{\dagger},\hspace{0.1cm} Wl^2\delta\delta^{\dagger}}$}\\
			\hline
			
			$\Psi^2\Phi X$&
			\multicolumn{2}{l|}{$\color{blue}{\boxed{\frac{1}{2}\left(N_f^2-N_f\right)(el^\dagger)^2\delta^\dagger Bl}}$}\\
			\hline
			
			$\Psi^2\Phi^2 \mathcal{D}$&
			\multicolumn{2}{l|}{$(N_f^2)QQ^{\dagger}\delta\delta^{\dagger}\mathcal{D},\hspace{0.1cm} (N_f^2)uu^{\dagger}\delta\delta^{\dagger}\mathcal{D},\hspace{0.1cm} (N_f^2)dd^{\dagger}\delta\delta^{\dagger}\mathcal{D},\hspace{0.1cm} (N_f^2)LL^{\dagger}\delta\delta^{\dagger}\mathcal{D},\hspace{0.1cm} (N_f^2)elel^{\dagger}\delta\delta^{\dagger}\mathcal{D},\hspace{0.1cm}\color{blue}{(N_f^2)\delta H^{\dagger}elL\mathcal{D}}$}\\
			\hline
			
			\multirow{2}{*}{$\Psi^2\Phi^3$}&
			\multicolumn{2}{l|}{$\color{blue}{(N_f^2)\delta\delta^{\dagger} H L^{\dagger}el,\hspace{0.1cm} (N_f^2)\delta\delta^{\dagger} H dQ^{\dagger},\hspace{0.1cm} (N_f^2)\delta\delta^{\dagger} H^{\dagger} uQ^{\dagger},\hspace{0.1cm}
			\frac{1}{2}\left(N_f^2+N_f\right)L^2\delta( H^{\dagger})^2,}$}\\
		
		    &
		    \multicolumn{2}{l|}{$\color{blue}{\frac{1}{2}\left(N_f^2+N_f\right)el^2\delta^2\delta^{\dagger},\hspace{0.1cm} \frac{1}{2}\left(N_f^2+N_f\right)el^2\delta H H^{\dagger}}$}\\
			\hline
			
			$\Phi^4\mathcal{D}^2$&
			\multicolumn{2}{l|}{$\delta^2(\delta^{\dagger})^2\mathcal{D}^2,\hspace{0.1cm} 2 H H^{\dagger}\delta\delta^{\dagger}\mathcal{D}^2$}\\
			\hline		    
	\end{tabular}}
	\caption{\small SM + Doubly Charged Scalar\,($\delta^{++}$): Operators of mass dimension-6 excluding pure SM operators. There are no mass dimension-5 operators except pure SM operators. Here, $\delta\rightarrow\delta^{++}$ and $\delta^{\dagger}\rightarrow\delta^{--}$.}
	\label{table:model2-operators}
\end{table} 

\subsubsection*{\underline{SM + Complex Triplet Scalar}}

Unlike the previous two scenarios, an entire multiplet may be light enough to be included as an IR DOF. The $SU(2)_L$ complex triplet $(\Delta)$ leading to neutrino mass generation through the Type-II seesaw is one such case \cite{Konetschny:1977bn,Senjanovic:1978ev,Schechter:1980gr}. The effective operators involving $\Delta$ have been catalogued in Table~\ref{table:model3-operators}.

\begin{table}[h]
	\centering
	\renewcommand{\arraystretch}{1.8}
	{\tiny\begin{tabular}{|c|c|c|}
			\hline
			\multicolumn{3}{|c|}{\textbf{Mass Dimension-5}}\\
			\hline
			\textbf{Operator Class}&
			\multicolumn{2}{c|}{\textbf{Operators (in non-covariant form)}}\\
			\hline
			
			$\Phi^5$&
			\multicolumn{2}{l|}{$\color{blue}{ H^3 H^{\dagger}\Delta^{\dagger},\hspace{0.1cm} 2 H^2\Delta(\Delta^{\dagger})^2}$}\\
			\hline
			
			$\Psi^2\Phi^2$&
			\multicolumn{2}{l|}{$\color{blue}{(N_f^2)\Delta H^{\dagger}L^{\dagger}el,\hspace{0.1cm} (N_f^2)\Delta H^{\dagger}dQ^{\dagger},\hspace{0.1cm} (N_f^2)\Delta^{\dagger} H uQ^{\dagger},\hspace{0.1cm} \frac{1}{2}\left(N_f^2+N_f\right)el^2\Delta^2
			}$}\\
			\hline
			\hline
			\multicolumn{3}{|c|}{\textbf{Mass Dimension-6}}\\
			\hline
			\textbf{Operator Class}&
			\multicolumn{2}{c|}{\textbf{Operators (in non-covariant form)}}\\
			\hline
			$\Phi^6$&
			\multicolumn{2}{l|}{$2\Delta^3(\Delta^{\dagger})^3,\hspace{0.1cm} 3\Delta^2(\Delta^{\dagger})^2 H H^{\dagger},\hspace{0.1cm} 3\Delta\Delta^{\dagger} H^2( H^{\dagger})^2,\hspace{0.1cm} \color{blue}{ H^4(\Delta^{\dagger})^2}$}\\
			\hline
			
			$\Phi^2X^2$&
			\multicolumn{2}{l|}{$\color{blue}{Bl^2\Delta^{\dagger}\Delta,\hspace{0.1cm} Gl^2\Delta^{\dagger}\Delta,\hspace{0.1cm} 2Wl^2\Delta^{\dagger}\Delta,\hspace{0.1cm} Bl\,Wl\Delta^{\dagger}\Delta} $}\\
			\hline
			
			$\Psi^2\Phi X$&
			\multicolumn{2}{l|}{$\color{blue}{(N_f^2)L^2\Delta Wl,\hspace{0.1cm} \boxed{\frac{1}{2}\left(N_f^2-N_f\right)L^2\Delta Bl}}$}\\
			\hline
			
			\multirow{2}{*}{$\Psi^2\Phi^2 \mathcal{D}$}&
			\multicolumn{2}{l|}{$(N_f^2)uu^{\dagger}\Delta\Delta^{\dagger}\mathcal{D},\hspace{0.1cm} (N_f^2)dd^{\dagger}\Delta\Delta^{\dagger}\mathcal{D},\hspace{0.1cm} (N_f^2)elel^{\dagger}\Delta\Delta^{\dagger}\mathcal{D},\hspace{0.1cm} (2N_f^2)QQ^{\dagger}\Delta\Delta^{\dagger}\mathcal{D},$}\\
			
			&
			\multicolumn{2}{l|}{$(2N_f^2)LL^{\dagger}\Delta\Delta^{\dagger}\mathcal{D},\hspace{0.1cm}
			\color{blue}{(N_f^2)\Delta H Lel\mathcal{D}}$}\\
			\hline
			
			\multirow{2}{*}{$\Psi^2\Phi^3$}&
			\multicolumn{2}{l|}{$\color{blue}{(2N_f^2)\Delta\Delta^{\dagger} H L^{\dagger}el,\hspace{0.1cm} (2N_f^2)\Delta\Delta^{\dagger} H dQ^{\dagger},\hspace{0.1cm} (2N_f^2)\Delta\Delta^{\dagger} H^{\dagger} uQ^{\dagger},\hspace{0.1cm} }$}\\
			
			&
			\multicolumn{2}{l|}{$\color{blue}{\left(N_f^2+N_f\right)L^2\Delta^2\Delta^{\dagger},\hspace{0.1cm} \frac{1}{2}\left(3N_f^2+N_f\right)L^2\Delta H H^{\dagger},\hspace{0.1cm} \frac{1}{2}\left(N_f^2+N_f\right)el^2 H^2\Delta}$}\\
			\hline
			
			$\Phi^4\mathcal{D}^2$&
			\multicolumn{2}{l|}{$3\Delta^2(\Delta^{\dagger})^2\mathcal{D}^2,\hspace{0.1cm} 4\Delta\Delta^{\dagger} H H^{\dagger}\mathcal{D}^2$}\\
			\hline		    
	\end{tabular}}
	\caption{\small SM + Complex Triplet Scalar\,($\Delta$): Operators of mass dimensions-5 and -6 excluding pure SM operators.}
	\label{table:model3-operators}
\end{table} 

\subsubsection*{\underline{SM + $SU(2)$ Quadruplet Scalar}}

$SU(2)$ quadruplet scalars often appear in the study of fermionic dark matter candidates which themselves are quadruplets under $SU(2)$ \cite{Tait:2016qbg}. They have also appeared in discussions of the Type-III Seesaw where they contribute to the Dirac mass of a lepton triplet \cite{Ren:2011mh}. Lastly, they also furnish a doubly charged scalar and provide an avenue to study the relevant phenomenology \cite{Chakrabortty:2015zpm}. Here, we have considered a $SU(2)$ quadruplet scalar with hypercharge $3/2$ and provided the effective operators of mass dimensions-5 and -6 in Table~\ref{table:model4-operators}.

\begin{table}[h]
	\centering
	\renewcommand{\arraystretch}{1.8}
	{\tiny\begin{tabular}{|c|c|c|}
			\hline
			\multicolumn{3}{|c|}{\textbf{Mass Dimension-5}}\\
			\hline
			\textbf{Operator Class}&
			\multicolumn{2}{c|}{\textbf{Operators (in non-covariant form)}}\\
			\hline
			$\Psi^2\Phi^2$&
			\multicolumn{2}{l|}{$\color{blue}{\frac{1}{2}\left(N_f^2+N_f\right) H^{\dagger}\Theta L^2}$}\\
			\hline
			\hline
			\multicolumn{3}{|c|}{\textbf{Mass Dimension-6}}\\
			\hline
			\textbf{Operator Class}&
			\multicolumn{2}{c|}{\textbf{Operators (in non-covariant form)}}\\
			\hline
			
			$\Phi^6$&
			\multicolumn{2}{l|}{$3\Theta^3(\Theta^{\dagger})^3,\hspace{0.1cm}  4HH^{\dagger}\Theta^2(\Theta^{\dagger})^2,\hspace{0.1cm} 3H^2(H^{\dagger})^2\Theta\Theta^{\dagger},\hspace{0.1cm} \color{blue}{H^4 H^{\dagger}\Theta^{\dagger},\hspace{0.1cm} 2H^3\Theta(\Theta^{\dagger})^2}$}\\
			\hline	
					
			$\Phi^2X^2$&
			\multicolumn{2}{l|}{$\color{blue}{Bl^2\Theta\Theta^{\dagger},\hspace{0.1cm} Gl^2\Theta\Theta^{\dagger},\hspace{0.1cm} 2Wl^2\Theta\Theta^{\dagger},\hspace{0.1cm} Bl\,Wl\Theta\Theta^{\dagger} } $}\\
			\hline
			
			$\Psi^2\Phi^2 \mathcal{D}$&
			\multicolumn{2}{l|}{$(N_f^2)uu^{\dagger}\Theta\Theta^{\dagger}\mathcal{D},\hspace{0.1cm} (N_f^2)dd^{\dagger}\Theta\Theta^{\dagger}\mathcal{D},\hspace{0.1cm} (N_f^2)elel^{\dagger}\Theta\Theta^{\dagger}\mathcal{D},\hspace{0.1cm} (2N_f^2)QQ^{\dagger}\Theta\Theta^{\dagger}\mathcal{D},\hspace{0.1cm} (2N_f^2)LL^{\dagger}\Theta\Theta^{\dagger}\mathcal{D}$}\\
			\hline
			
			\multirow{2}{*}{$\Psi^2\Phi^3$}&
			\multicolumn{2}{l|}{$\color{blue}{(N_f^2)L^{\dagger}el\Theta(H^{\dagger})^2,\hspace{0.1cm} (N_f^2)Q^{\dagger}d\Theta(H^{\dagger})^2,\hspace{0.1cm} (N_f^2)Q^{\dagger}u\Theta^{\dagger}H^2,\hspace{0.1cm} (2N_f^2)L^{\dagger}el H\Theta\Theta^{\dagger}, } $}\\
			
			&
			\multicolumn{2}{l|}{$\color{blue}{(2N_f^2)Q^{\dagger}d H\Theta\Theta^{\dagger},\hspace{0.1cm} (2N_f^2)Q^{\dagger}u H^{\dagger}\Theta\Theta^{\dagger}} $}\\
			\hline
			
			$\Phi^4\mathcal{D}^2$&
			\multicolumn{2}{l|}{$4HH^{\dagger}\Theta\Theta^{\dagger}\mathcal{D}^2,\hspace{0.1cm}4\Theta^2(\Theta^{\dagger})^2 $}\\
			\hline		    
	\end{tabular}}
	\caption{\small SM + $SU(2)$ Quadruplet Scalar\,($\Theta$): Operators of mass dimensions-5 and -6 excluding pure SM operators.}
	\label{table:model4-operators}
\end{table} 

\subsubsection*{\underline{SM + $SU(2)$ Quintuplet Scalar}}

$SU(2)$ quintuplet or higher $n$-plet scalars are commonly studied in the context of Minimal Dark Matter (MDM) models \cite{Cai:2017fmr}, where the neutral component is usually proposed to be the candidate particle. At the same time, a $SU(2)$ quintuplet can also furnish a doubly charged scalar. So any discussion of such multiplets brings forth a discussion of the phenomenology of doubly charged scalars \cite{Chakrabortty:2015zpm}. We provide the effective operators of mass dimensions-5 and -6 in Table~\ref{table:model5-operators}.

\begin{table}[h]
	\centering
	\renewcommand{\arraystretch}{1.7}
	{\tiny\begin{tabular}{|c|c|c|}
			\hline
			\multicolumn{3}{|c|}{\textbf{Mass Dimension-5}}\\
			\hline
			\textbf{Operator Class}&
			\multicolumn{2}{c|}{\textbf{Operators (in non-covariant form)}}\\
			\hline
			$\Phi^5$&
			\multicolumn{2}{l|}{$\Omega^3 H^{\dagger} H,\hspace{0.1cm} \Omega( H^{\dagger})^2 H^2,\hspace{0.1cm} 
			\Omega^5$}\\
			\hline
			$\Phi X^2$&
			\multicolumn{2}{l|}{$\color{blue}{Wl^2\Omega}$}\\
			\hline
	        \hline
	        \multicolumn{3}{|c|}{\textbf{Mass Dimension-6}}\\
	        \hline
	        \textbf{Operator Class}&
	        \multicolumn{2}{c|}{\textbf{Operators (in non-covariant form)}}\\
	        \hline
			
			$\Phi^6$&
			\multicolumn{2}{l|}{$2\Omega^6,\hspace{0.1cm}
			2\Omega^2 H^2( H^{\dagger})^2,\hspace{0.1cm} \Omega^4 H H^{\dagger}$}\\
			\hline
			
			$\Phi^2X^2$&
			\multicolumn{2}{l|}{$\color{blue}{Bl^2\Omega^2,\hspace{0.1cm} Gl^2\Omega^2,\hspace{0.1cm}
			2Wl^2\Omega^2}$}\\
			\hline
			
			$\Psi^2\Phi^2 \mathcal{D}$&
			\multicolumn{2}{l|}{$(N_f^2)QQ^{\dagger}\Omega^2\mathcal{D},\hspace{0.1cm} (N_f^2)LL^{\dagger}\Omega^2\mathcal{D}$}\\
			\hline
			
			$\Psi^2\Phi^3$&
			\multicolumn{2}{l|}{$\color{blue}{\frac{1}{2}\left(N_f^2+N_f\right)\Omega H^2L^2,\hspace{0.1cm}
			(N_f^2)\Omega^2 H elL^{\dagger},\hspace{0.1cm}
		    (N_f^2)\Omega^2 H dQ^{\dagger},\hspace{0.1cm}
	        (N_f^2)\Omega^2 H^{\dagger} uQ^{\dagger}}$}\\
			\hline
			
			$\Phi^4\mathcal{D}^2$&
			\multicolumn{2}{l|}{$2\Omega^4\mathcal{D}^2,\hspace{0.1cm} 2\Omega^2 H H^{\dagger}\mathcal{D}^2$}\\
			\hline		    
	\end{tabular}}
	\caption{\small SM + $SU(2)$ Quintuplet Scalar\,($\Omega$): Operators of mass dimensions-5 and -6 excluding pure SM operators.}
	\label{table:model5-operators}
\end{table} 

\subsubsection*{\underline{SM + Left-Handed Triplet Fermion}}

Extensions of SM through the addition of a triplet fermion leads to the generation of neutrino mass through Type-III seesaw mechanism. This has also been discussed in the context of dark matter model building where the neutral component of $\left(\Sigma^{+},\Sigma^{0},\Sigma^{-}\right)$ is proposed as the DM candidate \cite{Ma:2008cu}. We have enlisted the effective operators of mass dimensions-5 and -6 in Table~\ref{table:model6-operators}.

\begin{table}[h]
	\centering
	\renewcommand{\arraystretch}{1.7}
	{\tiny\begin{tabular}{|c|c|c|}
			\hline
			\multicolumn{3}{|c|}{\textbf{Mass Dimension-5}}\\
			\hline
			\textbf{Operator Class}&
			\multicolumn{2}{c|}{\textbf{Operators (in non-covariant form)}}\\
			\hline
			$\Psi^2\Phi^2$&
			\multicolumn{2}{l|}{$\color{blue}{(N_f^2)\Sigma^2 HH^{\dagger}, \hspace{0.1cm} (N_f^2)\Sigma^{\dagger}el H^2}$}\\
			\hline
			$\Psi^2 X$&
			\multicolumn{2}{l|}{$\color{blue}{\boxed{\frac{1}{2}(N_f^2-N_f)Bl\Sigma^2}, \hspace{0.1cm} \frac{1}{2}(N_f^2+N_f)Wl\Sigma^2}$}\\
			\hline
			\hline
			\multicolumn{3}{|c|}{\textbf{Mass Dimension-6}}\\
			\hline
			\textbf{Operator Class}&
			\multicolumn{2}{c|}{\textbf{Operators (in non-covariant form)}}\\
			\hline
			
			$\Psi^2\Phi^2 \mathcal{D}$&
			\multicolumn{2}{l|}{$(2N_f^2)HH^{\dagger}\Sigma\Sigma^{\dagger}\mathcal{D}$}\\
			\hline
			
			$\Psi^2\Phi X$&
			\multicolumn{2}{l|}{$\color{blue}{(N_f^2) Bl L\Sigma H,\hspace{0.1cm} (2N_f^2)WlL\Sigma H}$}\\
			\hline
			
			$\Psi^2\Phi^3$&
			\multicolumn{2}{l|}{$\color{blue}{(2N_f^2)L\Sigma H^2H^{\dagger}}$}\\
		 	\hline
		 	
		 	\multirow{3}{*}{$\Psi^4$}&
		 	\multicolumn{2}{l|}{$\color{blue}{(2N_f^4)LQd^{\dagger}\Sigma, \hspace{0.1cm} (N_f^4)LQ^{\dagger}u\Sigma, \hspace{0.1cm} (N_f^4)L^2el^{\dagger}\Sigma, \hspace{0.1cm} \boxed{\frac{1}{2}(N_f^4-N_f^3)Q^2d\Sigma^{\dagger}}, \hspace{0.1cm} \frac{1}{4}(N_f^4+3N_f^2)\Sigma^4,}$}\\
		 	
		 	&
		 	\multicolumn{2}{l|}{$(N_f^4)uu^{\dagger}\Sigma\Sigma^{\dagger},\hspace{0.1cm}  (N_f^4)dd^{\dagger}\Sigma\Sigma^{\dagger},\hspace{0.1cm} (N_f^4)elel^{\dagger}\Sigma\Sigma^{\dagger},\hspace{0.1cm} (2N_f^4)QQ^{\dagger}\Sigma\Sigma^{\dagger},\hspace{0.1cm} (2N_f^4)LL^{\dagger}\Sigma\Sigma^{\dagger},$}\\
		 	
		 	&
		 	\multicolumn{2}{l|}{$ (\frac{3}{4}N_f^4+\frac{1}{2}N_f^3+\frac{3}{4}N_f^2)\Sigma^2(\Sigma^{\dagger})^2$}\\    
		 	\hline
	\end{tabular}}
	\caption{\small SM + Left-Handed Triplet Fermion\,($\Sigma$): Operators of mass dimensions-5 and -6 excluding pure SM operators.}
	\label{table:model6-operators}
\end{table}

\subsubsection*{\underline{SM + Right-Handed Singlet Fermion}}

\begin{table}[h]
	\centering
	\renewcommand{\arraystretch}{1.7}
	{\tiny\begin{tabular}{|c|c|c|}
			\hline
			\multicolumn{3}{|c|}{\textbf{Mass Dimension-5}}\\
			\hline
			\textbf{Operator Class}&
			\multicolumn{2}{c|}{\textbf{Operators (in non-covariant form)}}\\
			\hline
			$\Psi^2\Phi^2$&
			\multicolumn{2}{l|}{$\color{blue}{\frac{1}{2}(N_f^2+N_f)\mathcal{N}^2 HH^{\dagger}}$}\\
			\hline
			$\Psi^2 X$&
			\multicolumn{2}{l|}{$\color{blue}{\boxed{\frac{1}{2}(N_f^2-N_f)Bl~(\mathcal{N^\dagger})^2}}$}\\
			\hline
			\hline
			\multicolumn{3}{|c|}{\textbf{Mass Dimension-6}}\\
			\hline
			\textbf{Operator Class}&
			\multicolumn{2}{c|}{\textbf{Operators (in non-covariant form)}}\\
			\hline
			
			$\Psi^2\Phi^2 \mathcal{D}$&
			\multicolumn{2}{l|}{$(N_f^2)HH^{\dagger}\mathcal{N}\mathcal{N}^{\dagger}\mathcal{D}, \hspace{0.1cm} \color{blue}(N_f^2){H^2\mathcal{N}^{\dagger}el~\mathcal{D}}$}\\
			\hline
			
			$\Psi^2\Phi X$&
			\multicolumn{2}{l|}{$\color{blue}{(N_f^2) Bl~ L\mathcal{N}^{\dagger} H,\hspace{0.1cm} (N_f^2) Wl~ L\mathcal{N}^{\dagger} H}$}\\
			\hline
			
			$\Psi^2\Phi^3$&
			\multicolumn{2}{l|}{$\color{blue}{(N_f^2)L^{\dagger}\mathcal{N} H(H^{\dagger})^2}$}\\
			\hline
			
			\multirow{3}{*}{$\Psi^4$}&
			\multicolumn{2}{l|}{$\color{blue}{ \boxed{\frac{1}{12}(N_f^4-N_f^2)~\mathcal{N}^4}, \hspace{0.1cm} \frac{1}{2}(N_f^4+N_f^3)Q^2d\mathcal{N}, \hspace{0.1cm} (N_f^4)d^2u\mathcal{N},\hspace{0.1cm} (N_f^4)(L^{\dagger})^2el~\mathcal{N},\hspace{0.1cm}(N_f^4)u^{\dagger}d~el^{\dagger}\mathcal{N},\hspace{0.1cm} }$}\\
			
			&
			\multicolumn{2}{l|}{$\color{blue}{(N_f^4)LQ^{\dagger}u\mathcal{N}^{\dagger},\hspace{0.1cm} (2N_f^4)L^{\dagger}Q^{\dagger}d\mathcal{N},\hspace{0.1cm}} \color{black}{  (N_f^4)uu^{\dagger}\mathcal{N}\mathcal{N}^{\dagger},\hspace{0.1cm}  (N_f^4)dd^{\dagger}\mathcal{N}\mathcal{N}^{\dagger},\hspace{0.1cm} (N_f^4)elel^{\dagger}\mathcal{N}\mathcal{N}^{\dagger},} $}\\
			
			&
			\multicolumn{2}{l|}{$ (N_f^4)QQ^{\dagger}\mathcal{N}\mathcal{N}^{\dagger},\hspace{0.1cm} (N_f^4)LL^{\dagger}\mathcal{N}\mathcal{N}^{\dagger}, (\frac{1}{4}N_f^4+\frac{1}{2}N_f^3+\frac{1}{4}N_f^2)\mathcal{N}^2(\mathcal{N}^{\dagger})^2$}\\    
			\hline
	\end{tabular}}
	\caption{\small SM + Right-Handed Singlet Fermion\,($\mathcal{N}$): Operators of mass dimensions-5 and -6 excluding pure SM operators.}
	\label{table:model7-operators}
\end{table}

\begin{figure}
    \centering
	\includegraphics[scale=0.6]{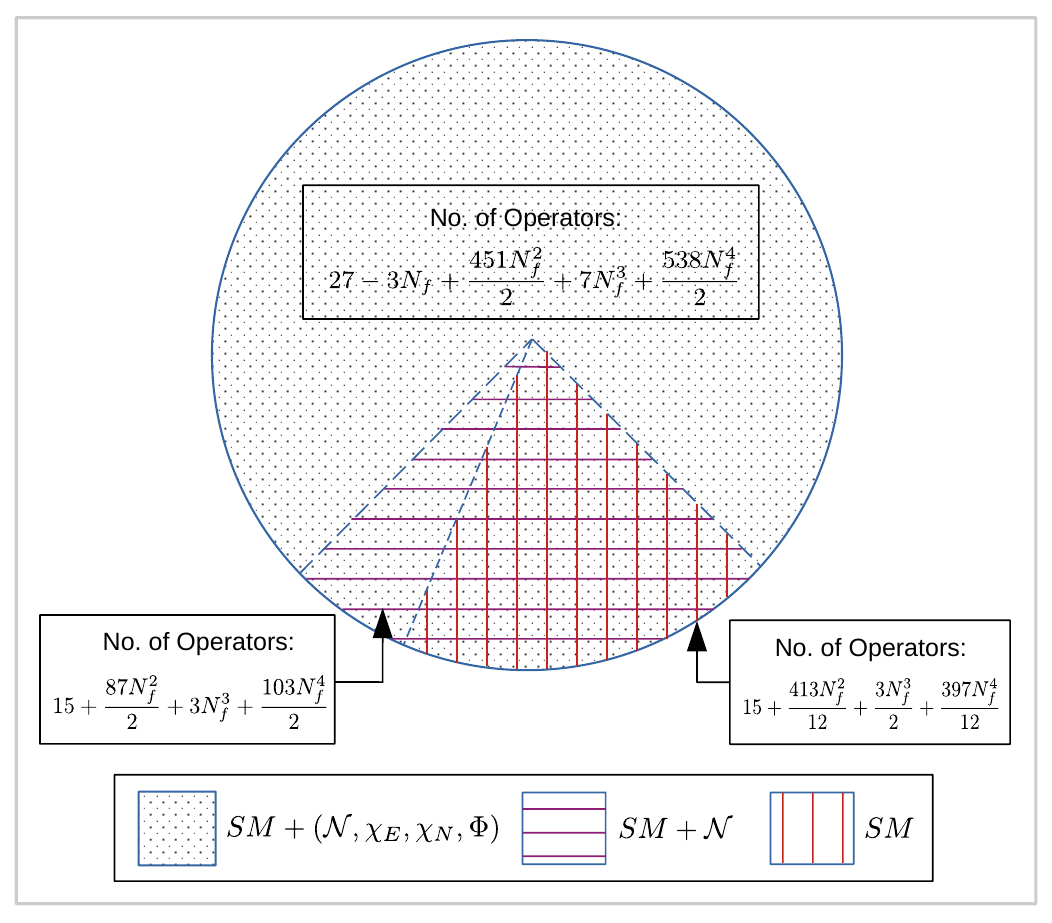}
    \caption{Pie-chart showing number of operators including their conjugates for 3 distinct scenarios.}
    \label{fig:pie-chart}
\end{figure}

Inclusion of a heavy right handed singlet fermion attempts to describe neutrino mass by way of the Type-I seesaw mechanism \cite{delAguila:2008ir}. But recently, the discussion has shifted towards the extension of the SM degrees of freedom by a light singlet fermion \cite{Liao:2016qyd} and construction of effective operators. One such scenario was considered in \cite{Chala:2020vqp} where the full particle content was comprised of SM fields, a right handed fermion singlet ($\mathcal{N}$), a couple of vector-like fermions ($\chi_E, \chi_N$) and a scalar ($\Phi$). Operators of dimensions-5 and -6, comprised of SM fields and the right handed singlet fermion are obtained when the vector-like fermions and the heavy scalar are integrated out. We have categorized all the dimensions-5 and -6 operators highlighting interactions between the SM fields and the right handed fermion as well as self-interactions of the fermion in Table~\ref{table:model7-operators}. 
It must be remarked that the operator set given in \cite{Chala:2020vqp} contained redundancies w.r.t. equations of motion of the fields. Such redundancies do not exist in the operator set provided in Table~\ref{table:model7-operators}. We have also drawn a pie-chart showing the number of effective operators of mass dimension-6 for $N_f$ number of fermion flavours and distinguishing the three cases: (i) Operators composed only of SM fields (ii) Operators composed of SM fields + $\mathcal{N}$ and (iii) Operators composed of SM fields + $\mathcal{N}$ + $\chi_E$ + $\chi_N$ + $\Phi$ in Fig.~\ref{fig:pie-chart}.

\subsubsection*{\underline{Extensions of the SM through addition of Lepto-Quarks}}

We now shift our attention towards the extension of the SM by including lepto-quarks, i.e., particles having both baryon and lepton numbers \cite{Buchmuller:1986zs}. We expect to observe baryon and lepton number violation among the operators constituted by them. Lepto-quarks attract a lot of attention not just because they act as mediators between the quark and lepton sectors \cite{Dey:2017ede,Bar-Shalom:2018ure,Bandyopadhyay:2018syt,Das:2017kkm} but also because they can lead to the breaking of the $SU(3)$ color symmetry as well. We have considered the models discussed in \cite{Dey:2017ede,Bar-Shalom:2018ure}. We have listed the lepto-quark fields and their transformation properties under the gauge groups $SU(3)_C\otimes SU(2)_L\otimes U(1)_Y$ and the Lorentz group as well as their baryon and lepton numbers in Table~\ref{table:leptoquark-quantum-no}. The effective operators of dimensions-5 and -6 have been neatly categorized and given in Tables~\ref{table:leptoquark-model1-operators}-\ref{table:leptoquark-model4-operators}.

\begin{table}[h]
	\centering
	\renewcommand{\arraystretch}{1.8}
	{\tiny\begin{tabular}{|c|c|c|c|c|c|c|c|}
			\hline
			\textbf{Model No.}&
			\textbf{Lepto-quark}&
			$SU(3)_C$&
			$SU(2)_L$&
			$U(1)_Y$&
			\textbf{Spin}&
			\textbf{Baryon No.}&
			\textbf{Lepton No.}\\
			\hline
			
			1&
			$\chi_1$&
			3&
			2&
			1/6&
			0&
			1/3&
			-1\\
			\hline
			
			2&
			$\chi_2$&
			3&
			2&
			7/6&
			0&
			1/3&
			-1\\
			\hline
			
			3&
			$\Phi_1$&
			3&
			1&
			2/3&
			0&
			1/3&
			-1\\
			\hline
			
			4&
			$\Phi_2$&
			3&
			1&
			-1/3&
			0&
			1/3&
			-1\\
			\hline
	\end{tabular}}
	\caption{\small Quantum numbers of various lepto-quark fields under the SM gauge groups, their spins and baryon and lepton numbers.}
	\label{table:leptoquark-quantum-no}
\end{table} 

\begin{table}[h]
	\centering
	\renewcommand{\arraystretch}{1.8}
	{\tiny\begin{tabular}{|c|c|c|}
			\hline
			\multicolumn{3}{|c|}{\textbf{Mass Dimension-5}}\\
			\hline
			\textbf{Operator Class}&
			\multicolumn{2}{c|}{\textbf{Operators (in non-covariant form)}}\\
			\hline
			
			\multirow{2}{*}{$\Psi^2\Phi^2$}&
			\multicolumn{2}{l|}{$\color{blue}{(N_f^2)LQ\chi_1^2, \hspace{0.1cm} (N_f^2)el u\chi_1^2, \hspace{0.1cm} (N_f^2)ud(\chi_1^{\dagger})^2, \hspace{0.1cm} (N_f^2+N_f)Q^2(\chi_1^{\dagger})^2, \hspace{0.1cm} (2N_f^2)LQH\chi_1^{\dagger}, \hspace{0.1cm} (N_f^2)el u H\chi_1^{\dagger},}$}\\
			
			&
			\multicolumn{2}{l|}{$\color{blue}{ (N_f^2)udH^{\dagger}\chi_1, \hspace{0.1cm} (N_f^2)Q^2H^{\dagger}\chi_1, \hspace{0.1cm} \boxed{\frac{1}{2}(N_f^2-N_f)d^2H\chi_1}}$}\\
			\hline
			\hline
			\multicolumn{3}{|c|}{\textbf{Mass Dimension-6}}\\
			\hline
			\textbf{Operator Class}&
			\multicolumn{2}{c|}{\textbf{Operators (in non-covariant form)}}\\
			\hline
			$\Phi^6$&
			\multicolumn{2}{l|}{$2\chi_1^3(\chi_1^{\dagger})^3, \hspace{0.1cm} 3\chi_1^2(\chi_1^{\dagger})^2HH^{\dagger}, \hspace{0.1cm} 2\chi_1\chi_1^{\dagger}H^2(H^{\dagger})^2 $}\\
			\hline
			
			$\Phi^2X^2$&
			\multicolumn{2}{l|}{$\color{blue}{Bl^2\chi_1^{\dagger}\chi_1, \hspace{0.1cm} 2Gl^2\chi_1^{\dagger}\chi_1, \hspace{0.1cm} BlGl\chi_1^{\dagger}\chi_1, \hspace{0.1cm} BlWl\chi_1^{\dagger}\chi_1, \hspace{0.1cm} GlWl\chi_1^{\dagger}\chi_1, \hspace{0.1cm} Wl^2\chi_1^{\dagger}\chi_1 } $}\\
			\hline
			
			$\Psi^2\Phi X$&
			\multicolumn{2}{l|}{$\color{blue}{(N_f^2)BlLd^{\dagger}\chi_1, \hspace{0.1cm} (N_f^2)Gl Ld^{\dagger}\chi_1, \hspace{0.1cm} (N_f^2)Wl Ld^{\dagger}\chi_1}$}\\
			\hline
			
			\multirow{2}{*}{$\Psi^2\Phi^2 \mathcal{D}$}&
			\multicolumn{2}{l|}{$(2N_f^2)uu^{\dagger}\chi_1\chi_1^{\dagger}\mathcal{D}, \hspace{0.1cm} (2N_f^2)dd^{\dagger}\chi_1\chi_1^{\dagger}\mathcal{D}, \hspace{0.1cm} (4N_f^2)QQ^{\dagger}\chi_1\chi_1^{\dagger}\mathcal{D}, \hspace{0.1cm} (2N_f^2)LL^{\dagger}\chi_1\chi_1^{\dagger}\mathcal{D}, $}\\
			
			&
			\multicolumn{2}{l|}{$(N_f^2)elel^{\dagger}\chi_1\chi_1^{\dagger}\mathcal{D}, \hspace{0.1cm}  \color{blue}{(N_f^2)el d^{\dagger}H\chi_1\mathcal{D}, \hspace{0.1cm} (2N_f^2)L Q^{\dagger}H\chi_1\mathcal{D}}$}\\
			\hline
			
			\multirow{2}{*}{$\Psi^2\Phi^3$}&
			\multicolumn{2}{l|}{$\color{blue}{(4N_f^2) Q^{\dagger}d H\chi_1\chi_1^{\dagger}, \hspace{0.1cm} (4N_f^2) Q^{\dagger}u H^{\dagger}\chi_1\chi_1^{\dagger}, \hspace{0.1cm} (N_f^2) Q^{\dagger}d \chi_1^3, \hspace{0.1cm} (N_f^2) Qu^{\dagger}\chi_1^3, \hspace{0.1cm} (N_f^2) Q^{\dagger}el  H^2\chi_1,}$}\\
			
			&
			\multicolumn{2}{l|}{$\color{blue}{(2N_f^2) L^{\dagger}el H\chi_1\chi_1^{\dagger}, \hspace{0.1cm} (2N_f^2) L^{\dagger}d \chi_1^{\dagger}HH^{\dagger}, \hspace{0.1cm} (N_f^2) L^{\dagger}d H^{\dagger}\chi_1^2, \hspace{0.1cm} (2N_f^2) L^{\dagger}d (\chi_1^{\dagger})^2\chi_1, \hspace{0.1cm} (N_f^2) Lu^{\dagger}H^2\chi_1}$}\\
			\hline
			
			$\Phi^4\mathcal{D}^2$&
			\multicolumn{2}{l|}{$4\chi_1^2(\chi_1^{\dagger})^2\mathcal{D}^2, \hspace{0.1cm} 4HH^{\dagger}\chi_1\chi_1^{\dagger}\mathcal{D}^2, \hspace{0.1cm} \color{blue}{\chi_1^3H^{\dagger}\mathcal{D}^2}$ }\\
			\hline		    
	\end{tabular}}
	\caption{\small Lepto-Quark Model 1\,($\chi_1$): Operators of mass dimensions-5 and -6 excluding pure SM operators.}
	\label{table:leptoquark-model1-operators}
\end{table}

\begin{table}[h]
	\centering
	\renewcommand{\arraystretch}{1.8}
	{\tiny\begin{tabular}{|c|c|c|}
			\hline
			\multicolumn{3}{|c|}{\textbf{Mass Dimension-5}}\\
			\hline
			\textbf{Operator Class}&
			\multicolumn{2}{c|}{\textbf{Operators (in non-covariant form)}}\\
			\hline
			
			$\Psi^2\Phi^2$&
			\multicolumn{2}{l|}{$\color{blue}{\boxed{\frac{1}{2}(N_f^2-N_f)d^2H^{\dagger}\chi_2} }$}\\
			\hline
			\hline
			\multicolumn{3}{|c|}{\textbf{Mass Dimension-6}}\\
			\hline
			\textbf{Operator Class}&
			\multicolumn{2}{c|}{\textbf{Operators (in non-covariant form)}}\\
			\hline
			$\Phi^6$&
			\multicolumn{2}{l|}{$2\chi_2^3(\chi_2^{\dagger})^3, \hspace{0.1cm} 3\chi_2^2(\chi_2^{\dagger})^2HH^{\dagger}, \hspace{0.1cm} 2\chi_2\chi_2^{\dagger}H^2(H^{\dagger})^2$}\\
			\hline
			
			$\Phi^2X^2$&
			\multicolumn{2}{l|}{$\color{blue}{Bl^2\chi_2^{\dagger}\chi_2, \hspace{0.1cm} 2Gl^2\chi_2^{\dagger}\chi_2, \hspace{0.1cm} BlGl\chi_2^{\dagger}\chi_2, \hspace{0.1cm} BlWl\chi_2^{\dagger}\chi_2, \hspace{0.1cm} GlWl\chi_2^{\dagger}\chi_2, \hspace{0.1cm} Wl^2\chi_2^{\dagger}\chi_2 } $}\\
			\hline
			
			$\Psi^2\Phi X$&
			\multicolumn{2}{l|}{$\color{blue}{(N_f^2)BlQ el^{\dagger}\chi_2^{\dagger}, \hspace{0.1cm} (N_f^2)GlQ el^{\dagger}\chi_2^{\dagger}, \hspace{0.1cm} (N_f^2)WlQ el^{\dagger}\chi_2^{\dagger}, \hspace{0.1cm} (N_f^2)BlL u^{\dagger}\chi_2, \hspace{0.1cm} (N_f^2)GlL u^{\dagger}\chi_2, \hspace{0.1cm} (N_f^2)WlL u^{\dagger}\chi_2}$}\\
			\hline
			
			\multirow{2}{*}{$\Psi^2\Phi^2 \mathcal{D}$}&
			\multicolumn{2}{l|}{$(2N_f^2)uu^{\dagger}\chi_2\chi_2^{\dagger}\mathcal{D}, \hspace{0.1cm} (2N_f^2)dd^{\dagger}\chi_2\chi_2^{\dagger}\mathcal{D}, \hspace{0.1cm} (4N_f^2)QQ^{\dagger}\chi_2\chi_2^{\dagger}\mathcal{D}, \hspace{0.1cm} (N_f^2)elel^{\dagger}\chi_2\chi_2^{\dagger}\mathcal{D}, \hspace{0.1cm} (2N_f^2)LL^{\dagger}\chi_2\chi_2^{\dagger}\mathcal{D},  $}\\
			
			&
			\multicolumn{2}{l|}{$\color{blue}{(N_f^2)el d^{\dagger}H^{\dagger}\chi_2\mathcal{D}, \hspace{0.1cm} (N_f^2)el u^{\dagger}H\chi_2\mathcal{D}, \hspace{0.1cm} (2N_f^2)LQ^{\dagger}H^{\dagger}\chi_2\mathcal{D}}$}\\
			\hline
			
			\multirow{2}{*}{$\Psi^2\Phi^3$}&
			\multicolumn{2}{l|}{$\color{blue}{(4N_f^2) Q^{\dagger}dH\chi_2\chi_2^{\dagger}, \hspace{0.1cm} (4N_f^2) Q^{\dagger}uH^{\dagger}\chi_2\chi_2^{\dagger}, \hspace{0.1cm} (2N_f^2) Q^{\dagger}el \chi_2HH^{\dagger}, \hspace{0.1cm} (2N_f^2) Q^{\dagger}el \chi_2^2\chi_2^{\dagger}, \hspace{0.1cm} (2N_f^2) Lu^{\dagger} \chi_2^2\chi_2^{\dagger},}$}\\
			
			&
			\multicolumn{2}{l|}{$\color{blue}{(2N_f^2) L^{\dagger} el H\chi_2\chi_2^{\dagger}, \hspace{0.1cm} (N_f^2) L^{\dagger}d H^2\chi_2^{\dagger}, \hspace{0.1cm} (2N_f^2) Lu^{\dagger} \chi_2HH^{\dagger}}$}\\
			\hline
			
			$\Phi^4\mathcal{D}^2$&
			\multicolumn{2}{l|}{$4\chi_2^2(\chi_2^{\dagger})^2\mathcal{D}^2, \hspace{0.1cm} 4HH^{\dagger}\chi_2\chi_2^{\dagger}\mathcal{D}^2$ }\\
			\hline		    
	\end{tabular}}
	\caption{\small Lepto-Quark Model 2\,($\chi_2$): Operators of mass dimensions-5 and -6 excluding pure SM operators.}
	\label{table:leptoquark-model2-operators}
\end{table}

\begin{table}[h]
	\centering
	\renewcommand{\arraystretch}{1.8}
	{\tiny\begin{tabular}{|c|c|c|}
			\hline
			\multicolumn{3}{|c|}{\textbf{Mass Dimension-5}}\\
			\hline
			\textbf{Operator Class}&
			\multicolumn{2}{c|}{\textbf{Operators (in non-covariant form)}}\\
			\hline
			
			$\Psi^2\Phi^2$&
			\multicolumn{2}{l|}{$\color{blue}{ \frac{1}{2}(N_f^2+N_f)u^2(\Phi_1^{\dagger})^2, \hspace{0.1cm} (N_f^2)Ld^{\dagger}H^{\dagger}\Phi_1, \hspace{0.1cm} (N_f^2)Q^{\dagger}el H\Phi_1, \hspace{0.1cm} (N_f^2)Lu^{\dagger}H\Phi_1}$}\\
			\hline
			\hline
			\multicolumn{3}{|c|}{\textbf{Mass Dimension-6}}\\
			\hline
			\textbf{Operator Class}&
			\multicolumn{2}{c|}{\textbf{Operators (in non-covariant form)}}\\
			\hline
			$\Phi^6$&
			\multicolumn{2}{l|}{$\Phi_1^3(\Phi_1^{\dagger})^3, \hspace{0.1cm} \Phi_1^2(\Phi_1^{\dagger})^2HH^{\dagger}, \hspace{0.1cm} \Phi_1\Phi_1^{\dagger}H^2(H^{\dagger})^2$}\\
			\hline
			
			$\Phi^2X^2$&
			\multicolumn{2}{l|}{$\color{blue}{Bl^2\Phi_1^{\dagger}\Phi_1, \hspace{0.1cm} 2Gl^2\Phi_1^{\dagger}\Phi_1, \hspace{0.1cm} BlGl\Phi_1^{\dagger}\Phi_1, \hspace{0.1cm} Wl^2\Phi_1^{\dagger}\Phi_1}$}\\
			\hline
			
			$\Psi^2\Phi X$&
			\multicolumn{2}{l|}{$\color{blue}{\frac{1}{2}(N_f^2+N_f)Brd^2\Phi_1, \hspace{0.1cm} (N_f^2)Grd^2\Phi_1 }$}\\
			\hline
			
			\multirow{2}{*}{$\Psi^2\Phi^2 \mathcal{D}$}&
			\multicolumn{2}{l|}{$(2N_f^2)uu^{\dagger}\Phi_1\Phi_1^{\dagger}\mathcal{D}, \hspace{0.1cm} (2N_f^2)dd^{\dagger}\Phi_1\Phi_1^{\dagger}\mathcal{D}, \hspace{0.1cm} (2N_f^2)QQ^{\dagger}\Phi_1\Phi_1^{\dagger}\mathcal{D}, \hspace{0.1cm} (N_f^2)LL^{\dagger}\Phi_1\Phi_1^{\dagger}\mathcal{D},$}\\
			
			&
			\multicolumn{2}{l|}{$(N_f^2)elel^{\dagger}\Phi_1\Phi_1^{\dagger}\mathcal{D}, \hspace{0.1cm}   \color{blue}{(N_f^2)QdH^{\dagger}\Phi_1\mathcal{D}, \hspace{0.1cm} (N_f^2)LuH\Phi_1^{\dagger}\mathcal{D}}$}\\
			\hline
			
			\multirow{2}{*}{$\Psi^2\Phi^3$}&
			\multicolumn{2}{l|}{$\color{blue}{(2N_f^2)Q^{\dagger}dH\Phi_1\Phi_1^{\dagger}, \hspace{0.1cm} (2N_f^2)Q^{\dagger}uH^{\dagger}\Phi_1\Phi_1^{\dagger}, \hspace{0.1cm} (N_f^2)L^{\dagger}elH\Phi_1\Phi_1^{\dagger}, \hspace{0.1cm} (N_f^2)LQH^2\Phi_1^{\dagger}, \hspace{0.1cm}}$}\\
			
			&
			\multicolumn{2}{l|}{$\color{blue}{\boxed{\frac{1}{2}(N_f^2-N_f)d^2\Phi_1^2\Phi_1^{\dagger}}, \hspace{0.1cm} \boxed{\frac{1}{2}(N_f^2-N_f)d^2\Phi_1HH^{\dagger}}, \hspace{0.1cm} \boxed{\frac{1}{2}(N_f^2-N_f)Q^2\Phi_1(H^{\dagger})^2}}$}\\
			\hline
			
			$\Phi^4\mathcal{D}^2$&
			\multicolumn{2}{l|}{$2\Phi_1^2(\Phi_1^{\dagger})^2\mathcal{D}^2, \hspace{0.1cm} 2HH^{\dagger}\Phi_1\Phi_1^{\dagger}\mathcal{D}^2 $}\\
			\hline		    
	\end{tabular}}
	\caption{\small Lepto-Quark Model 3\,($\Phi_1$): Operators of mass dimensions-5 and -6 excluding pure SM operators.}
	\label{table:leptoquark-model3-operators}
\end{table}
\clearpage
\begin{table}[h]
	\centering
	\renewcommand{\arraystretch}{1.8}
	{\tiny\begin{tabular}{|c|c|c|}
			\hline
			\multicolumn{3}{|c|}{\textbf{Mass Dimension-5}}\\
			\hline
			\textbf{Operator Class}&
			\multicolumn{2}{c|}{\textbf{Operators (in non-covariant form)}}\\
			\hline
			
			$\Psi^2\Phi^2$&
			\multicolumn{2}{l|}{$\color{blue}{\frac{1}{2}(N_f^2+N_f)d^2(\Phi_2^{\dagger})^2, \hspace{0.1cm} (N_f^2)Ld^{\dagger}H\Phi_2 }$}\\
			\hline
			\hline
			\multicolumn{3}{|c|}{\textbf{Mass Dimension-6}}\\
			\hline
			\textbf{Operator Class}&
			\multicolumn{2}{c|}{\textbf{Operators (in non-covariant form)}}\\
			\hline
			$\Phi^6$&
			\multicolumn{2}{l|}{$\Phi_2^3(\Phi_2^{\dagger})^3, \hspace{0.1cm} \Phi_2^2(\Phi_2^{\dagger})^2HH^{\dagger}, \hspace{0.1cm} \Phi_2\Phi_2^{\dagger}H^2(H^{\dagger})^2$}\\
			\hline
			
			$\Phi^2X^2$&
			\multicolumn{2}{l|}{$\color{blue}{Bl^2\Phi_2^{\dagger}\Phi_2, \hspace{0.1cm} 2Gl^2\Phi_2^{\dagger}\Phi_2, \hspace{0.1cm} BlGl\Phi_2^{\dagger}\Phi_2, \hspace{0.1cm} Wl^2\Phi_2^{\dagger}\Phi_2 } $}\\
			\hline
			
			\multirow{2}{*}{$\Psi^2\Phi X$}&
			\multicolumn{2}{l|}{$\color{blue}{(N_f^2)Gl\Phi_2Q^2, \hspace{0.1cm} \boxed{\frac{1}{2}(N_f^2-N_f)Bl\Phi_2Q^2}, \hspace{0.1cm} \frac{1}{2}(N_f^2+N_f)Wl\Phi_2Q^2, \hspace{0.1cm} (2N_f^2)Gl\Phi_2^{\dagger} u^{\dagger}d^{\dagger}, \hspace{0.1cm} (N_f^2)Bl\Phi_2^{\dagger} u^{\dagger}d^{\dagger},}$}\\
			
			&
			\multicolumn{2}{l|}{$\color{blue}{(N_f^2)Bl\Phi_2^{\dagger}LQ, \hspace{0.1cm} (N_f^2)Wl\Phi_2^{\dagger}LQ, \hspace{0.1cm} (N_f^2)Gl\Phi_2^{\dagger}LQ, \hspace{0.1cm} (N_f^2)Bl\Phi_2el^{\dagger}u^{\dagger}, \hspace{0.1cm} (N_f^2)Gl\Phi_2el^{\dagger}u^{\dagger}}$}\\
			\hline
			
			\multirow{3}{*}{$\Psi^2\Phi^2 \mathcal{D}$}&
			\multicolumn{2}{l|}{$(2N_f^2)uu^{\dagger}\Phi_2\Phi_2^{\dagger}\mathcal{D}, \hspace{0.1cm} (2N_f^2)dd^{\dagger}\Phi_2\Phi_2^{\dagger}\mathcal{D}, \hspace{0.1cm} (2N_f^2)QQ^{\dagger}\Phi_2\Phi_2^{\dagger}\mathcal{D}, \hspace{0.1cm} (N_f^2)elel^{\dagger}\Phi_2\Phi_2^{\dagger}\mathcal{D}, \hspace{0.1cm} (N_f^2)LL^{\dagger}\Phi_2\Phi_2^{\dagger}\mathcal{D}, $}\\
			
			&
			\multicolumn{2}{l|}{$\color{blue}{(N_f^2)QdH\Phi_2\mathcal{D}, \hspace{0.1cm} (N_f^2)QuH^{\dagger}\Phi_2\mathcal{D}, \hspace{0.1cm} (N_f^2)Q el H\Phi_2^{\dagger}\mathcal{D}, \hspace{0.1cm} (N_f^2)LdH\Phi_2^{\dagger}\mathcal{D}, \hspace{0.1cm} (N_f^2)LuH^{\dagger}\Phi_2^{\dagger}\mathcal{D}, \hspace{0.1cm}}$}\\
			
			&
			\multicolumn{2}{l|}{$\color{blue}{(N_f^2)L^{\dagger}Q\Phi_2^2\mathcal{D}, \hspace{0.1cm} (N_f^2)el^{\dagger}d\Phi_2^2\mathcal{D}}$}\\
			\hline
			
			\multirow{3}{*}{$\Psi^2\Phi^3$}&
			\multicolumn{2}{l|}{$\color{blue}{(2N_f^2)Q^{\dagger}dH\Phi_2\Phi_2^{\dagger}, \hspace{0.1cm} (2N_f^2)Q^{\dagger}uH^{\dagger}\Phi_2\Phi_2^{\dagger}, \hspace{0.1cm} (N_f^2)L^{\dagger}elH\Phi_2\Phi_2^{\dagger}, \hspace{0.1cm} (N_f^2)LQ\Phi_2(\Phi_2^{\dagger})^2, \hspace{0.1cm} (N_f^2)el\,u \Phi_2(\Phi_2^{\dagger})^2, \hspace{0.1cm}}$}\\
			
			&
			\multicolumn{2}{l|}{$\color{blue}{ \frac{1}{2}(N_f^2+N_f)Q^2\Phi_2^2\Phi_2^{\dagger}, \hspace{0.1cm} (N_f^2)u\,d\Phi_2^2\Phi_2^{\dagger}, \hspace{0.1cm} (2N_f^2)LQ\Phi_2^{\dagger}HH^{\dagger}, \hspace{0.1cm} (N_f^2)el\,u\Phi_2^{\dagger}HH^{\dagger}, \hspace{0.1cm} (N_f^2)u\,d\Phi_2HH^{\dagger},}$}\\
			
			&
			\multicolumn{2}{l|}{$\color{blue}{(N_f^2)Q^2\Phi_2HH^{\dagger}}$}\\
			\hline
			
			$\Phi^4\mathcal{D}^2$&
			\multicolumn{2}{l|}{$2\Phi_2^2(\Phi_2^{\dagger})^2\mathcal{D}^2, \hspace{0.1cm} 2HH^{\dagger}\Phi_2\Phi_2^{\dagger}\mathcal{D}^2$ }\\
			\hline		    
	\end{tabular}}
	\caption{\small Lepto-Quark Model 4\,($\Phi_2$): Operators of mass dimensions-5 and -6 excluding pure SM operators.}
	\label{table:leptoquark-model4-operators}
\end{table}

\subsubsection*{\underline{Interactions of a light dark matter candidate with SM fields}}

Recently, models built of the lighter SM fermions, after having integrated out the heavy particles, i.e., the $SU(2)$ gauge bosons, the Higgs and the top quark, with the gauge group $SU(3)_C\otimes U(1)_{em}$ and the corresponding gauge fields the gluons ($G^a_{\mu \nu}$) and the photon ($F_{\mu\nu}$) have sparked some interest. Also known as Low Energy Effective Field Theory (LEFT), \cite{Jenkins:2017dyc,Jenkins:2017jig,Aebischer:2017gaw,Dekens:2019ept} these below electroweak scale models provide a nice platform to study interactions between weakly interacting light dark matter and the SM fermions \cite{Brod:2017bsw}. We consider two distinct scenarios \cite{Brod:2017bsw}, one where the DM candidate is a complex scalar and another where it is a fermion. Since we are working in the Weyl basis we separately define the left and right chiral parts. In each of these cases, there is an extra $U(1)_D$ charge assigned to the dark matter candidate, see Table~\ref{table:DM_extension_of_broken_SM}. We have catalogued the effective operators for the two cases in Tables~\ref{table:dm_model1-operators} and \ref{table:dm_model2-operators} respectively.

\begin{table}[h]
	\centering
	\renewcommand{\arraystretch}{1.8}
	{\tiny\begin{tabular}{|c|c|c|c|c|c|}
			\hline
			\textbf{Model No.}&
			\textbf{DM-Candidate}&
			$SU(3)_C$&
			$U(1)_{em}$&
			$U(1)_{D}$&
			\textbf{Spin}\\
			\hline
			
			1&
			$\phi$&
			1&
			0&
			1&
			0\\
			\hline
			\multirow{2}{*}{2}&
			$\chi_L$&
			1&
			0&
			1&
			1/2\\
			
			&
			$\chi_R$&
			1&
			0&
			1&
			1/2\\
			
			\hline
	\end{tabular}}
	\caption{\small Quantum numbers of various Dark-Matter fields under the broken SM gauge groups extended by $U(1)_D$.}
	\label{table:DM_extension_of_broken_SM}
\end{table}

\begin{table}[h]
	\centering
	\renewcommand{\arraystretch}{1.7}
	{\tiny\begin{tabular}{|c|c|c|}
			\hline
			\multicolumn{3}{|c|}{\textbf{Mass Dimension-5}}\\
			\hline
			\textbf{Operator Class}&
			\multicolumn{2}{c|}{\textbf{Operators (in non-covariant form)}}\\
			\hline
			\multirow{4}{*}{$\Phi^2\Psi^2$}&
			\multicolumn{2}{l|}{\color{blue}{$\phi^\dagger\phi \nu_{{e}_{L}}^2,\;\phi^\dagger \phi \nu_{{\mu}_{L}}^2,\;\phi^\dagger \phi \nu_{{\tau}_{L}}^2,\;\phi^\dagger\phi \nu_{{e}_{L}}\nu_{{\mu}_{L}},\; \phi^\dagger\phi \nu_{{e}_{L}}\nu_{{\tau}_{L}},\phi^\dagger\phi\nu_{{\mu}_{L}}\nu_{{\tau}_{L}},\;\phi^\dagger\phi e_R e_L^\dagger,\;\phi^\dagger\phi \mu_R \mu_L^\dagger,$}}\\
			& \multicolumn{2}{l|}{\color{blue}{$\phi^\dagger\phi \tau_R \tau_L^\dagger,\;\phi^\dagger\phi \mu_R e_L^\dagger,\; \phi^\dagger\phi \tau_R  e_L^\dagger,\;\phi^\dagger\phi \tau_R \mu_L^\dagger,\; \phi^\dagger\phi \mu_L e_R^\dagger,\;\phi^\dagger\phi \tau_L e_R^\dagger,\; \phi^\dagger\phi \tau_L  \mu_R^\dagger,\;\phi^\dagger\phi e_L \tau_R^\dagger,$}}\\
			& 
			\multicolumn{2}{l|}{\color{blue}{$\phi^\dagger\phi b_R b_L^\dagger,\;\phi^\dagger\phi s_R b_L^\dagger,\;\phi^\dagger\phi d_R b_L^\dagger,\;\phi^\dagger\phi d_L  b_R^\dagger,\;\phi^\dagger\phi s_L b_R^\dagger,\;\phi^\dagger\phi c_R c_L^\dagger,\;\phi^\dagger\phi s_L d_R^\dagger,\;\phi^\dagger\phi u_R c_L^\dagger, $}}\\
			& 
			\multicolumn{2}{l|}{\color{blue}{$\phi^\dagger\phi u_L c_R^\dagger,\;\phi^\dagger\phi d_R d_L^\dagger,\; \phi^\dagger\phi s_R  d_L^\dagger,\;\phi^\dagger\phi s_R s_L^\dagger,\;\phi^\dagger\phi s_L d_R^\dagger,\; \phi^\dagger\phi u_R u_L^\dagger $}}\\
			\hline
			\multicolumn{3}{|c|}{\textbf{Mass Dimension-6}}\\
			\hline
			\textbf{Operator Class}&
			\multicolumn{2}{c|}{\textbf{Operators (in non-covariant form)}}\\
			\hline
			
			$\Phi^2X^2$&
			\multicolumn{2}{l|}{$Fl^2\phi^\dagger\phi,\; Gl^2\phi^\dagger\phi$}\\
			\hline
			$\Phi^6$&
			\multicolumn{2}{l|}{$(\phi^\dagger)^3\phi^3$}\\
			\hline
			$\Phi^4\mathcal{D}^2$&
			\multicolumn{2}{l|}{$(\phi^\dagger)^2\phi^2\mathcal{D}$}\\
			\hline
			\multirow{6}{*}{$\Phi^2\Psi^2\mathcal{D}$}&
			\multicolumn{2}{l|}{\color{blue}{$\phi^\dagger\phi b_L b{}^\dagger_L\mathcal{D},\;\phi^\dagger\phi c_L c_L^\dagger\mathcal{D},\;\phi^\dagger\phi c_R c_R^\dagger\mathcal{D},\;\phi^\dagger\phi d_L d_L^\dagger\mathcal{D},\;\phi^\dagger\phi d_R d_R^\dagger\mathcal{D},\;\phi^\dagger\phi s_L s_L^\dagger\mathcal{D},\;\phi^\dagger\phi u_L u_L^\dagger\mathcal{D}, $}} \\
			&
			\multicolumn{2}{l|}{\color{blue}{$\phi^\dagger\phi s_R s_R^\dagger\mathcal{D},\;\phi^\dagger\phi u_R u_R^\dagger\mathcal{D},\;\phi^\dagger\phi s_L b_L{^\dagger}\mathcal{D},\;\phi^\dagger\phi d_R b_R^\dagger\mathcal{D},\;\phi^\dagger\phi u_L c_L^\dagger\mathcal{D},\;\phi^\dagger\phi d_L b_L^\dagger\mathcal{D},\;\phi^\dagger\phi \mu_L e_L^\dagger\mathcal{D},$}}\\
			&
			\multicolumn{2}{l|}{\color{blue}{$\phi^\dagger\phi \tau_L e_L^\dagger\mathcal{D},\;\phi^\dagger\phi \mu_R e_R^\dagger\mathcal{D},\;\phi^\dagger\phi \tau_R e_R^\dagger\mathcal{D},\;\phi^\dagger\phi \tau_L \mu_L^\dagger\mathcal{D},\; \phi^\dagger\phi \tau_R \mu_R^\dagger\mathcal{D},\;\phi^\dagger\phi \nu_{\mu_L} \nu_{e_L}^\dagger\mathcal{D},\;\phi^\dagger\phi \nu_{\tau_L} \nu_{\mu_L}^\dagger\mathcal{D},$}}\\
			& 
			\multicolumn{2}{l|}{$\phi^\dagger\phi s_R d_R^\dagger\mathcal{D},\;\phi^\dagger\phi s_L d_L^\dagger\mathcal{D},\;\phi^\dagger\phi u_R c_R^\dagger\mathcal{D},\;\phi^\dagger\phi s_R b_R^\dagger\mathcal{D},\;\phi^\dagger\phi b_R b_R^\dagger\mathcal{D},\;\phi^\dagger\phi e_L e_L^\dagger\mathcal{D},\;\phi^\dagger\phi e_R e_R^\dagger\mathcal{D},$}\\
			& 
			\multicolumn{2}{l|}{$\phi^\dagger\phi \mu_L \mu_L^\dagger\mathcal{D},\;\phi^\dagger\phi \mu_R \mu_R^\dagger\mathcal{D},\;\phi^\dagger\phi \tau_L \tau_L^\dagger\mathcal{D},\;\phi^\dagger\phi \tau_R\tau_R^\dagger\mathcal{D},\;\phi^\dagger\phi \nu_{\mu_L} \nu_{\mu_L}^\dagger\mathcal{D},\;\phi^\dagger\phi \nu_{\tau_L} \nu_{\tau_L}^\dagger\mathcal{D},$}\\
			&
			\multicolumn{2}{l|}{$\phi^\dagger\phi\nu_{e_L} \nu_{e_L}^\dagger\mathcal{D},\;$\color{blue}{$\phi^\dagger\phi \nu_{\tau_L} \nu_{e_L}^\dagger\mathcal{D}$}}\\
			\hline
	\end{tabular}}
	\caption{\small Model 1\,($\phi$): Operators of mass dimensions-5 and -6 excluding pure SM operators.}
	\label{table:dm_model1-operators}
\end{table}

\begin{table}[h]
	\centering
	\renewcommand{\arraystretch}{1.7}
	{\tiny\begin{tabular}{|c|c|c|}
			\hline
			\multicolumn{3}{|c|}{\textbf{Mass Dimension-5}}\\
			\hline
			\textbf{Operator Class}&
			\multicolumn{2}{c|}{\textbf{Operators (in non-covariant form)}}\\
			\hline
			\multirow{1}{*}{$\Phi^2\Psi^2$}&
			\multicolumn{2}{l|}{\color{blue}{$Fr{\chi}_L^\dagger\chi_R,$}}\\
			\hline
			\multicolumn{3}{|c|}{\textbf{Mass Dimension-6}}\\
			\hline
			\textbf{Operator Class}&
			\multicolumn{2}{c|}{\textbf{Operators (in non-covariant form)}}\\
			\hline	
			\multirow{6}{*}{$\Psi^4$}&
			\multicolumn{2}{l|}{\color{blue}{$f^2\chi_L^\dagger\chi_R,\hspace{0.2cm}f_1\,f_2\chi_L^\dagger\chi_R, \hspace{1cm} f,f_1,f_2 \in \{\nu_{{e}_{L},\nu_{{\mu}_{L}},\nu_{{\tau}_{L}}} \}$}}\\
			
			&
			\multicolumn{2}{l|}{$q^{\dagger}_{L,R}q^{}_{L,R}\chi_{L,R}^\dagger\chi^{}_{L,R},\hspace{0.2cm} \hspace{1cm} q \in \{u,d,c,s,b \}$}\\
			
			&
			\multicolumn{2}{l|}{\color{blue}{$q^{\dagger}_{L,R}q^{}_{R,L}\chi_{L,R}^\dagger\chi^{}_{R,L},\hspace{0.2cm} \hspace{1cm} q \in \{u,d,c,s,b \}$}}\\
			
			&
			\multicolumn{2}{l|}{$l^{\dagger}_{L,R}l^{}_{L,R}\chi_{L,R}^\dagger\chi^{}_{L,R},\hspace{0.2cm} \hspace{1cm} l \in \{e,\mu, \tau \}$}\\
			
			&
			\multicolumn{2}{l|}{\color{blue}{$l^{\dagger}_{L,R}l^{}_{R,L}\chi_{L,R}^\dagger\chi^{}_{R,L},\hspace{0.2cm} \hspace{1cm} l \in \{e,\mu, \tau \}$}}\\
			
			&
			\multicolumn{2}{l|}{\color{blue}{$\chi_{L,R}^2\chi_{L,R}^2$}}\\
			
			\hline
	\end{tabular}}
	\caption{\small Model 2\,($\chi_{L,R}$): Operators of mass dimensions-5 and -6 excluding pure SM operators.}
	\label{table:dm_model2-operators}
\end{table} 
\noindent
In the second model, the broken SM is extended by fermions ($\chi_L, \chi_R$). The imposition of a $U(1)_D$ symmetry removes the terms having odd powers of $\chi_{L,R}$ and $\chi_{L,R}^{\dagger}$. The remaining terms contain one operator at mass dimension-5 and 275 operators at dimension-6. Since all the dimension-6 operators have a similar form we have only given a schematic form of these operators in Table~\ref{table:dm_model2-operators}.

%% file: FlavourStructure.tex
\section{Covariant form of operators and their explicit flavour dependence}\label{sec:flavour-structure}

Explicit flavour dependence in higher dimensional operators has always been of interest and drawn the attention of active researchers \cite{Ecker:1983uh,Buchmuller:1985jz,PhysRevD.40.1521,Pospelov:1996fq,Kovalenko:2009td,Chakrabortty:2015zpm,Guadagnoli:2010sd,Seon:2011ni,Barry:2013xxa,Chrzaszcz:2013uz,Hambye:2013jsa,Esteves:2010si,Helo:2010cw,Blanke:2011ry,Das:2012ii,Geng:2016auy,Silvestrini:2018dos,Davidson:2018kud,Hurth:2019ula}. Our program \textbf{GrIP} offers the user an avenue to gain familiarity with the possible ways in which the flavour and symmetry indices can be contracted to construct independent sets of operators. Among the models studied in this paper, the operators could always be categorized into distinct classes. Of those classes we can identify the following two classes where explicit flavour dependence becomes significant:
(i) $\Psi^4$ and (ii) $\Psi^2\Phi X$.

As the program output does not show the flavour and symmetry indices explicitly, it is difficult to explain why and how the number of operators varies as a function of the number of fermion flavours ($N_f$). To do so we are required to rewrite those operators ($\mathcal{O}_i$) in covariant form. 

\begin{itemize}
	\item $\mathcal{O}_i \in \Psi^4$: Here, we have focussed our attention on the $\Psi^4$ sector of Table~\ref{table:sm-dim-5-6-op}.
	
	Now, the fermions in this model can be divided into 4 categories based on their internal quantum numbers: 
	\begin{enumerate}[(a)]
		\item $SU(3)$ triplet and $SU(2)$ doublet: $Q$,  
		\item $SU(3)$ triplet and $SU(2)$ singlet: $u,\,d$,
		\item $SU(3)$ singlet and $SU(2)$ doublet: $L$,
		\item $SU(3)$ singlet and $SU(2)$ singlet: $el$.
	\end{enumerate}

	We can have three possible covariant structures for the $\Psi^4$ operators constituted of the above:  $(i)$ $(\bar{\Psi}\gamma_{\mu}\Psi)(\bar{\Psi}\gamma^{\mu}\Psi)$, $(ii)$ $(\bar{\Psi}\gamma_{\mu}\tau^{I}\Psi)(\bar{\Psi}\gamma^{\mu}\tau^{I}\Psi)$ and $(iii)$  $(\bar{\Psi}\gamma_{\mu}T^{a}\Psi)(\bar{\Psi}\gamma^{\mu}T^{a}\Psi)$\footnote{$\tau^{I}$ and $T^a$ are the generators of $SU(2)$ and $SU(3)$ group respectively.}. But all of these need not be independent. These may be connected through Fierz identities. If all three structures are possible as in the case of (a), then only two of them are independent. Thus, the independent structures for fermions of types (a)-(d) are:
	\begin{enumerate}[(1)]
		\item $(\bar{\Psi}\gamma_{\mu}\Psi)(\bar{\Psi}\gamma^{\mu}\Psi)$ for (b), (c), (d) and  
		\item $(\bar{\Psi}\gamma_{\mu}\Psi)(\bar{\Psi}\gamma^{\mu}\Psi)$,  $(\bar{\Psi}\gamma_{\mu}\tau^{I}\Psi)(\bar{\Psi}\gamma^{\mu}\tau^{I}\Psi)$ for (a).
	\end{enumerate}

    In case (1), if $\Psi$ transforms non-trivially under $SU(3)$ and $SU(2)$ then there are $\frac{1}{2}N_f^2(N_f^2+1)$ number of symmetric combinations. If $\Psi$ is an overall singlet under these non-abelian groups, then the number of symmetric combinations is\\ $[\frac{1}{2}N_f(N_f+1)]\times[\frac{1}{2}N_f(N_f+1)]$. For case (2), the counting gets doubled compared to case (1) as there exist two independent structures, i.e., the number becomes $2\times[\frac{1}{2}N_f^2(N_f^2+1)]$ as $\Psi$ transforms non-trivially under the non-abelian groups.
    
    We must also remark that when all four $\Psi$'s are different then we get a factor of $N_f^4$ or $2N_f^4$ depending on the number of independent structures.
    
	\item $\mathcal{O}_i \in \Psi^2\Phi X$: Within this class, we have considered three operators 
	$N_f^2Wl_{1}\Delta_1L_1^2$,
	$\frac{1}{2}\left(N_f^2+N_f\right)Wl_{2}\Delta_2L_1^2$ and $\frac{1}{2}\left(N_f^2-N_f\right)Bl\Delta_1L_1^2 $ from Table~\ref{table:mlrsm-dim-6-op}. Now, focussing on $Wl_{i}\Delta_iL_j^2$, we see that since $\Delta$ is Lorentz scalar and $Wl$ transforms as $(1,0)$ under $SU(2)_l\times SU(2)_r$, $L_j^2$ must be a symmetric product of $(\frac{1}{2},0)$ and $(\frac{1}{2},0)$, i.e., in Lorentz indices. 
	Now, if $i=j$, then $L_j^2$ can be yield both a symmetric and an  antisymmetric product in internal symmetry indices. Thus we must include both such structures which gives $\frac{1}{2}N_f(N_f+1)$ and $\frac{1}{2}N_f(N_f-1)$ symmetric and antisymmetric combinations respectively, i.e., a total number $N_f^2$.
	Now, if $i\neq j$, then $L_j^2$ must appear only as a symmetric combination in internal symmetry indices. Otherwise, the full term will be antisymmetric in internal indices and vanish identically. So, the total number of combinations is $\frac{1}{2}N_f(N_f+1)$.
	
	Following a similar line of reasoning, in the case of $Bl \Delta_1L_1^2$ the number of combinations is $\frac{1}{2}N_f(N_f-1)$ due to the antisymmetric nature of ($L_1\Delta_1L_1$) in flavour indices. 
\end{itemize}

\clearpage

%% file: conclusion.tex
\section{Conclusion and Remarks}

In this paper, our primary objective has been to elucidate the construction of group invariant polynomials of the quantum fields. The relevance of such invariants in high energy physics stems from the fact that most of the phenomenological models contain certain quantum fields as the degrees of freedom that are attributed unique transformation properties under the assigned symmetries. The phenomenological explorations within a proposed model rely on Feynman vertices and for that information of a complete Lagrangian is necessary. Recognition of the fact that the individual constituents of the Lagrangian are invariants under all the symmetries of the model (global, gauge as well as spacetime) motivates us to delve deeper into their construction. The fundamental ingredients for such development are the characters of different representations and Haar measure corresponding to the connected compact groups representing the symmetries of the adopted model.  In the first part of this paper, we have outlined the detailed mathematical steps in an algorithmic way to lay the platform for the  central part of this work which is based around developing the $\text{\texttt{Mathematica}}^{\tiny\textregistered}$  package, \textbf{GrIP}. We have explained all the necessary information such as how to install the package, and to create an input model file. We have also shown how to translate the output of the program into a working Lagrangian through some example scenarios. One of the most essential guiding principles behind the analysis has been to ensure that the required input be as minimal as possible and the output as simple as it can be. This is reflected in the character computation where the only input needed is the Dynkin label for the corresponding representation and again in the development of \textbf{GrIP} where the input only comprises of the quantum fields corresponding to the particles of a given model and their respective transformations under various symmetry groups. We would like to mention that we have also provided a separate sub-program \textbf{CHaar} which computes the character of representation using the suitable Dynkin label provided by the user.

\noindent
Adopting the output of \textbf{GrIP} for any phenomenological analysis is based on its identification as the Lagrangian. So, we have kept the provision to identify the operators at different mass dimensions, for different values of overall baryon and lepton numbers, and the different number of fermion flavours. All this has been exemplified through both non-supersymmetric as well as supersymmetric scenarios. We have shown how in addition to being a Lagrangian builder, \textbf{GrIP} can also act as a search tool for rare processes mediated by varying degrees of baryon and lepton number violations. At the same time, it can also enable the user to filter out the output based on the conservation or violation of externally imposed global symmetry. Both of these targets have been achieved through suitably defined functions within the package and their working principles have been emphasized through relevant examples. The contemporary state of particle physics has transcended past the Standard Model (SM). Keeping this in mind, we have employed our package in laying the groundwork for the first step towards a model-independent comparative study of different BSM scenarios. We have mentioned the need for an extension of the SM particle content through the addition of certain infrared degrees of freedom and paved the way for the study of BSM-EFT. We have further highlighted this by constructing and tabulating the higher dimensional effective operators for a plethora of extended SM scenarios ranging from $SU(2)$ scalars transforming in various multiplets to lepto-quarks, and some possible dark matter particles. Lastly, we have commented on the distinct flavour structures of a few classes of operators and shown how these can be understood based on group-theoretic principles.\\

\noindent
In the future, we will further explore the avenue of BSM-EFT. We will employ esoteric tools based on statistical methods and Effective Field Theory to pinpoint the new physics proposals, most favoured by experimental data. 

\section*{Acknowledgements}

We acknowledge the highly insightful discussions with Santosh Nadimpalli. We thank Arjun Bagchi, Shamik Banerjee, Diptarka Das, Nilay Kundu, Amitava Raychaudhuri for useful discussions, and their comments on the draft. This work is supported by the Science and Engineering Research Board, Government of India, under the agreements SERB/PHY/2016348 (Early Career Research Award) and SERB/PHY/2019501 (MATRICS) and Initiation Research Grant, agreement number IITK/PHY/2015077, by IIT Kanpur.

%% file: appendix_theory.tex
\section{Appendix}
\subsection{Non-supersymmetric Models}\label{appendix:nonsusy-models}
Below we shall discuss a couple of popular BSM scenarios and highlight the utility of the Hilbert Series in the construction of their operator sets.  
\subsubsection*{\underline{Minimal Left-Right Symmetric Model (MLRSM)}}\label{subsec:mlrsm}
The Minimal Left-Right Symmetric Model (MLRSM) is an intricate extension of the Standard Model where $P$ and $CP$ symmetries are on  par with the gauge symmetries \cite{Mohapatra:1974gc,Senjanovic:1975rk, Mohapatra:1979ia,Mohapatra:1980yp,Gunion:1989in,Deshpande:1990ip,Duka:1999uc,Senjanovic:1978ee,Grifols:1978wk,Olness:1985bg,Frank:1991sy,Chang:1992bg,Maalampi:1993tj,Gluza:1994ad,Bhattacharyya:1995nt,Boyarkina:2000bn,Barenboim:2001vu,Gogoladze:2003bb,Azuelos:2004mwa,Kiers:2005gh,Jung:2008pz,Guadagnoli:2010sd,Blanke:2011ry,Mohapatra:2013cia,Aydemir:2014ama,Maiezza:2015lza,Maiezza:2015qbd,Dev:2016dja,Chakrabortty:2010rq,Chakrabortty:2010zk,Maiezza:2016ybz,Deppisch:2017xhv,Dev:2018foq}. In this model the right handed neutrino is accommodated naturally. The field content and the transformation properties of those fields under the gauge groups $SU(3)_C\otimes SU(2)_L\otimes SU(2)_R\otimes U(1)_{B-L}$ and their spins are enlisted in Table~\ref{table:mlrsm-quantum-no}. We have also given the gauge group characters corresponding to each field in the table. The Lorentz characters can be conveniently obtained based on the spin of each particle. The Hilbert Series can be obtained by following a similar strategy as in the case of 2HDM. The Hilbert Series output and the corresponding covariant form of the renormalizable operators has been listed in Tables~\ref{table:mlrsm-renorm-output} and \ref{table:mlrsm-renorm-output-2}. The categorization of the non-covariant HS output of mass dimension-6 has been shown in Tables~\ref{table:mlrsm-dim-6-op} and \ref{table:mlrsm-dim-6-op-2}. See \cite{Anisha:2019nzx} for their covariant forms. All the operators are provided for general $N_f$. Some of the operators vanish for $N_f = 1$ and these have been highlighted by enclosing them in boxes in Tables~\ref{table:mlrsm-dim-6-op} and \ref{table:mlrsm-dim-6-op-2}. 

\begin{table}[h]
	\centering
	\renewcommand{\arraystretch}{1.8}
	{\tiny\begin{tabular}{|c|c|c|c|c|c|c|}
			\hline
			\textbf{MLRSM Fields}&
			$SU(3)_C$&
			$SU(2)_L$&
			$SU(2)_R$&
			$U(1)_{B-L}$&
			\textbf{Spin}&
			\textbf{Gauge Group Characters}\\
			\hline
			
			$\Phi$&
			1&
			2&
			2&
			0&
			0&
			$\chi_{({SU(2)_L})_{2}}\cdot \chi_{({SU(2)_R})_{2}}$\\
			
			$\Delta_L$&
			1&
			3&
			1&
			2&
			0&
			$\chi_{({SU(2)_L})_{3}}\cdot \chi_{({U(1)})_{2}}$\\
			
			$\Delta_R$&
			1&
			1&
			3&
			2&
			0&
			$\chi_{({SU(2)_R})_{3}}\cdot \chi_{({U(1)})_{2}}$\\
			
			$Q^p_L$&
			3&
			2&
			1&
			1/3&
			1/2&
			$\chi_{({SU(3)})_{3}}\cdot \chi_{({SU(2)_L})_{2}}\cdot \chi_{({U(1)})_{1/3}}$\\
			
			$Q^p_R$&
			3&
			1&
			2&
			1/3&
			1/2&
			$\chi_{({SU(3)})_{3}}\cdot \chi_{({SU(2)_R})_{2}}\cdot \chi_{({U(1)})_{1/3}}$\\
			
			$L^p_L$&
			1&
			2&
			1&
			-1&
			1/2&
			$\chi_{({SU(2)_L})_{2}}\cdot \chi_{({U(1)})_{\text{-}1}}$\\
			
			$L^p_R$&
			1&
			1&
			2&
			-1&
			1/2&
			$\chi_{({SU(2)_R})_{2}}\cdot \chi_{({U(1)})_{\text{-}1}}$\\
			
			$B_{\mu\nu}$&
			1&
			1&
			1&
			0&
			1&
			$1$\\
			
			$W^{I}_{L\mu\nu}$&
			1&
			3&
			1&
			0&
			1&
			$\chi_{({SU(2)_L})_{2}}$\\
			
			$W^{I}_{R\mu\nu}$&
			1&
			1&
			3&
			0&
			1&
			$\chi_{({SU(2)_R})_{2}}$\\
			
			$G^{a}_{\mu\nu}$&
			8&
			1&
			1&
			0&
			1&
			$\chi_{({SU(3)})_{8}}$\\
			\hline
			\hline
			
			$\mathcal{D}_\mu$&
			\multicolumn{6}{c|}{\textbf{Covariant Derivative}}\\
			\hline
			
	\end{tabular}}
	\caption{\small MLRSM: Quantum numbers of fields under the gauge groups, their spins and gauge group characters. Since MLRSM only contains fields with spins-0, -1/2, and -1, the relevant Lorentz characters are the ones given in Eq.~\eqref{eq:lorentz-char}. Here, $x$ parametrizes the $U(1)$ character. Also, $I$ = 1,2,3 denotes the  $SU(2)$ index, $a$ = 1,2,.....,8 the $SU(3)$ index and $p$ = 1,...,$N_f$ the flavor index. The color and isospin indices have been suppressed. $L$ and $R$ denote whether the fields transform non-trivially under $SU(2)_L$ or $SU(2)_R$. }
	\label{table:mlrsm-quantum-no}
\end{table}

\begin{table}[h]
	\centering
	\renewcommand{\arraystretch}{2.0}
	{\tiny\begin{tabular}{||c|c||c|c||c||}
			\hline
			\multicolumn{5}{|c|}{\textbf{Mass Dimension-2}}\\
			\hline
			
			\multirow{2}{*}{\textbf{HS Output}}&
			\multirow{2}{*}{\textbf{Covariant Form}}&
			\multirow{2}{*}{\textbf{HS Output}}&
			\multirow{2}{*}{\textbf{Covariant Form}}&
			\multirow{2}{*}{\textbf{\textbf{No. of Operators}}}\\
			
			&
			&
			&
			&
			\textbf{(including h.c.)}\\
			\hline
			$\phi^{\dagger}\phi$&
			$Tr[\Phi^{\dagger}\Phi]$&
			$\Delta_1^{\dagger}\Delta_1$&
			$Tr[\Delta_L^{\dagger}\Delta_L]$&
			\multirow{2}{*}{5}\\
			\cline{1-4}
			$\color{blue}{\phi^2}$&
			$\color{blue}{Tr[\tilde{\Phi}^{\dagger}\Phi]}$&
			$\Delta_2^{\dagger}\Delta_2$&
			$Tr[\Delta_R^{\dagger}\Delta_R]$&\\
			\hline
			\hline
			
			\multicolumn{5}{|c|}{\textbf{Mass Dimension-4}}\\
			\hline
			
			\multirow{2}{*}{\textbf{HS Output}}&
			\multirow{2}{*}{\textbf{Covariant Form}}&
			\multirow{2}{*}{\textbf{HS Output}}&
			\multirow{2}{*}{\textbf{Covariant Form}}&
			\multirow{2}{*}{\textbf{No. of Operators}}\\
			
			&
			&
			&
			&
			\textbf{(including h.c.)}\\
			\hline
			
			$Bl^2+Br^2$&
			$B_{\mu\nu}B^{\mu\nu},\hspace{0.1cm} B_{\mu\nu}\tilde{B}^{\mu\nu}$&
			$Gl^2+Gr^2$&
			$G^a_{\mu\nu}G^{a\mu\nu},\hspace{0.1cm} G^a_{\mu\nu}\tilde{G}^{a\mu\nu}$&
			\multirow{11}{*}{$14N_f^2+2N_f+22$}\\
			\cline{1-4}
			$Wl_{1}^2+Wr_{1}^2$&
			$W^I_{L\mu\nu}W_L^{I\mu\nu},\hspace{0.1cm} W^I_{L\mu\nu}\tilde{W}_L^{I\mu\nu}$&
			$Wl_{2}^2+Wr_{2}^2$&
			$W^J_{R\mu\nu}W_R^{J\mu\nu},\hspace{0.1cm} W^J_{R\mu\nu}\tilde{W}_R^{J\mu\nu}$&
			\\
			\cline{1-4}
			$\color{blue}{N_f^2Q_1^{\dagger}Q_2\phi}$&
			$\color{blue}{\overline{Q}_L^r\Phi Q_R^s}$&
			$\color{blue}{N_f^2Q_1^{\dagger}Q_2\phi^{\dagger}}$&
			$\color{blue}{\overline{Q}_L^r\tilde{\Phi} Q_R^s}$&
			\\
			\cline{1-4}
			
			$\color{blue}{N_f^2L_1^{\dagger}L_2\phi}$&
			$\color{blue}{\overline{L}_L^r\Phi L_R^s}$&
			$\color{blue}{N_f^2L_1^{\dagger}L_2\phi^{\dagger}}$&
			$\color{blue}{\overline{L}_L^r\tilde{\Phi} L_R^s}$&
			\\
			\cline{1-4}
			
			$\color{blue}{\frac{1}{2}\left(N_f^2+N_f \right)L_1^2\Delta_1}$&
			$\color{blue}{(L^r_L)^TCi\sigma_2\Delta_L L_L^s}$&
			$\color{blue}{\frac{1}{2}\left(N_f^2+N_f \right)L_2^2\Delta_2}$&
			$\color{blue}{(L^r_R)^TCi\sigma_2\Delta_R L_R^s}$&
			\\
			\cline{1-4}
			${}^{\textcolor{black}{\clubsuit}}N_f^2Q_1^{\dagger}Q_1\mathcal{D}$&
			$\overline{Q}_L^r\slashed{\mathcal{D}}Q_L^r$&
			${}^{\textcolor{black}{\clubsuit}}N_f^2Q_2^{\dagger}Q_2\mathcal{D}$&
			$\overline{Q}_R^r\slashed{\mathcal{D}}Q_R^r$&
			\\
			\cline{1-4}
			\multirow{2}{*}{$2\phi\phi^{\dagger}\Delta_2\Delta_2^{\dagger} $}&
			$Tr[\Phi^{\dagger}\Phi]Tr[\Delta_R^{\dagger}\Delta_R]$&
			${}^{\textcolor{black}{\clubsuit}}N_f^2L_1^{\dagger}L_1\mathcal{D}$&
			$\overline{L}_L^r\slashed{\mathcal{D}}L_L^r$&
			\\
			\cline{3-4}
			&
			$Tr[\Phi^{\dagger}\Delta_R^{\dagger}\Phi\Delta_R]$&
			${}^{\textcolor{black}{\clubsuit}}N_f^2L_2^{\dagger}L_2\mathcal{D}$&
			$\overline{L}_R^r\slashed{\mathcal{D}}L_R^r$&
			\\
			\cline{1-4}
			$\color{blue}{\phi^4}$&
			$\color{blue}{Tr[\tilde{\Phi}^{\dagger}\Phi]Tr[\tilde{\Phi}^{\dagger}\Phi]}$&
			$\color{blue}{\phi^3\phi^{\dagger}}$&
			$\color{blue}{Tr[\Phi^{\dagger}\Phi]Tr[\tilde{\Phi}^{\dagger}\Phi]}$&
			\\
			\cline{1-4}
			$\color{blue}{\phi^2\Delta_1\Delta_1^{\dagger}}$&
			$\color{blue}{Tr[\tilde{\Phi}^{\dagger}\Phi]Tr[\Delta_L^{\dagger}\Delta_L]}$&
			$\color{blue}{\phi^2\Delta_2\Delta_2^{\dagger}}$&
			$\color{blue}{Tr[\tilde{\Phi}^{\dagger}\Phi]Tr[\Delta_R^{\dagger}\Delta_R]}$&
			\\
			\cline{1-4}
			$\color{blue}{\phi^2\Delta_1\Delta_2^{\dagger}}$&
			$\color{blue}{Tr[\tilde{\Phi}^{\dagger}\Delta_R^{\dagger}\Phi\Delta_L]}$&
			$\color{blue}{\phi^2\Delta_2\Delta_1^{\dagger}}$&
			$\color{blue}{Tr[\tilde{\Phi}^{\dagger}\Delta_L^{\dagger}\Phi\Delta_R]}$&
			\\
			\cline{1-4}
			\hline
		\end{tabular}}
	\caption{\small MLRSM: Renormalizable operators as Hilbert Series output and their covariant form. The operators in blue have distinct hermitian conjugates which we have not written explicitly. Here, $I,J$ = 1,2,3 are $SU(2)_L$ and $SU(2)_R$ indices respectively; $a$ = 1,...,8 are $SU(3)$ indices and $r,s$ = $1,2,...,N_f$ are flavour indices which are summed over with the suitable coupling constants. ${\textcolor{black}{\clubsuit}}$ - In the Hilbert Series the fermion kinetic terms appear with a factor of $N_f^2$ but in the physical Lagrangian there is a flavour symmetry which forces the kinetic terms to be diagonal and the factor of $N_f^2$ is reduced to $N_f$.}
\label{table:mlrsm-renorm-output}
\end{table}

\begin{table}[h]
	\centering
	\renewcommand{\arraystretch}{2.0}
	{\tiny\begin{tabular}{||c|c||c|c||c||}
			\hline
			
			\multicolumn{5}{|c|}{\textbf{Mass Dimension-4}}\\
			\hline
			
			\multirow{2}{*}{\textbf{HS Output}}&
			\multirow{2}{*}{\textbf{Covariant Form}}&
			\multirow{2}{*}{\textbf{HS Output}}&
			\multirow{2}{*}{\textbf{Covariant Form}}&
			\multirow{2}{*}{\textbf{No. of Operators}}\\
			
			&
			&
			&
			&
			\textbf{(including h.c.)}\\
			\hline
			$\color{blue}{\phi\phi^{\dagger}\Delta_1\Delta_2^{\dagger}}$&
			$\color{blue}{Tr[\Phi^{\dagger}\Delta_R^{\dagger}\Phi\Delta_L]}$&
			$\color{blue}{(\Delta_1^{\dagger})^2\Delta_2^2}$&
			$Tr[\Delta_L^{\dagger}\Delta_L^{\dagger}]Tr[\Delta_R\Delta_R]$&
			\multirow{7}{*}{$16$}\\
			\cline{1-4}
			$\Delta_1\Delta_1^{\dagger}\Delta_2\Delta_2^{\dagger} $&
			$Tr[\Delta_L^{\dagger}\Delta_L]Tr[\Delta_R^{\dagger}\Delta_R]$&
			$\phi^{\dagger}\phi\mathcal{D}^2$&
			$Tr[(D_{\mu}\Phi)^{\dagger}(D^{\mu}\Phi)]$&
			\\
			\cline{1-4}
			$\Delta_1^{\dagger}\Delta_1\mathcal{D}^2$&
			$Tr[(D_{\mu}\Delta_L)^{\dagger}(D^{\mu}\Delta_L)]$&
			$\Delta_2^{\dagger}\Delta_2\mathcal{D}^2$&
			$Tr[(D_{\mu}\Delta_R)^{\dagger}(D^{\mu}\Delta_R)]$&
			\\
			\cline{1-4}
			\multirow{2}{*}{$2\phi^2(\phi^{\dagger})^2 $}&
			$Tr[\Phi^{\dagger}\Phi]Tr[\Phi^{\dagger}\Phi]$&
			\multirow{2}{*}{$2\Delta_1^2(\Delta_1^{\dagger})^2 $}&
			$Tr[\Delta_L^{\dagger}\Delta_L]Tr[\Delta_L^{\dagger}\Delta_L]$&
			\\
			&
			$Tr[\Phi^{\dagger}\tilde{\Phi}]Tr[\tilde{\Phi}^{\dagger}\Phi]$&
			&
			$Tr[\Delta_L^{\dagger}\Delta_L^{\dagger}]Tr[\Delta_L\Delta_L]$&
			\\
			\cline{1-4}
			\multirow{2}{*}{$2\Delta_2^2(\Delta_2^{\dagger})^2 $}&
			$Tr[\Delta_R^{\dagger}\Delta_R]Tr[\Delta_R^{\dagger}\Delta_R]$&
			\multirow{2}{*}{$2\phi\phi^{\dagger}\Delta_1\Delta_1^{\dagger} $}&
			$Tr[\Phi^{\dagger}\Phi]Tr[\Delta_L^{\dagger}\Delta_L]$&
			\\
			&
			$Tr[\Delta_R^{\dagger}\Delta_R^{\dagger}]Tr[\Delta_R\Delta_R]$&
			&
			$Tr[\Phi^{\dagger}\Delta_L^{\dagger}\Phi\Delta_L]$&
			\\
			\hline
	\end{tabular}}
	\caption{\small Table~\ref{table:mlrsm-renorm-output} continued.}
	\label{table:mlrsm-renorm-output-2}
\end{table}

\begin{table}[h]
	\centering
	\renewcommand{\arraystretch}{1.8}
	{\tiny\begin{tabular}{|c|l|}
			\hline
			
			\multicolumn{2}{|c|}{\textbf{Mass Dimension-6}}\\
			\hline
			
			\textbf{Operator Class}&
			\hspace{3.7cm}\textbf{Operators (in non-covariant form)}\\
			\hline
			
			\multirow{6}{*}{$\Phi^6$}&
			$2(\phi^{\dagger})^3(\phi)^3,\hspace{0.1cm}2(\Delta_1^{\dagger})^3(\Delta_1)^3,\hspace{0.1cm} 2(\Delta_2^{\dagger})^3(\Delta_2)^3,\hspace{0.1cm} 4(\phi^{\dagger})^2(\phi)^2\Delta_1^{\dagger}\Delta_1,\hspace{0.1cm} 4(\phi^{\dagger})^2(\phi)^2\Delta_2^{\dagger}\Delta_2,\hspace{0.1cm} 3\phi^{\dagger}\phi(\Delta_1^{\dagger})^2(\Delta_1)^2,$\\
			
			&
			$3\phi^{\dagger}\phi(\Delta_2^{\dagger})^2(\Delta_2)^2,\hspace{0.1cm} 
			4\phi^{\dagger}\phi\Delta_1^{\dagger}\Delta_1\Delta_2^{\dagger}\Delta_2,\hspace{0.1cm}2(\Delta_1^{\dagger})^2(\Delta_1)^2\Delta_2^{\dagger}\Delta_2,\hspace{0.1cm} 2(\Delta_2^{\dagger})^2(\Delta_2)^2\Delta_1^{\dagger}\Delta_1,\hspace{0.1cm}  
			\color{blue}{\phi^6,\hspace{0.1cm} \phi^5\phi^{\dagger},\hspace{0.1cm} 2\phi^4(\phi^{\dagger})^2,}$\\
			
			&
			$\color{blue}{\phi^4\Delta_1^{\dagger}\Delta_1,\hspace{0.1cm} \phi^4\Delta_2^{\dagger}\Delta_2,\hspace{0.1cm} \phi^4\Delta_1^{\dagger}\Delta_2,\hspace{0.1cm} \phi^4\Delta_2^{\dagger}\Delta_1,\hspace{0.1cm} 2\phi^3\phi^{\dagger}\Delta_1^{\dagger}\Delta_1,\hspace{0.1cm} 2\phi^3\phi^{\dagger}\Delta_2^{\dagger}\Delta_2,\hspace{0.1cm} 2\phi^3\phi^{\dagger}\Delta_1^{\dagger}\Delta_2,\hspace{0.1cm} 2\phi^3\phi^{\dagger}\Delta_2^{\dagger}\Delta_1,\hspace{0.1cm}}$\\
			
			&
			$\color{blue}{3(\phi^{\dagger})^2(\phi)^2\Delta_1^{\dagger}\Delta_2,\hspace{0.1cm} 2\phi^2\Delta_1^2(\Delta_1^{\dagger})^2,\hspace{0.1cm} 2\phi^2\Delta_2^2(\Delta_2^{\dagger})^2,\hspace{0.1cm} \phi^2\Delta_1^2(\Delta_2^{\dagger})^2,\hspace{0.1cm} \phi^2\Delta_2^2(\Delta_1^{\dagger})^2,\hspace{0.1cm} 2\phi^2\Delta_1\Delta_2(\Delta_1^{\dagger})^2,\hspace{0.1cm} }$\\
			
			&
			$\color{blue}{2\phi^2\Delta_1\Delta_2(\Delta_2^{\dagger})^2,\hspace{0.1cm} 2\phi^2\Delta_1^2\Delta_1^{\dagger}\Delta_2^{\dagger},\hspace{0.1cm} 2\phi^2\Delta_2^2\Delta_1^{\dagger}\Delta_2^{\dagger},\hspace{0.1cm} 2\phi^2\Delta_1\Delta_1^{\dagger}\Delta_2\Delta_2^{\dagger},\hspace{0.1cm} 2\phi\phi^{\dagger}\Delta_1\Delta_2(\Delta_1^{\dagger})^2,\hspace{0.1cm} 2\phi\phi^{\dagger}\Delta_1\Delta_2(\Delta_2^{\dagger})^2,}$\\
			
			&
			$\color{blue}{\phi\phi^{\dagger}\Delta_1^2(\Delta_2^{\dagger})^2,\hspace{0.1cm} \Delta_1^2\Delta_2(\Delta_2^{\dagger})^3,\hspace{0.1cm} \Delta_1\Delta_2^2(\Delta_1^{\dagger})^3}$\\
			\hline
			
			\multirow{4}{*}{$\Phi^2X^2$}&
			$\color{blue}{Bl^2\phi^2,\hspace{0.1cm} Bl^2(\phi^{\dagger})^2,\hspace{0.1cm} Gl^2\phi^2,\hspace{0.1cm} Gl^2(\phi^{\dagger})^2,\hspace{0.1cm} Bl^2\phi^{\dagger}\phi,\hspace{0.1cm}  Gl^2\phi^{\dagger}\phi,\hspace{0.1cm} Bl^2\Delta_1^{\dagger}\Delta_1,\hspace{0.1cm}  Gl^2\Delta_1^{\dagger}\Delta_1,\hspace{0.1cm} Bl^2\Delta_2^{\dagger}\Delta_2,\hspace{0.1cm}  Gl^2\Delta_2^{\dagger}\Delta_2,\hspace{0.1cm} }$\\
			
			&
			$\color{blue}{Wl_{1}^2\phi^2,\hspace{0.1cm} Wl_{1}^2(\phi^{\dagger})^2,\hspace{0.1cm} Wl_{1}^2\phi^{\dagger}\phi,\hspace{0.1cm}  
			Wl_{2}^2\phi^2,\hspace{0.1cm} Wl_{2}^2(\phi^{\dagger})^2,\hspace{0.1cm} 
			Wl_{2}^2\phi^{\dagger}\phi,\hspace{0.1cm} 2Wl_{1}^2\Delta_1^{\dagger}\Delta_1,\hspace{0.1cm}  Wl_{1}^2\Delta_2^{\dagger}\Delta_2,\hspace{0.1cm}  
			Wl_{2}^2\Delta_1^{\dagger}\Delta_1, } $\\
			
			&
			$\color{blue}{2Wl_{2}^2\Delta_2^{\dagger}\Delta_2,\hspace{0.1cm} Wl_{1}Wl_{2}\phi^2,\hspace{0.1cm} Wl_{1}Wl_{2}(\phi^{\dagger})^2,\hspace{0.1cm} Wl_{1}Wl_{2}\phi^{\dagger}\phi,\hspace{0.1cm} Wl_{1}Wl_{2}\Delta_1^{\dagger}\Delta_2,\hspace{0.1cm}  
				Wl_{1}Wl_{2}\Delta_2^{\dagger}\Delta_1, }$\\
			
			&
			$\color{blue}{BlWl_{1}\phi^{\dagger}\phi,\hspace{0.1cm} BlWl_{2}\phi^{\dagger}\phi,\hspace{0.1cm} BlWl_{1}\Delta_1^{\dagger}\Delta_1,\hspace{0.1cm}  
				BlWl_{2}\Delta_2^{\dagger}\Delta_2 }$\\
			\hline
			
			\multirow{4}{*}{$\Psi^2\Phi X$}&
			$\color{blue}{\boxed{\frac{1}{2}\left(N_f^2-N_f\right)Bl\Delta_1L_1^2},\hspace{0.1cm} \boxed{\frac{1}{2}\left(N_f^2-N_f\right)Bl\Delta_2^{\dagger}(L_2^{\dagger})^2},\hspace{0.1cm} N_f^2Bl\phi^{\dagger}L_1L_2^{\dagger}, \hspace{0.1cm}  N_f^2Bl\phi L_1L_2^{\dagger},\hspace{0.1cm}  N_f^2Wl_{1}\Delta_1L_1^2,\hspace{0.1cm}}$\\
			
			&
			$\color{blue}{\frac{1}{2}\left(N_f^2+N_f\right)Wl_{1}\Delta_1^{\dagger}(L_2^{\dagger})^2,\hspace{0.1cm} \frac{1}{2}\left(N_f^2+N_f\right)Wl_{2}\Delta_2L_1^2,\hspace{0.1cm} N_f^2Wl_{2}\Delta_2^{\dagger}(L_2^{\dagger})^2,\hspace{0.1cm} N_f^2Wl_{1}\phi^{\dagger}L_1L_2^{\dagger}, \hspace{0.1cm}  N_f^2Wl_{1}\phi L_1L_2^{\dagger},\hspace{0.1cm} }$\\
			
			&
			$\color{blue}{N_f^2Wl_{2}\phi^{\dagger}L_1L_2^{\dagger}, \hspace{0.1cm}  N_f^2Wl_{2}\phi L_1L_2^{\dagger},\hspace{0.1cm} N_f^2Bl\phi^{\dagger}Q_1Q_2^{\dagger}, \hspace{0.1cm} N_f^2Bl\phi Q_1Q_2^{\dagger}, \hspace{0.1cm} N_f^2Wl_{1}\phi^{\dagger}Q_1Q_2^{\dagger}\hspace{0.1cm},  N_f^2Wl_{1}\phi Q_1Q_2^{\dagger}\hspace{0.1cm}}$\\
			
			&
			$\color{blue}{N_f^2Wl_{2}\phi^{\dagger}Q_1Q_2^{\dagger}, \hspace{0.1cm}  N_f^2Wl_{2}\phi Q_1Q_2^{\dagger},\hspace{0.1cm}  N_f^2Gl\phi^{\dagger}Q_1Q_2^{\dagger}\hspace{0.1cm},  N_f^2Gl\phi Q_1Q_2^{\dagger}\hspace{0.1cm}}$\\
			\hline
			
			\multirow{4}{*}{$\Psi^2\Phi^2 \mathcal{D}$}&
			$2N_f^2\phi\phi^{\dagger}L_1L_1^{\dagger}\mathcal{D},\hspace{0.1cm} 2N_f^2\Delta_1\Delta_1^{\dagger}L_1L_1^{\dagger}\mathcal{D},\hspace{0.1cm} N_f^2\Delta_2\Delta_2^{\dagger}L_1L_1^{\dagger}\mathcal{D},\hspace{0.1cm} 2N_f^2\phi\phi^{\dagger}L_2L_2^{\dagger}\mathcal{D},\hspace{0.1cm} N_f^2\Delta_1\Delta_1^{\dagger}L_2L_2^{\dagger}\mathcal{D}, $\\
			
			&
			$2N_f^2\Delta_2\Delta_2^{\dagger}L_2L_2^{\dagger}\mathcal{D},\hspace{0.1cm} 2N_f^2\phi\phi^{\dagger}Q_1Q_1^{\dagger}\mathcal{D},\hspace{0.1cm} 2N_f^2\Delta_1\Delta_1^{\dagger}Q_1Q_1^{\dagger}\mathcal{D},\hspace{0.1cm} N_f^2\Delta_2\Delta_2^{\dagger}Q_1Q_1^{\dagger}\mathcal{D},\hspace{0.1cm} 2N_f^2\phi\phi^{\dagger}Q_2Q_2^{\dagger}\mathcal{D}, $\\
			
			&
			$N_f^2\Delta_1\Delta_1^{\dagger}Q_2Q_2^{\dagger}\mathcal{D},\hspace{0.1cm} 2N_f^2\Delta_2\Delta_2^{\dagger}Q_2Q_2^{\dagger}\mathcal{D},\hspace{0.1cm} \color{blue}{N_f^2\phi^2L_1L_1^{\dagger}\mathcal{D},\hspace{0.1cm} N_f^2\phi^2L_2L_2^{\dagger}\mathcal{D},\hspace{0.1cm} N_f^2\phi^2Q_1Q_1^{\dagger}\mathcal{D},\hspace{0.1cm} N_f^2\phi^2Q_2Q_2^{\dagger}\mathcal{D}, }$\\
			
			&
			$\color{blue}{N_f^2\phi\Delta_1L_1L_2\mathcal{D},\hspace{0.1cm} N_f^2\phi\Delta_2L_1L_2\mathcal{D},\hspace{0.1cm}
			N_f^2\phi^{\dagger}\Delta_1L_1L_2 \mathcal{D}, N_f^2\phi^{\dagger}\Delta_2L_1L_2 \mathcal{D}}$\\
			\hline
			\multirow{9}{*}{$\Psi^2\Phi^3$}&
			$\color{blue}{\frac{1}{2}\left(N_f^2+N_f\right)L_1^2\Delta_1\phi^2,\hspace{0.1cm} \frac{1}{2}\left(N_f^2+N_f\right)L_1^2\Delta_1(\phi^{\dagger})^2,\hspace{0.1cm} \frac{1}{2}\left(3N_f^2+N_f\right)L_1^2\Delta_1\phi\phi^{\dagger},\hspace{0.1cm} \left(N_f^2+N_f\right)L_1^2\Delta_1^{2}\Delta_1^{\dagger}, }$\\
			
			&
			$\color{blue}{\frac{1}{2}\left(N_f^2+N_f\right)L_1^2\Delta_1\Delta_2\Delta_2^{\dagger},\hspace{0.1cm} \frac{1}{2}\left(N_f^2+N_f\right)L_2^2\Delta_2\phi^2,\hspace{0.1cm} \frac{1}{2}\left(N_f^2+N_f\right)L_2^2\Delta_2(\phi^{\dagger})^2,\hspace{0.1cm} \frac{1}{2}\left(3N_f^2+N_f\right)L_2^2\Delta_2\phi\phi^{\dagger},}$\\
			
			&
			$\color{blue}{\frac{1}{2}\left(N_f^2+N_f\right)L_2^2\Delta_2\Delta_1\Delta_1^{\dagger},\hspace{0.1cm} \left(N_f^2+N_f\right)L_2^2\Delta_2^2\Delta_2^{\dagger},\hspace{0.1cm} \frac{1}{2}\left(N_f^2+N_f\right)L_1^2\Delta_2\phi^2,\hspace{0.1cm} \frac{1}{2}\left(N_f^2+N_f\right)L_1^2\Delta_2(\phi^{\dagger})^2,}$\\
			
			&
			$\color{blue}{\frac{1}{2}\left(N_f^2+N_f\right)L_1^2\Delta_2^2\Delta_1^{\dagger},\hspace{0.1cm} N_f^2L_1^2\Delta_2\phi\phi^{\dagger},\hspace{0.1cm} N_f^2L_2^2\Delta_1\phi\phi^{\dagger},\hspace{0.1cm} \frac{1}{2}\left(N_f^2+N_f\right)L_2^2\Delta_1(\phi^{\dagger})^2,\hspace{0.1cm} \frac{1}{2}\left(N_f^2+N_f\right)L_2^2\Delta_1\phi^2, }$\\
			
			&
			$\color{blue}{\frac{1}{2}\left(N_f^2+N_f\right)L_2^2\Delta_1^2\Delta_2^{\dagger},\hspace{0.1cm} N_f^2L_1^{\dagger}L_2\phi^3,\hspace{0.1cm} 2N_f^2L_1^{\dagger}L_2\phi^2\phi^{\dagger},\hspace{0.1cm} 2N_f^2L_1^{\dagger}L_2\phi(\phi^{\dagger})^2,\hspace{0.1cm} N_f^2L_1^{\dagger}L_2(\phi^{\dagger})^3,\hspace{0.1cm} 2N_f^2L_1^{\dagger}L_2\phi\Delta_1^{\dagger}\Delta_1,\hspace{0.1cm} }$\\
			
			&
			$\color{blue}{2N_f^2L_1^{\dagger}L_2\phi\Delta_2^{\dagger}\Delta_2,\hspace{0.1cm} 2N_f^2L_1^{\dagger}L_2\phi^{\dagger}\Delta_1^{\dagger}\Delta_1,\hspace{0.1cm} 2N_f^2L_1^{\dagger}L_2\phi^{\dagger}\Delta_2^{\dagger}\Delta_2,\hspace{0.1cm} N_f^2L_1^{\dagger}L_2\phi\Delta_1^{\dagger}\Delta_2,\hspace{0.1cm} N_f^2L_1^{\dagger}L_2\phi^{\dagger}\Delta_1^{\dagger}\Delta_2,\hspace{0.1cm}}$\\
			
			&
			$\color{blue}{ N_f^2L_1^{\dagger}L_2\phi\Delta_2^{\dagger}\Delta_1,\hspace{0.1cm} N_f^2L_1^{\dagger}L_2\phi^{\dagger}\Delta_2^{\dagger}\Delta_1,\hspace{0.1cm}} N_f^2Q_1^{\dagger}Q_2\phi^3,\hspace{0.1cm} 2N_f^2Q_1^{\dagger}Q_2\phi^2\phi^{\dagger},\hspace{0.1cm} 2N_f^2Q_1^{\dagger}Q_2\phi(\phi^{\dagger})^2,\hspace{0.1cm} N_f^2Q_1^{\dagger}Q_2(\phi^{\dagger})^3,\hspace{0.1cm} $\\
			
			&
			$\color{blue}{2N_f^2Q_1^{\dagger}Q_2\phi\Delta_1^{\dagger}\Delta_1,\hspace{0.1cm} 2N_f^2Q_1^{\dagger}Q_2\phi^{\dagger}\Delta_1^{\dagger}\Delta_1,\hspace{0.1cm} 2N_f^2Q_1^{\dagger}Q_2\phi\Delta_2^{\dagger}\Delta_2,\hspace{0.1cm} 2N_f^2Q_1^{\dagger}Q_2\phi^{\dagger}\Delta_2^{\dagger}\Delta_2,\hspace{0.1cm} N_f^2Q_1^{\dagger}Q_2\phi\Delta_1^{\dagger}\Delta_2,\hspace{0.1cm}}$\\
			
			&
			$\color{blue}{N_f^2Q_1^{\dagger}Q_2\phi^{\dagger}\Delta_1^{\dagger}\Delta_2,\hspace{0.1cm} N_f^2Q_1^{\dagger}Q_2\phi\Delta_2^{\dagger}\Delta_1,\hspace{0.1cm} N_f^2Q_1^{\dagger}Q_2\phi\Delta_2^{\dagger}\Delta_1\hspace{0.1cm}}$\\
			\hline				
			\multirow{3}{*}{$\Phi^4 \mathcal{D}^2$}&
			$4(\phi^{\dagger})^2\phi^2\mathcal{D}^2, \hspace{0.1cm} 3(\Delta_1^{\dagger})^2(\Delta_1)^2\mathcal{D}^2,\hspace{0.1cm} 3(\Delta_2^{\dagger})^2(\Delta_2)^2\mathcal{D}^2,\hspace{0.1cm} 4\phi^{\dagger}\phi\Delta_1^{\dagger}\Delta_1\mathcal{D}^2,\hspace{0.1cm} 4\phi^{\dagger}\phi\Delta_2^{\dagger}\Delta_2\mathcal{D}^2,\hspace{0.1cm} 2\Delta_1^{\dagger}\Delta_1\Delta_2^{\dagger}\Delta_2\mathcal{D}^2,$ \\
			
			&
			$\color{blue}{\phi^3\phi^{\dagger} \mathcal{D}^2,\hspace{0.1cm} 2\phi^2\Delta_1^{\dagger}\Delta_1 \mathcal{D}^2,\hspace{0.1cm} 2\phi^2\Delta_2^{\dagger}\Delta_2 \mathcal{D}^2,\hspace{0.1cm} \phi^2\Delta_1^{\dagger}\Delta_2 \mathcal{D}^2,\hspace{0.1cm} \phi^2\Delta_2^{\dagger}\Delta_1 \mathcal{D}^2,\hspace{0.1cm} 2\phi^2\Delta_1^{\dagger}\Delta_1 \mathcal{D}^2,\hspace{0.1cm} (\Delta_1^{\dagger})^2(\Delta_2)^2 \mathcal{D}^2,\hspace{0.1cm}}$\\
			
			&
			$\color{blue}{2\phi^{\dagger}\phi\Delta_1^{\dagger}\Delta_2\mathcal{D}^2,\hspace{0.1cm} \phi^4\mathcal{D}^2}$\\
			\hline	
	\end{tabular}}
\caption{\small MLRSM: Operators of mass dimension-6. Operators in blue have distinct hermitian conjugates and boxed operators vanish for $N_f=1$. There are no operators of mass dimension-5 in this case.}
\label{table:mlrsm-dim-6-op}
\end{table} 
\clearpage
\begin{table}[h]
	\centering
	\renewcommand{\arraystretch}{1.8}
	{\tiny\begin{tabular}{|c|l|}
			\hline
			
			\multicolumn{2}{|c|}{\textbf{Mass Dimension-6}}\\
			\hline
			
			\textbf{Operator Class}&
			\hspace{3.7cm}\textbf{Operators (in non-covariant form)}\\
			\hline			
			$X^3$&
			$ Wl_{1}^3,\hspace{0.2cm} Wr_{1}^3,\hspace{0.2cm} Wl_{2}^3,\hspace{0.2cm} Wr_{2}^3,\hspace{0.2cm} Gl^3,\hspace{0.2cm} Gr^3 \hspace{0.2cm}$\\
			\hline	
			
			\multirow{4}{*}{$\Psi^4$}&
			$\frac{1}{2}\left(N_f^4+N_f^2\right)L_1^2(L_1^{\dagger})^2,\hspace{0.1cm} \frac{1}{2}\left(N_f^4+N_f^2\right)L_2^2(L_2^{\dagger})^2,\hspace{0.1cm} N_f^4L_1L_1^{\dagger}L_2L_2^{\dagger},\hspace{0.1cm} \color{blue}{\frac{1}{2}\left(N_f^4+N_f^2\right)L_1^2(L_2^{\dagger})^2},\hspace{0.1cm}$\\
			
			&
			$\left(N_f^4+N_f^2\right)Q_1^2(Q_1^{\dagger})^2,\hspace{0.1cm} \left(N_f^4+N_f^2\right)Q_2^2(Q_2^{\dagger})^2,\hspace{0.1cm} 2N_f^4Q_1Q_1^{\dagger}Q_2Q_2^{\dagger},\hspace{0.1cm} \color{blue}{\left(N_f^4+N_f^2\right)Q_1^2(Q_2^{\dagger})^2,}$\\
			
			&
			$2N_f^4L_1L_1^{\dagger}Q_1Q_1^{\dagger},\hspace{0.1cm} 2N_f^4L_2L_2^{\dagger}Q_2Q_2^{\dagger},\hspace{0.1cm} N_f^4L_1L_1^{\dagger}Q_2Q_2^{\dagger},\hspace{0.1cm} N_f^4L_1L_1^{\dagger}Q_2Q_2^{\dagger},\hspace{0.1cm} \color{blue}{\frac{1}{3}\left(2N_f^4+N_f^2\right)L_1Q_1^3, }$\\
			
			&
			$\color{blue}{2N_f^4L_1^{\dagger}L_2Q_1^{\dagger}Q_2,\hspace{0.1cm} N_f^4L_2^{\dagger}L_1Q_1^{\dagger}Q_2,\hspace{0.1cm} \frac{1}{2}\left(N_f^4+N_f^3\right)L_1Q_1Q_2^2,\hspace{0.1cm} \frac{1}{2}\left(N_f^4+N_f^3\right)L_2Q_2Q_1^2,\hspace{0.1cm} \frac{1}{3}\left(2N_f^4+N_f^2\right)L_2Q_2^3}$\\
			\hline		    
	\end{tabular}}
	\caption{\small Table~\ref{table:mlrsm-dim-6-op} continued.}
	\label{table:mlrsm-dim-6-op-2}
\end{table} 

\subsubsection*{\underline{SU(5) Grand Unification - The Georgi Glashow Model}}\label{subsubsec:su5-gut}

The SM gauge group $SU(3)\otimes SU(2)\otimes U(1)$ can be successfully embedded in $SU(5)$ - a unified group \cite{Georgi:1974sy,Langacker:1980js,Buras:1977yy,Chakrabortty:2009xm,Ross:1985ai,Mohapatra:1986uf}. The field content of this model and their transformation properties are listed in Table~\ref{table:su5-quantum-no}. 
\begin{table}[h]
	\centering
	\renewcommand{\arraystretch}{1.8}
	{\scriptsize\begin{tabular}{|c|c|c|c|}
			\hline
			\textbf{Fields}&
			$SU(5)$&
			\textbf{Spin}&
			\textbf{Gauge Group Characters}\\
			\hline
			
			$\Phi_b^a$&
			24&
			0&
			$\chi_{({SU(5)})_{24}}$\\
			
			$\varphi^a$&
			5&
			0&
			$\chi_{({SU(5)})_{5}}$\\
			
			$\psi_{L,a}$&
			$\overline{5}$&
			1/2&
			$\chi_{({SU(5)})_{\overline{5}}}$\\
			
			$\Psi^{ab}_L$&
			10&
			1/2&
			$\chi_{({SU(5)})_{10}}$\\
			
			$\mathcal{A}^a_{b,\mu\nu}$&
			24&
			1&
			$\chi_{({SU(5)})_{24}}$\\
			
			\hline
			\hline
			$\mathcal{D}_\mu$&
			\multicolumn{3}{c|}{\textbf{Covariant Derivative}}\\
			\hline
	\end{tabular}}
	\caption{\small $SU(5)$ GUT: Quantum numbers of fields under the gauge groups, their spins and gauge group characters. Since the model only contains fields with spins-0, -1/2, and -1, the relevant Lorentz characters are the ones given in Eq.~\eqref{eq:lorentz-char}.  Here, $a$, $b$ are $SU(5)$ indices and $L$ denotes the chirality (i.e., left handedness of the field).}
	\label{table:su5-quantum-no}
\end{table} 

\vspace{-0.3cm} \noindent
We shall now utilize the strategy sketched in section~\ref{subsubsec:sun-haar-char} and show the explicit computation of the Haar measure and the characters of the relevant representations of $SU(5)$.\\
\vspace{-0.3cm}\\
\noindent \textbf{Haar Measure}: 
Using Eq.~\eqref{eq:eps-z-reln}, we can obtain for $SU(5)$, $\epsilon_1$ = $z_1$, $\epsilon_2$ = $z_1^{-1}z_2$, \\$\epsilon_3$ = $z_2^{-1}z_3$, $\epsilon_4$ = $z_3^{-1}z_4$ and $\epsilon_5$ = $z_4^{-1}$. The Vandermonde determinant for this case is: 

\vspace{-0.6cm}
{\scriptsize\begin{eqnarray}\label{eq:vandermonde-su5}
	\Delta\left(\epsilon\right) = \begin{vmatrix}
	\epsilon_1^{4}\hspace{0.1cm} \ \ \epsilon_1^{3}\hspace{0.1cm} \ \ \epsilon_1^2\hspace{0.1cm}\ \ \epsilon_1\hspace{0.1cm}\ \ 1\\
	\epsilon_2^{4}\hspace{0.1cm} \ \ \epsilon_2^{3}\hspace{0.1cm} \ \ \epsilon_2^2\hspace{0.1cm}\ \ \epsilon_2\hspace{0.1cm}\ \ 1\\
	\epsilon_3^{4}\hspace{0.1cm} \ \ \epsilon_3^{3}\hspace{0.1cm} \ \ \epsilon_3^2\hspace{0.1cm}\ \ \epsilon_3\hspace{0.1cm}\ \ 1\\
	\epsilon_4^{4}\hspace{0.1cm} \ \ \epsilon_4^{3}\hspace{0.1cm} \ \ \epsilon_4^2\hspace{0.1cm}\ \ \epsilon_4\hspace{0.1cm}\ \ 1\\
	\epsilon_5^{4}\hspace{0.1cm} \ \ \epsilon_5^{3}\hspace{0.1cm} \ \ \epsilon_5^2\hspace{0.1cm}\ \ \epsilon_5\hspace{0.1cm}\ \ 1\\
	\end{vmatrix} = \prod_{1\leq i<j\leq 5}\left(\epsilon_i-\epsilon_j\right).
	\end{eqnarray}}

\vspace{-0.5cm} \noindent $\Delta(\epsilon^{-1})$ is obtained by replacing $\epsilon_i$ by $\epsilon_i^{-1}$ in the above expression. Then using Eq.~\eqref{eq:haar-measure},

\vspace{-0.5cm}
{\scriptsize\begin{eqnarray}\label{eq:su5-haar-measure}
	d\mu_{SU(5)} &=& \frac{1}{5!\left(2\pi i\right)^{4}} \frac{dz_1}{z_1}\frac{dz_2}{z_2}\frac{dz_3}{z_3}\frac{dz_4}{z_4}\Delta\left(\epsilon\right)\Delta\left(\epsilon^{-1}\right)\nonumber\\
	&=& \frac{1}{120\left(2\pi i\right)^{4}} \frac{dz_1}{z_1}\frac{dz_2}{z_2}\frac{dz_3}{z_3}\frac{dz_4}{z_4}\left(1-\frac{z_1^2}{z_2}\right)\left(1-\frac{z_2}{z_1^2}\right)\left(1-\frac{z_1z_2}{z_3}\right)\left(1-\frac{z_3}{z_1z_2}\right)\left(1-\frac{z_1z_3}{z_4}\right)\left(1-\frac{z_4}{z_1z_3}\right) \nonumber\\
	& & \left(1-z_1z_4\right)\left(1-\frac{1}{z_1z_4}\right)\left(1-\frac{z_2^2}{z_1z_3}\right)\left(1-\frac{z_1z_3}{z_2^2}\right)\left(1-\frac{z_2z_3}{z_1z_4}\right)\left(1-\frac{z_1z_4}{z_2z_3}\right)\left(1-\frac{z_4^2}{z_3}\right)\left(1-\frac{z_3}{z_4^2}\right) \nonumber\\
	& &
	\left(1-\frac{z_2z_4}{z_1}\right)\left(1-\frac{z_1}{z_2z_4}\right)\left(1-\frac{z_3^2}{z_2z_4}\right)\left(1-\frac{z_2z_4}{z_3^2}\right)\left(1-\frac{z_3z_4}{z_2}\right)\left(1-\frac{z_2}{z_3z_4}\right).
	\end{eqnarray}}
\noindent
\textbf{Characters}
\begin{itemize}
	\item The Fundamental Representation: $5 \equiv (1,0,0,0)$ 
	
	Using Eq.~\eqref{eq:lambda-recursion}, we get $\vec{\lambda}$ = $(1,0,0,0,0)$. Now, since $\vec{\rho}$ = $(4,3,2,1,0)$ therefore, $\vec{r} = \vec{\lambda}+\vec{\rho} = (5,3,2,1,0)$ and the character is obtained as:
	
	\vspace{-0.8cm}
	{\scriptsize\begin{eqnarray}\label{eq:su5-5-char}
		\chi(\epsilon_1,\epsilon_2,\epsilon_3,\epsilon_4,\epsilon_5) &=& \frac{|\epsilon^5,\hspace{0.1cm}\epsilon^3,\hspace{0.1cm}\epsilon^2,\hspace{0.1cm}\epsilon,\hspace{0.1cm}1|} {|\epsilon^4,\hspace{0.1cm}\epsilon^3,\hspace{0.1cm}\epsilon^2,\hspace{0.1cm}\epsilon,\hspace{0.1cm}1|} = \frac{1}{\prod_{1\leq i<j\leq 5}\left(\epsilon_i-\epsilon_j\right)}
		\begin{vmatrix}
		\epsilon_1^{5}\hspace{0.1cm} \ \ \epsilon_1^{3}\hspace{0.1cm} \ \ \epsilon_1^{2}\hspace{0.1cm} \ \ \epsilon_1\hspace{0.1cm}\ \ 1\\
		\epsilon_2^{5}\hspace{0.1cm} \ \ \epsilon_2^{3}\hspace{0.1cm} \ \ \epsilon_2^{2}\hspace{0.1cm} \ \ \epsilon_2\hspace{0.1cm}\ \ 1\\
		\epsilon_3^{5}\hspace{0.1cm} \ \ \epsilon_3^{3}\hspace{0.1cm} \ \ \epsilon_3^{2}\hspace{0.1cm} \ \ \epsilon_3\hspace{0.1cm}\ \ 1\\
		\epsilon_4^{5}\hspace{0.1cm} \ \ \epsilon_4^{3}\hspace{0.1cm} \ \ \epsilon_4^{2}\hspace{0.1cm} \ \ \epsilon_4\hspace{0.1cm}\ \ 1\\
		\epsilon_5^{5}\hspace{0.1cm} \ \ \epsilon_5^{3}\hspace{0.1cm} \ \ \epsilon_5^{2}\hspace{0.1cm} \ \ \epsilon_5\hspace{0.1cm}\ \ 1\\
		\end{vmatrix}\nonumber\\
		&=& \epsilon_1 + \epsilon_2 + 
		\epsilon_3 + \epsilon_4 + \epsilon_5. \nonumber\\
		\therefore\;\chi_{({SU(5)})_{5}}(z_1, z_2, z_3, z_4) &=& z_1 + \frac{z_2}{z_1} + \frac{z_3}{z_2} + \frac{z_4}{z_3} + \frac{1}{z_4}.
		\end{eqnarray}} 
	
	\vspace{-0.8cm}
	\item The Anti-fundamental Representation: $\overline{5} \equiv (0,0,0,1)$ 
	
	With $\vec{\lambda}$ = $(1,1,1,1,0)$ and $\vec{r} = \vec{\lambda}+\vec{\rho} = (5,4,3,2,0)$, the character is obtained as:
	
	\vspace{-0.7cm}
	{\scriptsize\begin{eqnarray}\label{eq:su5-5-bar-char}
		\chi(\epsilon_1,\epsilon_2,\epsilon_3,\epsilon_4,\epsilon_5) &=& \frac{|\epsilon^5,\hspace{0.1cm}\epsilon^4,\hspace{0.1cm}\epsilon^3,\hspace{0.1cm}\epsilon^2,\hspace{0.1cm}1|} {|\epsilon^4,\hspace{0.1cm}\epsilon^3,\hspace{0.1cm}\epsilon^2,\hspace{0.1cm}\epsilon,\hspace{0.1cm}1|} = \frac{1}{\prod_{1\leq i<j\leq 5}\left(\epsilon_i-\epsilon_j\right)}
		\begin{vmatrix}
		\epsilon_1^{5}\hspace{0.1cm} \ \ \epsilon_1^{4}\hspace{0.1cm} \ \ \epsilon_1^{3}\hspace{0.1cm} \ \ \epsilon_1^{2}\hspace{0.1cm}\ \ 1\\
		\epsilon_2^{5}\hspace{0.1cm} \ \ \epsilon_2^{4}\hspace{0.1cm} \ \ \epsilon_2^{3}\hspace{0.1cm} \ \ \epsilon_2^{2}\hspace{0.1cm}\ \ 1\\
		\epsilon_3^{5}\hspace{0.1cm} \ \ \epsilon_3^{4}\hspace{0.1cm} \ \ \epsilon_3^{3}\hspace{0.1cm} \ \ \epsilon_3^{2}\hspace{0.1cm}\ \ 1\\
		\epsilon_4^{5}\hspace{0.1cm} \ \ \epsilon_4^{4}\hspace{0.1cm} \ \ \epsilon_4^{3}\hspace{0.1cm} \ \ \epsilon_4^{2}\hspace{0.1cm}\ \ 1\\
		\epsilon_5^{5}\hspace{0.1cm} \ \ \epsilon_5^{4}\hspace{0.1cm} \ \ \epsilon_5^{3}\hspace{0.1cm} \ \ \epsilon_5^{2}\hspace{0.1cm}\ \ 1\\
		\end{vmatrix}\nonumber\\ 
		&=& \epsilon_1\epsilon_2\epsilon_3\epsilon_4 + \epsilon_1\epsilon_2\epsilon_3\epsilon_5 + \epsilon_1\epsilon_2\epsilon_4\epsilon_5 + \epsilon_1\epsilon_3\epsilon_4\epsilon_5 + \epsilon_2\epsilon_3\epsilon_4\epsilon_5.\nonumber\\
		\therefore\;\chi_{({SU(5)})_{\overline{5}}}(z_1, z_2, z_3, z_4) 
		&=& z_4 + \frac{z_3}{z_4} + \frac{z_2}{z_3} + \frac{z_1}{z_2} + \frac{1}{z_1}.
		\end{eqnarray}}
	
	\vspace{-0.8cm}
	\item The Decuplet Representation: $10 \equiv (0,1,0,0)$ 
	
	With $\vec{\lambda}$ = $(1,1,0,0,0)$ and $\vec{r} = \vec{\lambda}+\vec{\rho} = (5,4,2,1,0)$, the character is obtained as:
	
	\vspace{-0.7cm}
	{\scriptsize\begin{eqnarray}\label{eq:su5-10-char}
		\chi(\epsilon_1,\epsilon_2,\epsilon_3,\epsilon_4,\epsilon_5) &=& \frac{|\epsilon^5,\hspace{0.1cm}\epsilon^4,\hspace{0.1cm}\epsilon^2,\hspace{0.1cm}\epsilon,\hspace{0.1cm}1|} {|\epsilon^4,\hspace{0.1cm}\epsilon^3,\hspace{0.1cm}\epsilon^2,\hspace{0.1cm}\epsilon,\hspace{0.1cm}1|} = \frac{1}{\prod_{1\leq i<j\leq 5}\left(\epsilon_i-\epsilon_j\right)}
		\begin{vmatrix}
		\epsilon_1^{5}\hspace{0.1cm} \ \ \epsilon_1^{4}\hspace{0.1cm} \ \ \epsilon_1^{2}\hspace{0.1cm} \ \ \epsilon_1\hspace{0.1cm}\ \ 1\\
		\epsilon_2^{5}\hspace{0.1cm} \ \ \epsilon_2^{4}\hspace{0.1cm} \ \ \epsilon_2^{2}\hspace{0.1cm} \ \ \epsilon_2\hspace{0.1cm}\ \ 1\\
		\epsilon_3^{5}\hspace{0.1cm} \ \ \epsilon_3^{4}\hspace{0.1cm} \ \ \epsilon_3^{2}\hspace{0.1cm} \ \ \epsilon_3\hspace{0.1cm}\ \ 1\\
		\epsilon_4^{5}\hspace{0.1cm} \ \ \epsilon_4^{4}\hspace{0.1cm} \ \ \epsilon_4^{2}\hspace{0.1cm} \ \ \epsilon_4\hspace{0.1cm}\ \ 1\\
		\epsilon_5^{5}\hspace{0.1cm} \ \ \epsilon_5^{4}\hspace{0.1cm} \ \ \epsilon_5^{2}\hspace{0.1cm} \ \ \epsilon_5\hspace{0.1cm}\ \ 1\\
		\end{vmatrix}\nonumber\\
		&=& \epsilon_1\epsilon_2 + \epsilon_1\epsilon_3 + \epsilon_1\epsilon_4 + \epsilon_1\epsilon_5 + \epsilon_2\epsilon_3 + \epsilon_2\epsilon_4 + \epsilon_2\epsilon_5 + \epsilon_3\epsilon_4 + \epsilon_3\epsilon_5 + \epsilon_4\epsilon_5. \nonumber\\
		\therefore\;\chi_{({SU(5)})_{10}}(z_1, z_2, z_3, z_4) &=& \frac{z_2 z_4}{z_1 z_3}+\frac{z_1 z_3}{z_2}+\frac{z_2}{z_1 z_4}+\frac{z_1 z_4}{z_3}+\frac{z_3}{z_1}+\frac{z_1}{z_4}+\frac{z_3}{z_2 z_4}+\frac{z_4}{z_2}+z_2+\frac{1}{z_3}.
		\end{eqnarray}}

	\vspace{-0.8cm}
	\item The Adjoint Representation: $24 \equiv (1,0,0,1)$ 
	
	With $\vec{\lambda}$ = $(2,1,1,1,0)$ and $\vec{r} = \vec{\lambda}+\vec{\rho} = (6,4,3,2,0)$, the character is obtained as:
	
	\vspace{-0.7cm}
	{\scriptsize\begin{eqnarray}\label{eq:su5-24-char}
		\chi(\epsilon_1,\epsilon_2,\epsilon_3,\epsilon_4,\epsilon_5) &=& \frac{|\epsilon^6,\hspace{0.1cm}\epsilon^4,\hspace{0.1cm}\epsilon^3,\hspace{0.1cm}\epsilon^2,\hspace{0.1cm}1|} {|\epsilon^4,\hspace{0.1cm}\epsilon^3,\hspace{0.1cm}\epsilon^2,\hspace{0.1cm}\epsilon,\hspace{0.1cm}1|} = \frac{1}{\prod_{1\leq i<j\leq 5}\left(\epsilon_i-\epsilon_j\right)}
		\begin{vmatrix}
		\epsilon_1^{6}\hspace{0.1cm} \ \ \epsilon_1^{4}\hspace{0.1cm} \ \ \epsilon_1^{3}\hspace{0.1cm} \ \ \epsilon_1^{2}\hspace{0.1cm}\ \ 1\\
		\epsilon_2^{6}\hspace{0.1cm} \ \ \epsilon_2^{4}\hspace{0.1cm} \ \ \epsilon_2^{3}\hspace{0.1cm} \ \ \epsilon_2^{2}\hspace{0.1cm}\ \ 1\\
		\epsilon_3^{6}\hspace{0.1cm} \ \ \epsilon_3^{4}\hspace{0.1cm} \ \ \epsilon_3^{3}\hspace{0.1cm} \ \ \epsilon_3^{2}\hspace{0.1cm}\ \ 1\\
		\epsilon_4^{6}\hspace{0.1cm} \ \ \epsilon_4^{4}\hspace{0.1cm} \ \ \epsilon_4^{3}\hspace{0.1cm} \ \ \epsilon_4^{2}\hspace{0.1cm}\ \ 1\\
		\epsilon_5^{6}\hspace{0.1cm} \ \ \epsilon_5^{4}\hspace{0.1cm} \ \ \epsilon_5^{3}\hspace{0.1cm} \ \ \epsilon_5^{2}\hspace{0.1cm}\ \ 1\\
		\end{vmatrix}\nonumber\\
		&=& \epsilon_1^2\epsilon_2\epsilon_3\epsilon_4 + \epsilon_1\epsilon_2^2\epsilon_3\epsilon_4 + \epsilon_1\epsilon_2\epsilon_3^2\epsilon_4 + \epsilon_1\epsilon_2\epsilon_3\epsilon_4^2 +
		\epsilon_1^2\epsilon_2\epsilon_3\epsilon_5 + \epsilon_1\epsilon_2^2\epsilon_3\epsilon_5 + \epsilon_1\epsilon_2\epsilon_3^2\epsilon_5 \nonumber\\
		& &  + \epsilon_1\epsilon_2\epsilon_3\epsilon_5^2 + \epsilon_1^2\epsilon_2\epsilon_4\epsilon_5 + \epsilon_1\epsilon_2^2\epsilon_4\epsilon_5 + \epsilon_1\epsilon_2\epsilon_4^2\epsilon_5 + \epsilon_1\epsilon_2\epsilon_4\epsilon_5^2 + \epsilon_1^2\epsilon_3\epsilon_4\epsilon_5 + \epsilon_1\epsilon_3^2\epsilon_4\epsilon_5 \nonumber\\
		& &  + \epsilon_1\epsilon_3\epsilon_4^2\epsilon_5 + \epsilon_1\epsilon_3\epsilon_4\epsilon_5^2 + \epsilon_2^2\epsilon_3\epsilon_4\epsilon_5 + \epsilon_2\epsilon_3^2\epsilon_4\epsilon_5 + \epsilon_2\epsilon_3\epsilon_4^2\epsilon_5 + \epsilon_2\epsilon_3\epsilon_4\epsilon_5^2 + 4\epsilon_1\epsilon_2\epsilon_3\epsilon_4\epsilon_5.\nonumber\\
		\therefore\;\chi_{({SU(5)})_{24}}(z_1, z_2, z_3, z_4) 
		&=&  \frac{z_1^2}{z_2}+\frac{z_2}{z_1^2}+\frac{z_1 z_3}{z_2^2}+\frac{z_2^2}{z_1 z_3}+\frac{z_1 z_4}{z_2 z_3}+\frac{z_2 z_3}{z_1 z_4}+\frac{z_1 z_2}{z_3}+\frac{z_3}{z_1 z_2}+\frac{z_1}{z_2 z_4}+\frac{z_2 z_4}{z_1}+\frac{z_1 z_3}{z_4} \nonumber\\
		& & +\frac{z_4}{z_1 z_3}+z_1 z_4+\frac{1}{z_1 z_4}+\frac{z_2 z_4}{z_3^2}+\frac{z_3^2}{z_2 z_4}+\frac{z_3 z_4}{z_2}+\frac{z_2}{z_3 z_4}+\frac{z_4^2}{z_3}+\frac{z_3}{z_4^2}+4.
		\end{eqnarray}}
\end{itemize}
\noindent
These results can be used to obtain the Hilbert Series for this model. We have tabulated the Hilbert Series as well as the covariant form of the operators that constitute the renormalizable Lagrangian in Table~\ref{table:su5-renorm-lagrangian}.

\begin{table}[h]
	\centering
	\renewcommand{\arraystretch}{1.8}
	{\tiny\begin{tabular}{|c|c|c|c|}
			\hline
			\multicolumn{4}{|c|}{\textbf{Renormalizable Operators}}\\
			\hline
			
			\multirow{2}{*}{\textbf{Mass Dimension}}&
			\multirow{2}{*}{\textbf{HS Output}}&
			\multirow{2}{*}{\textbf{Covariant Form}}&
			\multirow{2}{*}{\textbf{\textbf{No. of Operators}}}\\
			
			&
			&
			&
			\textbf{(including h.c.)}\\
			\hline
			2&
			$\Phi^2,\hspace{0.1cm} \varphi^{\dagger}\varphi$&
			$(\Phi^a_b\Phi_a^b),\hspace{0.1cm} (\varphi^{\dagger}_a\varphi^a)$&
			2\\
			\hline
			
			3&
			$\Phi^3,\hspace{0.1cm} \Phi\varphi^{\dagger}\varphi$&
			$(\Phi^a_b\Phi_c^b\Phi_a^c),\hspace{0.1cm} (\varphi^{\dagger}_a\Phi_b^a\varphi^b)$&
			2\\
			\hline
			
			\multirow{5}{*}{4}&
			$2\Phi^4$&
			$(\Phi^a_b\Phi_a^b)^2,\hspace{0.1cm} (\Phi^a_b\Phi_c^b\Phi^c_d\Phi_a^d)$&
			\multirow{5}{*}{14}\\
			\cline{2-3}
			
			&
			$2\Phi^2\varphi^{\dagger}\varphi,\hspace{0.1cm} (\varphi^{\dagger})^2\varphi$&
			$(\Phi^a_b\Phi_a^b)(\varphi^{\dagger}_c\varphi^c),\hspace{0.1cm} (\Phi^a_b\Phi_c^b\varphi^{\dagger}_a\varphi^c),\hspace{0.1cm} (\varphi^{\dagger}_a\varphi^a)^2$&
			\\
			\cline{2-3}

			&
			$\color{blue}{\Psi_L^2\varphi,\hspace{0.1cm} \Psi_L\psi_L\varphi^{\dagger}}$&
			$\color{blue}{\epsilon_{abcde}(\Psi_L^{ab})^T C\Psi_L^{cd}\varphi^e,\hspace{0.1cm} \Psi_L^{ab}\psi_{L,a}\varphi^{\dagger}_b}$&
			\\
			\cline{2-3}
			
			&
			$\Psi_L^{\dagger}\mathcal{D}\Psi_L,\hspace{0.1cm} \psi_L^{\dagger}\mathcal{D}\psi_L$&
			$\overline{\Psi}_L\slashed{\mathcal{D}}\Psi_L,\hspace{0.1cm} \overline{\psi}_L\slashed{\mathcal{D}}\psi_L$&
			\\
			\cline{2-3}
			
			&
			$\varphi^{\dagger}\varphi\mathcal{D}^2,\hspace{0.1cm}\mathcal{A}l^2+\mathcal{A}r^2$&
			$(\mathcal{D}_{\mu}\varphi)^{\dagger}(\mathcal{D}^{\mu}\varphi),\hspace{0.1cm} \mathcal{A}^{\mu\nu}\mathcal{A}_{\mu\nu},\hspace{0.1cm} \tilde{\mathcal{A}}^{\mu\nu}\mathcal{A}_{\mu\nu}$&
			\\
			\hline			
	\end{tabular}}
	\caption{\small Georgi-Glashow Model: Operators that comprise the renormalizable Lagrangian (for $N_f=1$). The operators in blue have distinct hermitian conjugates which we have not written explicitly. Here, $a,b,c$ are $SU(5)$ indices. We have suppressed the $SU(5)$ indices while writing the kinetic terms.}
	\label{table:su5-renorm-lagrangian}
\end{table}

\subsection{Supersymmetric Models}\label{appendix:susy-models}
We shall now discuss model building in the context of supersymmetric scenarios. For each of the following models we will enlist the chiral superfields and tabulate operators of various canonical dimensions constituted of these superfields. It must be noted that for each of these scenarios one can introduce vector superfields and with a suitable use of \textbf{GrIP} obtain the full operator set at different canonical dimensions.
\subsubsection*{\underline{Next to Minimal Supersymmetric Standard Model - NMSSM}}
The Next to Minimal Supersymmetric Standard Model extends MSSM by addition of a gauge group singlet chiral superfield $\mathcal{S}$.
We have noted the transformation properties of the chiral superfields of this model under the gauge groups $SU(3)_C\otimes SU(2)_L\otimes U(1)_Y$ in Table~\ref{table:nmssm-quantum-no}. We have collected the operators constituted only of chiral superfields of canonical dimensions-1, -2, -3, and -4 comprised of beyond MSSM interactions in Table~\ref{table:nmssm-output}, see Ref.~\cite{Ellwanger:2009dp}.

\begin{table}[h]
	\centering
	\renewcommand{\arraystretch}{1.8}
	{\tiny\begin{tabular}{|c|c|c|c|c|c|}
			\hline
			\textbf{Superfields}&
			$SU(3)_C$&
			$SU(2)_L$&
			$U(1)_Y$\\
			\hline
			
			$H_u$&
			1&
			2&
			1/2\\
			
			$H_d$&
			1&
			2&
			-1/2\\
			
			$Q^i$&
			3&
			2&
			1/6\\
			
			$U^i$&
			$\overline{3}$&
			1&
			-2/3\\
			
			$D^i$&
			$\overline{3}$&
			1&
			1/3\\
			
			$L^i$&
			1&
			2&
			-1/2\\
			
			$E^i$&
			1&
			1&
			1\\
			
			$\mathcal{S}$&
			1&
			1&
			0\\
			\hline
	\end{tabular}}
	\caption{\small Next to Minimal Supersymmetric Standard Model: Quantum numbers of superfields under the gauge groups. Internal symmetry indices have been suppressed.  $i=1,2,..,N_f$ is the flavour index.}
	\label{table:nmssm-quantum-no}
\end{table} 

\begin{table}[h]
	\centering
	\renewcommand{\arraystretch}{2.0}
	{\tiny\begin{tabular}{|c|c|c|}
			\hline
			
			\textbf{Canonical Dim.}&
			\textbf{Operators}&
			\textbf{No. of Operators}\\
			\hline
			
			1&
			$\mathcal{S}$&
			1\\
			\hline
			
			2&
			$\mathcal{S}^2$&
			1\\
			\hline
			
			3&
			$H_dH_u\mathcal{S},\hspace{0.1cm} H_uL\mathcal{S},\hspace{0.1cm} \mathcal{S}^3$&
			3\\
			\hline
			
		4&
		$LEH_d\mathcal{S},\hspace{0.1cm} QDH_d\mathcal{S},\hspace{0.1cm} QUH_u\mathcal{S},\hspace{0.1cm} \mathcal{S}^4, \hspace{0.1cm} LQD\mathcal{S},\hspace{0.1cm} H_uL\mathcal{S}^2,\hspace{0.1cm} H_dH_u\mathcal{S}^2$&
		7\\	
		\hline
	\end{tabular}}
	\caption{\small Next to Minimal Supersymmetric Standard Model: Operators constituted only of chiral superfields.}
	\label{table:nmssm-output}
\end{table}

\subsubsection*{\underline{The Supersymmetric Pati-Salam Model}}

The transformation properties of the superfields of the Supersymmetric Pati-Salam Model under the gauge groups $SU(4)_C\otimes SU(2)_L\otimes SU(2)_R$ are enlisted in Table~\ref{table:susypatisalam-quantum-no}. The results for this model are based on \cite{Ahmed:2018jlv}. We have listed the operators constituted only of chiral superfields of canonical dimensions-2, -3 and -4 in Table~\ref{table:susypatisalam-output}. The terms relevant to the phenomenology discussed in \cite{Ahmed:2018jlv} are highlighted using red colour.
\begin{table}[h]
	\centering
	\renewcommand{\arraystretch}{1.8}
	{\tiny\begin{tabular}{|c|c|c|c|c|c|}
			\hline
			\textbf{Superfields}&
			$SU(4)_C$&
			$SU(2)_L$&
			$SU(2)_R$\\
			\hline
			
			$\mathcal{F}$&
			4&
			2&
			1\\
			
			$\overline{\mathcal{F}}$&
			$\overline{4}$&
			1&
			2\\
			
			$\mathcal{H}$&
			$\overline{4}$&
			1&
			2\\
			
			$\overline{\mathcal{H}}$&
			4&
			1&
			2\\
			
			$h$&
			1&
			2&
			2\\
			
			$\Delta$&
			6&
			1&
			1\\
			
			$\Sigma$&
			15&
			1&
			1\\
			\hline
	\end{tabular}}
	\caption{\small Supersymmetric Pati-Salam Model: Quantum numbers of superfields under the gauge groups. Internal symmetry indices have been suppressed.}
	\label{table:susypatisalam-quantum-no}
\end{table}  

\begin{table}[h]
	\centering
	\renewcommand{\arraystretch}{2.0}
	{\tiny\begin{tabular}{|c|c|c|}
			\hline
			
			\textbf{Canonical Dim.}&
			\textbf{Operators}&
			\textbf{No. of Operators}\\
			\hline
			
			2&
			$\overline{\mathcal{F}}\,\overline{\mathcal{H}},\hspace{0.1cm} \color{purple}{h^2,\hspace{0.1cm} \Sigma^2,\hspace{0.1cm} \Delta^2,\hspace{0.1cm} \overline{\mathcal{H}}\mathcal{H}}$&
			5\\
			\hline
			
			3&
			$\overline{\mathcal{F}}\mathcal{F}h,\hspace{0.1cm} \mathcal{F}\mathcal{H}h,\hspace{0.1cm} \overline{\mathcal{F}}\mathcal{H}\Delta,\hspace{0.1cm} \overline{\mathcal{F}}\,\overline{\mathcal{H}}\Sigma, \hspace{0.1cm} \color{purple}{\Delta\mathcal{H}^2,\hspace{0.1cm} \Delta\overline{\mathcal{H}}^2,\hspace{0.1cm} \overline{\mathcal{H}}\mathcal{H}\Sigma,\hspace{0.1cm} \Sigma^3}$&
			8\\
			\hline

			\multirow{3}{*}{4}&
			$\Delta^4,\hspace{0.1cm} \overline{\mathcal{F}}^2\mathcal{F}^2,\hspace{0.1cm} \Delta^2h^2,\hspace{0.1cm} \overline{\mathcal{F}}\mathcal{F}^2\mathcal{H},\hspace{0.1cm}  \color{purple}{h^4,\hspace{0.1cm} h^2\overline{\mathcal{H}}\mathcal{H},\hspace{0.1cm} 2\overline{\mathcal{H}}^2\mathcal{H}^2,\hspace{0.1cm}
			2\overline{\mathcal{H}}\mathcal{H}\Sigma^2,\hspace{0.1cm} 2\Sigma^4,}$&
			\multirow{3}{*}{31}\\
			
			&
			$ \Delta^2\overline{\mathcal{F}}\,\overline{\mathcal{H}},\hspace{0.1cm} \Delta h\overline{\mathcal{H}}\mathcal{F},\hspace{0.1cm} h^2\overline{\mathcal{F}}\,\overline{\mathcal{H}}, \Delta^2\overline{\mathcal{H}}\mathcal{H},\hspace{0.1cm} 2\overline{\mathcal{F}}\,\overline{\mathcal{H}}^2\mathcal{H},\hspace{0.1cm} \Delta\mathcal{F}^2\Sigma,\hspace{0.1cm} \Delta\overline{\mathcal{F}}^2\Sigma,\hspace{0.1cm}  2\overline{\mathcal{F}}\mathcal{H}\Delta\Sigma,$&
			\\
			
			&
			$\overline{\mathcal{F}}\mathcal{F}h\Sigma,\hspace{0.1cm} \mathcal{F}\mathcal{H}h\Sigma,\hspace{0.1cm} \mathcal{H}^2\Delta\Sigma,\hspace{0.1cm} \overline{\mathcal{H}}^2\Delta\Sigma,\hspace{0.1cm} 2\Delta^2\Sigma^2,\hspace{0.1cm} 2h^2\Sigma^2,\hspace{0.1cm} 2\overline{\mathcal{F}}\,\overline{\mathcal{H}}\Sigma^2$&
			\\
			
			\hline
	\end{tabular}}
	\caption{\small Supersymmetric Pati-Salam Model: Operators constituted only of chiral superfields. Terms in red correspond to the operators given in \cite{Ahmed:2018jlv}.}
	\label{table:susypatisalam-output}
\end{table}
\clearpage
\subsubsection*{\underline{Minimal Supersymmetric Left-Right Model}}\label{subsubsec:susylr}

The transformation properties of the superfields of the Minimal Supersymmetric Left-Right Model under the gauge groups $SU(3)_C\otimes SU(2)_L\otimes SU(2)_R\otimes U(1)_{B-L}$ are enlisted in Table~\ref{table:susylr-quantum-no}. We have reproduced the results of \cite{Aulakh:1998nn}, i.e., the operators constituted only of chiral superfields of canonical dimensions-2, -3, and -4. We also extend the operator set by adding terms of canonical dimension-5. We have tabulated these in Table~\ref{table:susylr-output}. 
\begin{table}[h]
	\centering
	\renewcommand{\arraystretch}{1.8}
	{\tiny\begin{tabular}{|c|c|c|c|c|c|}
			\hline
			\textbf{Superfields}&
			$SU(3)_C$&
			$SU(2)_L$&
			$SU(2)_R$&
			$U(1)_{B-L}$\\
			\hline
			
			$\Phi$&
			1&
			2&
			2&
			0\\
			
			$\Delta$&
			1&
			3&
			1&
			1\\
			
			$\overline{\Delta}$&
			1&
			3&
			1&
			-1\\
			
			$\Delta_c$&
			1&
			1&
			3&
			-1\\
			
			$\overline{\Delta}_c$&
			1&
			1&
			3&
			1\\
			
			$Q^i$&
			3&
			2&
			1&
			1/6\\
			
			$Q^i_c$&
			$\overline{3}$&
			1&
			2&
			-1/6\\
			
			$L^i$&
			1&
			2&
			1&
			-1/2\\
			
			$L^i_c$&
			1&
			1&
			2&
			1/2\\
			\hline
	\end{tabular}}
	\caption{\small Minimal Supersymmetric Left Right Model: Quantum numbers of superfields under the gauge groups. Internal symmetry indices have been suppressed. $i$=1,2,...,$N_f$ is the flavour index.}
	\label{table:susylr-quantum-no}
\end{table} 

\begin{table}[h]
	\centering
	\renewcommand{\arraystretch}{2.0}
	{\tiny\begin{tabular}{|c|c|}
			\hline
			\multicolumn{2}{|c|}{\textbf{Canonical Dimension-2}}\\
			\hline
			\hspace{2.2cm}\textbf{Operators}&
			\textbf{No. of Operators}\\
			
			\hline

			$\Delta\overline{\Delta}, \hspace{0.1cm} \Delta_c\overline{\Delta}_c, \hspace{0.1cm}
			\Phi^2$&
			3\\
			\hline
  			\hline
			\multicolumn{2}{|c|}{\textbf{Canonical Dimension-3}}\\
			\hline
			\hspace{2.2cm}\textbf{Operators}&
			\textbf{No. of Operators}\\
			
			\hline
			
			$\frac{1}{2}\left(N_f^2+N_f\right)L^2\Delta, \hspace{0.1cm} \frac{1}{2}\left(N_f^2+N_f\right)L_c^2\Delta_c, \hspace{0.1cm} N_f^2LL_c\Phi, \hspace{0.1cm} N_f^2QQ_c\Phi$&
			$3N_f^2+N_f$\\
			\hline
			\hline
			\multicolumn{2}{|c|}{\textbf{Canonical Dimension-4}}\\
			\hline
			\hspace{2.2cm}\textbf{Operators}&
			\textbf{No. of Operators}\\
			
			\hline
			
			$ \Phi^4,\hspace{0.1cm} 2(\Delta)^2(\overline{\Delta})^2,\hspace{0.1cm} 2(\Delta_c)^2(\overline{\Delta}_c)^2,\hspace{0.1cm}(\Delta)^2(\Delta_c)^2,\hspace{0.1cm} (\overline{\Delta})^2(\overline{\Delta}_c)^2,\hspace{0.1cm}
			\Delta\overline{\Delta}\Delta_c\overline{\Delta}_c,\hspace{0.1cm} \Delta\overline{\Delta}\Phi^2,\hspace{0.1cm}\Delta_c\overline{\Delta}_c\Phi^2,\hspace{0.1cm}  \Delta\Delta_c\Phi^2,\hspace{0.1cm}  \overline{\Delta}\overline{\Delta}_c\Phi^2,$&
			\multirow{2}{*}{$ \frac{29}{12}N_f^4 -\frac{1}{2}N_f^3 + \frac{1}{12}N_f^2 + 12$}\\

			$\frac{1}{2}\left(N_f^4+N_f^2\right)Q^2Q_c^2, \hspace{0.1cm} N_f^4LL_cQQ_c,\hspace{0.1cm} \boxed{\frac{1}{3}\left(N_f^4-N_f^2\right)LQ^3},\hspace{0.1cm} \boxed{\frac{1}{3}\left(N_f^4-N_f^2\right)L_cQ_c^3},\hspace{0.1cm} \boxed{\frac{1}{4}N_f^2\left(N_f-1\right)^2L^2L_c^2}$&
			\\
			
			\hline
			\hline
			\multicolumn{2}{|c|}{\textbf{Canonical Dimension-5}}\\
			\hline
			\hspace{2.2cm}\textbf{Operators}&
			\textbf{No. of Operators}\\
			
			\hline
			
			$N_f^2LL_c\Delta\Delta_c\Phi,\hspace{0.1cm} N_f^2LL_c\overline{\Delta}\overline{\Delta}_c\Phi,\hspace{0.1cm} N_f^2QQ_c\Delta\Delta_c\Phi,\hspace{0.1cm} N_f^2QQ_c\overline{\Delta}\overline{\Delta}_c\Phi,\hspace{0.1cm} 2N_f^2QQ_c\Delta\overline{\Delta}\Phi,\hspace{0.1cm} 2N_f^2QQ_c\Delta_c\overline{\Delta}_c\Phi,$&
			\multirow{5}{*}{$20N_f^2+6N_f$}\\
			
			$2N_f^2LL_c\Delta\overline{\Delta}\Phi,\hspace{0.1cm} 2N_f^2LL_c\Delta_c\overline{\Delta}_c\Phi,\hspace{0.1cm}N_f^2LL_c\Phi^3,\hspace{0.1cm} N_f^2QQ_c\Phi^3,\hspace{0.1cm} \frac{1}{2}\left(N_f^2+N_f\right)
			L^2\Delta\Phi^2,\hspace{0.1cm} \frac{1}{2}\left(N_f^2+N_f\right)
			L_c^2\Delta_c\Phi^2,$&
			\\
			
			$\left(N_f^2+N_f\right)
			L^2\Delta^2\overline{\Delta},\hspace{0.1cm} \frac{1}{2}\left(N_f^2+N_f\right)
			L^2\overline{\Delta}_c\Phi^2,\hspace{0.1cm} \frac{1}{2}\left(N_f^2+N_f\right)
			L_c^2\overline{\Delta}\Phi^2,\hspace{0.1cm} \frac{1}{2}\left(N_f^2+N_f\right)
			L^2\Delta\Delta_c\overline{\Delta}_c,$&
			\\
			
			$\frac{1}{2}\left(N_f^2+N_f\right)
			L_c^2\Delta_c\Delta\overline{\Delta},\hspace{0.1cm} 
			\frac{1}{2}\left(N_f^2+N_f\right)
			L^2\overline{\Delta}(\overline{\Delta}_c)^2,\hspace{0.1cm} 
			\frac{1}{2}\left(N_f^2+N_f\right)
			L_c^2\overline{\Delta}_c(\overline{\Delta})^2,\hspace{0.1cm}\left(N_f^2+N_f\right)
			L_c^2\Delta_c^2\overline{\Delta}_c$&
			\\
			\hline
	\end{tabular}}
	\caption{\small Minimal Supersymmetric Left-Right Model: Operators constituted only of chiral superfields.}
	\label{table:susylr-output}
\end{table}
\clearpage

%% file: appendix_code.tex
\subsection{Dimensions and Dynkin labels for few representations of $SU(N)$}\label{appendix:grip-supporting-file}
%%%%%%%%%%%%%%%%%%%%%%%%%%%%%%%%%%%%%%%%%%%%%%%%%%%%%%%%%%%%%%%%%%%%%%%%%

\subsubsection*{$SU(2)$}

\begin{table}[h]
\begin{minipage}{.33\linewidth}	
	\centering
		\renewcommand{\arraystretch}{1.8}
				{\scriptsize\begin{tabular}[h]{|c|c|}
			\hline
			\textbf{Dimension} & \textbf{Dynkin label} \\
			\hline
			1 & $\{0\}$ \\
			\hline
			2 & $\{1\}$\\
			\hline
			3 & $\{2\} $\\
			\hline
			4 & $\{3\}$\\
			\hline
			5 & $\{4\}$\\
			\hline
			6 & $\{5\}$\\
			\hline
			7 & $\{6\}$\\
			\hline
			\end{tabular}}
	
	\end{minipage}%
		\begin{minipage}{.33\linewidth}
			\centering
			\renewcommand{\arraystretch}{1.8}
			{\scriptsize\begin{tabular}[h]{|c|c|}
			\hline
			\textbf{Dimension} & \textbf{Dynkin label} \\
			\hline
			8 & $\{7\}$ \\
			\hline
			9 & $\{8\}$\\
			\hline
			10 & $\{9\}$ \\
			\hline
			11 & $\{10\}$\\
			\hline
			12 & $\{11\}$\\
			\hline
			13 & $\{12\}$\\
			\hline
			14 & $\{13\}$\\
			\hline
				\end{tabular}}
	
		\end{minipage}
		\begin{minipage}{.33\linewidth}
			\centering
			\renewcommand{\arraystretch}{1.8}
			{\scriptsize\begin{tabular}[h]{|c|c|}
				\hline
			\textbf{Dimension} & \textbf{Dynkin label} \\
			\hline
			15 & $\{14\} $\\
			\hline
			16 & $\{15\}$\\
			\hline
			17 & $\{16\} $\\
			\hline
			18 & $\{17\}$\\
			\hline
			19 & $\{18\}$\\
			\hline
			20 & $\{19\}$\\
			\hline
			21 & $\{20\}$\\
			\hline
				\end{tabular}}
	
		\end{minipage}%
		\caption{\small Dynkin labels corresponding to a few low dimensional representations of $SU(2)$.}
	\label{table:SU2}
	\end{table}

\subsubsection*{$SU(3)$}

	\begin{table}[h]
		\begin{minipage}{.35\linewidth}
			\centering
		\renewcommand{\arraystretch}{1.8}	
			{\scriptsize\begin{tabular}[h]{|c|c|}
				\hline
				\textbf{Dimension} & \textbf{Dynkin label} \\
				\hline
			1 & $\{0,0\} $\\
			\hline
			3 & $\{1,0\}$\\
			\hline
			$\overline{3}$ & $\{0,1\}$ \\
			\hline
			6 & $\{2,0\}$\\
			\hline
			$\overline{6}$ & $\{0,2\}$\\
			\hline
			8 & $\{1,1\}$\\
			\hline
		\end{tabular}}
	
			\end{minipage}%
		\begin{minipage}{.3\linewidth}
				\centering
			\renewcommand{\arraystretch}{1.8}
				{\scriptsize\begin{tabular}[h]{|c|c|}
					\hline
					\textbf{Dimension} & \textbf{Dynkin label} \\
					\hline
			10 & $\{3,0\}$ \\
			\hline
			$\overline{10}$ & $\{0,3\}$\\
			\hline
			15 & $\{2,1\}$ \\
			\hline
			$\overline{15}$ & $\{1,2\}$\\
			\hline
			$15^{\prime}$ & $\{4,0\}$\\
			\hline
			$\overline{15^{\prime}}$ & $\{0,4\}$ \\
			\hline
		\end{tabular}}

			\end{minipage}%
		\begin{minipage}{.35\linewidth}
				\centering
			\renewcommand{\arraystretch}{1.8}
			{\scriptsize\begin{tabular}[h]{|c|c|}
					\hline
					\textbf{Dimension} & \textbf{Dynkin label} \\
					\hline
			21 & $\{0,5\}$\\
			\hline
			$\overline{21}$ & $\{5,0\}$ \\
			\hline
			24 & $\{1,3\}$\\
			\hline
			$\overline{24}$ & $\{3,1\}$\\
			\hline
			27 & $\{2,2\}$\\
			\hline
			28 & $\{6,0\}$\\
			\hline
		\end{tabular}}
	
		\end{minipage}%
		\caption{\small Dynkin labels corresponding to a few low dimensional representations of $SU(3)$.}
	\label{table:SU3}
	\end{table}

\newpage

\subsubsection*{$SU(4)$}

	\begin{table}[h]
		\begin{minipage}{.35\linewidth}
		\centering
		\renewcommand{\arraystretch}{1.8}	
		{\scriptsize\begin{tabular}[h]{|c|c|}
				\hline
				\textbf{Dimension} & \textbf{Dynkin label} \\
				\hline
			1 & $\{0,0,0\}$ \\
			\hline
			4 & $\{1,0,0\}$\\
			\hline
			$\overline{4}$ & $\{0,0,1\}$ \\
			\hline
			6 & $\{0,1,0\}$\\
			\hline
			10 & $\{2,0,0\}$\\
			\hline
			$\overline{10}$ & $\{0,0,2\}$\\
			\hline
			15 & $\{1,0,1\}$\\
			\hline
				\end{tabular}}
		
		\end{minipage}%
		\begin{minipage}{.3\linewidth}
			\centering
			\renewcommand{\arraystretch}{1.8}	
			{\scriptsize\begin{tabular}[h]{|c|c|}
					\hline
					\textbf{Dimension} & \textbf{Dynkin label} \\
					\hline
			20 & $\{0,1,1\}$ \\
			\hline
			$\overline{20}$ & $\{1,1,0\}$\\
			\hline
			$20^{\prime}$ & $\{0,2,0\}$ \\
			\hline
			$20^{\prime\prime}$ & $\{0,0,3\}$\\
			\hline
			$\overline{20^{\prime\prime}}$  & $\{3,0,0\}$\\
			\hline
			35 & $\{4,0,0\}$\\
			\hline
			$\overline{35}$ & $\{0,0,4\}$ \\
			\hline
		\end{tabular}}
	
		\end{minipage}%
		\begin{minipage}{.35\linewidth}
		\centering
		\renewcommand{\arraystretch}{1.8}	
		{\scriptsize\begin{tabular}[h]{|c|c|}
				\hline
				\textbf{Dimension} & \textbf{Dynkin label} \\
				\hline
			36 & $\{2,0,1\}$\\
			\hline
			$\overline{36}$ & $\{1,0,2\}$ \\
			\hline
			45 & $\{2,1,0\}$\\
			\hline
			$\overline{45}$ & $\{0,1,2\}$\\
			\hline
			50 & $\{0,3,0\}$\\
			\hline
			56 & $\{5,0,0\}$\\
			\hline
			$\overline{56}$ & $\{0,0,5\}$ \\
			\hline
			\end{tabular}}
			
		\end{minipage}%
		\caption{\small Dynkin labels corresponding to a few low dimensional representations of $SU(4)$.}
	\label{table:SU4}
	\end{table}

\subsubsection*{$SU(5)$}

\begin{table}[h]
	\begin{minipage}{.35\linewidth}
		\centering
		\renewcommand{\arraystretch}{1.8}	
		{\scriptsize\begin{tabular}[h]{|c|c|}
				\hline
				\textbf{Dimension} & \textbf{Dynkin label} \\
				\hline
				1 & $\{0,0,0,0\}$ \\
				\hline
				5 & $\{1,0,0,0\}$\\
				\hline
				$\overline{5}$ & $\{0,0,0,1\}$\\
				\hline
				10 & $\{0,1,0,0\}$ \\
				\hline
				$\overline{10}$ & $\{0,0,1,0\}$\\
				\hline
				15 & $\{2,0,0,0\}$\\
				\hline
				$\overline{15}$ & $\{0,0,0,2\}$\\
				\hline
				24 & $\{1,0,0,1\}$\\
				\hline
				35 & $\{0,0,0,3\}$ \\
				\hline
				$\overline{35}$ & $\{3,0,0,0\}$\\
				\hline
				$40$ & $\{0,0,1,1\}$ \\
				\hline
				$\overline{40}$ & $\{1,1,0,0\}$\\
				\hline
				\end{tabular}}
		
	\end{minipage}%
	\begin{minipage}{.3\linewidth}
		\centering
		\renewcommand{\arraystretch}{1.8}	
		{\scriptsize\begin{tabular}[h]{|c|c|}
				\hline
				\textbf{Dimension} & \textbf{Dynkin label} \\
				\hline
				$45$ & $\{0,1,0,1\}$ \\
				\hline
				$\overline{45}$  & $\{1,0,1,0\}$\\
				\hline
				50 & $\{0,0,2,0\}$\\
				\hline
				$\overline{50}$ & $\{0,2,0,0\}$ \\
				\hline
				70 & $\{2,0,0,1\}$\\
				\hline
				$\overline{70}$ & $\{1,0,0,2\}$ \\
				\hline
				$70^{\prime}$ & $\{0,0,0,4\}$\\
				\hline
				$\overline{70^{\prime}}$ & $\{4,0,0,0\}$ \\
				\hline
				75 & $\{0,1,1,0\}$\\
				\hline
				105 & $\{0,0,1,2\}$\\
				\hline
				$\overline{105}$ & $\{2,1,0,0\}$\\
				\hline
				126 & $\{2,0,1,0\}$\\
				\hline
		\end{tabular}}
		
	\end{minipage}%
	\begin{minipage}{.35\linewidth}
		\centering
		\renewcommand{\arraystretch}{1.8}	
		{\scriptsize\begin{tabular}[h]{|c|c|}
				\hline
				\textbf{Dimension} & \textbf{Dynkin label} \\
				\hline
				$\overline{126}$ & $\{0,1,0,2\}$\\
				\hline
				$126^{\prime}$ & $\{5,0,0,0\}$\\
				\hline
				$\overline{126^{\prime}}$ & $\{0,0,0,5\}$\\
				\hline
				$160$ & $\{3,0,0,1\}$\\
				\hline
				$\overline{160}$ & $\{1,0,0,3\}$\\
				\hline
				$175$ & $\{1,1,0,1\}$\\
				\hline
				$\overline{175}$ & $\{1,0,1,1\}$\\
				\hline
				$175^{\prime}$ & $\{1,2,0,0\}$ \\
				\hline
				$\overline{175^{\prime}}$ & $\{0,0,2,1\}$\\
				\hline
				$175^{\prime\prime}$ & $\{0,3,0,0\}$\\
				\hline
				$\overline{175^{\prime\prime}}$ & $\{0,0,3,0\}$\\
				\hline
				$200$ & $\{2,0,0,2\}$\\
				\hline
		\end{tabular}}
		
	\end{minipage}%
	\caption{\small Dynkin labels corresponding to a few low dimensional representations of $SU(5)$.}
	\label{table:SU5}
\end{table}	
\clearpage